\documentclass{jpp}

\pdfoutput=1
\usepackage{graphicx}
\usepackage{epstopdf, epsfig}
\usepackage{color}
\usepackage{bbm}
\usepackage[pagebackref=true,colorlinks=true,linkcolor=blue,citecolor=blue,urlcolor=blue]{hyperref}

\usepackage{usebib}
\newcommand{\tens}[1]{\ensuremath\textbf{\textsf{#1}}}
\renewcommand{\vec}[1]{\protect\mathbf{#1}}
\newcommand{\grad}{\vec{\boldsymbol{\nabla}}}
\newcommand{\curl}{\vec{\boldsymbol{\nabla}}\mathbf{\times}}
\renewcommand{\div}{\vec{\boldsymbol{\nabla}}\cdot}
\newcommand{\f}[2]{\frac{#1}{#2}}
\newcommand{\dpart}[2]{\f{\partial #1}{\partial #2}}
\newcommand{\dpartshort}[2]{{\partial #1}/{\partial #2}}
\newcommand{\ddpart}[2]{\f{\partial^2 #1}{{\partial #2}^2}}
\newcommand{\ddpartmixed}[3]{\f{\partial^2 #1}{{\partial #2}{\partial #3}}}
\newcommand{\dlag}[2]{\f{\mathrm{D} #1}{\mathrm{D} #2}}
\newcommand{\dlagshort}[2]{\mathrm{D} #1/\mathrm{D} #2}
\newcommand{\deriv}[2]{\f{\mathrm{d} #1}{{\mathrm{d} #2}}}

\newcommand{\diff}{{\mathrm{d}}}
\newcommand{\vpar}{v_\parallel}
\newcommand{\vperp}{v_\perp}
\newcommand{\vths}{v_{\mathrm{th},s}}

\newcommand{\vthi}{v_{\mathrm{th},i}}

\newcommand{\lambdamfps}{\lambda_{\mathrm{mfp},s}}

\newcommand{\lambdamfpi}{\lambda_{\mathrm{mfp},i}}

\newcommand{\Tpars}{T_{\parallel,s}}
\newcommand{\Tperps}{T_{\perp,s}}

\newcommand{\ppar}{P_{\parallel}}
\newcommand{\pperp}{P_{\perp}}
\newcommand{\ppars}{P_{\parallel,s}}
\newcommand{\pperps}{P_{\perp,s}}

\newcommand{\ppari}{P_{\parallel,i}}
\newcommand{\pperpi}{P_{\perp,i}}

\newcommand{\qpars}{Q_{\parallel,s}}
\newcommand{\qperps}{Q_{\perp,s}}

\newcommand{\qpari}{Q_{\parallel,i}}
\newcommand{\qperpi}{Q_{\perp,i}}

\newcommand{\hatB}{\hat{B}}
\newcommand{\vA}{\vec{A}}
\newcommand{\vU}{\vec{U}}
\newcommand{\vB}{\vec{B}}
\newcommand{\hatvB}{\hat{\vec{B}}}
\newcommand{\vJ}{\vec{J}}
\newcommand{\vE}{\vec{E}}
\newcommand{\vW}{\vec{W}}
\newcommand{\vOmega}{\boldsymbol{\Omega}}
\newcommand{\vEMF}{\boldsymbol{\mathcal{E}}}
\newcommand{\mean}[1]{\overline{#1}}
\newcommand{\fluct}[1]{#1}

\newcommand{\fluctB}{\fluct{b}}
\newcommand{\fluctU}{\fluct{u}}

\newcommand{\fluctvE}{\fluct{\vec{e}}}
\newcommand{\fluctvA}{\fluct{\vec{a}}}
\newcommand{\fluctvB}{\fluct{\vec{b}}}
\newcommand{\fluctvU}{\fluct{\vec{u}}}
\newcommand{\fluctvJ}{\fluct{\vec{j}}}
\newcommand{\fluctvomega}{\fluct{\boldsymbol{\omega}}}

\newcommand{\meanB}{\mean{B}}
\newcommand{\meanU}{\mean{U}}

\newcommand{\meanEMF}{\mean{\mathcal{E}}}
\newcommand{\meanvA}{\mean{\vA}}
\newcommand{\meanvB}{\mean{\vB}}
\newcommand{\meanvU}{\mean{\vU}}
\newcommand{\meanvJ}{\mean{\vJ}}
\newcommand{\meanvE}{\mean{\vE}}
\newcommand{\meanvW}{\mean{\vW}}
\newcommand{\meanvEMF}{\mean{\vEMF}}

\newcommand{\equ}[1]{equation~(\ref{#1})}
\newcommand{\equs}[2]{equations~(\ref{#1})-(\ref{#2})}
\newcommand{\Equ}[1]{Equation~(\ref{#1})}

\newcommand{\fig}[1]{Fig.~\ref{#1}}
\newcommand{\Fig}[1]{Figure~\ref{#1}}
\newcommand{\sect}[1]{\S\ref{#1}}

\newcommand{\Sect}[1]{Section~\ref{#1}}
\newcommand{\app}[1]{App.~\ref{#1}}
\shorttitle{Dynamo theories}
\shortauthor{F. Rincon}

\newbibfield{booktitle}
\newcommand{\GG}[1]{}
\bibinput{dynamo}

\newcommand{\red}[1]{#1}

\title{Dynamo theories}

\author{Fran\c{c}ois Rincon\aff{1,2}
  \corresp{\email{francois.rincon@irap.omp.eu}}}

\affiliation{\aff{1}Universit\'e de Toulouse; UPS-OMP; IRAP: Toulouse, France
\aff{2} CNRS; IRAP; 14 avenue Edouard Belin, F-31400 Toulouse, France}

\begin{document}

\maketitle

\begin{abstract}
These lecture notes are based on a tutorial given in 2017 at 
a plasma physics winter school in Les Houches. Their aim is to
provide a self-contained graduate-student level
introduction to the theory and modelling of the dynamo effect
in turbulent fluids and plasmas, blended with a review of
current research in the field. The primary focus is on
the physical and mathematical concepts underlying different
(turbulent) branches of dynamo theory, with some astrophysical,
geophysical and experimental context disseminated throughout the
document. The text begins with an introduction to the rationale,
observational and historical roots of the subject, and to the basic
concepts of magnetohydrodynamics relevant to dynamo
theory. The next two sections discuss the fundamental
phenomenological and mathematical aspects of (linear and nonlinear)
small- and large-scale MHD dynamos. These sections are complemented by
an overview of a selection of current active research topics in the
field, including the numerical modelling of the geo- and solar
dynamos, shear dynamos driven by turbulence with zero net helicity,
and MHD-instability-driven dynamos such as the magnetorotational
dynamo. The difficult problem
of a unified, self-consistent statistical treatment of small and
large-scale dynamos at large magnetic Reynolds numbers is also
discussed throughout the text. Finally, an excursion is made into the
relatively new but increasingly popular realm of magnetic-field
generation in weakly-collisional plasmas. A short discussion of the
outlook and challenges for the future of the field concludes the
presentation.
\end{abstract}

\renewcommand{\contentsname}{\setcounter{footnote}{1}
  Contents\footnote{Subsections marked with asterisks contain some
    fairly advanced, technical or specialised material, and may be
    skipped on a first reading.}}

\setcounter{tocdepth}{3}
\tableofcontents

\section{Introduction}
\subsection{About these notes}
These lecture notes expand (significantly) on a two-hour tutorial 
given at the 2017 Les Houches school \textit{``From laboratories to
astrophysics: the expanding universe of plasma physics''}. 
Many excellent books  and reviews have already been written
on the subjects of dynamo theory, planetary and astrophysical
magnetism. Most of them, however, are either quite specialised,
or simply too advanced for non-specialists seeking a general
entry-point into the field. The multidisciplinary context of this
school, taking place almost a century after Larmor's original
idea of self-exciting fluid dynamos, provided an ideal opportunity to
craft a self-contained, wide-ranging, yet relatively accessible
introduction to the subject.

One of my central preoccupations in the writing process has been to
attempt to distill in clear and relatively concise
terms the essence of each of the problems covered, and to highlight to
the best of my abilities the successes, limitations and connections of
different lines of research in logical order. Although I may not have entirely
succeeded, my sincere hope is that this review will nevertheless turn
out to be generally useful to observers, experimentalists, theoreticians, PhD
students, newcomers and established researchers in the field alike,
and will foster new original research on dynamos of all kinds.
It is quite inevitable, though, that such an ideal can only be sought
at the expense of total exhaustivity and mathematical rigour, and
necessitates making difficult editorial choices.  To borrow
Keith Moffatt's wise words in the introduction of his 1973 Les Houches
lecture notes on fluid
dynamics and dynamos, \textit{``it will be evident that in the time
available I have had to skate over certain difficult topics with
indecent haste. I hope however that I have succeeded in conveying
something of the excitement of current research in dynamo theory and
something of the general flavour of the subject. Those already
acquainted with the subject will know that my account is woefully
one-sided''}. Suggestions for further reading on the many different
branches of dynamo research discussed in the text are provided
throughout the document and in \app{biblio} to mitigate
these limitations.

Finally, while the main focus of the notes is on the physical and
practical mathematical aspects of dynamo theory in general,
contextual information is provided throughout to connect the material
presented to astrophysical, geophysical and experimental
dynamo problems. In particular, a selection of astrophysical and
planetary dynamo research topics, including the geo-, solar, and
accretion-disc dynamos, is highlighted in the most advanced parts of
the review to give a flavour of the diversity of research and
challenges in the field.

\subsection{Observational roots of dynamo theory\label{observations}}
Dynamo theory finds its roots in the human observation of
the Universe, and in the quest to understand the origin of magnetic
fields observed or inferred in a variety of astrophysical systems. 
This includes planetary magnetism (the Earth, other planets and their
satellites), solar and stellar magnetism, and cosmic
magnetism (galaxies, clusters and the Universe as a whole).
We will therefore start with a brief overview of the main
features of astrophysical and planetary magnetism.

\begin{figure}
  \centerline{\includegraphics[width=0.8\textwidth]{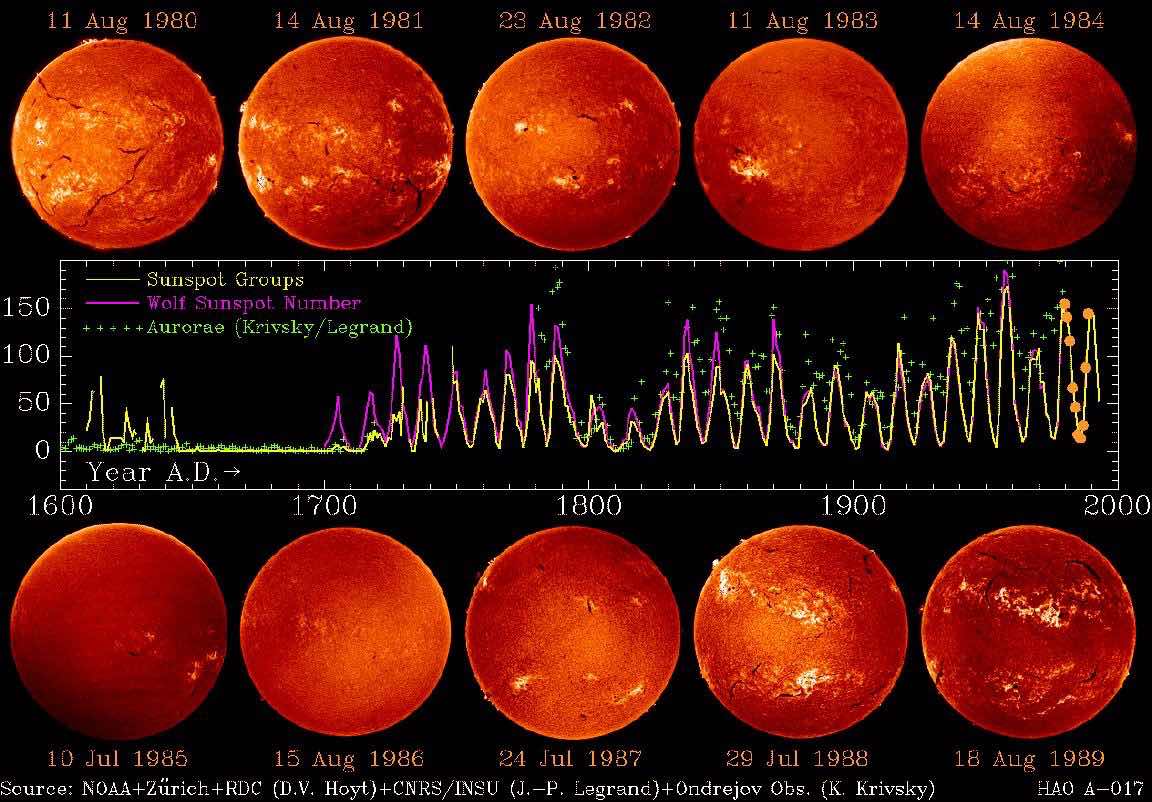}}
  \vspace{4mm}
  \centerline{\includegraphics[width=0.8\textwidth]{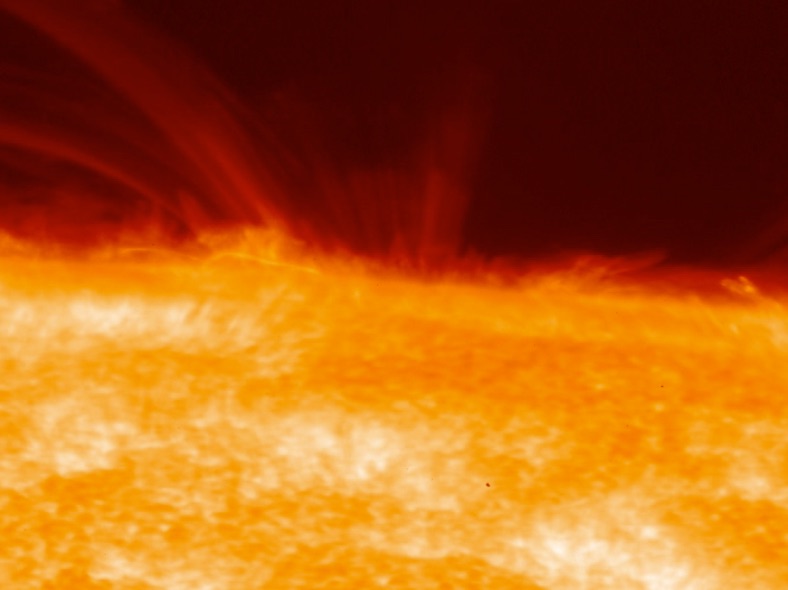}}
    \vspace{2mm}
\caption{Top: large-scale solar magnetism. The eleven-year
  magnetic solar cycle ($\#21-22$) observed in the chromosphere
  through the H$\alpha$ spectral line (full solar discs), and
  historical sunspot-number record (Credits:
  NOAA/Z\"urich/RDC/CNRS/INSU/Ondresjov Observatory/HAO). Bottom:
  local and global solar magnetic dynamics. The rapidly-evolving small-scale
  magnetic carpet, spicules and sunspot arches imaged near the limb in the lower
  chromosphere through the CaH spectral line (Credits: SOT/Hinode/JAXA/NASA).}
\label{figsun}
\end{figure}

Consider first solar magnetism, whose evolution on human timescales
and day-to-day monitoring make it a more intuitive dynamical
phenomenon to apprehend than other forms of astrophysical magnetism.
For the purpose of the discussion, we can single out two
``easily'' observable dynamical magnetic timescales on the Sun. The
first one is the eleven-year magnetic cycle timescale over which the
large-scale solar magnetic field reverses. The solar cycle shows
up in many different observational records, the most well-known of
which is probably the number of sunspots as a function of time, see
\fig{figsun} (top) (note that the eleven-year cycle is also
chaotically modulated on longer timescales). \red{Large-scale solar
magnetism is characterised by an average (mean) field of only a few
tens of Gauss \citep[see e.g. the review by][]{charbonneau14}, however the field itself can
exceed kiloGauss strengths in large-scale features like sunspots}\footnote{\red{1~Gauss$=10^{-4}$~Tesla
    is the most commonly encountered magnetic-field unit in astrophysics.
    Gaussian c.g.s. units are used throughout most of the text.}}. 
There is also a lot of dynamical, small-scale, disordered
 magnetism in the solar surface photosphere and chromosphere,
evolving on short time and spatial scales comparable to
those of thermal convective motions at the surface (from a minute to
an hour, and from a few kilometres to a few thousands of kilometres).
This so-called ``network'' and ``internetwork'' small-scale magnetism,
depicted in \fig{figsun} (bottom), was discovered much more recently
\citep{livingston71}. Its typical strength ranges from a few to a few 
hundred Gauss, and does not appear to be significantly modulated over
the course of the global solar cycle (see
e.g. \cite{solanki06,stenflo13} for reviews).
Large-scale stellar magnetic fields, including time-dependent ones,
have been detected on many other stars \citep[e.g.][]{donati09}, but
only for the Sun do we have accurate, temporally and
spatially-resolved direct measurements of small-scale stellar magnetism.

\begin{figure}
  \centerline{\includegraphics[width=0.8\textwidth]{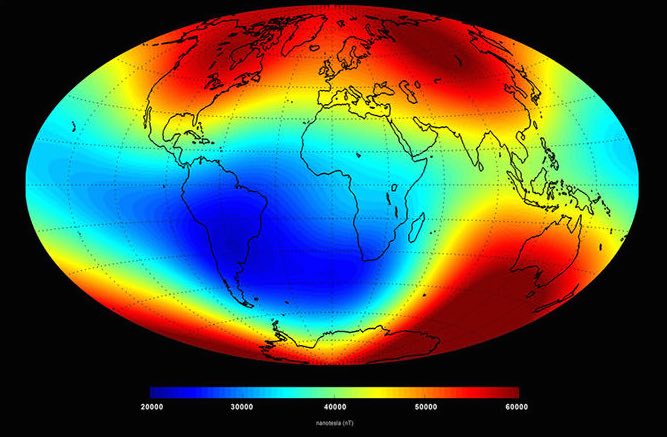}}
  \vspace{5mm}
  \centerline{\includegraphics[width=0.5\textwidth]{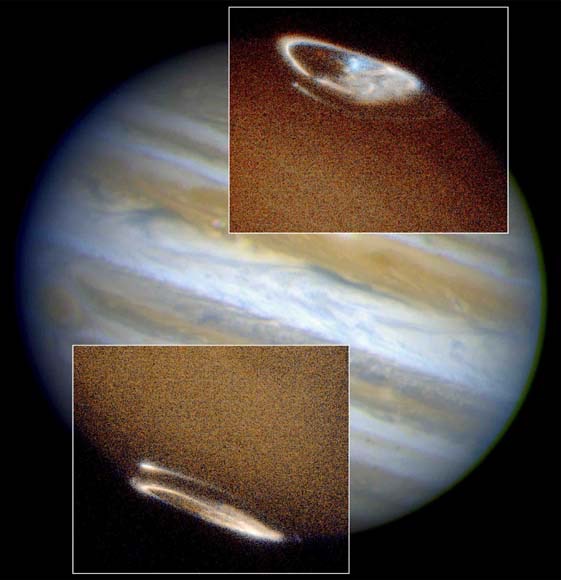}}
    \vspace{2mm}
\caption{Top: direct satellite measurements of the Earth's magnetic field
  strength (in nano Teslas) in 2014 at an altitude of 450 km (Credits:
  Swarm/CNES/ESA). Bottom: ultra-violet emission of a 1998 Jupiter
  aurora (Credits: J. Clarke/STIS/WFPC2/HST/NASA/ESA).\label{figplanets}} 
\end{figure}

The second major natural, human-felt phenomenon that inspired the
development of dynamo theory is of course the Earth's magnetic
field, whose strength at the surface of the Earth is of the order 
of 0.1 Gauss \red{\citep[$10^{-5}$~T, ][]{finlay10}}. The dynamical evolution and
structure of the field, including its many irregular reversals over a
hundred-thousand to million-year timescale, is established through
paleomagnetic and archeomagnetic records, marine navigation books, and
is now monitored with satellites, as shown in
\fig{figplanets} (top). While the terrestrial field is probably 
highly multiscale and multipolar in the liquid iron part of the core where
it is generated, it is primarily considered as a form of large-scale
dynamical magnetism involving a north and 
south magnetic pole. Several other planets of the solar system also
exhibit large-scale, low-multipole surface magnetic fields and
magnetospheres. \Fig{figplanets} (bottom) shows auroral
emissions on Jupiter, whose magnetic field has a typical surface
strength of a few Gauss \red{\citep[a few $10^{-4}$~T,][]{khurana04}}. Just
as in the Earth's case, the large-scale external field of the other magnetic
planets is almost certainly not representative of the
structure of the field in the interior.
\begin{figure}
\centerline{\includegraphics[height=0.8\textwidth,angle=-90]{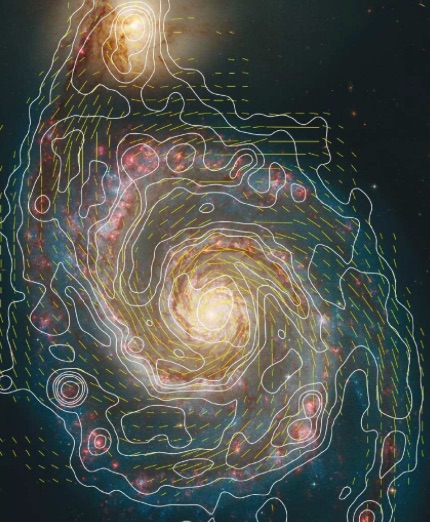}}
\vspace{4mm}
\centerline{\includegraphics[width=0.8\textwidth]{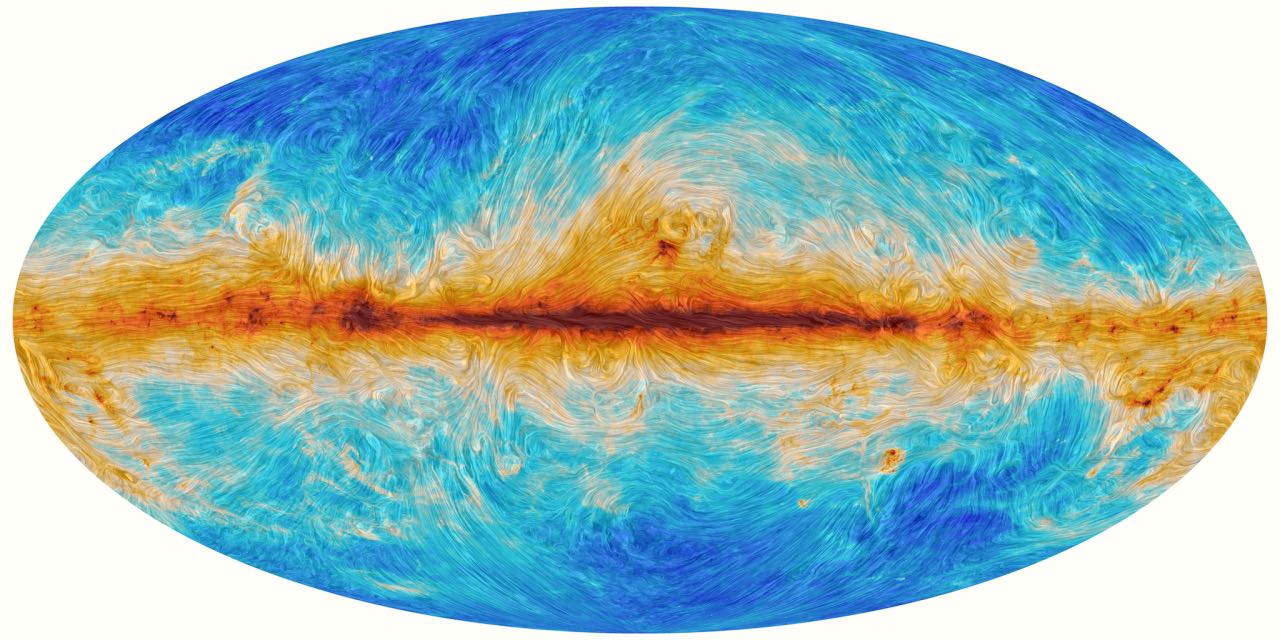}}
    \vspace{2mm}
\caption{Top: large-scale spiral magnetic structure (line segments) of the
  M51 galaxy established from radio observations of polarised synchrotron
  emission by cosmic rays (Credits: MPIfR Bonn and Hubble Heritage
  Team. Graphics: Sterne and Weltraum). Bottom: map of the microwave
  galactic dust emission convolved with galactic magnetic-field lines
  reconstructed from polarisation maps of the dust emission
  (Credits: M.~A. Miville-Desch\^enes/CNRS/ESA/Planck
  collaboration).\label{figgalaxy}}
\end{figure}

Moving further away from the Earth, we also learned in the second part 
of the twentieth century that galaxies,
including our own Milky Way, host magnetic
fields with a typical strength of the order of a few $10^{-5}$ Gauss \red{\citep{beck13}}.
For a long time, observations would only reveal the ordered
large-scale, global magnetic structure whose projection in the
galactic plane would often take the form of spirals, see
\fig{figgalaxy} (top). But recent high-resolution observations of
polarised dust emission in our galaxy, displayed in \Fig{figgalaxy}
(bottom), have now also established that the galactic magnetic field
has a very intricate multiscale structure, of which a large-scale
ordered field is just one component.
\begin{figure}
  \centerline{\includegraphics[width=0.6\textwidth]{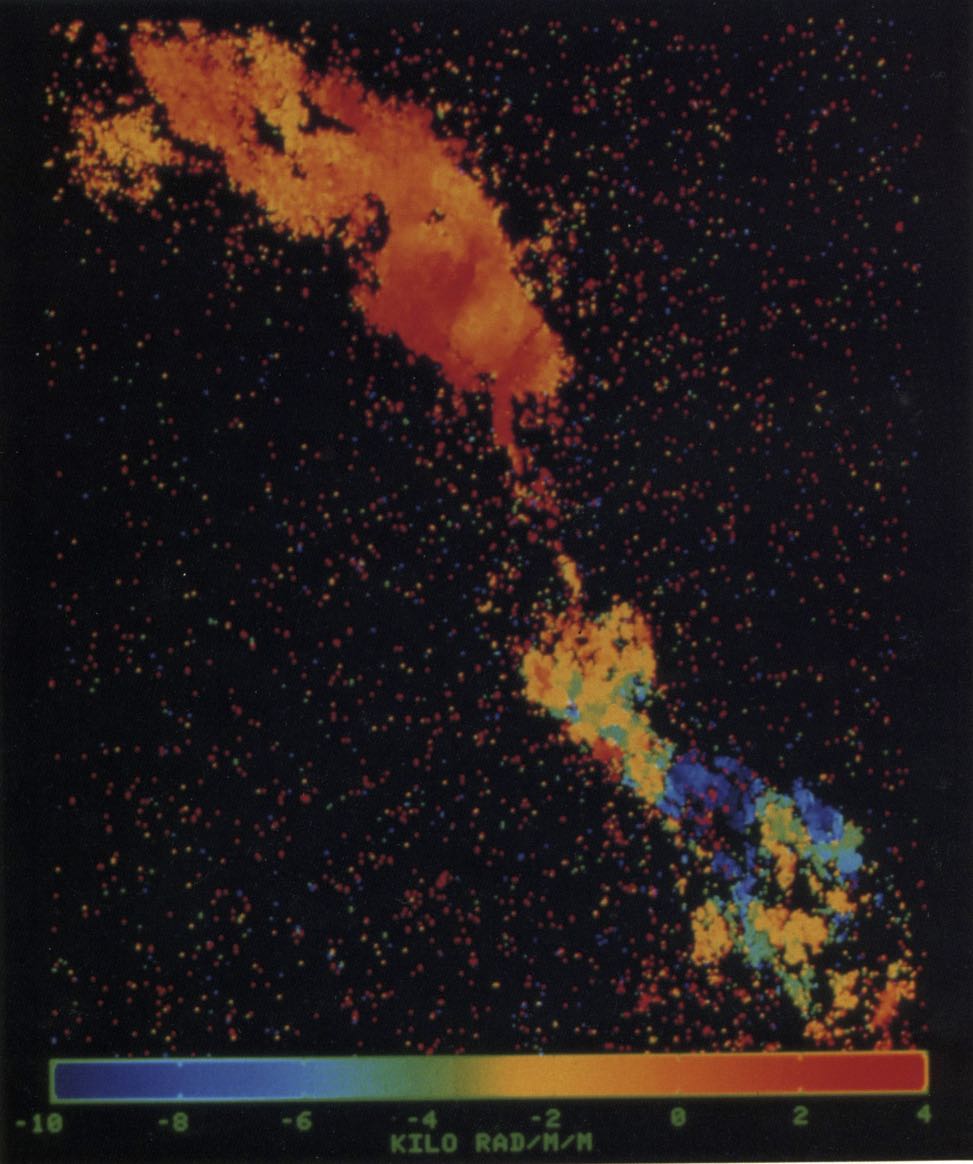}}
\vspace{5mm}
\centerline{\includegraphics[width=0.8\textwidth]{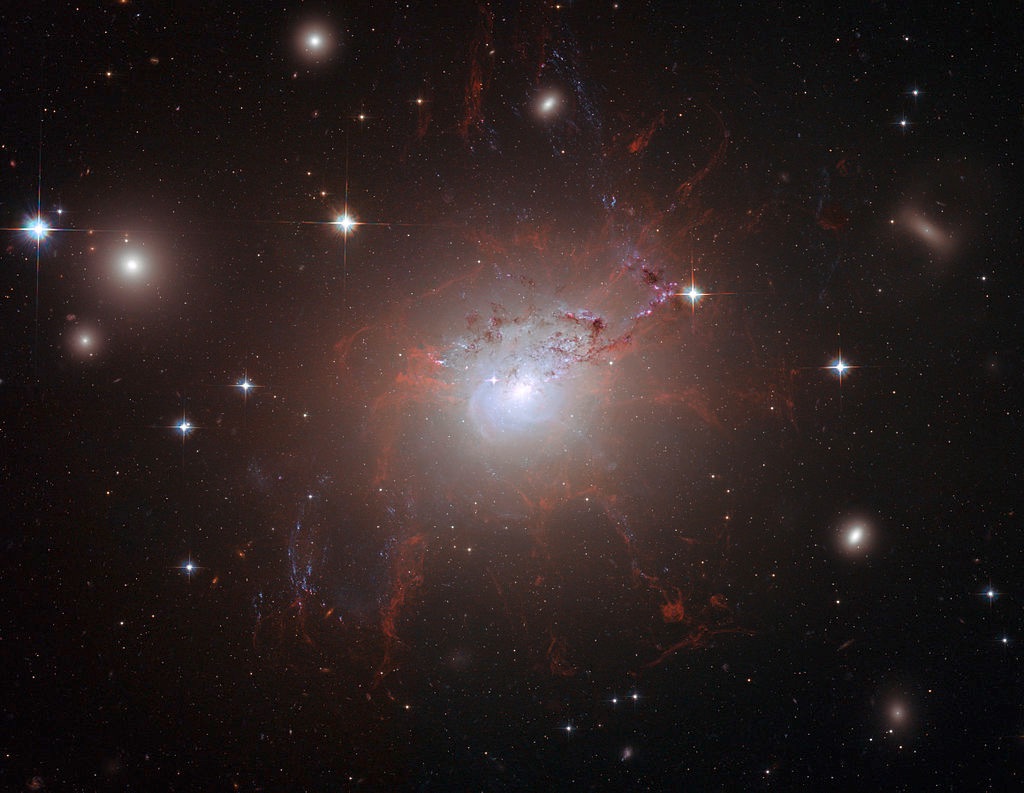}}
\vspace{2mm}
\caption{Top: Faraday rotation measure map (a proxy for the
  line-of-sight component of the magnetic field) in the
  synchrotron-illuminated radio-lobes of the Hydra A cluster (Credits:
  Taylor \& Perley/VLA/NRAO). Bottom: visible-light observations of
  magnetised filaments in the core of the Perseus cluster (Credits:
  Fabian et al./HST/ESA/NASA).\label{figclusters}}
\end{figure}

Magnetic fields of the order of a few $10^{-6}$~Gauss are also measured in 
the hot intracluster medium (ICM) of galaxy clusters
\red{\citep[e.~g.][]{carilli02,bonafede10}}.
The large-scale global structure and orientation of cluster fields, if
any, is not well-determined (it should be noted in this respect that global
differential rotation is not thought to be very important in clusters,
unlike in individual galaxies, stars and planets). On the other hand,
synchrotron polarimetry measurements in the radio-lobes of active
galactic nuclei (AGN), such as that shown in \fig{figclusters} (top),
suggest that there is a significant ``small-scale'', turbulent ICM
field component on scales comparable to or even smaller than a
kiloparsec \red{\citep{vogt05}}. Visible-light observations of the
ICM, including in the H$\alpha$ spectral line, also reveal the
presence of colder gas structured into magnetised filaments, see
\fig{figclusters} (bottom).

There has been as yet no direct detection of magnetic 
fields on even larger, cosmological scales. Magnetic fields in the
filaments of the cosmic web and intergalactic medium are thought to
be of the order of, but no larger than a few $10^{-9}$~Gauss at
Megaparsec scale. \red{This upper bound can be derived from a variety
of observational constraints, including on the cosmic microwave
background \citep{ade16}. Note however that a lower bound on the
typical intergalactic magnetic-field strength, of the order of
a few $10^{-16}$~Gauss, has been derived from high-energy
$\gamma$-ray observations \citep{neronov10}}. A detailed
discussion of the current observational bounds on the scales and
amplitudes of magnetic fields in the early Universe can 
be found in the review by \cite{durrer13}.

\subsection{What is dynamo theory about ?}
The dynamical nature, spatial structure and measured amplitudes of
astrophysical and planetary magnetic fields strongly suggest that they
must in most instances have been amplified to, and are further sustained
at significant levels by internal dynamical
mechanisms. In the absence of any such mechanism, calculations of
magnetic diffusion notably show that ``fossil'' fields present in the
early formation stages of different objects should decay over
cosmologically short timescales, see e.g. \cite{weiss2002,roberts13}
for geomagnetic estimates. Besides, even in relatively
high-conductivity environments such as stellar interiors, the
fossil-field hypothesis cannot easily explain the dynamical evolution
and reversals of large-scale magnetic fields over a timescale of the
order of a few years either. So, what are these field-amplifying and
field-sustaining mechanisms ? Most astrophysical objects (or at least some
subregions within them) are fluids/plasmas in a dynamical, turbulent
state. Even more importantly for the problem at hand, these
fluids/plasmas are electrically conducting. This
raises the possibility that internal flows create an 
electromotive force leading to the inductive self-excitation of
magnetic fields and electrical currents. This idea of self-exciting
fluid  dynamos was first put forward a century ago by
\cite{larmor1919} in the context  of solar (sunspot) magnetism.

From a fundamental physics perspective, dynamo theory therefore 
generally aims at describing the amplification and sustainment of
magnetic fields by flows of electrically conducting fluids and plasmas
-- most importantly turbulent ones. Important questions include
whether such an excitation and sustainment is possible at all, at
which rate the growth of initially very weak seed fields can proceed,
at what magnetic energy such processes saturate, and what the
time-dependence and spatial structure of dynamo-generated fields is in
different regimes. At the heart of these questions lies a variety of
difficult classical linear and nonlinear physics and applied mathematics
problems, many of which have a strong connection with more general
(open) problems in turbulence theory, including closure problems.

While fundamental theory is a perfectly legitimate object of
study on its own, there is also a strong demand for ``useful'' or
applicable mathematical models of dynamos. Obviously, researchers from
different backgrounds have very different conceptions of what a
useful model is, and even of what theory is. Astronomers for instance
are keen on phenomenological, low-dimensional models of large-scale
astrophysical magnetism with a few free parameters, as these provide an
intuitive framework for the interpretation of observations. Solar
and space physicists are interested in more quantitative and
fine-tuned versions of such models to predict solar activity in the
near future. Experimentalists need models that can help them minimise
the mechanical power required to excite dynamos in highly-customised
washing-machines filled with liquid sodium or plasma. Another 
major challenge of dynamo theory, then, is to build meaningful
bridges between these different communities by constructing
conceptual and mathematical dynamo models that are 
physically-grounded and rigorous, yet tractable and predictive. 
The overall task of dynamo theoreticians therefore appears to be
quite complex and multifaceted.

\subsection{Historical overview of dynamo research\label{history}}
Let us now give a very brief overview of the history of the subject 
as a matter of context for the main theoretical developments of the next
sections. More detailed historical accounts are available in different
reviews and books, including the very informative Encyclopedia of
Geomagnetism and Paleomagnetism \citep{gubbins07},  and the book by
\cite*{molokov07} on historical trends in magnetohydrodynamics.

Dynamo theory did not immediately take
off after the publication of Larmor's original ideas on solar
magnetism. Viewed from today's perspective, it is clear that the
intrinsic geometric and dynamical complexity of the problem was a
major obstacle to its development. This complexity was first hinted
by the demonstration by \cite{cowling33} that axisymmetric dynamo
action is not possible (\sect{cowling}). Cowling's conclusions
were not particularly encouraging\footnote{\textit{``The
theory proposed by Sir Joseph Larmor, that the magnetic field of a
sunspot is maintained by the currents it induces in moving matter,
is examined and shown to be faulty~; the same result also applies
for the similar theory of the maintenance of the general field of
Earth and Sun.''}} and apparently even led Einstein to voice a
pessimistic outlook on the subject \citep{krause93}. The first
significant positive developments only occurred after
the second world war, when \cite{elsasser46,elsasser47}, followed by
\cite{bullard54}, set about formulating a spherical theory of magnetic
field amplification by non-axisymmetric convective motions in the
liquid core of the Earth. In the same period, \cite{batchelor50} and
\cite{schluter50} started investigating the problem of magnetic
field amplification by generic three-dimensional turbulence from a
more classical statistical hydrodynamic perspective. In the wake of
Elsasser's and Bullard's work, \cite{parker55} published a seminal
semi-phenomenological article describing how differential
rotation and small-scale cyclonic motions could combine to excite
large-scale magnetic fields (\sect{largescalealpha}). Parker also
notably showed how such a mechanism could excite oscillatory dynamo
modes (now called Parker waves) reminiscent of the solar
cycle. The spell of Cowling's theorem was definitely broken
a few years later when \cite{herzenberg58} and \cite{backus58} 
found the first mathematical working examples of fluid dynamos.

The 1960s saw the advent of statistical dynamo theories. 
\cite{braginskii64,braginskii64b} first showed how an ensemble of
non-axisymmetric spiral wavelike motions could lead to the statistical
excitation of a large-scale magnetic field. Shortly after that,
\cite*{steenbeck66} published their mean-field theory of 
large-scale magnetic-field generation in flows lacking
parity/reflectional/mirror invariance
(\sect{largescalemeanfield}). These and a few other
pioneering studies \citep[e.g.][]{moffatt70,vainshtein70} 
put Parker's mechanism on a much stronger mathematical footing.
In the same period, \cite{kazantsev67} developed a quintessential
statistical model describing the dynamo excitation of small-scale
magnetic fields in non-helical (parity-invariant) random flows
(\sect{kazantsev}). Interestingly, Kazantsev's work predates the
observational detection of ``small-scale''  solar magnetic fields.
This golden age of dynamo research extended into the 1970s with 
further developments of the statistical theory, and the introduction 
of the concept of fast dynamos by \cite{vainshtein72}, which offered
a new phenomenological insight into the dynamics of turbulent
dynamo processes (\sect{fast}). ``Simple'' helical
dynamo flows that would later prove instrumental in the development of
experiments were also found in that period
\citep{roberts70,roberts72,ponomarenko73}.

It took another few years for the different theories to be
vindicated in essence by numerical simulations, as the essentially
three-dimensional nature of dynamos made the life of numerical people
quite hard at the time. In a very brief but results-packed 
article, \cite*{meneguzzi81} numerically demonstrated both the
excitation of a large-scale magnetic field in small-scale homogeneous
helical fluid turbulence, and that of small-scale magnetic
fields in non-helical turbulence. These results marked the beginnings
of a massive numerical business that is more than ever flourishing
today. Experimental evidence for dynamos,
on the other hand, was much harder to establish.
Magnetohydrodynamic (MHD) fluids are not easily available on tap in
the laboratory and the properties of liquid metals such
as liquid sodium create all kinds of power-supply, dissipation and
safety problems. Experimental evidence for helical dynamos 
was only obtained at the dawn the twenty-first
century in the Riga \citep{gailitis00} and Karlsruhe experiments
\citep{stieglitz01} relying upon very constrained flow geometries
designed after the work of \cite{ponomarenko73} and \cite{roberts70,roberts72}.
Readers are referred to an extensive review paper by
\cite{gailitis02} for further details. Further experimental
evidence of fluid dynamo action in a freer, more homogeneous turbulent 
setting has since been sought by several groups, but has so
far only been reported in the Von K\'arm\'an Sodium experiment
\citep[VKS,][]{monchaux07}. The decisive role of soft-iron
solid impellers in the excitation of a dynamo in this experiment
remains widely debated (see short discussion and references in
\sect{incompressible}). Overall, the VKS experiment provides a good
flavour of the current status, successes and difficulties of the
liquid metal experimental approach to exciting a turbulent dynamo.
For broader reviews and perspectives on experimental dynamo efforts,
readers are referred to \cite{stefani08,verhille10,lathrop11}.

\subsection{An imperfect dichotomy}
The historical development of dynamo theory 
has roughly proceeded along the lines of the seeming observational
dichotomy between large and small-scale magnetism, albeit not in a
strictly causal way. We usually refer to the processes
by which flows at a given scale statistically produce magnetic
fields at much larger scales as \textit{large-scale dynamo
  mechanisms}. Global rotation and/or large-scale shear usually
(though not always) plays an important role in this
context. \red{As we shall discover, large-scale dynamos also naturally
  produce a significant amount of small-scale magnetic field, however
  magnetic fields at scales comparable to or smaller than that of the
  flow can also be excited by independent \textit{small-scale dynamo
    mechanisms} if the fluid/plasma is sufficiently
  ionised. Importantly, the latter are usually much faster and
  can be excited even in the absence of system rotation or shear}.

The dichotomy between small- and large-scale dynamos
has the merits of clarity and simplicity, and will therefore be used
in this tutorial as a rough guide to organise the presentation.
However it is not as clear-cut and perfect as it looks at first
glance, for a variety of reasons. Most importantly, large-scale and
small-scale magnetic-field generation processes can take place
simultaneously in a given system, and the outcome of these
processes is entirely  up to one of the most dreaded words in physics:
\textit{nonlinearity}. In fact, most astrophysical and planetary
magnetic fields are in a \textit{saturated}, dynamical nonlinear
state: they can have temporal variations such as reversals or rapid
fluctuations, but their typical strength does not change by many orders of
magnitudes over long periods of time; their energy content is also
generally not small comparable to that of fluid motions, which
suggests that they exert dynamical feedback on these
motions. Therefore, dynamos in nature involve strong couplings
between multiple scales, fields, and dynamical processes, including
distinct dynamo processes. Nonlinearity significantly blurs the lines
between large and small-scale dynamos (and in some cases also other MHD
instabilities), and adds a whole new layer of dynamical complexity
to an already difficult subject. The small-scale/large-scale
``unification'' problem is currently one of the most important in
dynamo research, and will accordingly be a recurring theme in this
review.

\subsection{Outline}
The rest of the text is organised as follows. \Sect{stage}
introduces classic MHD material and dimensionless quantities and
scales relevant to the dynamo problem, as well as some important
definitions, fundamental results and ideas such as anti-dynamo
theorems and the concept of fast dynamos. The core of the presentation
starts in \sect{smallscale} with an introduction to the 
phenomenological and mathematical models of small-scale MHD
dynamos. The fundamentals of linear and nonlinear large-scale MHD
dynamo theory are then reviewed in \sect{largescale}. These two
sections are complemented in \sect{complexLS} by essentially
phenomenological discussions of a selection of advanced research
topics including large-scale stellar and planetary dynamos driven by
rotating convection, large-scale dynamos driven by sheared
turbulence with vanishing net helicity, and dynamos mediated by MHD
instabilities such as the magnetorotational instability.
\Sect{plasma} provides an introduction to the relatively new
but increasingly popular realm of dynamos in weakly-collisional
plasmas. The notes end with a concise discussion of perspectives
and challenges for the field in \sect{conclusion}. A selection of good
reads on the subject can be found in \app{biblio}.
Subsections marked with asterisks contain some fairly advanced,
technical or specialised material, and may be skipped on first
reading.

\section{Setting the stage for MHD dynamos\label{stage}}
\subsection{Magnetohydrodynamics}
Most of these notes, except \sect{plasma}, are about fluid dynamo theories
in the non-relativistic, collisional, isotropic, single fluid MHD regime
in which the mean free path of liquid, gas or plasma particles is
significantly smaller than any dynamical scale of interest, and than
the smallest of the particle gyroradii. We will also assume that the
dynamics takes place at scales larger than the ion inertial length, so
that the Hall effect can be discarded. The isotropic MHD regime is
applicable to liquid metals, stellar interiors and galaxies to some
extent, but not quite to the ICM for instance, as we will discuss
later. Accretion discs can be in a variety of plasma states ranging
from hot and weakly collisional to cold and multifluid.

\subsubsection{Compressible MHD equations}
Let us start from the equations of compressible,
viscous, resistive magnetohydrodynamics. First, we 
have the continuity (mass conservation) equation
\begin{equation}
  \label{eq:cont}
  \dpart{\rho}{t}+\div{\left(\rho\vU\right)}=0~,
\end{equation}
where $\rho$ is the gas density and $\vU$ is the fluid velocity
field, and the momentum equation
\begin{equation}
  \label{eq:mom}
\rho\left(\dpart{\vU}{t}+\vU\cdot\grad{\,\vU}\right)=-\grad{P}+\f{\vJ\times\vB}{c}+\div{{\boldsymbol{\tau}}}+\vec{F}(\vec{x},t)~,
\end{equation}
where $P$ is the gas pressure,
$\tau_{ij}=\mu\left(\nabla_i{U_j}+\nabla_j{U_i}-(2/3)\,\delta_{ij}\div{\vU}\right)$
is the viscous stress tensor ($\mu$ is the dynamical viscosity and
$\nu=\mu/\rho$ is the kinematic viscosity),
$\vec{F}$ \red{is aforce per unit volume representing any kind of
external stirring mechanism} (impellers, gravity, spoon, supernovae, meteors
etc.), $\vB$ is the magnetic field, $\vJ=(c/4\pi)\curl{\vB}$ is the current
density, and $\vJ\times\vB/c$, the Lorentz force, describes
the dynamical feedback exerted by the magnetic field on fluid
motions. The evolution of $\vB$ is governed by the induction equation
\begin{equation}
  \label{eq:induc}
  \dpart{\vB}{t}=\curl{\left(\vU\times\vB\right)}-\curl{\left(\eta\curl{\vB}\right)}~,
\end{equation}
supplemented with the solenoidality condition
\begin{equation}
\label{eq:divB}
\div{\vB}=0~.
\end{equation}
\Equ{eq:induc} is derived from the Maxwell-Faraday equation and a
simple, isotropic Ohm's law for collisional electrons,
\begin{equation}
\label{eq:ohmMHD}
\vJ=\sigma\left(\vE+\f{\vU\times\vB}{c}\right),
\end{equation}
where $\sigma$ is the electrical conductivity of the fluid.
The first term $\vEMF=\vU\times\vB$ on the r.h.s. of
\equ{eq:induc} is called the electromotive force (EMF) and describes
the induction of magnetic field by the flow of conducting fluid from
an Eulerian perspective. The second term describes the diffusion of
magnetic field in a non-ideal fluid of magnetic diffusivity
$\eta=c^2/(4\pi\sigma)$. Both the Lorentz force and EMF terms in
\equs{eq:mom}{eq:induc} play a very important role in the dynamo
problem, but so do viscous and resistive dissipation. Finally, we have
the internal energy, or entropy equation
\begin{equation}
\label{eq:entropy}
\rho T \left(\dpart {S}{t}+\vU\cdot\grad\, S\right)=
D_\mu+D_\eta+\div{\left(K{\grad}T\right)}~,
\end{equation}
where $T$ is the gas temperature,
\red{$S\propto P/\rho^\gamma$} is the entropy ($\gamma$ is the adiabatic index),
 $D_\mu$ and $D_\eta$ stand for the viscous and resistive dissipation,
$K$ is the thermal conductivity and the last term on the r.h.s. stands
for thermal diffusion (we could also have added an inhomogeneous heat
source, or explicit radiative transfer). \red{An equation of state for
the thermodynamic variables, like the ideal gas law $P=\rho \mathcal{R} T$,
is also required in order to close this system ($\mathcal{R}$ here
denotes the gas constant).}

The compressible MHD equations describe the dynamics of waves,
instabilities, turbulence and shocks in all kinds of astrophysical
fluid systems, including stratified and/or (differentially) rotating
fluids, and accommodate a large range of dynamical magnetic
phenomena including dynamos and (fluid) reconnection. \red{The ideal MHD
limit corresponds to $\nu=\eta=K=0$}. The reader is referred to the
astrophysical fluid dynamics lecture notes of \cite{ogilvie16},
published in this journal, \red{for a very tidy derivation and presentation
of ideal MHD.}

\subsubsection{Important conservation laws in ideal MHD\label{conservation}}
There are two particularly important conservation laws in the ideal MHD
limit that involve the magnetic field and are of primary
importance in the context of the dynamo problem. To obtain the first
one we combine the continuity and ideal induction equations into
\begin{equation}
  \label{eq:frozen}
  \dlag{}{t}\left(\f{\vB}{\rho}\right)=\f{\vB}{\rho}\cdot\grad{\vU}~,
\end{equation}
where $\dlagshort{}{t}=\dpartshort{}{}t+\vU\cdot\grad{}$ is the
Lagrangian derivative.
\Equ{eq:frozen} for $\vB/\rho$ has the same form as the equation
describing the evolution of the Lagrangian separation vector $\delta
\vec{r}$ between two fluid particles,
\begin{equation}
  \label{eq:materialline}
  \dlag{\delta\vec{r}}{t}=\delta\vec{r}\cdot\grad{\vU}~.
\end{equation}
Hence, magnetic-field lines in ideal MHD can be thought of as being
``frozen into'' the fluid just as material lines joining fluid
particles. This is called Alfv\'en's theorem. Using this equation
\red{and \equ{eq:divB}}, it is also possible to show that the magnetic flux
through material surfaces $\delta \vec{S}$ (deformable surfaces moving
with the fluid) is conserved in ideal MHD, 
\begin{equation}
  \label{eq:fluxconservation}
  \dlag{}{t}\left(\vB\cdot\delta \vec{S}\right)=0~.
\end{equation}
If a material surface $\delta \vec{S}$ is deformed under the effect of
either shearing or compressive/expanding motions, the magnetic
field threading it must change accordingly so that $\vB\cdot\delta
\vec{S}$ remains the same. Alfv\'en's theorem enables us to appreciate
the kinematics of the magnetic field in a flow in a more intuitive
geometrical way than by just staring at equations, as it is
relatively easy to visualise  magnetic-field lines advected and
stretched by the flow. This will prove very helpful to develop an
intuition of how small and large-scale dynamo processes work.

A second important conservation law in ideal MHD in the context of
dynamo  theory is the conservation of magnetic helicity
$\mathcal{H}_m=\int\vA\cdot\vB\,\diff^3\vec{r}$, where
$\vA$ is the magnetic vector potential. To derive it, we first 
write the Maxwell-Faraday equation for $\vA$,
\begin{equation}
  \label{eq:vectorpotential}
\f{1}{c}\dpart{\vA}{t}=-\vE-\grad{\varphi}~,
\end{equation}
where $\varphi$ is the electrostatic potential. Combining
\equ{eq:vectorpotential} with \equ{eq:induc} gives
\begin{equation}
\label{eq:helicitynonideal}
  \dpart{}{t}(\vA\cdot\vB)+\div{\vec{F}_{\mathcal{H}_m}}=-2\eta\,(\curl{\vB})\cdot\vB~,
\end{equation}
where
\begin{equation}
\label{eq:helicityflux}
\vec{F}_{\mathcal{H}_m}=c\left(\varphi\vB+\vE\times\vA\right)
\end{equation}
is the total magnetic-helicity flux. In the ideal case, we see that
\equ{eq:helicitynonideal} reduces to an explicitly conservative local
evolution equation for $\vA\cdot\vB$,
\begin{equation}
  \label{eq:helicity}
  \dpart{}{t}(\vA\cdot\vB)+\div{\left[c\varphi\vB+\vA\times\left(\vU\times\vB\right)\right]}=0~,
\end{equation}
where \equ{eq:ohmMHD} with $\eta=0$ has been used to express the magnetic-helicity flux.
Note that both $\mathcal{H}_m$ and $\vec{F}_{\mathcal{H}_m}$ depend on the
choice of electromagnetic gauge and are therefore not uniquely
defined. Qualitatively, magnetic helicity provides a measure of the
linkage/knottedness of the magnetic field within the domain
considered and the conservation of magnetic helicity in ideal MHD is
therefore generally understood as a conservation of magnetic linkages
in the absence of magnetic diffusion or reconnection 
(see e.g. \cite{hubbard11,miesch12,blackman15,bodo17} for discussions
of magnetic helicity dynamics in different astrophysical dynamo contexts).

\subsubsection{Magnetic-field energetics\label{energetics}}
What about the driving and energetics of the magnetic field ? An
enlighting equation in that respect is that describing the
local Lagrangian evolution of the magnetic-field strength $B$
associated with a fluid particle in ideal MHD~($\eta=\nu=0$),
\begin{equation}
\f{1}{B}\dlag{B}{t}=\hatvB\hatvB:\grad{}\vU-\div{\vU}~,
\label{eq:Bmag}
\end{equation}
where $\hatvB=\vB/{B}$ is the unit vector defining the orientation of
the magnetic field attached to the fluid particle, and we have used
the double dot-product notation
$\hatvB\hatvB:\grad{}\vU=\hatB_i\hatB_j\nabla_iU_j$. \Equ{eq:Bmag}
follows directly from \equs{eq:cont}{eq:frozen}, and shows that any
increase of $B$ results from either a stretching of the magnetic
field along itself by a flow, or from a compression, and that the rate
at which $\ln B$ changes is proportional to the local shearing or
compression rate of the flow. Note that incompressible motions with no
component parallel to the \red{local original/initial background
  field} do not affect the field strength at linear order, and only
generate magnetic curvature perturbations (these are shear Alfv\'en
waves).  Going back to full resistive MHD, the global evolution
equation for the total magnetic energy \red{within a fixed volume},
derived for instance in the classic textbook of \cite{roberts67}, is
\begin{equation}
  \label{eq:benergy}
  \deriv{}{t}\int
  \f{|\vB|^2}{8\pi}\diff
  V=-\int\vU\cdot\f{\left(\vJ\times\vB\right)}{c}\diff V-\f{c}{4\pi}\int\left(\vE\times\vB\right)\cdot \diff\vec{S}-\int\f{|\vJ|^2}{\sigma} \diff V~,
\end{equation}
\red{where the surface integral is taken over the boundary of the
  volume, oriented by an outward normal vector}.
The first term on the r.h.s. is a volumetric term equal to the opposite of
the work done by the Lorentz force on the flow, \red{the second
  surface term is the Poynting flux of electromagnetic energy through
  the boundaries of the domain under consideration},
and the last term quadratic in $\vJ$ corresponds to Ohmic dissipation
of electrical currents into heat.  In the absence of a Poynting term
(for instance in a periodic domain), we see that magnetic energy can only be
generated at the expense of kinetic (mechanical) energy. In other
words, we must put in mechanical energy in order to drive a dynamo.

\subsubsection{Incompressible MHD equations for dynamo theory\label{incompressible}}
Starting from compressible MHD enabled us to show that compressive
motions, which are relevant to a variety of astrophysical situations, 
can formally contribute to the dynamics and amplification of magnetic
fields. However, much of the essence of the
dynamo problem can be captured in the much simpler framework of
incompressible, viscous, resistive MHD, which we will therefore mostly
use henceforth (further assuming constant kinematic viscosity and magnetic
diffusivity). In the incompressible limit, $\rho$ is uniform and
constant, and the distinction between thermal and magnetic pressure
disappears. The magnetic tension part of the Lorentz force provides
the only relevant dynamical magnetic feedback on the flow in that case%
\footnote{In the compressible case,  magnetic pressure exerts
  a distinct dynamical feedback on the flow. This becomes important if
  the magnetic energy is locally amplified to a level comparable
  to the thermal pressure and can notably lead to density
  evacuation.}. Rescaling $\vec{F}$ and $P$ by $\rho$ and
  $\vB$ by $(4\pi\rho)^{1/2}$, \red{so that $\vB$ now stands for the
    Alfv\'en velocity
    \begin{equation}
      \label{eq:alfvenspeed}
    \vU_A=\f{\vB}{\sqrt{4\pi\rho}}~,
  \end{equation}
 }and introducing the total pressure $\Pi=P+B^2/2$,
we can write the incompressible momentum equation as 
\begin{equation}
\label{eq:momincompressible}
\dpart{\vU}{t}+\vU\cdot\grad{\,\vU}=-\grad{\Pi}+\vB\cdot\grad\,\vB+\nu\Delta{\vU}+\vec{F}(\vec{x},t)~.
\end{equation}
The induction equation is rewritten as
\begin{equation}
  \label{eq:inducincompressible}
\dpart{\vB}{t}+\vU\cdot\grad\,\vB=\vB\cdot\grad{\,\vU}+\eta\Delta\vB~.
\end{equation}
This form separates the physical effects of the electromotive force
into two parts: advection/mixing represented by
$\vU\cdot\grad{\,\vB}$ on the left, and induction/stretching
represented by $\vB\cdot\grad{\,\vU}$ on the right.
Magnetic-stretching by shearing motions is the only way to amplify
magnetic fields in an incompressible flow of conducting fluid. In
order to formulate the problem completely,
\equs{eq:momincompressible}{eq:inducincompressible} must be
supplemented with
\begin{equation}
\div{\vU}=0~,\quad \div{\vB}=0~,
\end{equation}
and paired with an appropriate set of initial conditions, and boundary
conditions in space. The latter can be a
particularly tricky business in the dynamo context. Periodic boundary
conditions, for instance, are a popular choice among theoreticians but
may be problematic in the context of the saturation of large-scale
dynamos (\sect{largescalesat}). Certain types of magnetic boundary
conditions are also problematic for the definition of magnetic
helicity. The choice of boundary conditions and global configuration of
dynamo problems is not just a problem for theoreticians either: as
mentioned earlier, the choice of soft-iron vs. steel propellers
has a drastic effect on the excitation of a dynamo effect in the VKS
experiment \citep{monchaux07}, raising the question of whether this
dynamo is a pure fluid effect or a fluid-structure interaction effect
\citep[see e.g.][]{gissinger08a,giesecke12,kreuzahler17,nore18}.

\subsubsection{Shearing sheet model of differential rotation\label{shearingsheet}}
Differential rotation is present in many systems that sustain
dynamos, but can take many different forms depending on the geometry
and internal dynamics of the system at hand. As we will discover
in \sect{numexplor}, working in global cylindrical or spherical geometry
is particularly valuable if we seek to understand how
large-scale dynamos like the solar or geodynamo operate at a global
level, because these systems happen to have fairly complex
differential rotation laws and internal shear layers. On the other hand,
we do not in general need all this geometric complexity to
understand how rotation and shear affect dynamo processes at a
fundamental physical level. In fact, any possible simplification is
most welcome in this context, as many of the basic statistical dynamical
processes that we are interested in are usually difficult enough
to understand at a basic level. In what follows, we will therefore
make intensive use of a local Cartesian representation of differential
rotation, known as the shearing sheet model \citep{goldreich65}, that
will make it possible to study some essential effects of shear and
rotation on dynamos in a very simple and systematic way. 

\begin{figure}
\centering\includegraphics[width=\textwidth]{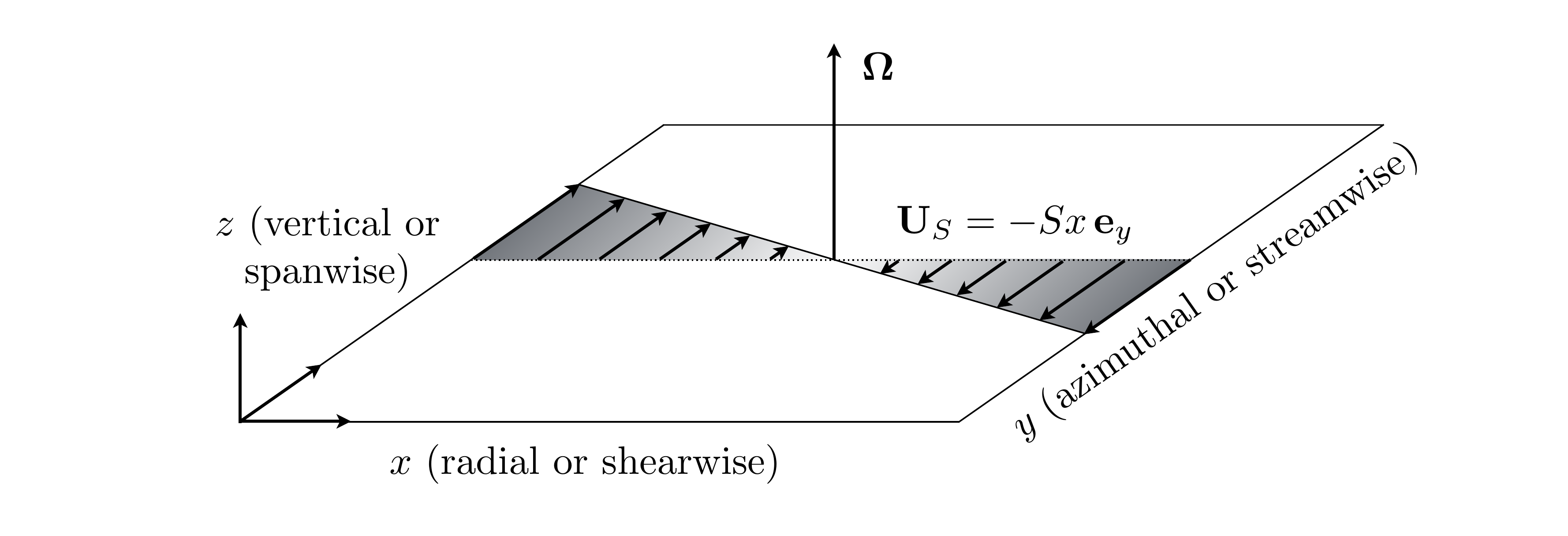}
\caption{The Cartesian shearing sheet  model of differentially rotating flows.\label{figshearingsheet}}
\end{figure}

Consider a simple cylindrical differential rotation law
$\vOmega=\Omega(R)\,\vec{e}_z$ in polar coordinates $(R,\varphi,z)$ 
(think of an accretion disc or a galaxy). To study the 
dynamics around a particular cylindrical radius $R_0$,
we can move to a frame of reference rotating at the local angular velocity,
$\Omega\equiv\Omega(R_0)$, and
solve the equations of rotating MHD locally (including Coriolis and
centrifugal accelerations) in a Cartesian coordinate system
($x,y,z$) centred on $R_0$, neglecting curvature effects (all of this
can be derived rigorously). Here, $x$ corresponds to the
direction of the local angular velocity gradient (the radial direction
in an accretion disc), and $y$ corresponds to the azimuthal
direction. In the rotating frame, the differential rotation around
$R_0$ reduces to a simple a linear shear flow
$\vU_S=-Sx\,\vec{e}_y$, where
$S\equiv-R_0\left.d\Omega/dR\right|_{R_0}$ is the local shearing rate
 (\fig{figshearingsheet}).

This model enables us to probe a variety of differential
rotation regimes by studying the individual or combined
effects of a pure rotation, parametrised by $\Omega$, and of a pure shear,
parametrised by $S$, on dynamos. For instance, we can study dynamos
in non-rotating shear flows by setting $\Omega=0$ and varying the
shearing rate $S$ with respect to the other timescales of the
problem, or we can study the effects of rigid rotation on
a dynamo-driving flow (and the ensuing dynamo) by varying $\Omega$
while setting $S=0$.
Cyclonic rotation regimes, for which the vorticity of the shear flow is
aligned with the rotation vector, have negative $\Omega/S$ in the
shearing sheet with our convention, while anticyclonic rotation
regimes correspond to positive $\Omega/S$.  In particular,
anticyclonic Keplerian rotation typical of accretion discs orbiting
around a central mass $M_*$, $\Omega (R)=\sqrt{\mathcal{G}M_*}/R^{3/2}$,
is characterised by $\Omega=(2/3)S$ in this model.

The numerical implementation of the local shearing sheet approximation
in finite domains is usually referred to as the ``shearing box'',
as it amounts to solving the equations in a  Cartesian box of dimensions
$(L_x,L_y,L_z)$ much smaller than the typical radius of curvature of the
system. In order to accommodate the linear shear in this
numerical problem, the $x$ coordinate is usually taken 
shear-periodic\footnote{A detailed description of a typical
implementation of shear-periodicity in the popular pseudospectral
numerical MHD code SNOOPY \citep{lesur07} can be found in App.~A of
\cite{riols13}.}, the $y$ coordinate is taken periodic, and the
choice of the boundary conditions in $z$ depends on whether some
stratification is incorporated in the modelling (if not, periodicity
in $z$ is usually assumed).

\subsection{Important scales and dimensionless numbers}
\subsubsection{Reynolds numbers}
Let us now consider some important scales and dimensionless numbers
in the dynamo problem based on
\equs{eq:momincompressible}{eq:inducincompressible}. First, we define
the scale of the system
under consideration as $L$, and the integral scale of the turbulence,
or the scale at which energy is injected into the flow, as
$\ell_0$. Depending on the problem under consideration, we will have
either $L~\sim \ell_0$, or $L\gg\ell_0$. Turbulent velocity field
fluctuations at scale $\ell_0$ are denoted by $u_0$. The kinematic
Reynolds number \begin{equation}
Re=\frac{u_0\ell_0}{\nu}
\end{equation}
measures the relative magnitude of inertial effects compared to viscous
effects on the flow. The Kolmogorov  scale $\ell_\nu\sim
Re^{-3/4}\ell_0$ is the scale at which kinetic
energy is dissipated in Kolmogorov turbulence, with $u_\nu\sim
Re^{-1/4}u_0$ the corresponding typical velocity at that scale.
The magnetic Reynolds number 
\begin{equation}
Rm=\frac{u_0\ell_0}{\eta}
\end{equation}
measures the relative magnitude of inductive (and mixing)
effects compared to resistive effects in \equ{eq:inducincompressible},
and is therefore a key number in dynamo theory.

\subsubsection{The magnetic Prandtl number landscape\label{pmlandscape}}
\begin{figure}
\includegraphics[width=\textwidth]{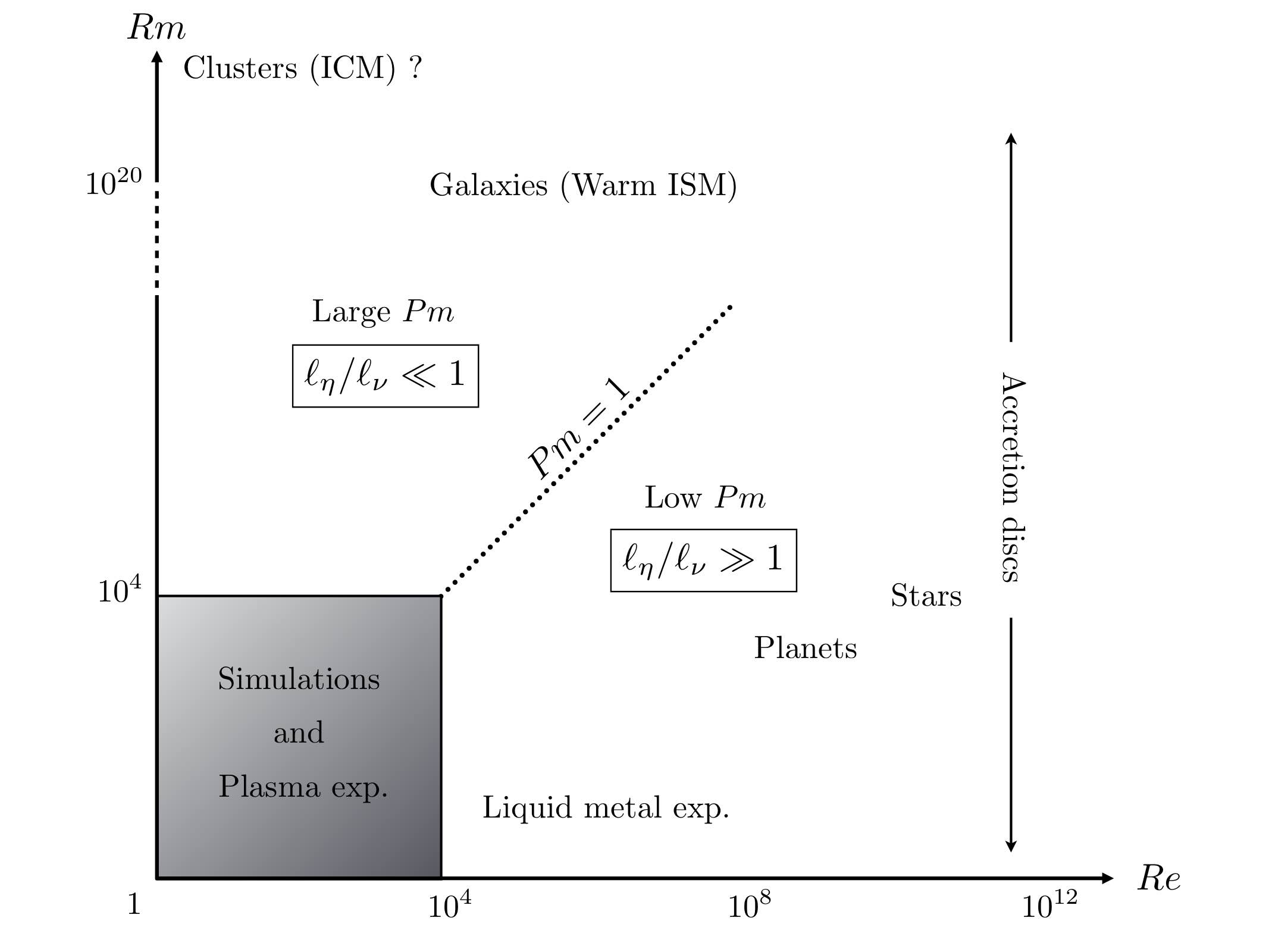}
  \caption{A qualitative representation of the magnetic Prandtl number
    landscape. The grey area depicts the range of $Re$ and $Rm$ 
    (based on r.m.s. velocities) thought to be accessible in the
    foreseeable future through either numerical simulations or plasma
    experiments. \label{figPmlandscape}}
\end{figure}

The ratio of the kinematic viscosity to the magnetic
diffusivity, the magnetic Prandtl number
\begin{equation}
Pm=\frac{\nu}{\eta}=\f{Rm}{Re}~,
\end{equation}
is also a key quantity in dynamo theory. Unlike
$Re$ and $Rm$, $Pm$ in a collisional fluid is an intrinsic property
of the fluid itself, not of the flow. Figure~\ref{figPmlandscape}
shows that conducting fluids and plasmas found in nature and in the lab have
a wide range of $Pm$. 
One reason for this wide distribution is that $Pm$
is very strongly dependent on both temperature and density. 
For instance, in a pure, collisional hydrogen plasma 
with equal ion and electron temperature,
\begin{equation}
\label{eq:Pm}
Pm\simeq 2.5\times 10^3 \f{T^4}{n\left(\ln \Lambda\right)^2}~,
\end{equation}
where $T$ is in Kelvin, $\ln\Lambda$ is the Coulomb logarithm and $n$
is the particle number density in $\mathrm{m}^{-3}$. This collisional
formula gives $Pm\sim 10^{25}$ or larger for
the very hot ICM of galaxy clusters (although it is probably not very
accurate in this context given the weakly collisional nature of the
ICM). The much denser and cooler plasmas in stellar interiors have
much lower $Pm$, for instance $Pm$ ranges approximately from $10^{-2}$
at the base of the solar convection zone to $10^{-6}$ below the
photosphere. Accretion-disc
plasmas can have all kinds of $Pm$, depending on the nature of
the accreting system, closeness to the central accreting object, 
and location with respect to the disc midplane.

Liquid metals like liquid iron in the Earth's core or liquid sodium
in dynamo experiments have very low $Pm$, typically $Pm\sim 10^{-5}$
or smaller. This has proven a major inconvenience for dynamo
experiments, as achieving even moderate $Rm$ in a very low 
$Pm$ fluid requires a very large $Re$ and therefore necessitates
a lot of mechanical input power, which in turns implies a lot of
heating. To add to the inconvenience, the turbulence generated at 
large $Re$ enhances the effective diffusion of the magnetic field,
which makes it even harder to excite interesting magnetic dynamics.
As a result, the experimental community has started to shift
attention to plasma experiments in which $Pm$ can in principle be
controlled and varied in the range $0.1<Pm<100$ by changing either the
temperature or density of the plasma, as illustrated by \equ{eq:Pm}.
Finally, due to computing power limitations implying finite numerical
resolutions, most virtual MHD fluids of computer simulations have
$0.1<Pm<10$ (with a few exceptions at large $Pm$). Hence, it is and
will remain impossible in a foreseeable future to simulate magnetic-field
amplification in any kind of regime found in nature. The best we can
hope for is that simulations of largish or smallish $Pm$
regimes can provide glimpses of the asymptotic dynamics.

The large and small $Pm$ MHD regimes are seemingly very different. 
To see this, consider first the ordering of the resistive scale
$\ell_\eta$, i.e. the typical scale at  which the magnetic field gets
dissipated in MHD, with respect to the viscous scale  $\ell_\nu$.

\begin{figure}
\includegraphics[width=\textwidth]{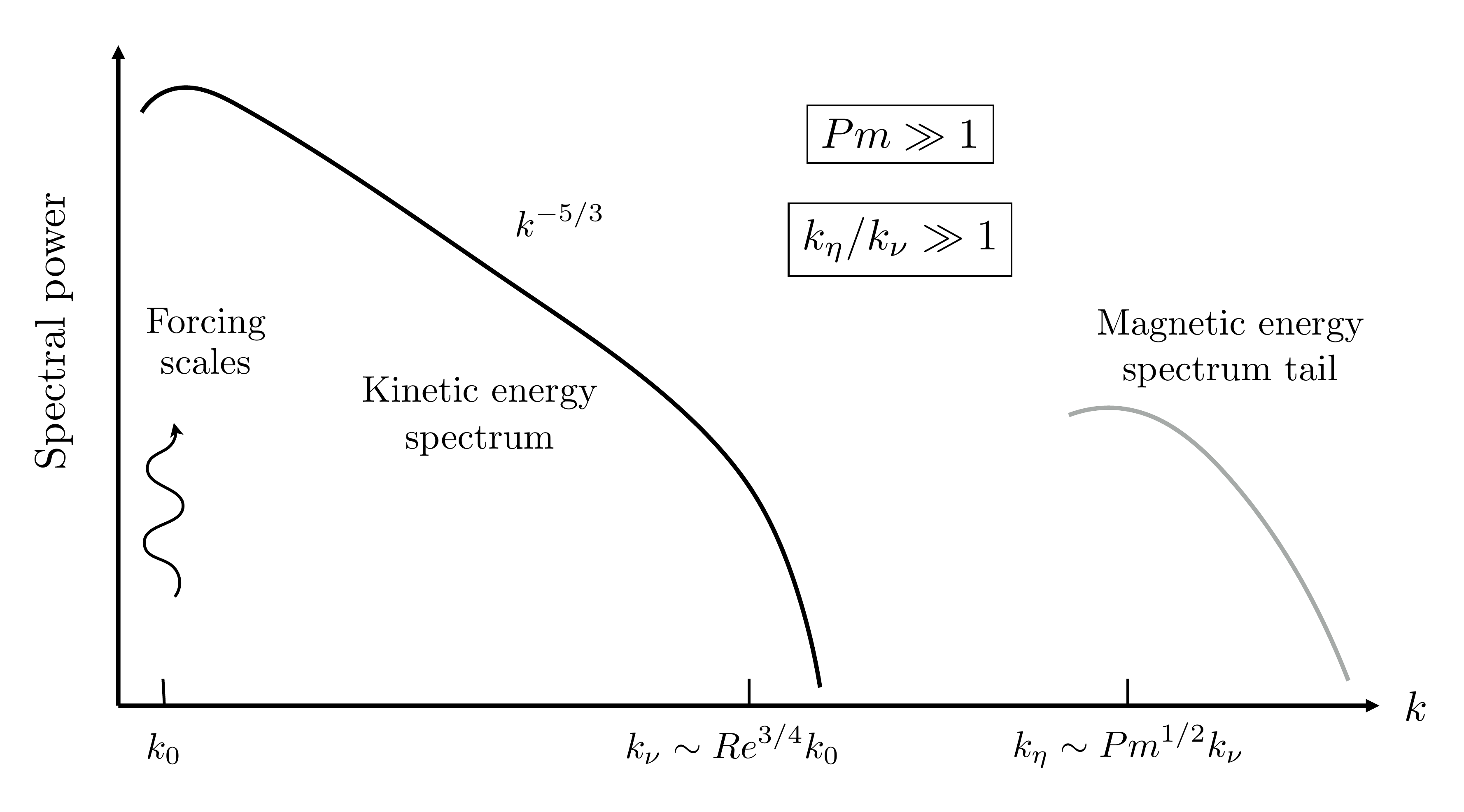}
  \caption{Ordering of scales and \red{qualitative representation} of the
    kinetic and magnetic energy spectra in $k$ (wavenumber) space at large
    $Pm$. \label{figlargePmspec}}
\end{figure}

\paragraph{\textit{Large magnetic Prandtl numbers.}}
For $Pm>1$, the resistive cut-off scale $\ell_\eta$ is smaller than the viscous
scale. This suggests that a lot of the magnetic energy resides at scales well
below any turbulent scale in the flow. \red{The situation is best
illustrated in \fig{figlargePmspec} by taking a spectral point of
view of the dynamics in wavenumber $k\sim 1/\ell$ space, introducing
the kinetic and magnetic energy spectra associated with the the
Fourier transforms in space of the velocity and magnetic field, and
the viscous and resistive wavenumbers $k_\nu\sim 1/\ell_\nu$ and
$k_\eta\sim 1/\ell_\eta$ (this kind of representation will be
frequently encountered in the rest of the review).}
To estimate $\ell_\eta$ more precisely in this regime, let us consider
the case of Kolmogorov turbulence for
which the rate of strain of eddies of size $\ell$
goes as $u_{\ell}/\ell\sim \ell^{-2/3}$. For this kind of turbulence,
the smallest viscous eddies are therefore also the fastest at
stretching the magnetic field. To estimate the resistive scale $\ell_\eta$,
we balance the stretching rate of these eddies $u_\nu/\ell_\nu\sim
Re^{1/2}u_0/\ell_0$ with the ohmic diffusion rate at the resistive scale
$\eta/\ell_\eta^2$. This gives
\begin{equation}
  \label{eq:letalargePm}
\ell_\eta\sim Pm^{-1/2}{\ell_\nu}~,\quad Pm \gg 1~.
\end{equation}

\begin{figure}
\includegraphics[width=\textwidth]{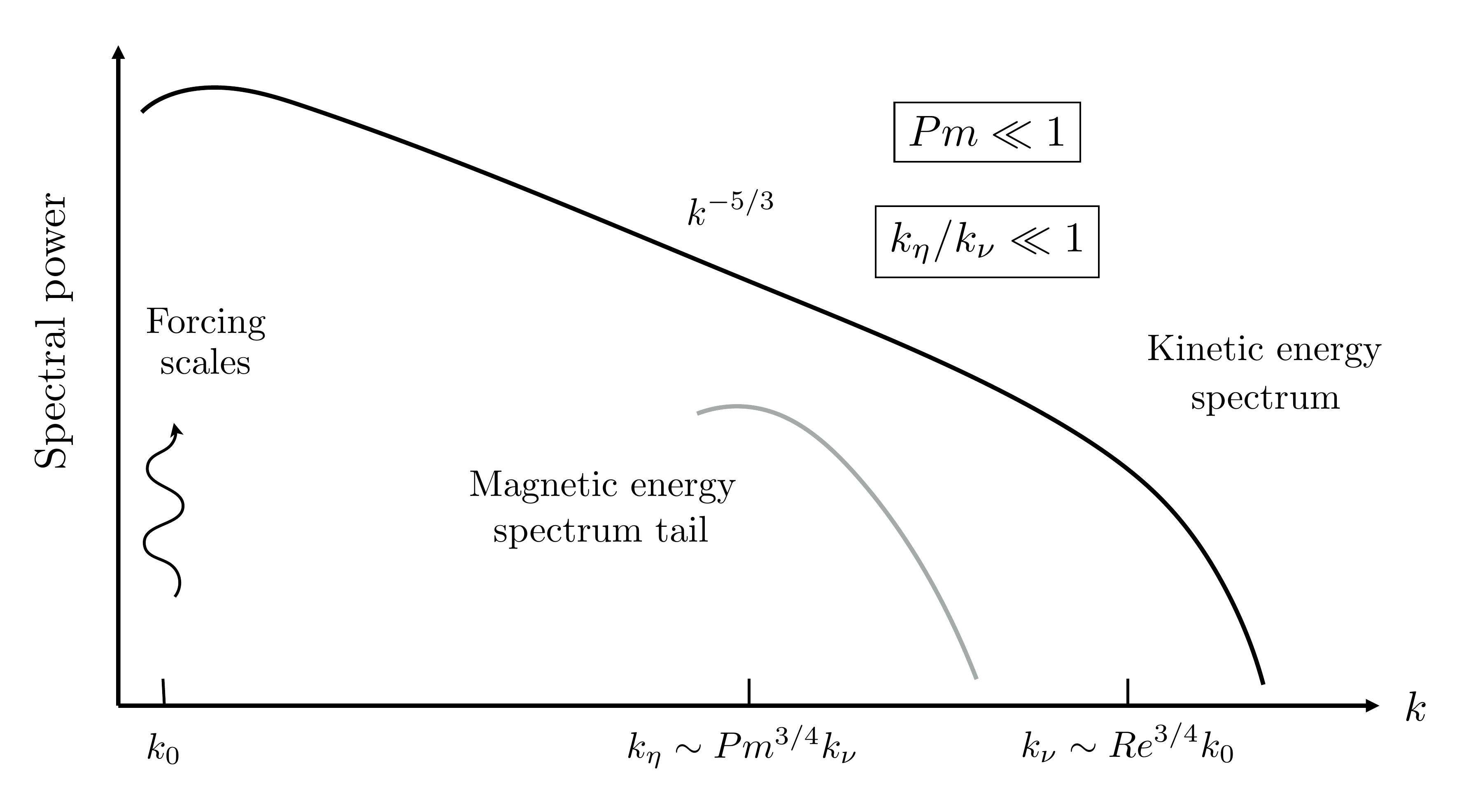}
  \caption{Ordering of scales and  \red{qualitative representation} of
    kinetic and magnetic energy spectra at low $Pm$. \label{figlowPmspec}}
\end{figure}

\paragraph{\textit{Low magnetic Prandtl numbers.}}
For $Pm<1$, we instead expect the resistive scale $\ell_\eta$ to fall
in the inertial range of the turbulence. This is illustrated
in spectral space in \fig{figlowPmspec}.
To estimate $\ell_\eta$ in this regime, we simply balance the
turnover/stretching rate
$u_\eta/\ell_\eta$ of the eddies at scale $\ell_\eta$  with the
magnetic diffusion rate $\eta/\ell_\eta^2$. Equivalently, this can be
formulated as $Rm(\ell_\eta)=u_{\ell_\eta}\ell_\eta/\eta\sim ~1$. The
result is
\begin{equation}
  \label{eq:letalowpm}
  \ell_\eta\sim Pm^{-3/4} \ell_\nu~,\quad Pm \ll 1~.
\end{equation}

Intuitively, the large-$Pm$ regime seems much more favourable to
dynamos. In particular, the fact that the magnetic field ``sees'' a
lot of turbulent activity at low $Pm$ could create many
complications. However, and contrary to what was for instance  argued
in the early days of dynamo theory by \cite{batchelor50}, we will see
in the next sections that dynamo action is possible at low
$Pm$. Besides, the large-$Pm$ regime has a lot of non-trivial dynamics
on display despite its seemingly simpler ordering of scales.

\subsubsection{Strouhal number}
Another important dimensionless quantity arising in dynamo theory is
the Strouhal number
\begin{equation}
St=\f{\tau_c}{\tau_{\mathrm{NL}}}~.
\end{equation}
This number measures the ratio between the correlation time $\tau_c$
and the nonlinear turnover time $\tau_{\mathrm{NL}}\sim\ell_u/u$ of an
eddy with a typical velocity $u$ at scale $\ell_u$. A similar number
appears in all dynamical fluid and plasma problems involving closures
and, despite being of order one in many physical systems worthy of
interest (including fluid turbulence), is usually used as a small
parameter to derive perturbative closures such as those
described in the next two sections. \cite{krommes02} 
offers an illuminating discussion of the potential problems of
perturbation theory applied to non-perturbative systems, many of
which are directly relevant to dynamo theory.

\subsection{Dynamo fundamentals}
Most of the material presented so far is relevant to a much broader
MHD context than just dynamo theory. We are now going to introduce
a few important definitions, and outline several general results and
concepts that are specific to this problem: anti-dynamo theorems and
fast/slow dynamos. A more in-depth and rigorous (yet accessible)
presentation of these topics can notably be found in Michael Proctor's
contribution to the collective book on ``Mathematical aspects of
Natural dynamos'' edited by \cite{dormy07}.

\subsubsection{Kinematic versus dynamical regimes}
The question of the amplification and further sustainment of
magnetic fields in MHD is fundamentally an \textit{instability
problem} with both linear and nonlinear aspects. The first
thing that we usually need to assess is whether the stretching of the
magnetic field by fluid motions can overcome its
diffusion. The magnetic Reynolds number $Rm$ provides a direct
measure of how these two processes compare, and is therefore the key
parameter of the problem. Most, albeit not all, dynamo flows
have a well-defined, analytically calculable or at least computable
$Rm_c$ above which magnetic-field generation becomes possible.

In the presence of an externally prescribed velocity field
(independent of $\vB$), the induction \equ{eq:inducincompressible}
is linear in $\vB$. The \textit{kinematic dynamo problem}
therefore consists in establishing what flows, or classes of flows,
can lead to exponential growth of magnetic energy starting from an
initially infinitesimal seed magnetic field, and in computing 
$Rm_c$ of the bifurcation and growing \textit{eigenmodes} of
\equ{eq:inducincompressible}.
The velocity field in the kinematic dynamo problem can be computed
numerically from the forced Navier-Stokes equation with
negligible Lorentz force\footnote{Or, in
  dynamo problems involving thermal convection, the Rayleigh-B\'enard or
  anelastic systems including \equ{eq:entropy}.}, or using simplified
numerical flow models, or prescribed analytically. This
\textit{linear} problem is relevant to the early
stages of magnetic-field amplification during which the magnetic
energy is small compared to the kinetic energy of the flow.

The \textit{dynamical, or nonlinear dynamo problem}, on the other
hand, consists in solving the full \textit{nonlinear} MHD system
consisting of \equs{eq:cont}{eq:entropy} (or the simpler
\equs{eq:momincompressible}{eq:inducincompressible} in the
incompressible case), including the magnetic back-reaction of the
Lorentz force on the flow. This problem is obviously
directly relevant to the \textit{saturation} of dynamos, but
it is more general than that. For instance, some systems
with linear dynamo bifurcations exhibit subcritical bistability,
i.e. they have pairs of nonlinear dynamo modes involving a
magnetically distorted version of the flow at $Rm$ smaller than the
kinematic $Rm_c$. There is also an important class of dynamical
magnetic-field-sustaining MHD processes, referred to as
instability-driven dynamos, which do not originate in a linear
bifurcation at all, and have no well-defined $Rm_c$. These different
mechanisms will be discussed in \sect{subcritical}.

\subsubsection{Anti-dynamo theorems\label{cowling}}
Are all flows of conducting fluids dynamos ? Despite the seemingly
simple nature of induction illustrated by \equ{eq:Bmag},
there are actually many generic cases in which magnetic fields
cannot be sustained by fluid motions in the limit of infinite times,
even at large $Rm$. Two of them are particularly important (and
annoying) for the development of theoretical models and experiments:
axisymmetric magnetic fields cannot be sustained by dynamo action
(\citeauthor{cowling33}'s
theorem, \citeyear{cowling33}), and planar, two-dimensional motions
cannot excite a dynamo (\citeauthor{zeldovich56}'s  theorem, \citeyear{zeldovich56}).

In order to give a general feel of the constraints that dynamos face,
let us sketch qualitatively how Cowling's theorem originates in 
an axisymmetric system in polar (cylindrical) geometry
$(R,\varphi,z)$. Assume that $\vB$ is an axisymmetric vector field
\begin{equation}
\vB=\curl{(\chi\vec{e}_\varphi/R)}+R\psi\,\vec{e}_\varphi
\end{equation}
where $\chi(R,z,t)$ is a poloidal flux function and \red{$R\psi$ is
  the toroidal magnetic field}. Similarly,  assume that $\vU$ is
axisymmetric with respect to the same axis of symmetry as $\vB$, i.e.
\begin{equation}
\vU=\vU_\mathrm{pol}+R\Omega\,\vec{e}_\varphi~,
\end{equation}
where $\vU_\mathrm{pol}(R,z,t),$ is an axisymmetric poloidal
velocity field in the $(R,z)$ plane and  $\Omega(R,z,t)\,\vec{e}_\varphi$
is an axisymmetric toroidal differential rotation. The poloidal and
toroidal components of \equ{eq:inducincompressible} respectively read
\begin{equation}
  \label{eq:poloidalchi}
\dpart{\chi}{t}+\vU_{\mathrm{pol}}\cdot\grad{\chi}=\eta\left(\Delta-\f{2}{R}\dpart{ }{R}\right)\chi
  \end{equation}
and
\begin{equation}
  \label{eq:toroidalpsi}
\dpart{\psi}{t}+\vU_\mathrm{pol}\cdot\grad{\psi}=\vB_\mathrm{pol}\cdot\grad{\Omega}+\eta\left(\Delta+\f{2}{R}\dpart{}{R}\right)\psi~.
\end{equation}
\Equ{eq:toroidalpsi} has a source term,
$\vB_\mathrm{pol}\cdot\grad{\Omega}$, which describes the
stretching of poloidal field into toroidal field by the
differential rotation, and is commonly referred to as 
the $\Omega$ effect in the astrophysical dynamo community (more on
this in \sect{largescaleOmega}).
However, there is no similar back-coupling between $\psi$ and $\chi$ in
\equ{eq:poloidalchi}, and therefore there is no converse way to
generate poloidal field out of toroidal field in such a system.
The problem is that there is no perennial source of poloidal flux $\chi$
in \equ{eq:poloidalchi}. The advection term on the l.h.s. describes the
redistribution/mixing of the  flux by the axisymmetric poloidal flow
in the $(R,z)$ plane and can only amplify the field locally and
transiently. The presence of resistivity on the r.h.s. then implies
that $\chi$ must ultimately decay, and therefore so must the
$\vB_\mathrm{pol}\cdot\grad{\Omega}$ source term in
\equ{eq:toroidalpsi}, and $\psi$. Overall, the constrained geometry of
this system therefore makes it impossible for the magnetic field to be
sustained\footnote{From a mathematical point of view, the linear
  induction operator for a pure shear flow is not self-adjoint. In a
  dissipative system, this kind of mathematical structure generically
  leads to transient secular growth followed by exponential or
  super-exponential decay, rather than simple exponential growth or
  decay of normal modes (see \cite{trefethen93} and \cite{livermore04}
  for a discussion in  the dynamo context).}. 

Cowling's theorem is one of the main reasons why the solar and geo- dynamo
problems are so complicated, as it notably shows that a magnetic dipole
strictly aligned with the rotation axis cannot be sustained by
a simple combination of axisymmetric differential rotation and meridional
circulation. Note however that axisymmetric flows like the
\cite{dudley89} flow or von K\'arm\'an flows \citep{marie03}, on which
the designs of several dynamo experiments are based, can excite
non-axisymmetric dynamo fields with dominant equatorial dipole
geometry ($m=1$ modes with respect to the axis of symmetry of the
flow). There are also mechanisms by which nearly-axisymmetric
magnetic fields can be generated in fluid flows with a strong
axisymmetric mean component \citep{gissinger08b}. We will find out in
\sect{largescale} how the relaxation of the assumption of flow
axisymmetry gives us the freedom to generate large-scale dynamos,
albeit generally at the cost of a much-enhanced dynamical complexity.

Many other anti-dynamo theorems have been proven using similar reasonings.
As mentioned above, the most significant one, apart from Cowling's
theorem, is Zel'dovich's
theorem that a two-dimensional planar flow (i.e. with only two
components), $\vU_{\mathrm{2D}}(x,y,t)$, cannot excite a dynamo. A
purely toroidal flow cannot excite a dynamo either, and a magnetic
field of the form $\vB(x,y,t)$ alone cannot be a dynamo field. All
these theorems are a consequence of the particular structure of the
vector induction equation, and imply that
a minimal geometric complexity is required for dynamos to work. But
what does ``minimal'' mean ? As computer simulations were still in
their infancy, a large number of applied mathematician brain hours
were devoted to tailoring flows with enough dynamical and geometrical
complexity to be dynamos, yet simple-enough mathematically to remain
tractable analytically or with a limited computing capacity.
2.5D (or 2D-3C) flows of the form $\vU(x,y,t)$ with three
non-vanishing components (i.e., including a $z$ component)
are popular configurations of this kind, that make it possible to
overcome anti-dynamo theorems at a minimal cost. Some well-known
examples are 2D-3C versions of the Roberts flow \citep{roberts72}, and
the Galloway-Proctor flow \citep[GP,][]{galloway92}. \red{The original
  version of the latter is periodic in time} and therefore has
relatively simple kinematic dynamo eigenmodes of the form
$\vB(x,y,z,t)=\mathcal{R}\left\{\vec{B}_{\mathrm{2D}-\mathrm{3C}}(x,y,t)\exp\left(st+ik_zz\right)\right\}$,
where $k_z$, the wavenumber of the magnetic perturbation along the $z$
direction, is a parameter of the problem,
$\vec{B}_{\mathrm{2D}-\mathrm{3C}}(x,y,t)$ \red{has the same
time-periodicity as the flow}, and $s$ is the (a priori
complex) growth rate of the dynamo for a given $k_z$.
While such flows are very peculiar in many respects,
they have been instrumental in the development of theoretical and
experimental dynamo research and have taught us a lot on the
dynamo problem in general. They retain some popularity nowadays
because they can be used to probe kinematic dynamos in higher-$Rm$
regimes than in the fully 3D problem by concentrating all the
numerical resolution  and computing power into just two spatial
dimensions. A contemporary example of this kind of approach will be
given in \sect{largeRmfrontier}.

\subsubsection{Slow versus fast dynamos\label{fast}}
Dynamos can be either \textit{slow} or \textit{fast}. Slow dynamos 
are dynamos whose existence hinges  on the spatial diffusion of the
magnetic field to couple different field components. These dynamos
therefore typically evolve on a large, system-scale Ohmic diffusion
timescale $\tau_{\eta,0}=\ell_0^2/\eta$ and their growth rate tends to
zero as $Rm\rightarrow \infty$, for instance (but not necessarily) as
some inverse power of $Rm$. For this reason, they are probably not
relevant to astrophysical systems with very large $Rm$ and  dynamical
magnetic timescales much shorter than $\tau_{\eta,0}$. A classic
example is the \cite{roberts70,roberts72} dynamo,
\red{which notably served as an inspiration for one of the first
experimental demonstrations of the dynamo effect in Karlsruhe
\citep{stieglitz01,gailitis02}}. Fast dynamos, on the other hand, are
dynamos whose growth rate
remains finite and becomes independent of $Rm$ as $Rm\rightarrow
\infty$. Although it is usually very hard to formally prove that a
dynamo is fast, most dynamo processes discussed in the next 
sections are thought to be fast, and so is for instance the 
previously mentioned \cite{galloway92} dynamo. The difficult
analysis of the Ponomarenko dynamo case illustrates the general
trickiness of this question \citep{gilbert88a}.
A more detailed comparative discussion of the characteristic
properties of classic examples of slow and fast dynamo flows is given
by Michael Proctor at p. 186 of the
\textit{\usebibentry{gubbins07}{title}} edited by \cite{gubbins07}.

\begin{figure}
\includegraphics[width=\textwidth]{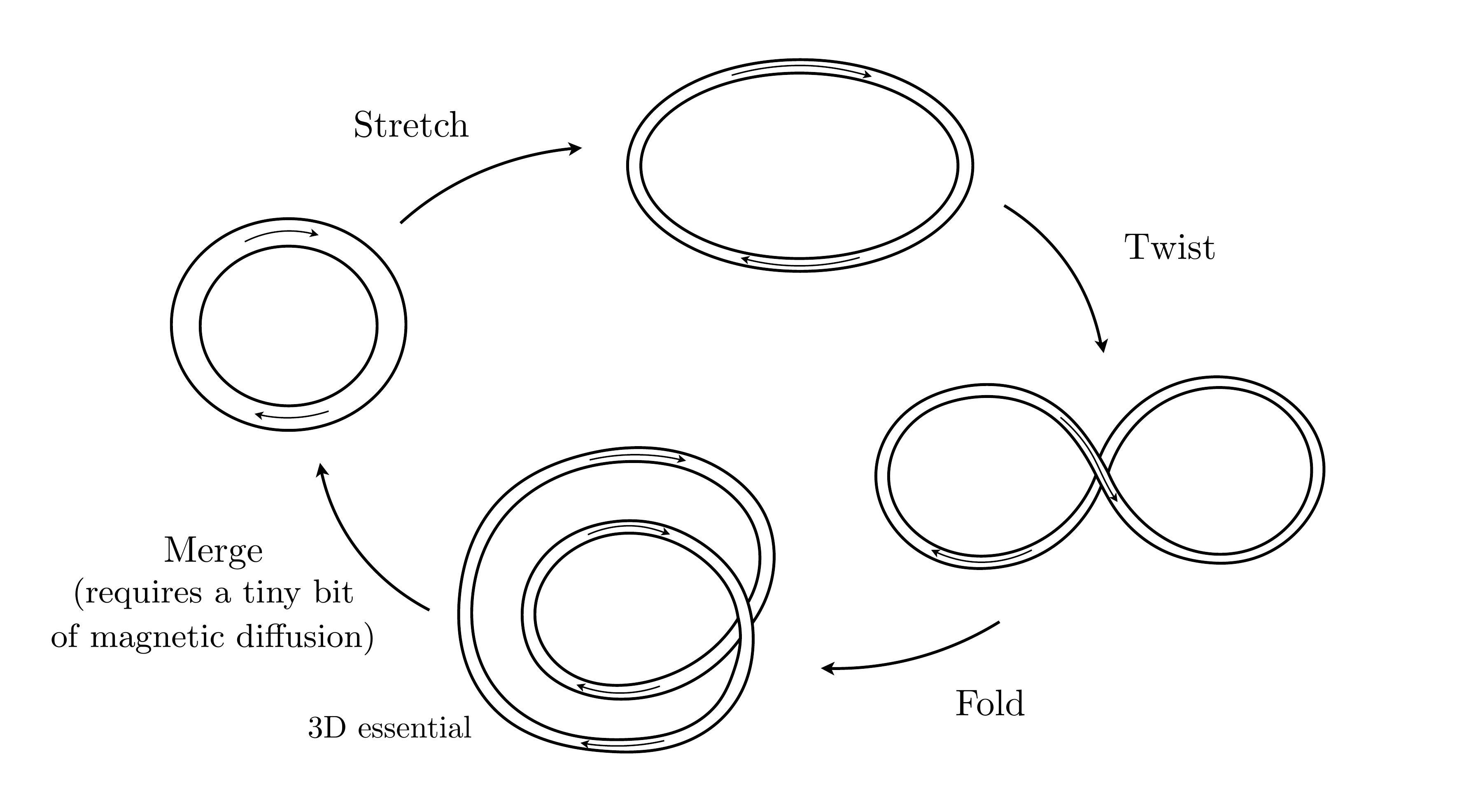}
\caption{The famous stretch-twist-fold dynamo cartoon, adapted from
  \cite{vainshtein72} and many others.\label{figstretchtwistfold}} 
\end{figure}

The standard physical paradigm of fast dynamos, originally due to
\cite{vainshtein72}, is the \textit{stretch, twist, fold} mechanism
pictured in numerous texts, including here in \fig{figstretchtwistfold}.
In this picture, a loop of magnetic field is stretched by shearing
fluid motions so that the field strength increases by a typical 
factor two over a turnover time through magnetic flux conservation. 
If the field is (subsequently or simultaneously) further twisted and
folded by out-of-plane motions, we obtain a fundamentally 3D ``double
tube'' similar to that shown at the bottom of \fig{figstretchtwistfold}.
In that configuration, the magnetic field in each flux tube has the same
orientation as in the neighbouring tube. The initial geometric
configuration can then be recovered by diffusive merging of two loops,
but with almost double magnetic field compared to the original
situation. If we think of this cycle as being a single
iteration of a repetitive discrete process (a discrete map), with each
iteration corresponding to a typical fluid eddy turnover, then we
have all the ingredients of a self-exciting process, whose growth
rate in the ideal limit of infinite $Rm$ is $\gamma_\infty=\ln 2$
(inverse turnover times). Only a tiny magnetic diffusivity is required
for the merging, as the latter can take place at arbitrary small
scale. The overall process is therefore not diffusion-limited.

One of the fundamental ingredients here is the stretching of the
magnetic field by the flow. More generally, it can be shown that an
essential requirement for fast dynamo action is that the flow exhibits
\textit{Lagrangian chaos} \citep{finn88}, i.e. trajectories of initially close
fluid particles must diverge exponentially, at least in some flow
regions. This key aspect of the problem, and its implications for the
structure of dynamo magnetic fields, will become more explicit in the
discussion of the small-scale dynamo phenomenology in the next section. 
Many fundamental mathematical aspects of fast kinematic dynamos have
been studied in detail using the original induction equation or
simpler idealised discrete maps that capture the essence of this
dynamics. We will not dive into this subject any further here, as 
it quickly becomes very technical, and has already extensively been
covered in dedicated reviews and books, including the monograph of
\cite{childress95} and a chapter by Andrew Soward in a collective
book of lectures on dynamos edited by \cite{proctor94}.

\section{Small-scale dynamo theory\label{smallscale}}
Dynamos processes exciting magnetic fields at scales smaller than the
typical integral or forcing scale $\ell_0$ of a flow are generically
referred to as small-scale dynamos, but can be very diverse in
practice. In this section, we will primarily be concerned with the
statistical theory of small-scale dynamos excited by turbulent,
non-helical velocity fluctuations $\vec{u}$ driven randomly by an
external artificial body force, or through natural
hydrodynamic instabilities (e.g. Rayleigh-B\'enard convection). We will
indistinctly refer to such dynamos as \textit{fluctuation} or small-scale
dynamos. The first question that we would like to address, of course,
is whether small-scale fluctuation dynamos are possible at all. We
know that ``small-scale'' fields and turbulence are present in
astrophysical objects, but is there actually a proper mechanism to
generate such fields from this turbulence ? In particular, can
vanilla non-helical homogeneous, isotropic incompressible fluid
turbulence drive a fluctuation dynamo in a conducting fluid, a
question first asked by \cite{batchelor50} ?

\subsection{Evidence for small-scale dynamos}
Direct experimental observations of small-scale fluctuation dynamos
have only recently been reported in laser experiments
\citep{meinecke15,tzeferacos18}, although the reported magnetic-field
amplification factor of $\sim 25$ is relatively small by
experimental standards. The most detailed evidence (and interactions
with theory) so far has been through numerical
simulations. In order to see what the basic evidence for small-scale
dynamos in a turbulent flow looks like, we will therefore simply have a look at
the original numerical study of \cite{meneguzzi81}, which served as a
template for many subsequent simulations\footnote{Pragmatic
  down-to-earth experimentalists feeling uneasy with a primarily
  numerical and theoretical perspective on physics problems may or may
  not find some comfort in the observation that essentially the same
  dynamo has since been reported in many MHD ``experiments'' performed
  with different resolutions and numerical methods.}. 

\begin{figure}
\centering\hbox{\includegraphics[height=0.29\textheight]{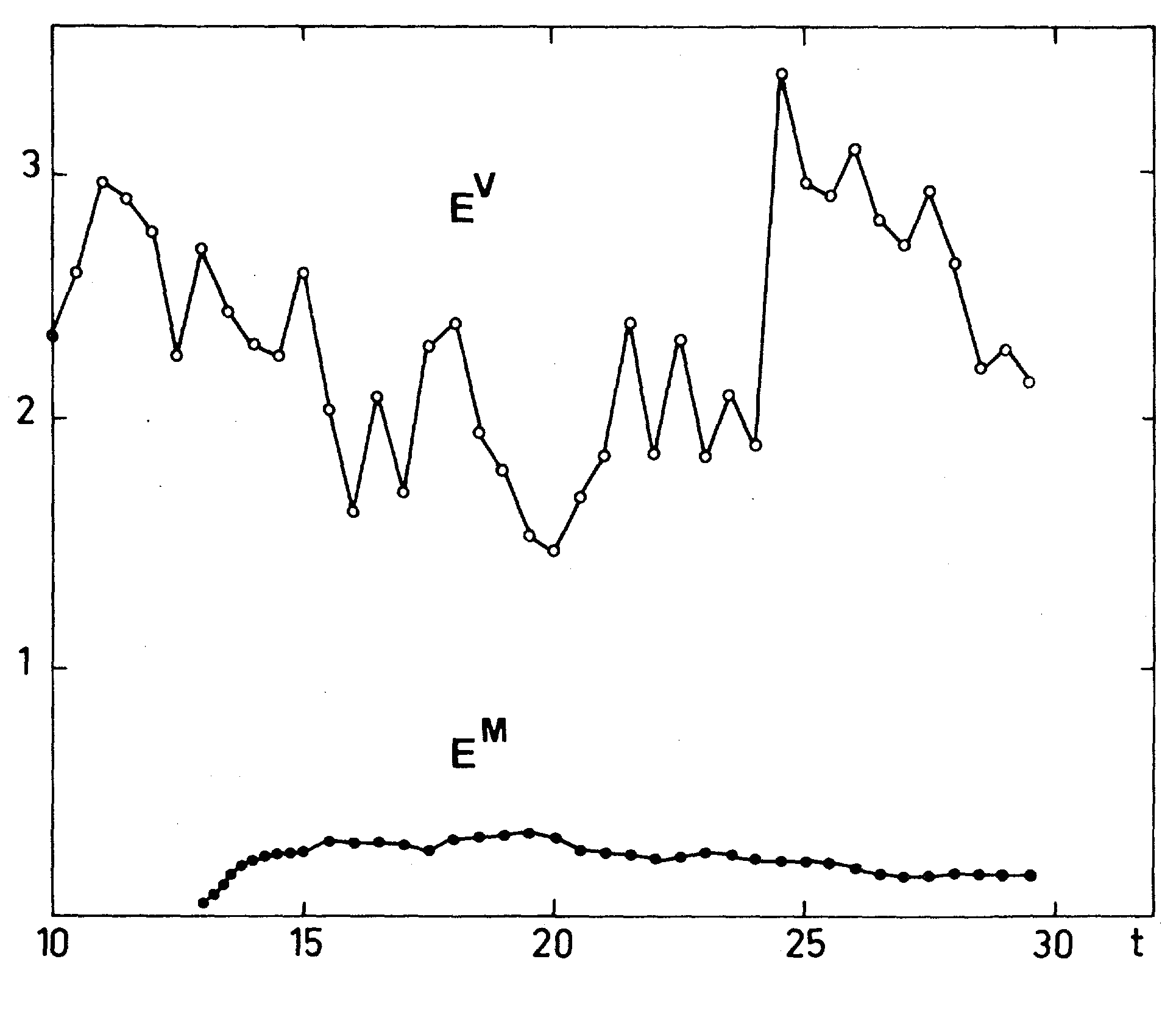}\includegraphics[height=0.29\textheight]{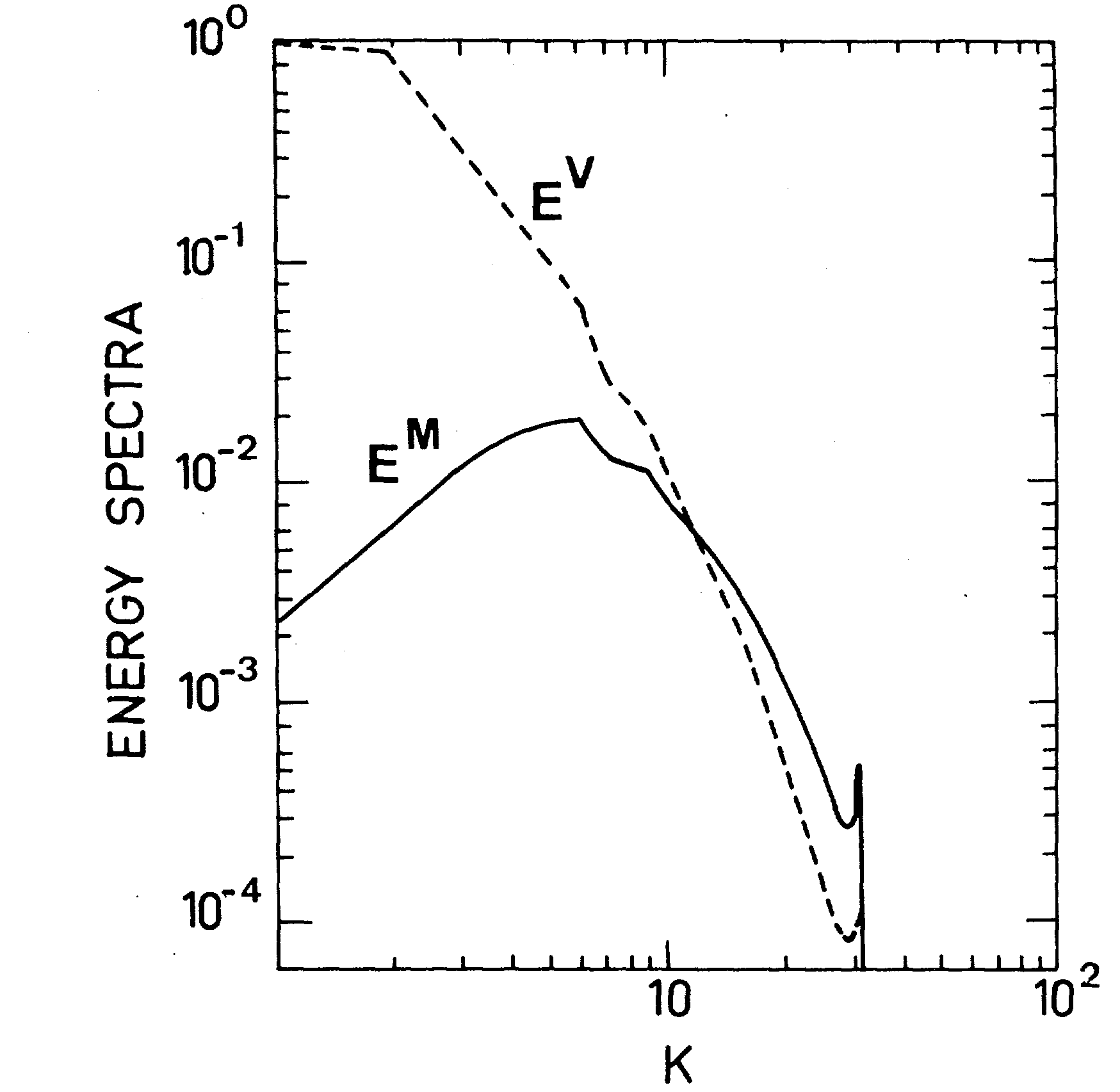}}
\caption{The first simulations of small-scale dynamo action,
  conducted with a pseudo-spectral MHD code and $64^3$ numerical
  resolution with dealiasing, $Rm=100$ and $Pm=1$ (time is measured
  in multiples of the turnover time $\ell_0/u_0$ at the turbulent
  forcing scale). Left: time-evolution of the kinetic
  $(E^V)$ and magnetic $(E^M)$ energies. Right: corresponding kinetic
  and magnetic energy spectra in the saturated stage
 \citep[adapted from][]{meneguzzi81}.\label{figMFP81small}}
\end{figure}

The \citeauthor{meneguzzi81} experiment starts with a
three-dimensional numerical simulation of incompressible, homogeneous,
isotropic, non-helical
Navier-Stokes hydrodynamic turbulence forced randomly at the scale of a
(periodic) numerical domain. This is done by direct numerical
integration of \equ{eq:momincompressible} at $Re=100$ with a
pseudo-spectral method. After a few turnover times ensuring that the
turbulent velocity field has reached a
statistically steady state, a small magnetic field seed is introduced
in the domain and both
\equs{eq:momincompressible}{eq:inducincompressible} are integrated
from there on ($Pm=1$ in the simulation). The
time-evolution of the total kinetic and magnetic energies during the
simulation is shown in \fig{figMFP81small} (left).  After the
introduction of the seed field, magnetic energy first grows,
and then saturates after a few turnover times by settling into a
statistically steady state. \Fig{figMFP81small} (right) shows the
kinetic and magnetic energy spectra in the saturated regime. 
The magnetic spectrum has a significant overlap with the velocity
spectrum, but peaks at a scale significantly smaller than the forcing
scale of the turbulence. Also, its shape is very different
from that of the velocity spectrum. We will discuss this later
in detail when we look at the theory.

To summarise, this simulation captures both the
kinematic and the dynamical regime of a small-scale dynamo effect
at $Rm=100$, $Pm=1$ (although the dynamical impact of the magnetic field
on  the flow in the saturated regime is not obvious in this particular
simulation), and it provides a glimpse of the statistical properties
of the magnetic dynamo mode through the shape of the magnetic
spectrum. Interestingly, it does not take much resolution to obtain
this result, as $64$ Fourier modes in each spatial direction
are enough to resolve the dynamo mode in this regime. Keep in mind,
though, that performing such a 3D MHD simulation was quite a technical
accomplishement in 1981, and required a massive allocation of CPU time
on a CRAY supercomputer !

\subsection{Zel'dovich-Moffatt-Saffman phenomenology\label{zeldo}}
Having gained some confidence that a small-scale dynamo
instability is possible, the next step is to understand
how it works. While the general stretch, twist, fold phenomenology
provides a qualitative flavour of how such a dynamo may proceed,
it would be nice to be able to make sense of it through a more
quantitative, yet physically transparent analysis.
Such an analysis was conducted by \cite{zeldovich84} for an idealised
time-dependent flow model consisting of a linear shear flow
\red{``renovated'' (refreshed)} randomly at regular time intervals, and
by \cite{moffatt64} for the simpler case of time-independent linear
shear, based on earlier hydrodynamic work by \cite{pearson59}.

Let us consider the incompressible, kinematic dynamo
problem~(\ref{eq:inducincompressible}) paired \red{with a simple
non-uniform but spatially ``smooth'' incompressible random
flow model $\vec{u}(\vec{r})=\mathbf{\sf{C}}\,\vec{r}$,
where $\sf{C}$ is a random matrix with $\mathrm{Tr}\,\sf{C}=0$},
and assume that the magnetic field at $t=0$,
$\vB(\vec{r},0)=\vB_0(\vec{r})$, has finite total
energy, no singularity and $\lim_{r\rightarrow \infty}
\vB_0(\vec{r})=0$. The evolution of the separation vector
$\delta \vec{r}$ connecting two fluid particles is given by
\begin{equation}
\deriv{\delta r_i}{t}={\sf{C}}_{ik}\delta r_k~.
\end{equation}

\begin{figure}
\centering\includegraphics[width=\textwidth]{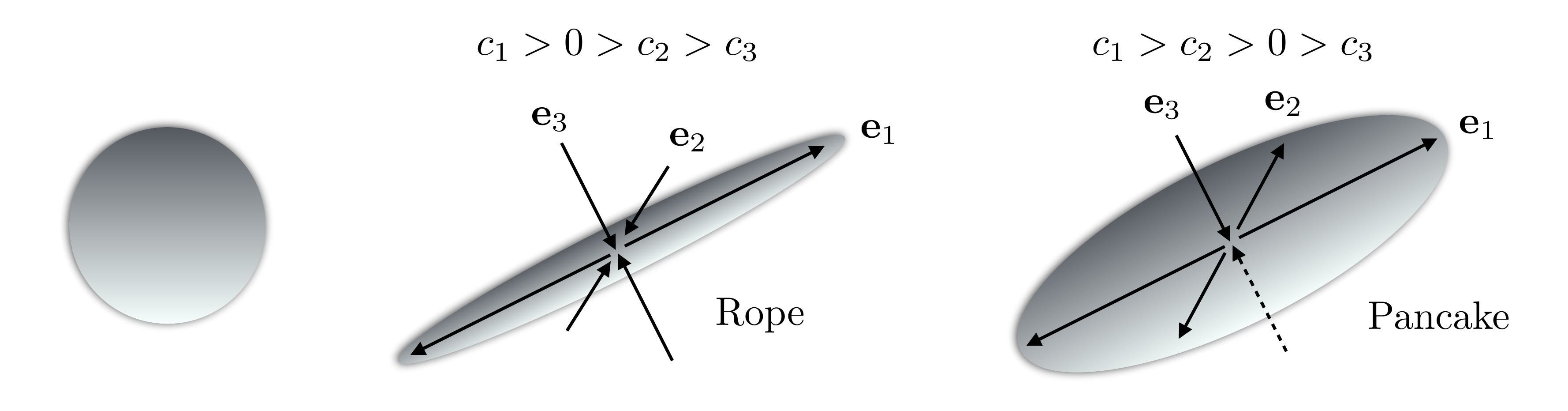}
\caption{Possible deformations of a fluid particle (or magnetic field
  lines) under an incompressible strain.\label{figropepancake}}
\end{figure}

We first analyse the situation considered by \cite{moffatt64}
where {\sf{C}} is constant. In  an appropriate basis
$(\vec{e}_1,\vec{e}_2,\vec{e}_3)$, ${\sf{C}}=\mathrm{diag}(c_1,c_2,c_3)$
with $c_1+c_2+c_3=0$. There are two possible ways to stretch and
squeeze the magnetic field, namely we can form magnetic ropes if
$c_1>0>c_2>c_3$, or magnetic pancakes if $c_1>c_2>0>c_3$.
Both cases are depicted in \fig{figropepancake}. 
We will only analyse the rope case in detail here, but will also give
the results for the pancake case. In both cases, we expect the
stretching of the magnetic field along $\vec{e}_1$ to lead
to magnetic amplification as $B^2\sim \exp (2 c_1t)$ in ideal MHD.
However, the perpendicular squeezing implies that even a tiny magnetic
diffusion matters. Is growth still possible in that case ?
To answer this question, we decompose $\vB$ into shearing Fourier modes
\begin{equation}
\label{Bdecomposed}
\vB(\vec{r},t)=\int\hat{\vB}_{\vec{k}_0}(t)\exp{\left(i\vec{k}(t)\cdot\vec{r}\right)}\diff^3\vec{k}_0~,
\end{equation}
where $\vec{k}_0$, the initial lagrangian wavenumbers, can be
thought of as labelling each evolving mode, and the
different $\vec{k}(t)\equiv\vec{k}(\vec{k}_0,t)$ are
time-evolving wavenumbers with $\vec{k}(t=0)=\vec{k}_0$
($\hat{\vB}_{\vec{k}_0}(t)$ in this context should not be confused with
the unit vector, introduced in  \sect{energetics}, defining the
orientation of $\vec{B}$). Replacing $\vB$ by this expression in the
induction equation, we have
\begin{equation}
\deriv{\hat{\vB}_{\vec{k}_0}}{t}={\sf{C}}\,\hat{\vB}_{\vec{k}_0}-\eta k^2\hat{\vB}_{\vec{k}_0}
\end{equation}
for each $\vec{k}_0$, and
\begin{equation}
\deriv{\vec{k}}{t}=-{\sf{C}}^T\vec{k}~.
\end{equation}
with $\vec{k}(t)\cdot\hat{\vB}_{\vec{k}_0}(t)=0$ at all times.
The diffusive part of the evolution goes as
\begin{equation}
  \exp\left(-\eta\int_0^tk^2(s)\diff s\right)
\end{equation}
and leads to the super-exponential decay of most Fourier modes because
$k_3 \sim k_{03}\exp(|c_3| t)$. At any given time $t$, the ``surviving''
wavenumbers live in an exponentially narrower cone of Fourier space
such that
\begin{equation}
  \eta\int_0^tk^2(s)\diff s=O(1)~.
\end{equation}
In the rope case, the initial wavenumber of the modes still surviving at time
$t$ is given by 
\begin{equation}
\label{eq:k0cone}
k_{02} \sim \exp(-|c_2| t),\quad k_{03} \sim \exp(-|c_3|
t)~.
\end{equation}
Accordingly, the initial magnetic field of these surviving modes
is
\begin{equation}
\left[\vB_{\vec{k}_0}(0)\cdot{\vec{e}_1}\right]\sim k_{02}/k_{01}\left[\vB_{\vec{k}_0}(0)\cdot{\vec{e}_2}\right]  \sim
\exp(-|c_2|t)
\end{equation} 
(from the solenoidality condition for $\vB$). As
the field is stretched along $\vec{e}_1$, we then find
that the amplitude of the surviving rope modes at time $t$ goes 
as
\begin{equation}
\label{eq:bc1c2}
\hat{\vB}_{\vec{k}_0}(t) \sim\exp{(c_1t)}\exp{(-|c_2|t)}.
\end{equation}
We can now estimate from \equ{Bdecomposed} the total magnetic field in
physical space. The first term in the integral goes as
$\exp{\left[(c_1-|c_2|)t\right]}$ from \equ{eq:bc1c2}, and the wavevector
space element as $\exp{\left[-(|c_2|+|c_3|) t\right]}$ from
\equ{eq:k0cone}, so that overall
\begin{equation}
\vB(\vec{r},t)\sim \exp\left(-|c_2|t\right)~.
\end{equation}
Hence the magnetic
field is stretched and squeezed into a rope that decays pointwise
asymptotically. But what about the total magnetic energy
\begin{equation}
  E_\mathrm{m}=\int\vB^2(t,\vec{r})\diff^3\vec{r}
\end{equation}
in the volume of fluid ? $|\vB|^2\sim \exp\left(-2|c_2|t\right)$,
but the volume occupied by the field goes as
$\exp\left(c_1t\right)$. Importantly,
there is no shrinking of the volume element along the second and third
axis because magnetic diffusion sets a minimum scale in these
directions.  Regrouping everything, we obtain
\begin{equation}
\label{eq:Emevol}
E_\mathrm{m}\sim \exp\left[(c_1-2|c_2|)t\right]\sim \exp\left[(|c_3|-|c_2|)t\right]~.
\end{equation}
The second twiddle equality only applies in 3D. Similar
conclusions hold for the pancake case, except that $E_\mathrm{m}\sim
\exp\left[(c_1-c_2)t\right]$.
Overall, we see that  the total magnetic energy of magnetic ropes
decays in 2D, because $|c_2|=c_1$ in that case. This is of course 
expected from Zel'dovich's anti-dynamo theorem. On the other hand, the
magnetic energy grows in 3D because $|c_3|>|c_2|$ and the volume
occupied by the magnetic field grows faster than the pointwise decay
rate of the field itself \citep{moffatt64}.

\begin{figure}
\centering\includegraphics[width=\textwidth]{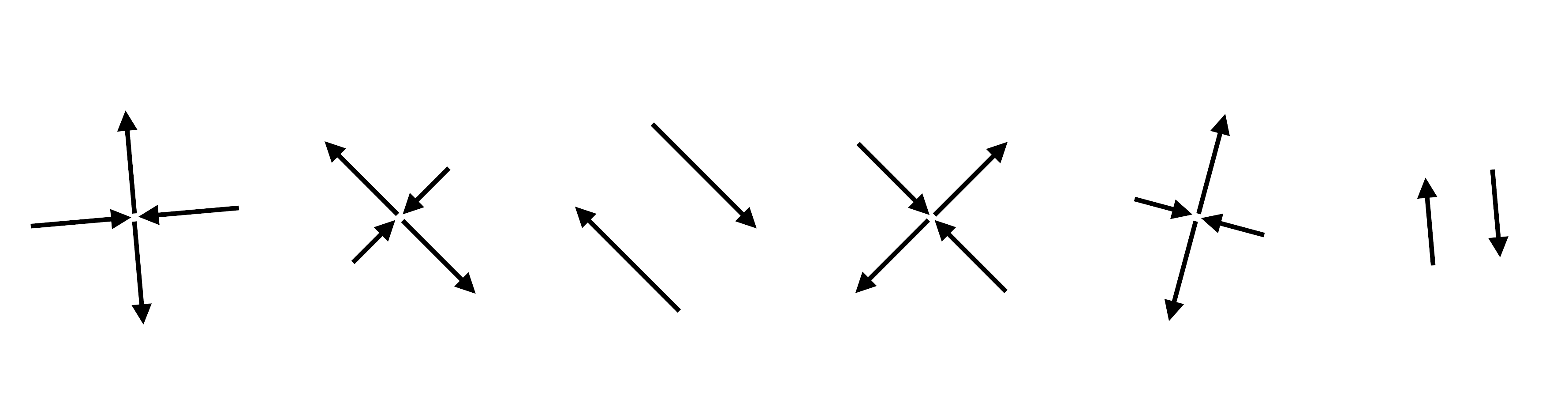}
\caption{A sequence of random linear shearing events 
(to be thought of in 3D).\label{figrandomshears}}
\end{figure}

\cite{zeldovich84} generalised these results 
to random, time-dependent shears. They considered a shear flow
``renovating'' every time-interval $\tau$, such as shown in
\fig{figrandomshears}.
This generates a succession of random area-preserving stretches and
squeezes, which can be described in \textit{multiplicative matrix
  form}.  More precisely, the matrix ${\sf{T}}_t\equiv
{\sf{T}}(t_0,t)$ such that $\vec{k}(\vec{k}_0,t)={\sf{T}}_t\vec{k}_0$
is put in Volterra multiplicative integral form
\begin{equation}
  {\sf{T}}_t=\prod_{s=0}^t\left[\tens{I}-{\sf{C}}^{\sf{T}}(s)\diff
    s\right]~,
\end{equation}
where $\tens{I}$ is the unit tensor. From there, the
formal solution of the linear induction equation is 
\begin{equation}
\vB(\vec{r},t)=\int\exp{\left[i{\sf{T}}_t\left(\vec{k}_0\cdot\vec{r}\right)-\eta\int_0^t({\sf{T}}_s\vec{k}_0)^2\diff
    s\right]}({\sf{T}}_t^{\sf{T}})^{-1}\vec{B}_{\vec{k}_0}(0)\diff^3\vec{k}_0~.
\end{equation}
The hard technical work lies in the calculation of the properties of the
multiplicative integral. \citeauthor{zeldovich84} managed to show that the
cumulative effects of any random sequence of shear can be reduced to
diagonal form. In particular, they proved the surprising result that 
for any such infinite sequence there is always a net positive Lyapunov
exponent $\gamma_1$ corresponding to a stretching direction
\begin{equation}
\lim_{n\rightarrow\infty} \f{1}{n\tau}\ln k(\vec{k}_0,n\tau) \equiv
\gamma_1 > 0~.
\end{equation}
The underlying Lyapunov basis is a function of the full random
sequence, but it is independent of time. This is a form of spontaneous
symmetry breaking: while the system has no
privileged direction overall, any particular infinite sequence of
random shears will generate a particular eigenbasis. Moreover, as a
particular sequence of random shears unfolds, this Lyapunov basis
crystallises exponentially fast in time, while the exponents converge
as $1/t$ \citep*{goldhirsch87}.

The random problem therefore reduces to that of the constant
strain matrix described earlier. This establishes that magnetic 
energy growth is possible in a smooth, 3D chaotic velocity field 
even in the presence of magnetic diffusion, and shows that
the exponential separation of initially nearby fluid trajectories
is critical to the dynamo process. 
The linear shear assumption can be relaxed to accommodate the 
case where the flow has large but finite size. The main difference in
that case is that magnetic field can also be constantly reseeded in
wavenumbers outside of the cone described by \equ{eq:k0cone} through
wavenumber couplings/scattering associated with the
$\vec{u}\times\vB$ induction term, and this effect facilitates 
the dynamo.

\begin{figure}
\centering\includegraphics[width=\textwidth]{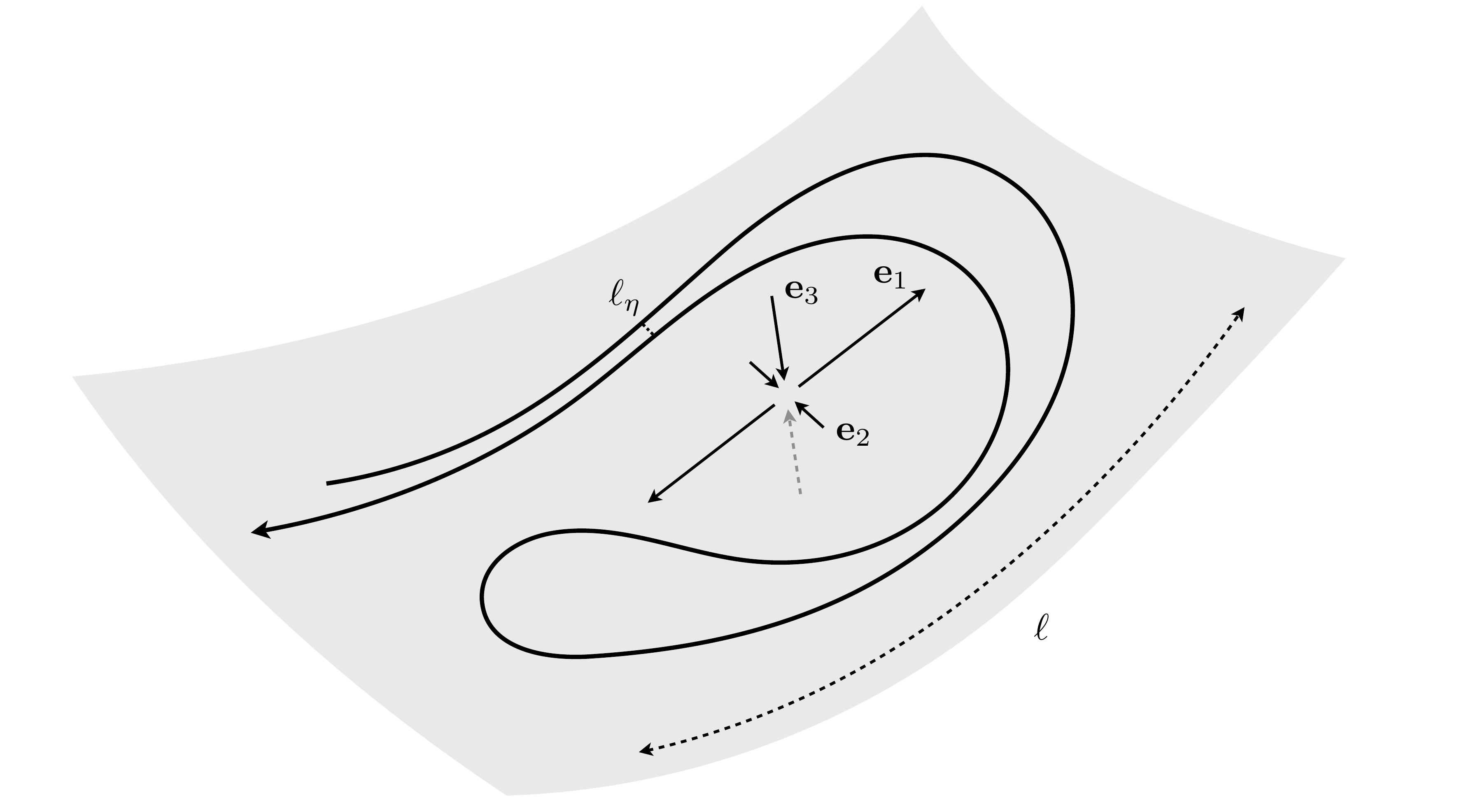}
\caption{Organisation of the magnetic field in folds perpendicular 
  to the local compressive direction $\vec{e}_3$ of a shearing velocity
  field ($c_2<0$ case). The typical flow scale over which the folds
  develop is denoted by $\ell$ here.\label{figfoldneutral}}
\end{figure}

Overall, what makes this dynamo possible in 3D but
not in 2D is the existence of an (almost) ``neutral'' direction
$\vec{e}_2$ in 3D. In 2D, $c_1+c_2=0$ and the field must be squeezed as much
along $\vec{e}_2$ as it is stretched along $\vec{e}_1$. In that case,
decays ensues according to the first twiddle inequality in \equ{eq:Emevol}.
In 3D, on the other hand, this exact constraint disappears and
some particular field configurations can survive the competition
between stretching and diffusion. More precisely, the surviving fields
are organised into folds in $(\vec{e}_1,\vec{e}_2)$ planes
perpendicular to the most compressive direction $\vec{e}_3$, with
reversals occurring along $\vec{e}_2$ at the resistive scale
$\ell_\eta$. This is illustrated in \fig{figfoldneutral}.

\subsection{Magnetic Prandtl number dependence of small-scale dynamos}
\subsubsection{Small-scale dynamo fields at $Pm>1$\label{largePmsection}}
A linear shear flow has a spatially uniform gradient and is therefore
the ultimate example of a large-scale shear flow. The magnetic mode
that results from this kind of dynamo, on the other hand, has typical
reversals at the resistive scale $\ell_\eta$, which of course becomes 
very small at large $Rm$. The problem described above is therefore
implicitly typical of the large-$Pm$ MHD regime introduced in
\sect{pmlandscape}. The fastest shearing eddies at large $Pm$ in
Kolmogorov turbulence are spatially-smooth, yet chaotic viscous
eddies, and take on the role of the random linear shear flow in the
Zel'dovich model. Interestingly, because this dynamo only requires a
smooth, chaotic flow to work, there should be no problem with exciting 
it down to $Re=O(1)$ (random Stokes flow). On the other
hand, $Rm$ must be large enough for stretching to win over
diffusion. There is therefore always a minimal requirement to resolve
resistive-scale reversals in numerical simulations (typically the 64
Fourier modes per spatial dimension of the \citeauthor{meneguzzi81}
simulation). 

\begin{figure}
  {\centering \hbox{\includegraphics[width=0.475\textwidth]{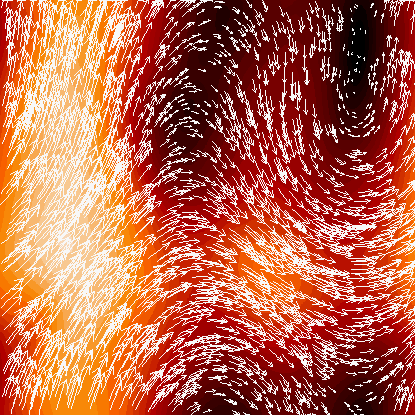}\hspace{0.05\textwidth}\includegraphics[width=0.475\textwidth]{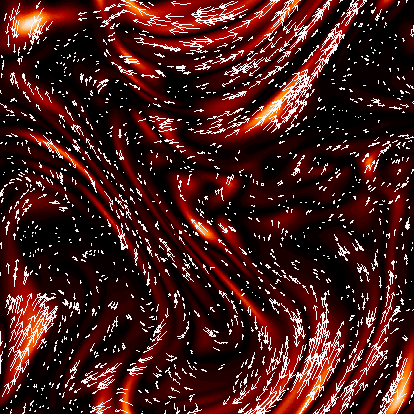}}}
\caption{Left: 2D snapshot of $|\vec{u}|$ in a 3D simulation
  of non-helical, homogeneous, isotropic smooth random flow
  forced at the box scale $\ell_0$ for $Re=1$
  ($\ell_\nu=\ell_0$). Right: snapshot of the strength
  $|\vB|$ of the kinematic dynamo magnetic field 
  generated by this flow for $Rm=Pm=1250$, and corresponding magnetic
  field directions (arrows). The field in this large-$Pm$ regime 
  has a strongly folded geometric structure: it is almost uniform along
  itself, but reverses on the very fine scale $\ell_\eta/\ell_0\sim
  Pm^{-1/2}$, $\sim 0.03$ in that example. The brighter regions 
  correspond to large field strengths
  \citep[adapted from][]{schekochihin04}.\label{figvizlargepm}}
\end{figure}

\begin{figure}
\centering\includegraphics[width=0.6\textwidth]{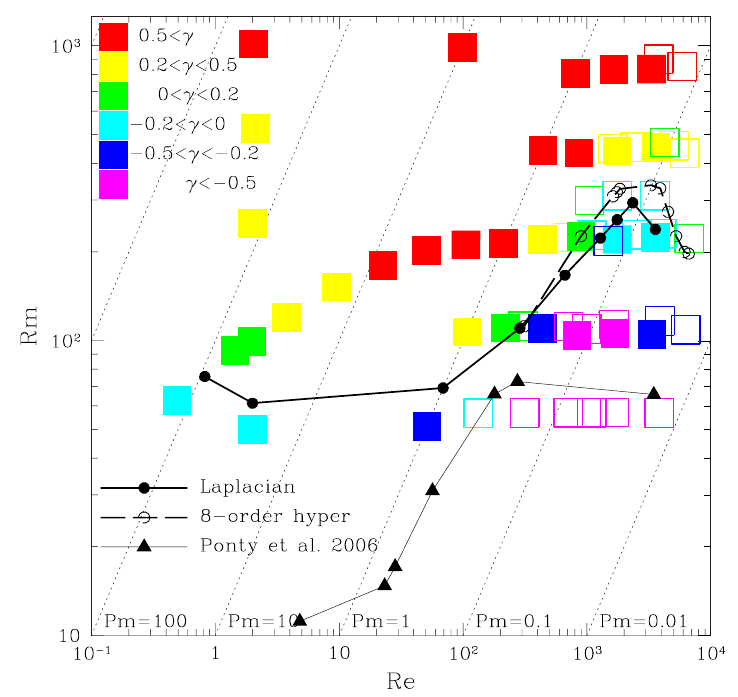}
\caption{Critical magnetic Reynolds number  $Rm_{c,\mathrm{ssd}}$
  (black solid line with full circles) and growth
  rates (colour squares) of the kinematic small-scale dynamo excited by
  non-helical, homogeneous, isotropic turbulence forced at the box
  scale, as a function of $Re$. The parameter range of the plot
  corresponds approximately to the grey box in
  \fig{figPmlandscape}. $Rm_{c,\mathrm{ssd}}$ increases by a factor 
  almost four for $Pm<1$ \citep[adapted
  from][]{schekochihin07}\label{figgrowthratePm}.}
\end{figure}

\begin{figure}
\centering{\hbox{\includegraphics[width=0.31\textwidth]{BlargePm}\hspace{0.03\textwidth}\includegraphics[width=0.31\textwidth]{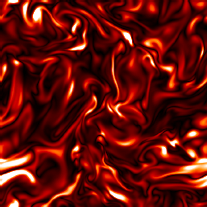}\hspace{0.03\textwidth}\includegraphics[width=0.31\textwidth]{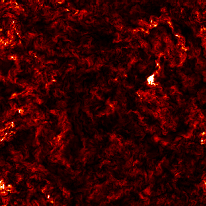}}}
\caption{2D snapshots of the strength
  $|\vB|$ of the kinematic dynamo magnetic field for 3D simulations of
  non-helical, homogeneous, isotropic turbulence forced at the box scale.
  Left: $Pm=Rm=1250$, $Re=1$. Centre: $Pm=1$, $Re=Rm=440$. Right:
  $Pm=0.07$, $Rm=430$, $Re=6200$ \red{(because this particular
    simulation uses hyperviscous dissipation only, the kinetic
    Reynolds number in this case is defined using an effective
    viscosity determined statistically from the  simulation
    data)}. Note the very different magnetic field structures
    between the $Pm=1$ and $Pm=0.07$ cases, despite $Rm$ being
    essentially the same in both simulations
    \citep[adapted from][]{schekochihin07}.\label{figvizlowpm}}
\end{figure}

Many \red{three-dimensional} direct numerical simulations (DNS) of the kind
conducted by \citeauthor{meneguzzi81} have now been performed, that
essentially confirm the Zel'dovich phenomenology and folded field
structure of the small-scale dynamo in the $Pm>1$ regime. Snapshots of
the smooth velocity field and particularly clean folded magnetic field structures
in the relatively asymptotic large-$Pm$ regime $Re=1$, $Pm=1250$, are
shown in \fig{figvizlargepm}. The Fourier spectra of these two images
(not shown) are obviously very different, which is of course
reminiscent of the \citeauthor{meneguzzi81} results. In fact, all
simulations down to $Pm=O(1)$, including the \citeauthor{meneguzzi81}
experiment, essentially produce a dynamo of the kind described above. 
\Fig{figgrowthratePm} provides a map in the $(Re,Rm)$ plane of the
dynamo growth rate $\gamma$ of the small-scale dynamo, and a plot of
the critical magnetic Reynolds number
$Rm_{c,\mathrm{ssd}}$ above which the dynamo is excited ($Rm$ here and
in \fig{figvizlargepm} and \fig{figvizlowpm} is defined as
$Rm=u_\mathrm{rms}\ell_0/(2\pi\eta)$). $Rm_{c,\mathrm{ssd}}$
is found to be almost independent of $Pm$ 
and approximately equal to 60 for $Pm>1$. As $Pm$ decreases from
large values to unity, the folded field structure gradually becomes
more intricate, but for instance we can always spot very fast field
reversals perpendicular to the field itself. This gradual change can
be seen on the two leftmost 2D snapshots of \fig{figvizlowpm}
representing  $|\vB|$ in simulations at $Pm=1250$ and $Pm=1$.

\subsubsection{Small-scale dynamo fields at $Pm<1$\label{lowPmsection}}
What about the $Pm<1$ case ? \cite{batchelor50} argued
based on an analogy between the induction equation and the vorticity
equation, that there could be no dynamo at all for $Pm<1$ (a concise
account of Batchelor's arguments on the small-scale dynamo can be
found in \cite{davidson13}, \S~18.3). As explained in
\sect{pmlandscape}, the magnetic field sees a very different, and much
more irregular velocity field in the low-$Pm$ regime, and we would
naturally expect this to have a negative impact on the dynamo. Whether
the dynamo survives in this regime remained an open and somewhat
controversial theoretical and numerical question for many years
\citep{vainshtein82,novikov83,vainshtein86,rogachevskii97,vincenzi02,haugen04,schekochihin04c,boldyrev04,schekochihin05}.

The first conclusive numerical demonstrations of kinematic dynamos at
low $Pm$ in non-helical isotropic homogeneous turbulence were only
performed a few years ago by \cite{iskakov07}. While the question of
the optimal numerical configuration to reach the low-$Pm$ dynamo
regime is not entirely settled, the main results of these
Meneguzzi-like simulations of homogeneous, isotropic turbulence are
that the critical $Rm$ of the dynamo increases significantly as $Pm$
becomes smaller than one (see \fig{figgrowthratePm}), and that the
nature of the low-$Pm$ dynamo is quite different from 
its large-$Pm$ counterpart. The rightmost snapshot of
\fig{figvizlowpm} shows for instance that the structure of the
magnetic field at $Pm<1$ is radically altered in comparison to even
the $Pm=1$ case.
The disappearance of the folded field structure is perhaps not that
surprising, given that we are completely outside of the domain of
applicability of Zel'dovich's smooth flow phenomenology for
$Pm<1$. Unfortunately, a clear physical understanding of the low-$Pm$
small-scale kinematic dynamo process comparable to that of the large
$Pm$ case is still lacking. As we will see in the next paragraph,
though, the increase in $Rm_{c,\mathrm{ssd}}$ at low $Pm$ can be
directly tied to the roughness of the velocity field at the resistive
scale, within the framework of the mathematical Kazantsev model.

Numerically, the problem with the low-$Pm$ regime is that one must
simultaneously ensure that $Rm$ is large enough to trigger the dynamo
($Rm_{c,\mathrm{ssd}}$ at low $Pm$ appears to be at least a factor two larger than at
large $Pm$ depending on how the problem is set-up), and that
$Re$ is significantly larger than $Rm$ ! In practical terms, a
resolution of $512^3$ is required to simulate such high $Re$
turbulence in pseudo-spectral numerical simulations with
explicit laplacian dissipation. Only now is this kind of MHD 
simulation becoming routine in computational fluid dynamics.
Note finally that the excitation of small-scale dynamos at both $Pm>1$
and $Pm<1$ appears to be quite independent of the hydrodynamic
turbulent-forcing mechanism, and even of the details of the
turbulent flow. For instance, results similar to \fig{figgrowthratePm}
have been obtained using hyperviscosity in DNS,  MHD shell-models
\citep{stepanov06,buchlin11} and DNS and large-eddy simulations of the
(turbulent) Taylor-Green flow \citep{ponty05}. Of importance to 
astrophysics, small-scale fluctuation dynamo action is also known to
be effective in $Pm>1$ simulations of turbulent thermal convection,
Boussinesq and stratified alike
\citep{nordlund92,cattaneo99,vogler07,graham10,moll11,bushby14}.
Low-$Pm$ turbulent convection is also widely thought to be the main
driver of small-scale solar-surface magnetic fields, although
clean, conclusive DNS simulations of a fluctuation dynamo driven by
turbulent convection at $Pm$ significantly smaller than one have still
not been conducted.

\subsection{Kinematic theory: the Kazantsev model\label{kazantsev}}
Would it not be nice if we could calculate analytically the growth
rate, energy spectrum, or probability density function of small-scale
dynamo fields for different kinds of velocity fields ? Despite all the
numerical evidence and data available on the kinematic small-scale
dynamo problem, there is still no general quantitative statistical
theory of the problem, for reasons that will soon become
clear. \cite{kazantsev67}, however, derived a solution 
to the problem under the assumption that the velocity
field is a random $\delta$-correlated-in-time (white-noise) Gaussian
variable. This particular statistical ensemble of velocity fields is
commonly referred to as the Kraichnan ensemble, after \cite{kraichnan68}
independently introduced it in his study of the structure of passive
scalars advected by turbulence.

At first glance, the Kazantsev-Kraichnan assumptions do not seem
very fitting to solve transport problems involving Navier-Stokes
turbulence, as the latter is intrinsincally non-Gaussian and has
a scale-dependent correlation time of the order of the eddy turnover
time. The Kazantsev model has however proven extremely useful to
calculate and even predict the kinematic properties of small-scale
dynamos,  and many of its results appear to be in very good quantitative
agreement with Navier-Stokes simulations. The same can be said of
the Kraichnan model for the passive scalar problem. It is also a very
elegant piece of applied mathematics that provides a nice playground
to acquaint oneself with turbulent closure problems, and offers a
different perspective on the physics of small-scale dynamos. We will
therefore go through the key points of the derivation of the Kazantsev
model for the simplest three-dimensional, incompressible, non-helical
case, and discuss some particularly important results that can be
derived from the model. More detailed derivations of the model,
including different variations in different MHD regimes, including
compressible ones, can notably be found in the work of
\cite{kulsrud92}, \cite{vincenzi02}, \cite*{schekochihin02},
\cite{boldyrev04}, and \cite*{tobias11b}, all of which have largely
inspired the following presentation.

\subsubsection{Kazantsev-Kraichnan assumptions on the velocity field}
We consider a three-dimensional, statistically steady and
homogeneous fluctuating incompressible velocity field with two-point,
two-time correlation function
\begin{equation}
  \label{eq:corrbase}
  \left<u^i(\vec{x},t)u^j(\vec{x}',t')\right>=R^{ij}(\vec{x}'-\vec{x},t'-t)~.
\end{equation}
We assume that $\vec{u}$ has \textit{Gaussian statistics},
\begin{equation}
  \label{eq:pdfu}
  P\left[\vec{u}\right]=C\exp\left[-\frac{1}{2}\int \diff t\int \diff t'\int \diff^3\vec{x}\int \diff^3\vec{x}'D_{ij}(t'-t,\vec{x}'-\vec{x})u^i(\vec{x},t)u^j(\vec{x}',t')\right]~,
\end{equation}
where $C$ is a normalisation factor and the covariance matrix $D_{ij}$
is related to $R^{ij}$ by
\begin{equation}
  \label{eq:covar}
\int \diff\tau\int \diff^3\vec{y} D_{ik}(t'-\tau,\vec{x}'-\vec{y})R^{kj}(\tau-t,\vec{y}-\vec{x})=\delta^j_i\delta(t'-t)\delta(\vec{x}'-\vec{x})~.
\end{equation}
We further assume that $\vec{u}$ is \textit{$\delta$-correlated in time}, 
\begin{equation}
\label{eq:deltacorrelated}
\left<u^i(\vec{x},t)u^j(\vec{x}',t')\right>=R^{ij}(\vec{x}'-\vec{x},t'-t)=\kappa^{ij}(\vec{r})\delta(t'-t)~,
\end{equation}
where $\vec{r}=\vec{x'}-\vec{x}$ is the spatial correlation vector.
We restrict the calculation to the isotropic, non-helical case for the time
being (the helical case is also interesting in the context of
large-scale dynamo theory and will be discussed in
\sect{kazantsevhelical}). In the absence of 
a particular axis of symmetry in the system and of helicity,
we are only allowed to use  $\delta^{ij}$ and $\vec{r}$ to construct
$\kappa^{ij}(\vec{r})$\footnote{If the turbulence is made helical (but remains isotropic),
\equ{eq:kappaij} must be supplemented by an extra term proportional to the
fully anti-symmetric Levi-Cevita tensor $\varepsilon^{ijk}$ (see
\sect{kazantsevhelical}).}, and the most general expression that we can
form is
\begin{equation}
  \label{eq:kappaij}
  \kappa^{ij}(\vec{r})=\kappa_N(r)\left(\delta^{ij}-\frac{r^ir^j}{r^2}\right)+\kappa_L(r)\f{r^ir^j}{r^2}~.
\end{equation}
where $r=|\vec{r}|$ and $\kappa_N$ and $\kappa_L$ are the tangential
and longitudinal velocity correlation functions. For an
incompressible/solenoidal vector field, we have
\begin{equation}
  \label{eq:corrsolenoi}
  \kappa_N=\kappa_L+(r\kappa_L')/2~.
\end{equation}
\subsubsection{Equation for the magnetic field correlator}
Our goal is to derive a \textit{closed} equation for the two-point,
single-time magnetic correlation function \red{(or, equivalently, for
the magnetic energy spectrum, as the two are related by a Fourier transform)}
\begin{equation}
  \label{eq:magneticcorrelation}  
 \left<B^i(\vec{x},t)B^j(\vec{x}',t)\right>=H^{ij}(\vec{r},t)~.
\end{equation}
For the same reasons as above, we can write
\begin{equation}
  \label{eq:Hij}
H^{ij}(\vec{r},t)=H_N(r,t)\left(\delta^{ij}-\frac{r^ir^j}{r^2}\right)+H_L(r,t)\frac{r^ir^j}{r^2},
\end{equation}
\begin{equation}
H_N=H_L+(rH_L')/2~.
\end{equation}
Taking the $i$-th component of the induction equation at
point $(\vec{x},t)$ and multiplying it by $\vB^j(\vec{x}',t)$,
then taking the $j$-th component at point $(\vec{x}',t)$
and multiplying it by $\vB^i(\vec{x},t)$ we find, after adding
the two results, the evolution equation for $H^{ij}$,
\begin{eqnarray}
\dpart{H^{ij}}{t} & = & 
\dpart{}{{x'^k}}\left(\left<B^i(\vec{x},t)B^k(\vec{x}',t)u^j(\vec{x'},t)\right>
-\left<B^i(\vec{x},t)B^j(\vec{x}',t)u^k(\vec{x'},t)\right>\right)\label{eq:unclosed} \\ \nonumber
& + & 
\dpart{}{x^k}\left(\left<B^k(\vec{x},t)B^j(\vec{x}',t)u^i(\vec{x},t)\right>
-\left<B^i(\vec{x},t)B^j(\vec{x}',t)u^k(\vec{x},t)\right>\right) \\ 
& + & \eta\left(\ddpart{}{x^k} + \ddpart{}{{x'^k}}\right)H^{ij}~,\nonumber
\end{eqnarray}
where $\dpartshort{}{{x'^k}}\left<\cdot \right>=-\dpartshort{}{x^{k}}\left<\cdot
\right>=\dpartshort{}{r^{k}}\left<\cdot \right>$ because of statistical
homogeneity. \Equ{eq:unclosed} is exact, but we are now faced with an important
difficulty: the time-derivative of the second-order
magnetic correlation function depends on mixed third-order correlation
functions, and we do not have explicit expressions for these correlators.
We could write down evolution equations for them too, but their
r.h.s. would then involve fourth-order correlation functions, and so
on. This is a familiar \textit{closure} problem. 

\subsubsection{Closure procedure in a nutshell\label{closure}*}
In order to make further progress, we have to find a (hopefully
physically) way to truncate the system of equations by replacing the
higher-order correlation functions with lower-order ones.
This is where the Kazantsev-Kraichnan assumptions of a random,
$\delta$-correlated-in-time Gaussian velocity field come into play.
The  assumption of Gaussian statistics implies that $n$th-order mixed
correlation functions involving $\vec{u}$ can be expressed in terms of
$(n-1)$th-order correlation functions using the Furutsu-Novikov
(Gaussian integration) formula:
\begin{equation}
  \label{eq:furutsu}
  \left<u^i(\vec{x},t)F[\vec{u}]\right>=\int \diff t''\int \diff^3\vec{x}''\left<u^i(\vec{x},t)u^l(\vec{x}'',t'')\right>\left<\frac{\delta F[\vec{u}]}{\delta u^l(\vec{x}'',t'')}\right>~,
\end{equation}
where $F[\vec{u}]$ stands for any functional of $\vec{u}$, and the
$\delta F/\delta u^i$ are functional derivatives.
We can use this formula in \equ{eq:unclosed} to replace
all the third-order moments appearing in the r.h.s. by integrals of
products of second-order moments. To illustrate how the closure
procedure proceeds, let us isolate just one of these terms,
\begin{equation}
\label{eq:furutsuapplied}
\left<u^i(\vec{x},t)B^k(\vec{x},t)B^j(\vec{x}',t)\right>=\int_0^t
\diff t''\!\!\int \diff ^3\vec{x}''\left<u^i(\vec{x},t)u^l(\vec{x}'',t'')\right>\left<\frac{\delta \left[B^k(\vec{x},t)B^j(\vec{x}',t)\right]}{\delta u^l(\vec{x}'',t'')}\right>.
\end{equation}
Applying the Gaussian integration formula is a critical first step,
but more work is needed. In particular, \equ{eq:furutsuapplied}
involves a time-integral encapsulating the effects of flow memory. For
a generic turbulent flow for which the correlation time is not small compared 
the relevant dynamical timescales of the problem, the problem is
non-perturbative and there is no known method to calculate such an
integral exactly. However, as a first step we could still assume that it
is small, and perform the integral perturbatively (the expansion parameter
will be the Strouhal number). The Kazantsev-Kraichnan assumption of
zero correlation-time corresponds to the lowest-order
calculation. Using \equ{eq:deltacorrelated} in \equ{eq:furutsuapplied}
removes the time-integral and leaves us with the task of calculating
the equal-time functional derivative $\left<\delta
  \left[B^k(\vec{x},t)B^j(\vec{x}',t)\right]/\delta
  u^l(\vec{x}'',t)\right>$. This expression can be explicitly
calculated using the expression of the formal solution of the
induction equation,
\begin{equation}
  \label{eq:formalinduc}
  B^k(\vec{x},t)=\int^t \diff t' \left[B^m(\vec{x},t')\dpart{u^k(\vec{x},t')}{x^m}-u^m(\vec{x},t')\dpart{B^k(\vec{x},t')}{x^m}+\eta\Delta B^k(\vec{x},t')\right]~.
\end{equation}
Functionally differentiating this equation (and that for
$B^j(\vec{x}',t)$) with respect to $\delta u^l(\vec{x''},t)$ introduces
$\delta(\vec{x}'-\vec{x}'')$ and $\delta(\vec{x}-\vec{x}'')$, which 
makes the space-integral in \equ{eq:furutsuapplied} trivial and
completes the closure procedure.

The end result of the full calculation outlined above are expressions
of all the mixed third-order correlation functions in terms of a
combination of products of the two-point second-order correlation
function of the magnetic field with the (prescribed) second-order
correlation function of the velocity field (and their spatial
derivatives). 

\subsubsection{Closed equation for the magnetic correlator}
From there, it can be shown easily using appropriate projection
operators and isotropy that the original, complicated unclosed
\equ{eq:unclosed} reduces to the much simpler closed scalar equation
for $H_L(r,t)$,
\begin{equation}
  \label{eq:HLequation}
\dpart{H_L}{t}=\kappa H_L''+\left(\frac{4}{r}\kappa+\kappa'\right)H_L'+\left(\kappa''+\frac{4}{r}\kappa'\right)H_L~,  
\end{equation}
where 
\begin{equation}
\label{eq:kappaequation}
\kappa(r)=2\eta+\kappa_L(0)-\kappa_L(r)
\end{equation} 
can be interpreted as (twice) the ``turbulent diffusivity''. If we now
perform the change of variables
\begin{equation}
  H_L(r,t)=\f{\psi(r,t)}{r^{2}\kappa(r)^{1/2}}~,
\end{equation}
we find that
\equ{eq:HLequation} reduces to a Schr\"odinger equation with imaginary
time
\begin{equation}
  \label{eq:schrodinger}
  \dpart{\psi}{t}=\kappa(r)\psi''-V(r)\psi~,
\end{equation}
which describes the evolution of the wave function of a quantum
particle of variable mass 
\begin{equation}
m(r)=\f{1}{2\kappa(r)}
\end{equation}
 in the potential
\begin{equation}
V(r)=\frac{2}{r^2}\kappa(r)-\frac{1}{2}\kappa''(r)-\frac{2}{r}\kappa'(r)-\f{\kappa'(r)^2}{4\kappa(r)}~.
\end{equation}
\subsubsection{Solutions}
We can then look for solutions of \equ{eq:schrodinger} of the form
\begin{equation}
  \label{eq:psischrod}
\psi=\psi_E(r)e^{-Et}~.  
\end{equation}
Keeping in mind that $\psi$ stands for the magnetic correlation
function, we see that exponentially growing dynamo modes correspond to
discrete $E<0$ bound states. The existence of such modes
depends on the shape of the Kazantsev potential, which is entirely
determined by the statistical properties of the velocity field
encapsulated in the function $\kappa(r)$. The variational result for
$E$ is
\begin{equation}
\label{eq:variational}
E=\f{\displaystyle{\int 2 mV{\psi_E}^2\diff r+\int {\psi_E'}^2\diff r}}{\displaystyle{\int
  2m{\psi_E}^2 \diff r}}~.
\end{equation}
If we are just interested in the question of whether a dynamo is
possible, we can equivalently solve the equation
\begin{equation}
  \label{eq:psiE}
\psi''_E+\left[E-V_\mathrm{eff}(r)\right]\psi_E=0~.
\end{equation}
where
\begin{equation}
  \label{eq:Veff}
  V_\mathrm{eff}(r)=\f{V(r)}{\kappa(r)}~.
\end{equation}
The ground state describes the fastest growing mode, and
therefore the long-time asymptotics in the kinematic regime of the
dynamo. 

\subsubsection{Different regimes\label{kazantsevregimes}}
\begin{figure}
\centering\includegraphics[width=\textwidth]{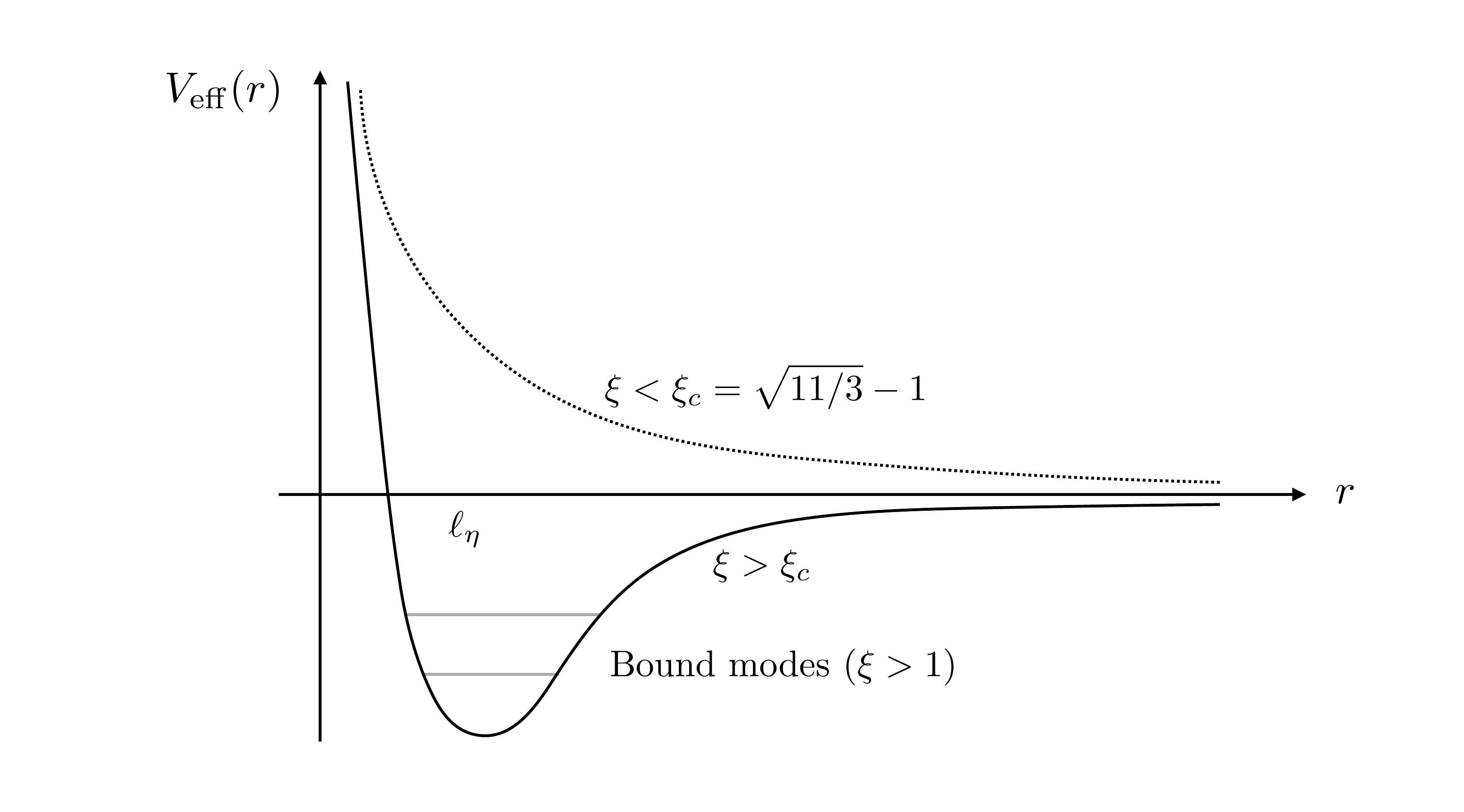}
\caption{Kazantsev potential as a function of $r$ for different
  roughness exponents $\xi$. An attractive potential forms at $\xi_c$.
  Bound (growing) dynamo modes require
  $\xi>1$. \label{figkazantsevpotential}}
\end{figure}

After all this hard work, we are now almost in a position to answer
much more specific questions. For instance, we would like to predict
whether dynamo action is possible in a smooth flow (something we
already know from \sect{zeldo}), and in a turbulent flow with
Kolmogorov scalings (something the \citeauthor{zeldovich84} model does
not tell us). Let us assume that the velocity field has power-law
scaling,
\begin{equation}
\kappa_L(0)-\kappa_L(r)\sim r^{\xi}~,
\end{equation}
where $\xi$ is called the roughness exponent. Introducing the
resistive scale $\ell_\eta=(2\eta)^{1/\xi}$, the effective Kazantsev
potential as a function of $\xi$ is 
\begin{eqnarray}
V_\mathrm{eff}(r) & = & \f{2}{r^2}\quad \mathrm{for}\quad r\ll \ell_\eta~,\\
& = & \f{2-3\,\xi/2-3\,\xi^2/4}{r^2}\quad\mathrm{for}\quad r\gg \ell_\eta~.
\end{eqnarray}
The overall shape of the potential is illustrated in \fig{figkazantsevpotential}
for different values of $\xi$. The potential becomes attractive at
$\xi=\xi_c=\sqrt{11/3}-1$, but growing bound modes only exist for $\xi>1$.

In order to find out which kind of flows are dynamos according to 
the Kazantsev model, we have to make a small handwaving
argument in order to relate the scaling properties of flows with
finite correlation times to those of Kazantsev-Kraichnan velocity fields. Recalling that
$\left<u^i(\vec{x},t)u^j(\vec{x}',t')\right>=\kappa^{ij}(\vec{r})\delta(t'-t)$,
and that a $\delta$ function is dimensionally the inverse of a time,
we write
\begin{equation}
  \label{eq:kappaturb}
  \kappa(r)\sim \delta u(r)^2\tau(r)\sim r\delta u(r)~,
\end{equation}
where $\delta u(r)$ is the typical velocity difference
between two points separated by $r$, and we have assumed that 
the relevant timescale in the dimensional analysis is the
scale-dependent eddy turnover time $\tau(r)\sim r/\delta u(r)$
(this twiddle-algebra analysis also clarifies the association of
$\kappa$ with a turbulent diffusivity).
Using \equ{eq:kappaturb}, we find that  $\xi=2$
for a smooth flow whose velocity increments scale as $\delta u\sim r$ 
(the linear shears of the Zel'dovich phenomenology). Qualitatively, it
is tempting to associate this $\xi=2$ case with a large-$Pm$ regime because all
the magnetic field sees at large $Pm$ is a large-scale, smooth viscous flow.
On the other hand, for a flow with Kolmogorov scaling, $\delta u \sim
r^{1/3}$ and $\xi=1+1/3=4/3$. This case would instead correspond to a low-$Pm$
regime in which the magnetic field sits in the middle of the inertial range.

According to the Kazantsev model, \red{a necessary condition for
small-scale dynamo action to be possible is that $\xi>1$. This
condition is satisfied} in both smooth flows ($\xi=2$,
$Pm\gg 1$ regime) and rough flows, including Kolmogorov-like
turbulence ($\xi=4/3$, $Pm\ll 1$ regime). The study of the
low-$Pm$-like case was not part of the
original \cite{kazantsev67} article, and only appeared in later work by
various authors (\cite{vainshtein82,vainshtein86,vincenzi02,boldyrev04},
  see also \cite{eyink10,eyink10b} for a Lagrangian perspective on the problem).
In this respect, we should also mention the work of \cite{kraichnan67}
contemporary to that of
\citeauthor{kazantsev67}. \citeauthor{kraichnan67} arrived
at a closed equation for the evolution of the magnetic-energy
spectrum similar to that derived by Kazantsev using a different
closure procedure, and found exponential growth of magnetic
energy by solving this equation numerically as an initial value
problem for a prescribed Kolmogorov velocity spectrum.

\subsubsection{Critical $Rm$}
\begin{figure}
\centering\includegraphics[width=\linewidth]{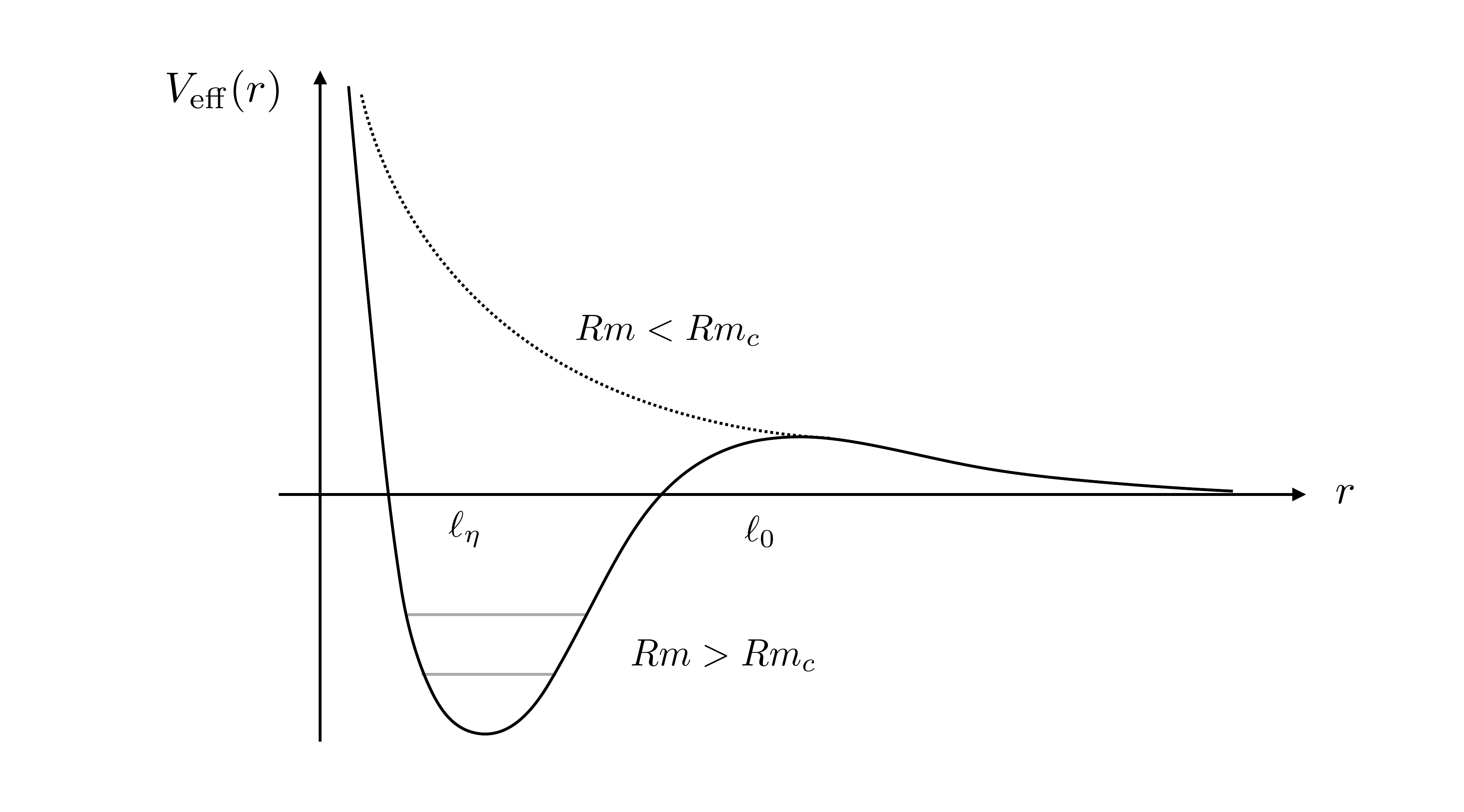}
\caption{Critical $Rm$ effect in the Kazantsev model. 
  The existence of an attractive potential (growing dynamo modes) requires
  a large-enough scale separation between the integral scale $\ell_0$
  and the resistive scale $\ell_\eta$. \label{kazantsevpotentialintegral}}
\end{figure}

Another interesting result that can be derived from the Kazantsev
model is the existence of a critical $Rm$ \citep{ruzmaikin81}.
A finite scale separation $\ell_0/\ell_\eta$ is introduced in the
problem by assuming the existence of an integral scale $\ell_0$ beyond
which the velocity field decorrelates. The shape of the 
Kazantsev potential in that case is shown in
\fig{kazantsevpotentialintegral}, and the asymptotic form is
\begin{eqnarray}
V_\mathrm{eff}(r) &= & \f{2}{r^2}\quad\mathrm{for}\quad r\ll \ell_\eta,\\
& = & \f{2-3\,\xi/2-3\,\xi^2/4}{r^2}\quad\mathrm{for}\quad
\ell_\eta\ll r\ll \ell_0~,\\
& =& \f{2}{r^2}\quad\mathrm{for}\quad \ell_0\ll r~.
\end{eqnarray}
The potential is now repulsive at both small and large scales.
As a result, the existence of an attractive range of scales and bound
modes now depends on the existence of a large-enough scale separation
$\ell_0/\ell_\eta$, or large enough $Rm$. If the scale separation is
too low, magnetic diffusion, whose stabilizing effects translate as a
repulsive potential at small-scales, always wins over stretching, and
no dynamo action is possible. This argument is independent of $\xi$, and
therefore predicts the existence of a critical $Rm$ in both large- and
low-$Pm$ regimes.

\subsubsection{\red{Selected results in the large-$Pm$ regime*}\label{largepmselection}}
The Kazantsev model can be used to derive a variety of quantitative
results when the magnetic field is excited at scales much smaller than
the spatial correlation scale of the flow. One particularly
interesting such case is the large-$Pm$ regime encountered earlier, for
which the dynamo is driven by a smooth viscous-scale flow. This is
called the Batchelor regime, in reference to
\citeauthor*{batchelor59}'s \citeyear{batchelor59} work 
on the transport of passive scalars. Note that this problem is not of mere
applied mathematics interest, and has notably been at the core of important
developments on the theory of astrophysical galactic dynamos
\citep{kulsrud92}.

In the Batchelor regime, the velocity field
correlator~(\ref{eq:kappaij})-(\ref{eq:corrsolenoi}) can be expanded
in Taylor series by writing $\kappa_L=\kappa_0-\kappa_2r^2/4+\cdots$,
$\kappa_N=\kappa_0-\kappa_2r^2/2+\cdots$, resulting in
\begin{equation}
  \label{eq:kappaexpand}
  \kappa^{ij}(r)=\kappa_0\delta^{ij}-{\kappa_2}\f{r^2}{2}\left(\delta^{ij}-\f{1}{2}\f{r^i r^j}{r^2}\right)+\cdots~,
\end{equation}

A technical digression is in order: while it is possible to
make quantitative calculations in physical space
starting from \equ{eq:kappaexpand}, the easiest and most popular route
to calculate magnetic eigenfunctions and the magnetic energy spectrum 
goes through Fourier-space. This was in fact the primary method used
by \cite{kazantsev67}. We will take that route in the next 
paragraphs, and find that the results can be expressed in terms of
the coefficients of the Taylor expansion~(\ref{eq:kappaexpand}). The
correspondence between the two descriptions can be found in
\cite{schekochihin02}, App. A. 
\smallskip

\paragraph{\textit{Fokker-Planck equation.}}
At scales much smaller than the viscous scale (small-scale
approximation), the spectral equivalent of the Schr\"odinger
\equ{eq:schrodinger} for the magnetic correlation function 
is a Fokker-Planck equation for the one-dimensional magnetic
energy spectrum 
\begin{equation}
M(k,t)= \f{1}{2}\int k^2 |\hat{\vB}_{\vec{k}}(t)|^2 \diff\Omega_\vec{k}~,
\end{equation}
where $\hat{\vB}_{\vec{k}}(t)$ denotes the Fourier transform in space
of $\vB$ (once again not to be confused in this context with the unit
vector in the direction of $\vec{B}$). Namely,
\begin{equation}
  \label{eq:fokkermagnetic}
  \dpart{M}{t}=\f{\gamma}{5}\left(k^2\ddpart{M}{k}-2k\dpart{M}{k}+6M\right)-2\eta k^2 M~,
\end{equation}
with
\begin{equation}
\label{eq:gammaideal}
\gamma=\f{5}{4}\kappa_2=\f{5}{2}|\kappa_L''(0)|\sim \delta u_\nu/\ell_\nu~,
\end{equation}
the typical growth rate of the magnetic energy, of the order
of the viscous shearing rate represented by the $\kappa_2$ 
parameter in the smooth flow expansion~(\ref{eq:kappaexpand}). 
These results are only quantitatively valid in the incompressible 3D case. 

\begin{figure}
\centering\includegraphics[width=\linewidth]{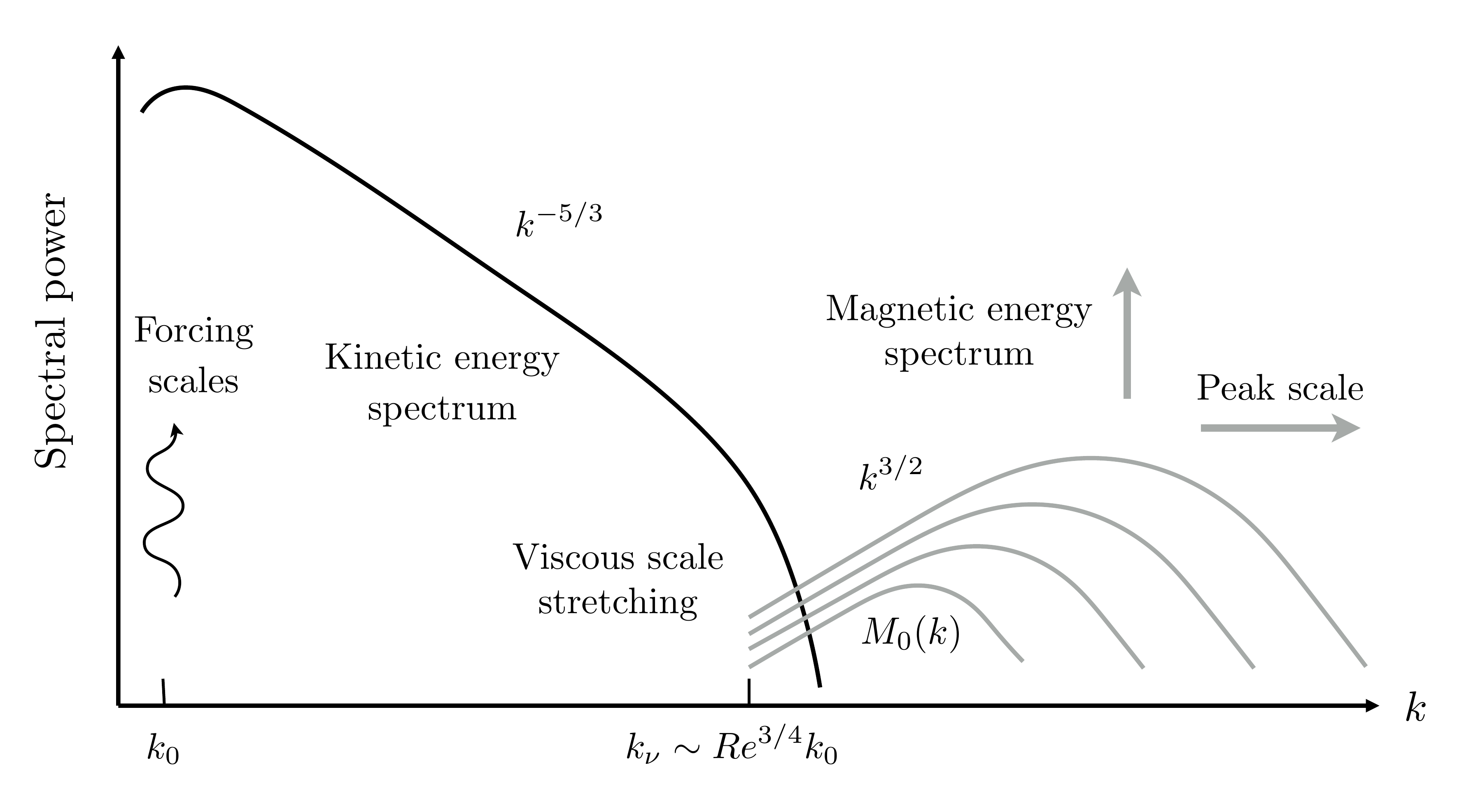}
  \caption{Evolution of the magnetic energy spectrum in the kinematic,
    diffusion-free large-$Pm$ regime, starting from an initial magnetic
    spectrum $M_0(k)$. The magnetic energy in each wavenumber increases,
    and so does the the peak wavenumber as the spectrum
    spreads.\label{diffusionfreeregime}}
\end{figure}

\paragraph{\textit{Kazantsev spectrum and growth rate.}}
\Equ{eq:fokkermagnetic} has a diffusion term in wavenumber space.
If we start with magnetic perturbations at scales much larger than the
resistive scale, the magnetic spectrum will both grow in time and spread
in wavenumber space until it hits the resistive scale. This early
evolution during which all $k\ll k_\eta$ is called the
diffusion-free regime and is illustrated in \fig{diffusionfreeregime}.
Assuming that we initially excite a spectrum of magnetic modes $M_0(k)$,
the evolution of the spectrum can be computed exactly
using Green's function of \equ{eq:fokkermagnetic} with negligible
magnetic diffusion,
\begin{equation}
  \label{eq:diffusionfreeM}
  M(k,t)=e^{3/4\gamma t}\int_0^{\infty}\f{\diff k'}{k'}M_0(k')\left(\f{k}{k'}\right)^{3/2}\sqrt{\f{5}{4\pi\gamma t}}\exp\left[-\f{5\,\ln^2\left(k/k'\right)}{4\gamma t}\right]~.
\end{equation}
On the large-scale side, the magnetic energy spectrum grows
as $k^{3/2}$. This result was first derived in \citeauthor{kazantsev67}'s
(\citeyear{kazantsev67}) paper and is therefore called the Kazantsev
spectrum. Interestingly, the energy in each wavenumber grows at the rate
$3\gamma/4$, but the total energy integrated over wavenumber space 
grows at the rate $2\gamma$ because the number of excited Fourier
modes also grows in time. 

\begin{figure}
\centering\includegraphics[width=\linewidth]{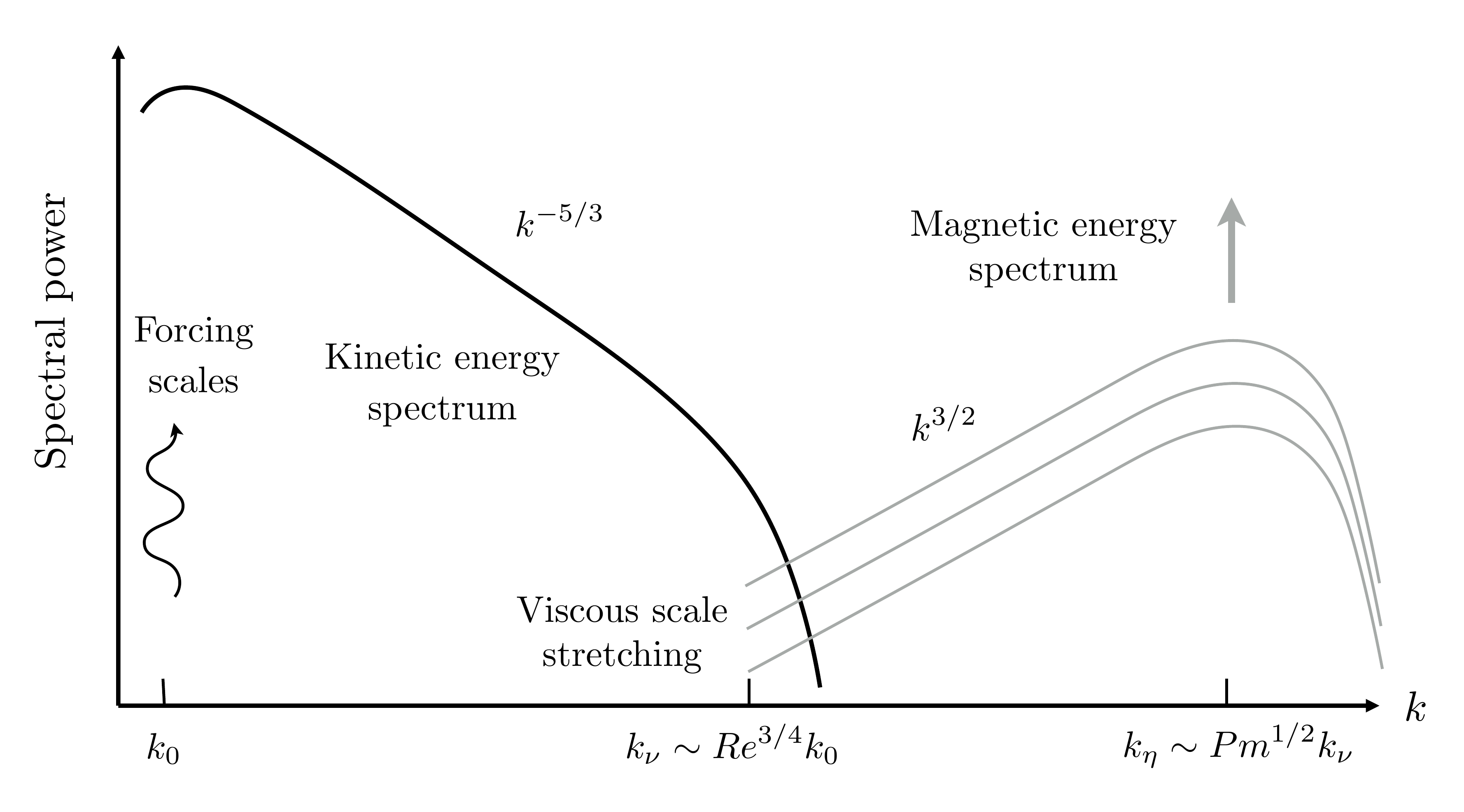}
  \caption{Evolution of the magnetic energy spectrum in the kinematic,
    diffusive large-$Pm$ regime. The shape of the magnetic
    spectrum is now fixed in time and peaks at the resistive scale
    $\ell_\eta$, but the magnetic energy continues to grow
    exponentially.\label{resistiveregime}}
\end{figure}

At the end of the diffusion-free regime, the spectrum hits $k\sim
k_\eta$ and we enter the diffusive regime, for which the long-time
asymptotic form is
\begin{equation}
  \label{eq:resistiveM}
M(k,t)\propto \left(\f{k}{k_\eta}\right)^{3/2}K_0\left(\f{k}{k_\eta}\right)e^{\lambda\gamma t}~,
\end{equation}
where $K_0$ is a Macdonald function, 
$k_\eta=\sqrt{\gamma/(10\eta)}\sim Pm^{1/2}k_\nu$, \red{and $\lambda$
here is a non-dimensional growth-rate prefactor}. The spectrum at
large scales is still $k^{3/2}$, but now peaks at the resistive
scale and falls off exponentially at even smaller scales.
While the exact value of $\lambda$ depends weakly on  $Pm$ and on 
the boundary condition imposed on the viscous side at low $k$, the
asymptotic total energy growth rate is now essentially $3\gamma/4$,
as the number of excited modes remains constant. The evolution
of the spectrum in this regime is illustrated in
\fig{resistiveregime}.

\red{Interestingly, this general spectral shape is
reminiscent of the numerical results shown in \fig{figMFP81small}}. In
fact, and despite all the assumptions behind the Kazantsev derivation,
all small-scale dynamo
simulations at $Pm>1$ exhibit a spectrum \red{with a positive slope at
wavenumbers larger than but close to $k_\nu$. Observing a clean
$k^{3/2}$ scaling, though, requires to go to fairly large
$Pm$ to reach a proper scale separation between $k_\nu$ and $k_\eta$.}

\paragraph{\textit{Magnetic probability density function and moments.}\label{pdfformalism}}
Probability density functions (p.d.f.) and statistical moments of
different orders can diagnose subtle dynamical features in both
experiments and numerical simulations and as such provide important
statistical tools to tackle many turbulence problems. To complete our
overview of the Kazantsev theory, we will therefore finally outline
a mathematical procedure to calculate the magnetic p.d.f. and
moments in the Kazantsev description in the diffusion-free regime.
The spatial dependence of the field is ignored for simplicity in this
particular derivation: this removes turbulent mixing effects from the
model, but not inductive ones. The key point of the derivation \citep[see,
e.g.,][]{schekochihin01,boldyrev01} is to introduce the characteristic
function,
\begin{equation}
  \label{eq:characteristic}
  Z(\vec{\mu},t)=\left<\exp\left[i\mu_iB^i(t)\right]\right>=\left<\tilde{Z}\right>
\end{equation}
where  $\tilde{Z}=\exp\left[i\mu_iB^i(t)\right]$ and  the
bracket average is over all the realisations of the velocity field.
This function is interesting because it is the Fourier transform in
$\vB$ of the p.d.f. of $\vB$, 
\begin{equation}
  \label{eq:fourierpdf}
  Z(\vec{\mu},t)=\int P\left[\vB\right]\exp\left[i\mu_i B^i(t)\right]\diff^3\vB~.
\end{equation}
Using the simplified induction equation
\begin{equation}
  \dpart{B^i}{t}=\sigma^i_kB^k~,
\end{equation}
where $\sigma^i_k=\dpartshort{u^i}{x^k}$ is the velocity strain
tensor, we obtain an evolution equation for $Z$,
\begin{equation}
  \label{eq:evolZ}
  \dpart{Z}{t}=\mu_i\dpart{}{\mu_k}\left<\sigma^i_k\tilde{Z}\right>~.
\end{equation}
This equation can be closed in the Kazantsev model using the same
Gaussian integration trick as in \sect{closure}. The result is
\begin{equation}
  \label{eq:Zclosed}
  \dpart{Z}{t}=\f{\kappa_2}{2}\,T^{ij}_{kl}\, \mu_i\dpart{}{\mu_k}\mu_j\dpart{}{\mu_l} Z~,
\end{equation}
where
\begin{equation}
  \label{eq:straincorrelator}
  T^{ij}_{kl}=-\f{1}{\kappa_2}\frac{\partial^2 \kappa^{ij}}{\partial r^k\partial r^l}=\delta^{ij}\delta_{kl}-\f{1}{4}\left(\delta^i_k\delta^j_l+\delta^i_l\delta^j_k\right)
\end{equation}
is the strain correlator for a 3D, incompressible, \red{isotropic} velocity
field. Performing an inverse Fourier transform from $\mu$ to $\vB$
variables and using the transformation
\begin{equation}
\mu_i\dpart{}{\mu_k}\left(\right)\rightarrow\dpart{}{B^i}\left[B^k\left(\right)\right]~,  
\end{equation}
we obtain a Fokker-Planck equation for the p.d.f.
\begin{equation}
  \label{eq:fokkerpdf}
\dpart{}{t}P\left[\vB\right]=\f{\kappa_2}{2}T^{ij}_{kl}B^k\dpart{}{B^i}{B^l}\dpart{}{B^j} P\left[\vB\right].
\end{equation}
\red{Thanks to the isotropy assumption}, this equation simplifies further as a 1D
diffusion equation with a drift for the p.d.f. $P_B[B](t)$ of the
field strength $B$,
\begin{equation}
  \label{eq:fokkerpdfiso}
  \dpart{}{t}P_B\left[B\right]=\f{\kappa_2}{4}\f{1}{B^2}\dpart{}{B}B^4\dpart{}{B}
P_B\left[B\right]~,
\end{equation}
which has the log-normal solution
\begin{equation}
  \label{eq:lognormal}
  P_B[B](t)=\f{1}{\sqrt{\pi\kappa_2 t}}\int_0^\infty \f{\diff B'}{B'}
  P_B\left[B'\right](t=0)\exp\left(-\f{\left[\ln(B/B')+(3/4)\kappa_2
        t\right]^2}{\kappa_2 t}\right)~,
\end{equation}
i.e. the statistics of the logarithm of the field strength
are Gaussian. Therefore, the magnetic field structure is strongly
intermittent, despite the fact that the velocity field itself is Gaussian.
This is actually expected from the central limit theorem on a
general basis, not just in the Kazantsev formalism, as
the induction equation is linear in $\vB$ and involves the
multiplicative random variable $\grad{\vec{u}}$ on the r.h.s. 
The magnetic moments of different order $n$, defined here as 
\begin{equation}
  \label{eq:magneticmoments}
\left<B^n(t)\right>=4\pi\int_0^\infty \diff B B^{2+n} P_B\left[B\right]~,
\end{equation}
grow as
\begin{equation}
  \label{eq:magneticmomentskazantsev}
  \left<B^n(t)\right>\propto \exp\left[\f{1}{4}n(n+3)\kappa_2 t\right]~.
\end{equation}
There is currently no similar general result for the
magnetic-field p.d.f. in the diffusive regime, although expressions
for the $n> 2$ moments of the field have been derived in this regime
by \cite{chertkov99} using a different mathematical technique. Their
results show that the structure of the field remains strongly
intermittent after the folds' reversal scale hits the
magnetic-diffusion scale.
\red{
\subsubsection{Miscellaneous observations}
\paragraph{\textit{Instability threshold subtleties.}}
\Equ{eq:magneticmomentskazantsev} indicates that magnetic moments of
different order have different ideal growth rates. Accordingly,
we would find that each moment becomes unstable at a different
$Rm$ in the resistive case. This behaviour can be attributed to the
multiplicative stochastic nature of the inductive term in the
linear induction equation. Which moment, then, provides the correct
prediction for the overall dynamo threshold ? It has been argued that
a proper instability threshold in this context is only well-defined
after taking into account nonlinear saturation terms that
tend to suppress the large-deviations events responsible for the field
intermittency. The dynamo onset of the full nonlinear problem then appears
to be given by the linear threshold derived from the statistical
average of the logarithm of the magnetic field, not of the magnetic
energy (second-order moment) as calculated in the Kazantsev model
\citep{sesha18}.}

\paragraph{\textit{Connection with finite-correlation-time turbulence.}}
Statistics of the magnetic field in the kinematic,
large-$Pm$ regime have been extensively studied in numerical
simulations of the MHD equations. While the log-normal shape of the
magnetic p.d.f. appears to be a robust feature of the simulations,
intermittency corrections imprinted in magnetic moments,
such as predicted by \equ{eq:magneticmomentskazantsev}, cannot 
be easily diagnosed due to subtle finite-size effects. An
in-depth-discussion of this thorny aspect of the
problem can be found in \cite{schekochihin04}. 
More generally, and despite its crude closure assumptions, the
predictions of the Kazantsev model (dynamo growth rate and statistics) are
in good overall agreement with the results of DNS at $Pm>1$. Non-zero
correlation-time effects appear to be of relatively minor qualitative
importance (see e.g. \cite{vainshtein80,chandran97,bhat15} for theoretical
arguments, and \cite{mason11} for a detailed numerical comparison with
turbulence DNS), and intermittent statistics of the fluctuation dynamo
fields are predicted by the model despite the gaussianity of the velocity
field. This suggests that the kinematic small-scale fluctuation dynamo
process is relatively insensitive to the details of the flow driving 
the dynamo, as long as the former is chaotic.

\subsection{Dynamical theory\label{NLss}}
We have found that a small-scale magnetic field seed can grow
exponentially on fast eddy-turnover timescales in a generic
chaotic/turbulent flow in the
kinematic dynamo regime. The dynamo field, however small initially,
will therefore inevitably become ``sufficiently large'' (soon to be
discussed) after a few turbulent turnover times for the
back-reaction of Lorentz force on the flow to become dynamically
significant. Understanding the evolution of small-scale
dynamo fields in this nonlinear dynamical regime is at least as
important as understanding the kinematic regime, not least because all
observed magnetic fields in the Universe are thought to be in a nonlinear
dynamical state involving a strong small-scale component. Definitive
quantitative theoretical results  remain scarce though, in spite of
the guidance now provided by high-resolution numerical simulations. 
We will therefore mostly discuss this problem from a qualitative,
phenomenological perspective.

\subsubsection{General phenomenology}
The first and most natural question to ask about saturation 
is arguably that of the dynamo efficiency: how much magnetic energy
should we expect relative to kinetic energy in the statistically
steady saturated state of the dynamo ? Two answers were historically
given to this question, one by \cite{batchelor50}  and the other by
\cite{schluter50}. \citeauthor{batchelor50}
observed that the induction equation for the magnetic field has
the same form as the evolution equation for hydrodynamic
vorticity. As the vorticity peaks at the viscous scale in
hydrodynamic turbulence, he then argued that the
dynamo should saturate when $\left<|\vB|^2\right>\sim u_\nu^2$. 
This argument predicts that very weak fields well below equipartition
are sufficient to saturate the dynamo in Kolmogorov turbulence
($u_\nu\sim Re^{-1/4}u_0$) with $Re\gg1$,
\begin{equation}
  \label{eq:subeq}
  \left<|\vB|^2\right> \sim Re^{-1/2} \left<|\vec{u}|^2\right>~,
\end{equation}
admittedly not a very interesting astrophysical prospect !
\citeauthor{schluter50}, on the other hand, argued that the outcome of
saturation should be global equipartition between kinetic and magnetic
energy,
\begin{equation}
\label{eq:equipart}
\left<|\vB|^2\right> \sim \left<|\vec{u}|^2\right>~,
\end{equation}
as a result of a gradual build-up of scale-by-scale equipartition 
from the smallest, least energetic scales. Which one (if either) is
correct ?  Batchelor's analogy between vorticity and magnetic field 
was shown to be flawed a long time ago, and numerical simulations
exhibit much higher saturation levels than predicted by his
model. However, scale-by-scale equipartition does not appear to 
hold either, as can be seen for instance by inspecting the 
saturated spectra in \fig{figMFP81small}.

In order to get a better handle on the issue, we must have a
detailed look at how magnetic growth is quenched physically. Geometry,
not just energetics, is very important in this problem, particularly
in the large-$Pm$ case. First, not all velocity fields are created
equal for the magnetic field, as
\equ{eq:Bmag} clearly illustrates. Only certain types of motions
relative to the magnetic field can make the field change, and it is
only by quenching such motions that the magnetic field can
saturate. Second, the magnetic tension
force $\vB\cdot\grad{\,\vB}$ mediating the dynamical feedback
(in incompressible flows) is proportional to the variation of the
magnetic field along itself, i.e. to magnetic curvature. A quick look
at the folded fields structure in \fig{figvizlargepm} reveals the
geometric complexity of the problem. Magnetic curvature 
is very small along the regions of straight magnetic field,
and very large in regions where the magnetic field bends to
reverse, and therefore so is the Lorentz force back-reaction.

\begin{figure}  
\centering\includegraphics[width=0.475\textwidth]{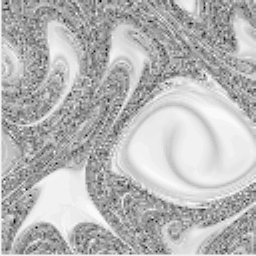}\hspace{0.04\textwidth}\includegraphics[width=0.475\textwidth]{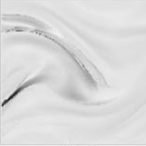}
  \caption{Spatial distribution of the finite-time Lyapunov stretching
    exponent in a dynamo simulation of the  GP
    flow. Light shades correspond to integrable regions with little or no
    exponential stretching, dark shades to chaotic regions with strong
    stretching. Left: kinematic regime map exhibiting fractal regions of chaotic
    dynamics and stability islands. Right: dynamical regime. Strongly
    chaotic regions have almost disappeared 
    \cite[adapted from][]{cattaneo96}.\label{satsmallscale}}
\end{figure}

\begin{figure}  
\centering \includegraphics[width=0.9\textwidth]{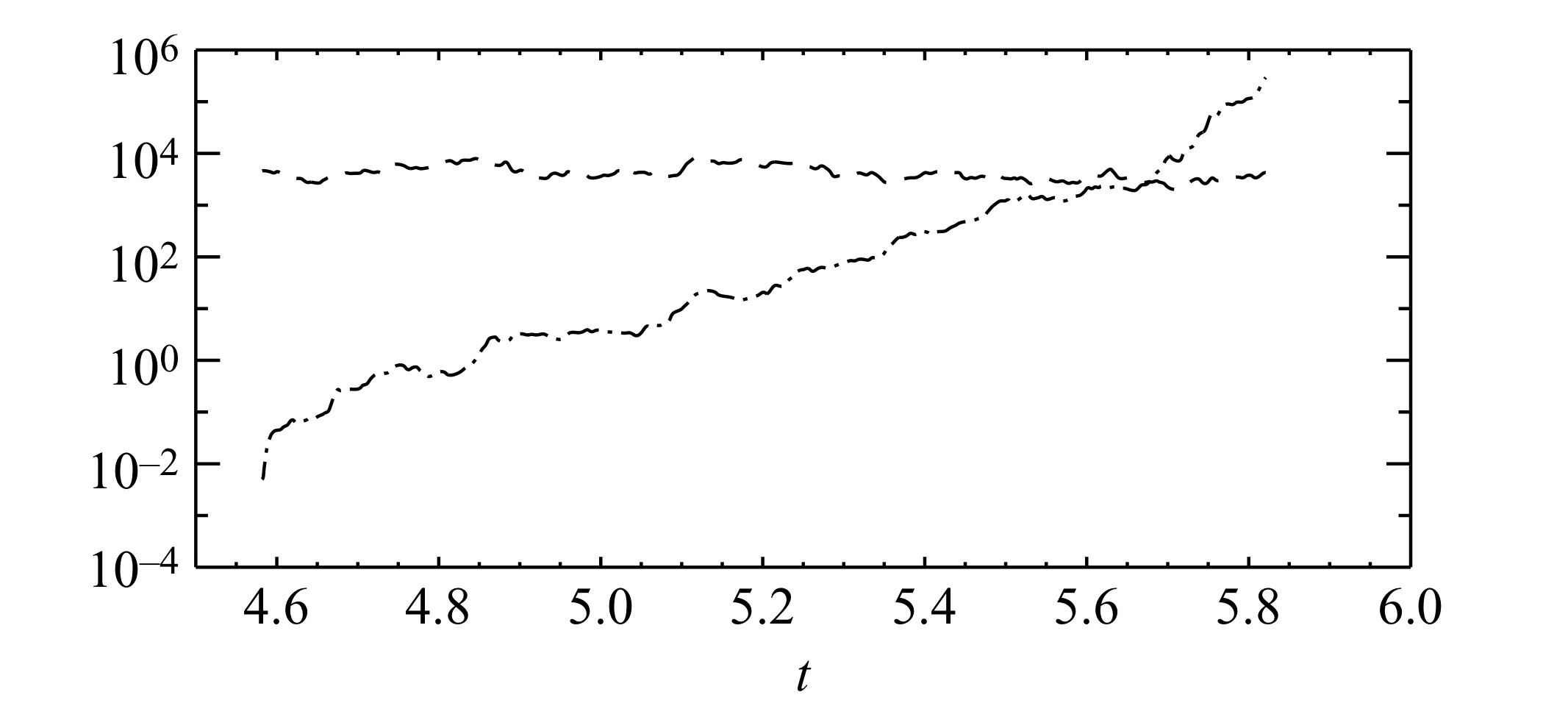}
  \caption{Magnetic energy in a simulation of small-scale dynamo
    action driven by turbulent thermal convection. Dashed-line: energy
    of the saturated magnetic field in the
    dynamical regime. Dash-dotted line: energy of an independent dummy
    magnetic field seeded in the course of the saturated phase. The
    dummy magnetic field grows exponentially, even though the true
    magnetic field has already saturated the velocity field 
    \citep[adapted from][]{cattaneo09}.\label{satsmallscaledummy}}
\end{figure}

These two observations are not independent though
because the energy gained by the field through the anisotropic
induction term must always be equal and opposite to that lost by the
velocity field through the effects of the Lorentz force 
(\sect{energetics}). In practice, the Lorentz force effectively reduces
Lagrangian chaos in the flow in comparison to the kinematic regime,
thereby quenching the inductive stretching of the magnetic field. This
effect is illustrated  in \fig{satsmallscale} by two maps of
the stretching Lyapunov exponent of the GP flow in the
kinematic and saturated regimes for $Pm=4$. The Lyapunov exponent
reduction is strongly correlated spatially with the structure of the
particular realisation of the dynamo magnetic field
being saturated. The velocity field saturated by a given small-scale
dynamo field realisation therefore remains a kinematic dynamo flow for
other independent field realisations/dummy magnetic fields governed by
an induction equation \citep[][see also \cite{branden18} for a more
detailed discussion]{tilgner08,cattaneo09}. A numerical illustration of
this effect is provided in \fig{satsmallscaledummy}.

\subsubsection{Nonlinear growth\label{NLgrowth}}
Now that we have a better appreciation of the effects of the 
Lorentz force on the flow, let us come back to the dynamical 
history of saturation. Various numerical simulations
\citep[e.g.][]{schekochihin04,ryu08,beresnyak12},
suggest that the kinematic stage is followed by a nonlinear 
growth phase during which magnetic energy grows linearly in 
time instead of exponentially. A possible scenario for such nonlinear
growth at large $Pm$ is as follows \citep{schekochihin02d}. In this
parameter regime, the end of the kinematic phase occurs when the
stretching by the most dynamo-efficient viscous scales gets suppressed,
\begin{eqnarray}
  \label{eq:viscoussupp}
\vB\cdot\grad{\vB} & \sim & \vec{u}\cdot\grad{\vec{u}}\sim
u_\nu^2/\ell_\nu~,\\
& \sim & B^2/\ell_\nu \quad \mathrm{(magnetic\,folds\, at\, viscous\, scale)}.
\end{eqnarray}
This corresponds to the magnetic ``Batchelor level'' $\left<|\vB|^2\right>
\sim {Re}^{-1/2} \left<|\vec{u}|^2\right>$. However, this need not be the final
saturation level. Once the fastest viscous motions are quenched, 
the baton of magnetic amplification can be passed to increasingly 
slower but more energetic field-stretching motions at increasingly
larger scales. To see how this process can lead to nonlinear growth,
let us denote the scale of the smallest, weakest, but fastest as yet
unsaturated motions at time $t$ by $\ell^*(t)$, and their stretching
rate by $\gamma^*(t)=u_{\ell^*}/\ell^*$.
These motions are the most efficient at amplifying the field at this
particular time, but by definition their kinetic energy is also such
that $\left<|\vB|^2\right>(t)\sim u_{\ell^*}^2$. From the induction
equation, we can then estimate that
\begin{eqnarray}
\deriv{\left<|\vB|^2\right>}{t} & \sim & \gamma^*(t)\left<|\vB|^2\right>~,\\ 
& \sim & u_{\ell^*}^3/\ell^*~.\label{secular1}
\end{eqnarray}
Now, $u_{\ell^*}^3/\ell^*$ is proportional to the rate of turbulent
transfer of kinetic energy $\varepsilon$ at scale $\ell^*$, but note
that this quantity is actually time- and scale-independent in Kolmogorov
turbulence. \Equ{secular1} therefore shows that for this kind of
turbulence, magnetic energy should grow linearly in time at a rate
corresponding to a fraction $\zeta$ of $\varepsilon$,
\begin{equation}
  \label{eq:secular2}
\left<|\vB|^2\right> \propto \zeta \varepsilon t~.
\end{equation}
In this regime, the magnetic field quenches stretching
motions at increasingly large $\ell^*(t)$ until its energy becomes
comparable to the kinetic energy of the largest, most energetic
shearing motions $u_0$ at the integral scale of the flow
$\ell_0$, i.e. the dynamo saturates in a state of global equipartition
\begin{equation}
\left<|\vB|^2\right> \sim \left<|\vec{u}|^2\right>~.
\end{equation}
As the scale $\ell^*(t)$ of the motions driving the dynamo slowly
increases, there should be a corresponding slow increase in the length
of magnetic folds. Also, note that  the gradual decrease of the maximum
instantaneous stretching rate $\gamma^*(t)$ leaves the diffusion of
the magnetic field unbalanced at increasingly larger scales in the course
of the process. This secular increase of magnetic energy should therefore
be accompanied by a corresponding slow increase of the resistive scale as
$\ell_\eta\sim (\eta/\gamma^*(t))^{1/2}\sim (\eta t)^{1/2}$ for Kolmogorov
turbulence. Perpendicular magnetic reversals associated with
the spectral peak at $\ell_\eta$ do not play an obvious active role in
the problem though, as they do not contribute to the Lorentz force.
Overall, we therefore expect that the folded structure of the
field and general shape of the large-$Pm$ magnetic spectrum should 
be preserved in this scenario.
The particular form of the Lorentz force implies that saturation
is possible without the need for scale-by-scale equipartition.
Spectra derived from numerical simulations of the saturated regime 
appear to support the latter conclusion qualitatively
\citep{schekochihin04}. 

The previous argument is non-local in the sense that the magnetic
field, which has reversals at the resistive scale, appears to
strongly interact with the velocity field at scales much larger than
that \citep*[that saturated isotropic MHD turbulence has a non-local
  character remains controversial though, see e.g.][]{aluie10}.
  A different nonlinear dynamo-growth scenario can be
  constructed if one instead assumes that nonlinear interactions
  between the velocity and magnetic field are predominantly
local \citep{beresnyak12}, as could for instance (but not necessarily
exclusively) be the case for $Pm$ of order one or smaller. Imagine a
situation in which the magnetic spectrum at time $t$ now peaks at a
scale (also denoted by
$\ell^*(t)$ for simplicity) where there is a local equipartition
between kinetic and magnetic energy. The dynamics at scales larger
than $\ell^*(t)$ is hydrodynamic and the turbulent rate of energy transfer
from these scales to $\ell^*(t)$ is therefore the standard hydrodynamic
transfer rate $\varepsilon$. At scales below $\ell^*(t)$, the magnetic
field becomes dynamically significant and the turbulent cascade turns
into a joint MHD cascade of magnetic and kinetic energy characterised
by a different constant turbulent transfer rate $\varepsilon_{\mathrm{MHD}}$.
Locality of interactions and energy conservation in scale space
then lead to the prediction that the magnetic energy should grow as
$(\varepsilon-\varepsilon_\mathrm{MHD})t$. For Kolmogorov turbulence,
$u_\ell\sim \ell^{1/3}$, and therefore a linear in time increase of 
the magnetic energy is associated with an increase
of the equipartition scale as $\ell^*(t)\sim t^{3/2}$. Numerical
results obtained by \cite{beresnyak12} for $Pm=1$ suggest that 
nonlinear magnetic-energy growth in this scenario may be
relatively inefficient, $(\varepsilon-\varepsilon_\mathrm{MHD})/\varepsilon\sim
0.05$. It has been argued recently that this nonlinear growth regime
may be relevant to galaxy clusters accreting through cosmic
time \citep{miniati15}. As the turbulence injection scale $\ell_0$ also
grows in such systems, the relative inefficiency of the dynamo in this
nonlinear regime would suggest that the ratio between $\ell^*$
and $\ell_0$ remains small as both scales grow, so that a
statistically steady saturated state is never achieved.

\subsubsection{Saturation at large and low $Pm$\label{satPm}}
It remains unclear which of the previous scenarios, if any, applies to
the large-$Pm$ limit and, more generally, whether a unique, universal
nonlinear growth scenario applies to different $Pm$ regimes.
Numerical simulations of the $Pm>1$ regime
\cite[e.g.][]{haugen04,schekochihin04,beresnyak12,bhat13}
do not as yet provide a definitive answer to these questions. As far as
the $Pm\gg 1$ regime is concerned, it has so far proven impossible to
explore numerically the astrophysically relevant asymptotic limit
$Rm\gg Re\gg 1$ in which the viscous and integral scales of the dynamo
flow are properly separated. 

The physics of saturation at low $Pm$ also very much remains \textit{terra
  incognita} at the time of writing of these notes. The only nonlinear
results available in this regime are due to \cite{sahoo11} and
\cite{branden11}. One of the positive conclusions of the latter study
is that dynamical simulations at low $Pm$ appear to be slightly less
demanding in terms of numerical resolution than their kinematic counterpart
because the saturated magnetic field diverts a significant fraction of
the cascaded turbulent energy before it reaches the viscous
scale. \citeauthor{branden11}'s simulations at largest $Rm$ also
suggest that the efficiency of the dynamo should be quite large even
at low $Pm$, with magnetic energy of the order of at least $30\%$ of
the kinetic energy.

\subsubsection{Reconnecting dynamo fields\label{reconnect}}
\begin{figure}
\includegraphics[width=\textwidth]{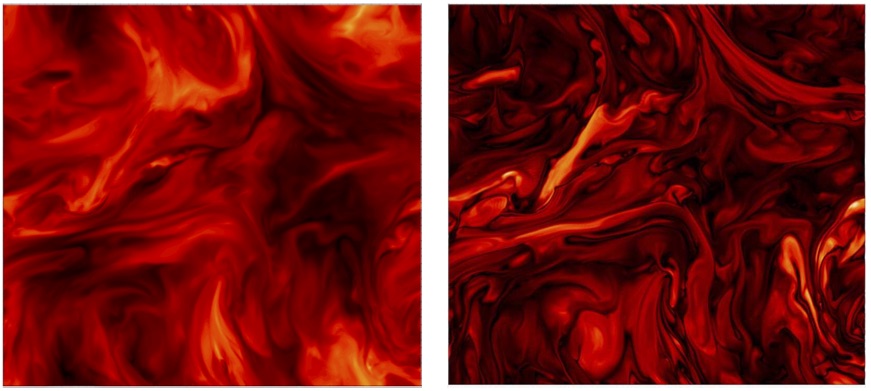}
\caption{2D snapshots of $|\vec{u}|$ (left)  and $|\vB|$ (right)
  in a nonlinear simulation of small-scale dynamo driven by
  turbulence forced at the box scale at $Re=290$
  $Rm=2900$, $Pm=10$ (the magnetic energy spectra for this $512^3$
  spectral simulation suggest that it is reasonably well resolved). 
  At such large $Rm$, the dynamo field becomes weakly supercritical to
  a secondary fast-reconnection instability in regions of reversing
  field polarities associated with strong electrical currents. The
  instability generates magnetic plasmoids and outflows, leaving a
  small-scale dynamical imprint on the velocity field (unpublished
  figure courtesy of A.~Iskakov and A.~Schekochihin).\label{plasmoids}}
\end{figure}

Another potentially very important problem in the nonlinear regime is
that of the stability of small-scale dynamo fields (in particular
magnetic folds at large $Pm$) to fast MHD reconnection for
Lundquist numbers $Lu=U_AL/\eta$ of the order of a few thousands
\red{(or equivalently $Rm$ of a few thousands in an equipartitioned
nonlinear dynamo regime characterised by a typical Alfv\'en velocity $U_A$
of the order of the r.m.s. flow velocity).} This regime has only become
accessible to numerical simulations in the last few years, but
plasmoid chains typical of this process \citep{loureiro07,loureiro12}
have now been observed in different high-resolution simulations of
small-scale dynamos by Andrey Beresnyak and by Alexei~Iskakov and
Alexander~Schekochihin (\fig{plasmoids}), as well as in simulations of
MHD turbulence driven by the magnetorotational instability \citep{kadowaki18}.
The relevance and implications of fast, stochastic reconnection processes
for the saturation of dynamos in general, and the small-scale dynamo
in particular, are currently not well understood, although there is some
nascent theoretical activity on the problem \citep*[][]{eyink11,eyink11b}.
A tentative phenomenological model of the reconnecting small-scale dynamo,
inspired by similar results on MHD turbulence in a guide field, is due
in an upcoming review by \cite{schekochihin19}.

\subsubsection{Nonlinear extensions of the Kazantsev model*\label{nlkazantsev}}
Let us finally quickly review a few nonlinear extensions of
the Kazantsev model that readers begging for more
quantitative mathematical derivations may enjoy exploring further. 
The general idea of these models is to solve \equ{eq:unclosed} for the
magnetic correlator for a dynamical velocity field consisting of
the original stochastic kinematic Kazantsev-Kraichnan field, plus 
a nonlinear magnetic-field-dependent dynamical correction
$\vec{u}_{\mathrm{NL}}$
accounting for the effect of the Lorentz force on the flow. By
massaging the new unclosed mixed correlators associated with
$\vec{u}_{\mathrm{NL}}$ in \equ{eq:unclosed}, one picks up new
quasilinear or nonlinear terms in Kazantsev's
\equ{eq:HLequation}. Stationary solutions of this modified equation
for the saturated magnetic correlator are then sought. 
Of course, in the absence of a good analytical procedure to solve the
Navier-Stokes equation with a magnetic back-reaction, the exact form
of $\vec{u}_{\mathrm{NL}}$ must be postulated on phenomenological
grounds. \cite{subramanian98},
for instance, introduced an ambipolar drift proportional to the
Lorentz force, $\vec{u}_{\mathrm{NL}}=a\,\vJ\times\vB$,
where $a$ is a dimensional constant. The closure procedure leads to a
simple dynamical renormalisation of the magnetic diffusivity
$\eta\rightarrow \eta_\mathrm{NL}=\eta+2a H_L(0,t)$, and
therefore to a renormalisation of the magnetic Reynolds number,
i.e. $Rm_\mathrm{NL}=u_0\ell_0/\eta_\mathrm{NL}$ \red{($H_L$ is
the longitudinal magnetic correlator introduced in \equ{eq:Hij})}.
The saturated state of the model is the neutral eigenmode of the
marginally-stable linear problem, scaled by
an amplitude fixed by the condition $Rm_{\mathrm{NL}}=Rm_c$. This
model predicts a total saturated magnetic energy much smaller than
the kinetic energy (by a factor $Rm_c$) with locally strong
equipartition fields organised into non-space-filling  narrow rope
structures.

Another model, put forward by \cite{boldyrev01}, postulates 
a velocity strain tensor of the form
\begin{equation}
\dpart{u_\mathrm{NL}^i}{x^{k}}=-\f{1}{\nu}\left(B^iB^k-\f{1}{3}\delta^{ik}B^2\right)~,
\end{equation}
motivated by the dynamical feedback of the Lorentz force on viscous
eddies expected in the early phases of saturation in the large
$Pm$ regime. This prescription neatly leads to a modified
version of the Fokker-Planck \equ{eq:fokkerpdf} for the p.d.f. of
magnetic-field strength, whose stationary solution is a Gaussian.
Simulations suggest that the magnetic  p.d.f. in the saturated regime
of the dynamo is exponential, but the determination of the exact shape
of the p.d.f. appears to bear some subtle dependence on rare but intense
stretching events. Phenomenological considerations can be used to
fine-tune a very similar model to generate exponential p.d.f.s
\citep{schekochihin02b}.  Yet another possibility, explored by
\cite{schekochihin04b}, is to postulate a local
magnetic-field-orientation dependent anisotropic
correction $\kappa^{a}$ to the correlation tensor $\kappa^{ij}$ of the
Kazantsev velocity field to model the effects of magnetic tension on
the flow. The generalised correlator can be expressed in spectral
space as
\begin{eqnarray*}\kappa^{ij}(\vec{k}) &=
  &\kappa^{(\mathrm{i})}(k,|\mu|)\left(\delta^{ij}-\hat{k}_i
    \hat{k}_j\right) 
  \label{eq:kazantsevaniso}\\
& + &  \kappa^{a}(k,|\mu|)\left(\hatB^i\hatB^j+\mu^2\hat{k}_i\hat{k}_j-\mu\hatB^i \hat{k}_j-\mu\hat{k}_i\hatB^j\right)~,\nonumber
\end{eqnarray*}
where $\mu=\hat{\vec{k}}\cdot\hatvB$ and $\hat{\vec{k}}=\vec{k}/k$.
The Kazantsev equation derived from this model is more complicated and
must be solved numerically. The results appear to reproduce the
evolution of the magnetic spectra of  simulations of the saturated
large-$Pm$ regime of the small-scale dynamo reasonably well. 

Overall, we see that multiple nonlinear extensions of the Kazantsev
formalism are possible. The potential of this kind of
semi-phenomenological approach to saturation has probably not yet been
fully explored.

\section{Fundamentals of large-scale dynamo theory\label{largescale}}
In the previous section, we studied the problem of small-scale dynamo action
at scales comparable to, or smaller than the integral scale of the
flow $\ell_0$. We are now going to consider large-scale dynamo
mechanisms by which the magnetic field is amplified at system scales
$L$ much larger than $\ell_0$. At first glance, large-scale dynamos
appear to be quite different from small-scale ones: while the latter
rely on dynamical mechanisms that do not particularly care whether the system
is globally isotropic or not, the former appear to generically 
require an element of large-scale symmetry-breaking such as rotation
or shear. In particular, we will see that flow helicity is a major
facilitator of large-scale dynamos. In the absence of any such
ingredient, it would seem extremely difficult (although not
necessarily impossible) to maintain the spatial and temporal
coherence of a large-scale magnetic-field component over many
turbulent ``eddies'' and turnover times. We should therefore not be
surprised to discover in this section that the classical theory of
large-scale dynamos looks very different from what we have
encountered so far. However, we will also find that the spectre of
small-scale dynamos looms over large-scale dynamos, and that both are
in fact seemingly inextricable
in large-$Rm$ regimes of astrophysical interest.  But, before we dive
into such theoretical considerations, let us once again
comfortably acquaint ourselves with the problem at hand by reviewing
some straightforward numerical results and a bit of phenomenology.

\subsection{Evidence for large-scale dynamos\label{LSevidence}}
As explained in \sect{observations}, observational measurements of
dynamical large-scale solar, stellar and planetary magnetic fields
provide the best empirical evidence for large-scale MHD dynamo
mechanisms. Reverse-engineering observations to understand
the detailed underlying physics, however, is a very challenging task. 
Basic experimental evidence for the kind of physical mechanisms
underlying large-scale dynamos that we will study in this section
has also been incredibly long and difficult to obtain
\citep{stieglitz01}, and has so far unfortunately only
provided relatively limited insights into the problem in generic
turbulent flows, especially at high $Rm$. Therefore, despite their own
flaws and limitations, numerical simulations have long been, and will
for the foreseeable future, remain the most powerful tool available to
theoreticians to make progress on this particular problem.

\begin{figure}
\centering\includegraphics[height=0.35\textwidth]{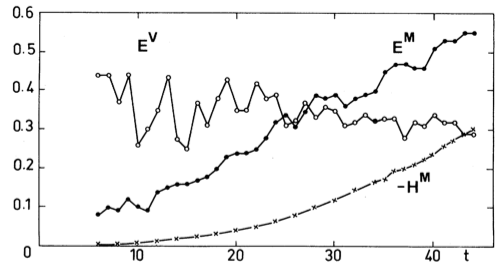}
  \caption{Numerical evidence for a kinematic large-scale dynamo 
    driven by ``small-scale'' 3D homogeneous, pseudo-isotropic,
    helical turbulence forced at $\ell_0$ corresponding to one fifth of
    the box size ($Rm\simeq 40$, $Pm=4$). Time-evolution of the kinetic energy
    $(E^V)$, magnetic energy $(E^M)$, and magnetic
    helicity $(H^M)$. Time is measured in multiples of an $O(1)$
    fraction of the turnover time $\ell_0/u_0$ at the injection scale 
    \citep{meneguzzi81}.\label{figMFP81large}}    
\end{figure}

\begin{figure}
\centering\includegraphics[height=\textwidth]{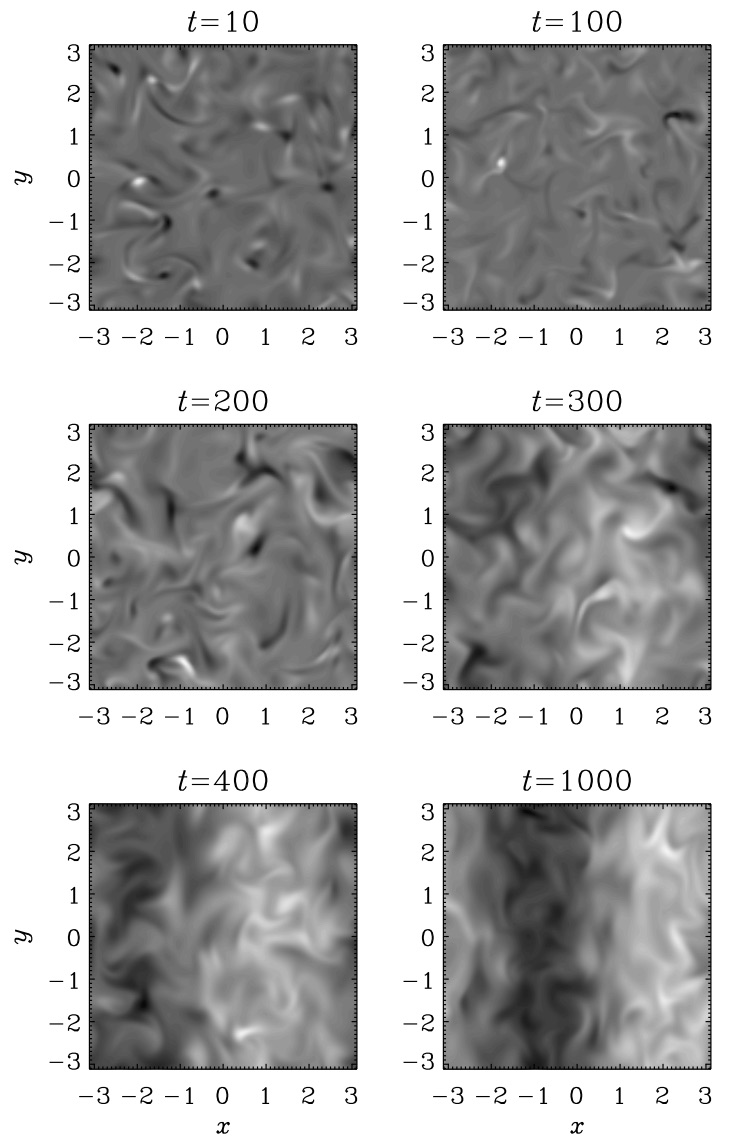}
  \caption{Time-evolution of the $x$-component of the magnetic field
    in a plane in a 3D simulation of large-scale
    dynamo action driven by homogeneous, pseudo-isotropic
    helical turbulence forced at one fifth of the box size ($Rm=180$,
    $Pm=1$). Time is measured in multiples of an $O(1)$ fraction of the
eddy turnover time \citep[adapted from][]{branden01}. \label{figbranden01large}}    
\end{figure}

The first published numerical evidence for a growing large-scale
dynamo is again due  to \cite{meneguzzi81}, and it was presented in the
same paper as the first simulation of the small-scale dynamo. In the second
simulation of their study, turbulent velocity fluctuations $\fluctvU$
\textit{with a net volume-averaged kinetic helicity}
$\left<\fluctvU\cdot\curl\fluctvU\right>_V$
are \textit{driven at an intermediate scale} $\ell_0<L$ of the (spatially
periodic) numerical domain: magnetic energy grows exponentially 
and ends up being predominantly concentrated at
the box scale $L$ (\fig{figMFP81large}). Comparing \fig{figMFP81large}
to \fig{figMFP81small}, we see
that large-scale helical dynamo growth takes place at a much
slower pace than the small-scale, non-helical dynamo. While $Rm$ in
these particular simulations is close to 40, the critical $Rm$ for
a large-scale, maximally helical dynamo in this configuration is
actually $O(1)$ \citep[see e. g.][]{branden01}. This is significantly
lower than $Rm_{c,\mathrm{ssd}}$ for the small-scale dynamo, and
explains why proof-of-principle simulations of this problem could be
done at relatively low resolution ($32^3$ in a spectral
representation). What does the magnetic field of this dynamo look
like ? \Fig{figbranden01large}, reproduced from the work
of \cite{branden01}, illustrates the
multiscale nature of the problem and the build-up of a large-scale
magnetic-field component at scale $L$, on top of magnetic-field
fluctuations at scales comparable to or smaller than $\ell_0$.
The simple conclusion of these selected controlled numerical
experiments (and many others) is that there appears to be such a 
thing as large-scale statistical dynamo action. Now, how does this
kind of dynamo work ? And why does flow helicity appear to be 
important in this context ? 

\subsection{Some phenomenology\label{phenom}}
In order to answer these questions and to introduce the phenomenology
of the problem in an intuitive way, let us travel back in time to
the historical roots of dynamo theory.

\subsubsection{Coherent large-scale shearing: the $\Omega$
  effect\label{largescaleOmega}}
\begin{figure}
\centering\includegraphics[width=\textwidth]{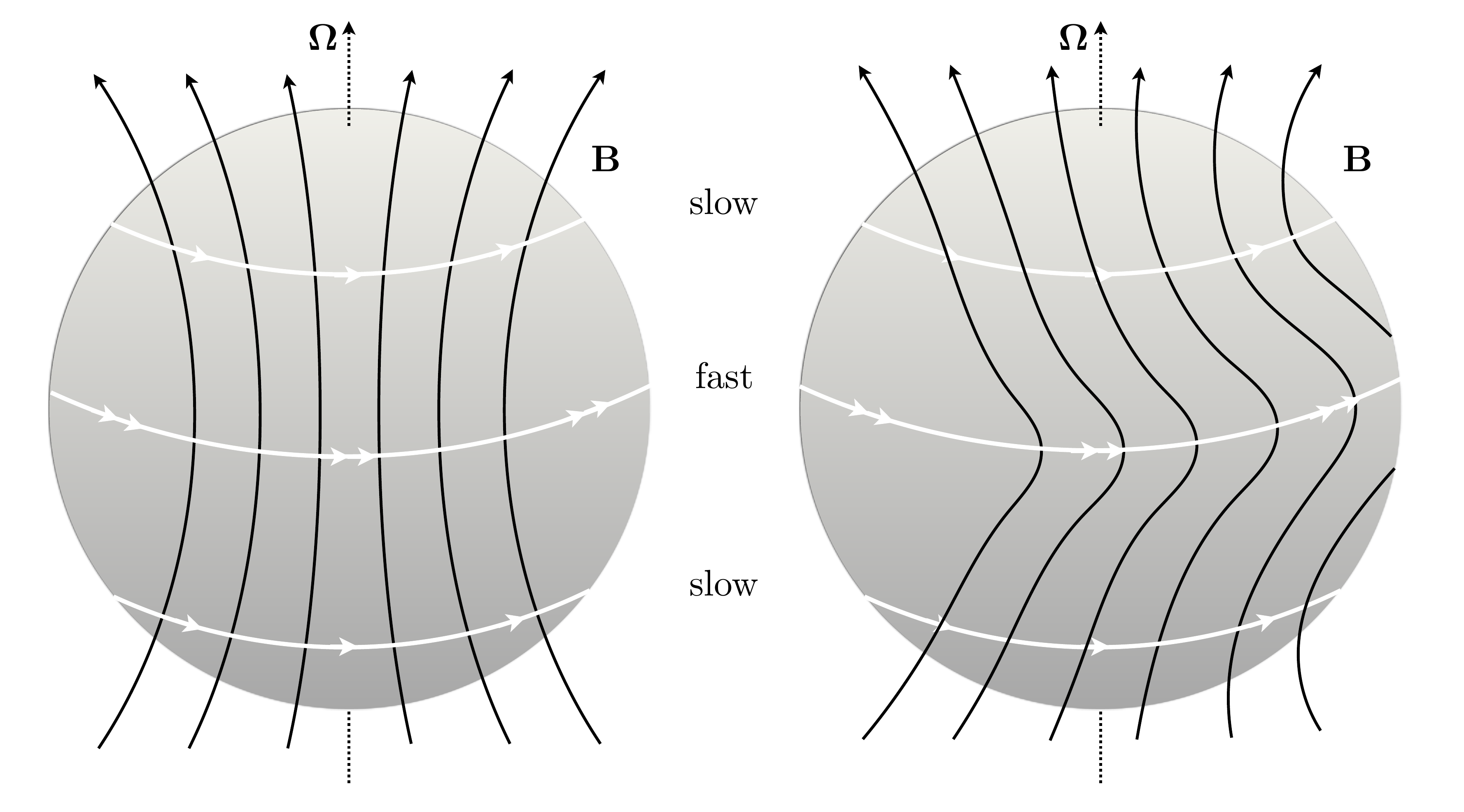}
  \caption{A poor perspective drawing of the $\Omega$ effect in a
    spherical fluid system with latitudinal differential rotation
    $\Omega(\theta)\vec{e}_z$ (maximum at the equator in this example): the
    latitudinal shear winds up an initial axisymmetric  poloidal field
    into a stronger axisymmetric toroidal field in the
    regions of fastest rotation. In the absence of resistivity and any
    other dynamical effect, the growth of the toroidal field is
    linear, not exponential, in time. In the resistive case, and
    in the absence of further three-dimensional dynamical effects, the
    field as a whole is ultimately bound to decay (Cowling's theorem).\label{figomegaeffect}}
\end{figure}

The winding-up of a weak magnetic-field component parallel
to the direction of the velocity gradient into a stronger
magnetic-field component parallel to the direction of the velocity
field itself, the $\Omega$ effect is undoubtedly a very important inductive
process in any shearing or differentially rotating system
\citep{cowling53}. This effect is illustrated in \fig{figomegaeffect}
in spherical geometry. As discussed in \sect{cowling}, however, this
effect is only good at producing toroidal field out of poloidal field,
and cannot by itself sustain a dynamo\footnote{A poloidal/toroidal and
  axisymmetric/non-axisymmetric
  decomposition and terminology is used here because the large-scale dynamo
problem is most commonly discussed in the context of cylindrical or
spherical systems such as accretion discs, ``washing machines'' filled
with liquid sodium, or stars. The same phenomenology applies in
Cartesian geometry though, as the physics discussed does not owe
its existence to curvature effects or geometric constraints.}.

\subsubsection{Helical turbulence: Parker's
  mechanism and the $\alpha$ effect\label{largescalealpha}}
\begin{figure}
\includegraphics[width=\textwidth]{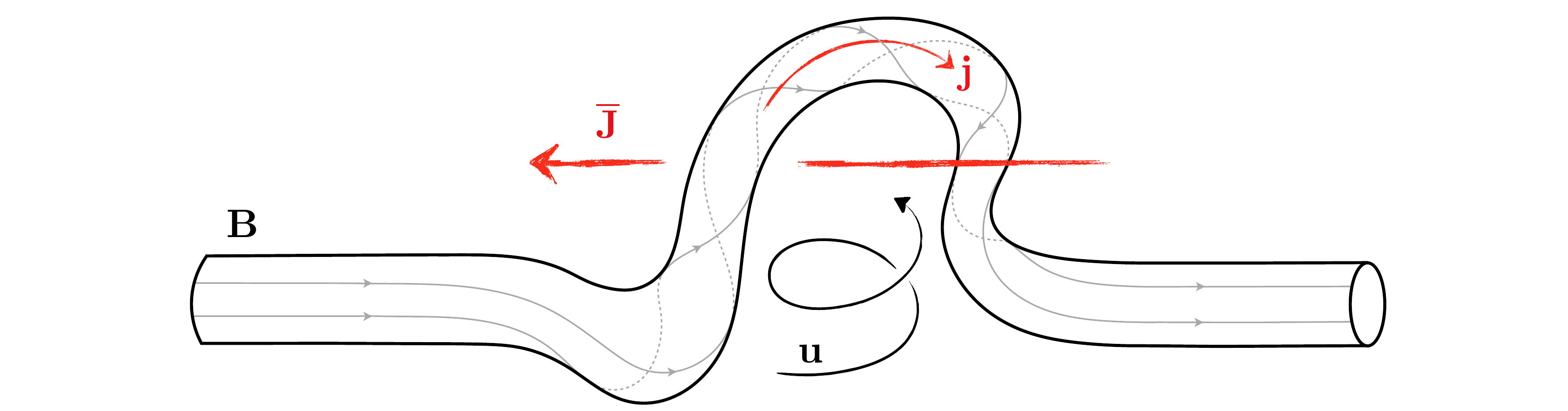}
\vspace{0.7cm}

\centering{\includegraphics[height=0.3\textwidth]{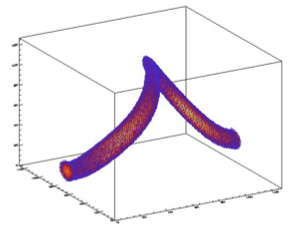}\hspace{1cm}\includegraphics[height=0.3\textwidth]{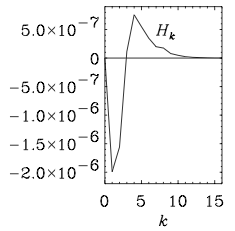}}
\caption{Top: sketch of the dynamics of a magnetic flux tube in
  Parker's mechanism for a right-handed helical velocity fluctuation $\fluctvU$,
  showing a left-handed large-scale magnetic writhe associated with a
  large-scale current $\meanvJ$, and a right-handed
  internal twist associated with a small-scale current
  $\fluctvJ$. This particular configuration is generally thought
  to be representative of the dynamics in the southern hemisphere of
  rotating stars with a strongly-stratified convection zone, where
  motions have a net cyclonic bias (\sect{alphastrat}). Bottom:
  computation of the Cauchy
  solution of an initially straight magnetic flux tube in a cyclonic
  velocity field (left), and corresponding magnetic-helicity spectrum
  (right) \citep[adapted from][]{yousef03}.\label{figtwisting}}  
\end{figure}

After \citeauthor{cowling33}'s (\citeyear{cowling33}) antidynamo work,
it became clear that more complex, three-dimensional,
non-axisymmetric physical mechanisms were required to sustain the
poloidal field against resistive decay. One such mechanism
 was first identified in a landmark publication by
\cite{parker55}, and is commonly referred to as the $\alpha$ effect
after the theoretical work of \cite{steenbeck66,steenbeck66b} that
we will introduce in \sect{largescalemeanfield}. In his work, Parker
considered the effects of helical fluid motions, typical of rotating
convection in the Earth's core or in the solar convection zone, on an
initially straight magnetic field perpendicular to the axis of rotation
(\fig{figtwisting} top). Field lines rising with the hot convecting
fluid also get twisted in the process, and thereby acquire a component
perpendicular to the original field. A statistical version of this
mechanism involving an ensemble of localised small-scale swirls, all
with the same sign of helicity, would effectively couple the
large-scale toroidal and poloidal field components, which could then
lead to effective large-scale dynamo action. 
Parker's mechanism can equally turn toroidal field into poloidal field, and
poloidal field into toroidal field (the initial horizontal orientation
of the magnetic field drawn in \fig{figtwisting} (top) is
arbitrary). It should  therefore be sufficient to excite a dynamo even
in the absence of large-scale shearing.

In practice, there are several issues with this simple picture,
the most important of which will be discussed in \sect{MFcoeff}
and in \sect{largescalesat}. Besides, as we will discover
later in this section and in \sect{complexLS}, the $\alpha$ effect is
not the only statistical effect capable of exciting a large-scale dynamo.
Parker's idea, however, was seminal in the development of large-scale
dynamo theory and remains one of its central pillars, not least
because it is directly connected to rotating dynamics, and rotation
and large-scale magnetism in the Universe always seem to go hand-in-hand.

\subsubsection{Writhe, twist, and magnetic helicity\label{writhetwist}}
An important aspect of the kinematic evolution of a magnetic
field in a helical velocity field is the associated dynamics 
of \textit{magnetic} helicity $\mathcal{H}_m$ subject to
the constraint of total magnetic helicity conservation
introduced in \sect{conservation}. To illustrate this, consider again the
simple three-dimensional evolution of an initially straight magnetic flux
tube in a steady right-handed helical swirl  
depicted in \fig{figtwisting} (top). As the swirl acts upon the flux
tube and makes it rise, a ``Parker loop'', or large-scale writhe, is
created. The current associated with this loop is
anti-parallel to the large-scale field
direction (and so is the vector potential in the Coulomb gauge),
i.e. the process generates negative magnetic helicity on scales larger
than that of the loop. But the same motion simultaneously twists the
magnetic-field lines around the flux tube, thereby generating a local
current within the tube with a positive projection along the local
field, i.e. positive local magnetic helicity. If we ignore resistive
effects and assume that there are no helicity losses out of the domain
under consideration, \equ{eq:helicity} shows that the total magnetic
helicity of the system is conserved, so that the negative
helicity/left-handed writhe generated on large
scales must be exactly balanced by the positive helicity/right-hand
twist generated on small scales if $\mathcal{H}_m=0$ initially (for
a left-handed helical swirl, the sign of large-scale and small-scale
magnetic helicity is opposite). Overall, Parker's kinematic mechanism
therefore tends to generate a partition of large- and small-scale
magnetic helicities of opposite signs.

This effect can be illustrated in a more quantitative way by computing 
for instance the magnetic helicity spectrum of a twisted magnetic flux
tube (\fig{figtwisting} bottom). 
Similar computations for fully turbulent helically-driven systems 
lead to the same results and conclusions \citep[e.g.][]{mininni11}.
In the period preceding the very first simulations of helical MHD
turbulence, it was suggested that such dynamics may be a consequence 
of the existence of an inverse ``cascade''  of magnetic helicity
associated with the magnetic helicity conservation constraint
\citep{frisch75}. This interpretation can be broadly justified
in the framework of a three-scale interaction model, however numerical
simulations suggest that the magnetic helicity transfer process in
helical MHD turbulence is very different in nature from hydrodynamic
cascades, as it does not proceed on turbulent dynamical timescales and
is non-local in spectral space \citep*{branden02,branden05}.

Finally, let us point out that the magnetic tension associated with 
the small-scale curvature of twisted magnetic-field lines should 
be an important source of back-reaction of the field on the flow in
the dynamical stages of the dynamo. Accordingly, we will discover in
\sect{largescalesat} that the dynamics of magnetic helicity is a key
ingredient of nonlinear theories of large-scale helical dynamos driven
by an $\alpha$ effect.

\subsection{Kinematic theory: mean-field electrodynamics\label{largescalemeanfield}}
How can we turn this kind of phenomenology into a mathematical theory of
large-scale dynamos ? The classical approach to this problem is called
\textit{mean-field} dynamo theory, or mean-field electrodynamics
\citep{steenbeck66,vainshtein70,moffatt70}. Its mathematical
machinery, and applications, have now been covered by a countless
number of authors. We will therefore only provide a superficial and
rather casual presentation of it in this tutorial, with the main
objective being to emphasise the main underlying ideas and to
frame the theory into a broader discussion. Readers are referred
to the classic textbooks by \cite{moffatt78} and \cite{krause80}, and
to recent dedicated reviews by \cite{hughes18} and \cite{branden18}
for more exhaustive presentations.

\subsubsection{Two-scale approach}
The general idea behind mean-field theory is to split 
the magnetic field, velocity field and the corresponding MHD equations
into mean and fluctuating parts in order to isolate the net r.m.s. effect
of the fluctuations on the mean (``large-scale'') fields. Depending on
the problem, the mean can be defined as an average over a statistical
ensemble of realisations of the flow, as an average over one dimension
(e.g., the toroidal direction for problems in spherical or cylindrical
geometry), over two dimensions, or, if there is a proper spatial and
temporal dynamical scale-separation in the problem, as an average over
the small-scale dynamics. \red{In what follows, we will assume
that any such averaging procedure satisfies the so-called Reynolds rules}
(for a more detailed discussion of averaging in the large-scale dynamo
problem, see \cite{branden05}, Chap.~6.2, and \cite{moffatt78},
Chap. 7.1), and will essentially consider a formulation based on a
two-scale decomposition of the dynamics. Namely, the magnetic and
velocity fields are split into large-scale, \textit{mean-field} parts
($\ell\gg\ell_0$, denoted by an overline) and small-scale, 
fluctuating parts ($\ell\le \ell_0$, denoted by $\fluctvB$ and
$\fluctvU$),  
\begin{eqnarray}
  \label{eq:split}
  \vU & = &\meanvU+\fluctvU~, \\
  \vB & = & \meanvB+\fluctvB~.
\end{eqnarray}
We will also occasionally interpret \equ{eq:split} as a
decomposition into axisymmetric and non-axisymmetric field 
components in systems with a rotation axis\footnote{In
  the general case, we should formally distinguish between
  azimuthal averages and small-scale or ensemble averages.
  This distinction is not necessary in the context of
  this presentation, and is therefore ignored for the sake
  of mathematical and notational simplicity.}. In order to minimise
 the physical complexity, we also restrict the analysis to the
 incompressible formulation of the kinematic dynamo problem with a
 uniform magnetic diffusivity, \equ{eq:inducincompressible}.
 
Let us start with the induction equation averaged over scales larger
than $\ell_0$, 
\begin{eqnarray}
  \label{eq:Bbar}
  \dpart{\protect\meanvB}{t}+\meanvU\cdot\grad{\,\meanvB}=\meanvB\cdot\grad{\,\meanvU}+\curl{(\mean{\fluctvU\times\fluctvB})}+\eta\Delta\meanvB~.
\end{eqnarray}
To predict the evolution of $\meanvB$, we must
determine the \textit{mean electromotive force} (EMF)
$\meanvEMF=\mean{\fluctvU\times\fluctvB}$
driving the dynamo. This term involves the statistical cross-correlations
between small-scale magnetic and velocity fluctuations of the kind
that we encountered in Parker's phenomenology. The big question, of
course, is how do we calculate it ? Ideally, we would like to find an
expression for $\meanvEMF$ in terms of just
$\meanvB$ and the statistical properties of the fluctuating
velocity field $\fluctvU$. The logical next step is therefore to
attempt to solve the induction equation for $\fluctvB$ in terms of
$\fluctvU$ and $\meanvB$, and to substitute the
solution into the above expression for
$\meanvEMF$. This equation is obtained by
subtracting \equ{eq:Bbar} from the full
induction \equ{eq:inducincompressible},
\begin{equation}
\label{eq:Btilde}
  \dpart{\fluctvB}{t}=\curl{\left[\left(\fluctvU\times\meanvB\right)+
\left(\meanvU\times\fluctvB\right)+\left(\fluctvU\times\fluctvB-\mean{\fluctvU\times\fluctvB}\right)\right]}+\eta\Delta\fluctvB~.
\end{equation}
The term in the first pair of parentheses on the r.h.s. describes the
induction of small-scale magnetic fluctuations, and mixing of the
field, due to the tangling and shearing of the mean field by
small-scale velocity fluctuations. This is the term that we are most
interested in in order to
compute a mean-field dynamo effect. The term in the second pair of
parentheses describes the effect of a large-scale velocity field on
small-scale magnetic fluctuations. Such a velocity field usually has a
much slower turnover time and amplitude compared to the fluctuations, 
and we will therefore consider that it has a subdominant role in the
inductive dynamics of small-scale fluctuations for the purpose of this
discussion. Finally, we have a combination of two terms in the third
pair of parentheses, which will be henceforth referred to as the
``tricky term'' (also often referred
to as the ``pain in the neck'' term). This term is quadratic in
fluctuations, and there is no obvious way to simplify it without
making further assumptions: so we meet again, old closure foe !
\red{We will come back to this problem in \sect{MFcoeff}.}

\subsubsection{Mean-field ansatz\label{MFansatz}}
\Equ{eq:Btilde} is a linear relationship between
$\fluctvB$ and $\meanvB$ if $\vU$ is independent of
$\vB$ (as is the case in the kinematic regime), i.e.
\begin{equation}
\label{eq:linear}
\mathcal{L}_{\vU}(\fluctvB)=\curl{\left(\fluctvU\times\meanvB\right)}~,
\end{equation}
where $\mathcal{L}_{\vU}$ is a linear operator functionally dependent on
$\vU$ \citep{hughes10}. Despite its seemingly innocuous nature,
there is actually quite a lot of complexity hiding in this 
equation due to the presence of the tricky term on the l.h.s., but
for the time being we are going to ignore the presence of this term
and simply postulate that \equ{eq:linear} is indicative of a
straightforward linear relationship between the fluctuations
$\fluctvB$ and the mean field $\meanvB$. If this holds, then we may
expand the mean  EMF as
\begin{equation}
\label{eq:meanfieldexpansion}
  (\mean{\fluctvU\times\fluctvB})_i=a_{ij}\meanB_j+b_{ijk}\nabla_k\meanB_j+\cdots~,
\end{equation}
where the spatial derivative is with respect to the slow spatial
variables over which $\meanvB$ is non-uniform.
This is called the mean-field expansion. Terms involving higher-order
derivatives are discarded because the spatial derivative is slow.
In the general case of inhomogeneous, stratified, anisotropic,
differentially rotating flows, this expansion is usually recast in
terms of a broader combination of greek-letter tensors and vectors that
neatly isolate different symmetries \citep{krause80,raedler06}:
  \begin{equation}
    \meanEMF_i=
 \alpha_{ij}\meanB_j+\left(\boldsymbol{\gamma}\times\meanvB\right)_i-\beta_{ij}\left(\curl{\meanvB}\right)_j-\left[\boldsymbol{\delta}\times\left(\curl{\meanvB}\right)\right]_i-\f{\kappa_{ijk}}{2}\left(\nabla_j\meanB_k+\nabla_k\meanB_j\right)~.
    \label{eq:MFexpansiongeneral}
  \end{equation}
  Here, $\boldsymbol{\alpha}$ and $\boldsymbol{\beta}$ are
  symmetric second-order tensors, $\boldsymbol{\gamma}$ and
  $\boldsymbol{\delta}$ are vectors, and $\boldsymbol{\kappa}$
  is a third-order tensor, namely
  \begin{equation}
    \label{eq:defalphabeta}
    \alpha_{ij}=\f{1}{2}\left(    a_{ij}+    a_{ji}\right)~,
\quad \beta_{ij}=\f{1}{4}\left(\varepsilon_{ikl}\,b_{jkl}+\varepsilon_{jkl}\,b_{ikl}\right)~,
  \end{equation}
  \begin{equation}
    \label{eq:defgammadelta}
\gamma_{i}=-\f{1}{2}\varepsilon_{ijk}\,a_{jk}~,\quad
\delta_i=\f{1}{4}\left(b_{jji}-b_{jij}\right)~,
\end{equation}
\begin{equation}
      \label{eq:defkappa}
  \kappa_{ijk}=-\f{1}{2}\left(b_{ijk}+b_{ikj}\right)~.
\end{equation}
All these quantities formally depend on $Rm$, $Pm$, and the statistics of
the velocity field (and magnetic field, in the dynamical regime).
The two most emblematic mean-field effects are those deriving from
$\boldsymbol{\alpha}$ and $\boldsymbol{\beta}$. The former is related
to Parker's mechanism and can drive a large-scale dynamo, while the
latter is easily interpreted as a turbulent magnetic diffusion (albeit
not necessarily a simple one).

\subsubsection{Symmetry considerations\label{LSsym}}
Symmetry considerations can be used to determine which of the previous
mean-field quantitities are in principle non-vanishing for a given
problem. In order to illustrate at a basic level how this is
generally done, and what kind of mean-field effects we might expect,
we will essentially discuss three particular symmetries in the rest of
this section: isotropy, parity, and homogeneity.

As mentioned in \sect{kazantsev}, there is no preferred direction
or axis of symmetry in an isotropic three-dimensional system.
Tensorial dynamical quantities can only be constructed
from $\delta_{ij}$ and the antisymmetric Levi-Cevita tensor
$\varepsilon_{ijk}$ in this case. On the other hand, if a fluid
system has one or several particular directions, as is the case in
the presence of rotation, stratification, a non-uniform large-scale
flow $\meanvU$, or a strong, dynamical large-scale magnetic field,
then we are in principle also allowed to use a combination of
quantities such as the rotation vector $\vOmega$, the direction of
gravity (or inhomogeneity) $\vec{g}$, the mean-flow deformation tensor
$\mean{\tens{D}}=\left[\grad\meanvU+(\grad\meanvU)^T\right]/2$,
the mean-flow vorticity $\meanvW=\curl{\meanvU}$
or even $\meanvB$ in the dynamical case to construct mean-field tensors.

A second important class of symmetry is parity invariance. In three
dimensions, a parity transformation, or point
reflection, is a combination of a reflection through a plane (an
improper rotation) and  a proper rotation of $\pi$ around an axis
perpendicular to that plane. Parity symmetry is therefore connected to
mirror symmetry, although the two must be distinguished in principle in
three-dimensions \citep[see discussion in][Chap. 7, footnote
4]{moffatt77}. We essentially ignore this distinction in what follows
and will talk indiscriminately of mirror/parity/reflection symmetry
breaking. Under a point reflection, the position vector transforms as
$\vec{r} \rightarrow -\vec{r}$, and the velocity field transforms as
$\vU\rightarrow -\vU$. Now, if we look at some local,
homogeneous, isotropic hydrodynamic solution of the Navier-Stokes
equation driven by a non-helical force and with no rotation, it is
clear that the image of the velocity field under a parity
transformation is itself a solution of the equations. In the MHD
and/or rotating case, on the other hand, the image of a solution
is only a physical solution itself if we keep $\vB$ and / or
$\vOmega$ the same, i.e. $\vB\rightarrow\vB$, and
$\vOmega\rightarrow\vOmega$.
Accordingly, we say that the velocity field is a true vector,
while the magnetic field and rotation vectors are
pseudo-vectors\footnote{The terminology ``true'' and ``pseudo'' stems
from the transformation laws of these quantities under simpler
mirror transformations. Mirror-invariance of MHD requires that we
reflect all vectors (true and pseudo) under a mirror transformation,
but further flip the sign of pseudo-vectors such as $\vB$.}.
Scalars and higher-order tensors can also be divided into pseudo
and true quantities, depending on how they transform under
reflections. Pseudo-scalars simply change sign under reflection.
Let us finally introduce a few simple rules for the manipulation of vectors:
vector products of pairs of true vectors or pairs of pseudo-vectors
produce a pseudo-vector, while mixed vector products involving a true
and a pseudo-vector produce a true vector. Curl operators turn a
true vector into a pseudo-vector, and vice-versa (as illustrated by
Ampere's law, or Biot-Savart's law of magnetostatics between the
electric current, a true vector field constructed from the velocities
of charged particles, and the magnetic field). 
Let us now come back to \equ{eq:MFexpansiongeneral}. The mean EMF
on the l.h.s., being the vector product of a true vector and a
pseudo-vector, is a true vector, while the mean magnetic field on the
r.h.s. is a pseudo-vector. Thanks to the particular decomposition
used in the equation, it is then straightforward to see
that $\boldsymbol{\alpha}$ and $\boldsymbol{\kappa}$ must be
pseudo-tensors, while $\boldsymbol{\beta}$ is a true
tensor. Similarly, $\boldsymbol{\delta}$ must be a
pseudo-vector, and $\boldsymbol{\gamma}$ is a true vector.

Why is this important ? If a flow is parity-invariant, as is for
instance the case of standard homogeneous, isotropic non-helical
turbulence, then its image under a parity transformation has the
exact same statistical properties, and should produce the exact same
physical statistical effects (with the same sign). However, we have
identified several pseudo-quantities, most importantly
$\boldsymbol{\alpha}$, which by construction must change
sign under a parity transformation. These quantities must therefore
vanish for a parity-invariant flow. In order for them to be non-zero,
parity invariance must be broken one way or the other, and a particular
way to achieve this is by making the flow helical. Parker's big idea
was to suggest that this kind of effect can in turn lead to
large-scale dynamo action. We will find in the next paragraphs that
this can indeed be proven rigorously in some particular regimes.
Of course, it is also this argument that motivated the set-up
of the dynamo simulations of \cite{meneguzzi81} discussed in the
introduction of this section.

More generally, the previous discussion highlights that the different
large-scale statistical effects encapsulated by the greek-letter tensors
introduced in \equ{eq:MFexpansiongeneral} are only non-zero when
some particular symmetries are broken. For pedagogical purposes,
we will essentially concentrate on kinematic and dynamical problems
involving simple isotropic, homogeneous $\alpha$ and $\beta$ effects
in this section. Important variants of these effects, as well
as additional effects arising in stratified, rotating
and shearing flows will however also be touched upon in \sect{otherstat}.

\subsubsection{Mean-field equation for pseudo-isotropic homogeneous flows}
In what follows, we will more specifically be concerned with the
Cartesian version of the problem of large-scale dynamo
action driven by isotropic, homogeneous, statistically-steady, but
non-parity-invariant flows (also referred to as pseudo-isotropic
flows), the archetype of which is our now familiar helically-forced
isotropic turbulence. Although the helical nature of many flows in
nature is a consequence of rotation, we will not explicitly try to
capture this effect here, and instead assume that helical motions are
forced externally (a discussion of the $\alpha$ effect in rotating,
stratified flow will be provided in \sect{alphastrat}). On the other
hand, we retain a large-scale shear flow typical of differentially
rotating systems to accommodate the possibility of an $\Omega$ effect,
but neglect possible anisotropic statistical effects associated with
this shear for simplicity (the latter will be discussed in \sect{SCeffect}).
Under all these assumptions, $\alpha_{ij}=\alpha\,\delta_{ij}$ and
$\beta_{ij}=\beta\,\delta_{ij}$ (or $b_{ijk}=\beta\,\epsilon_{ijk}$)
are both finite and independent of space and time, and all the other
mean-field effects in \equ{eq:MFexpansiongeneral} vanish. Substituting
\equ{eq:meanfieldexpansion} into \equ{eq:Bbar}, we obtain a closed
evolution equation for the mean field,
\begin{equation}
\label{eq:MFequation}
\dpart{\protect\meanvB}{t}+\meanvU\cdot\grad{\,\meanvB}
=\meanvB\cdot\grad{\,\meanvU}+\alpha\curl{\meanvB}+(\eta+\beta)\Delta\meanvB~.
\end{equation}
The term on the r.h.s. proportional to the curl of
$\meanvB$, the $\alpha$ effect, is the only one, \red{apart from the
  shear}, that can couple the different components of the mean field for the
simple system under consideration. The last term on the r.h.s., the
$\beta$ effect, acts as an effective turbulent magnetic diffusivity
operating on the mean field.

The presence of the $\alpha$ and $\beta$ terms in the linear
\equ{eq:MFequation} implies that we can now in principle 
obtain exponentially growing mean-field dynamo solutions if the 
statistical properties of the flow allow it\footnote{If $\alpha=0$,
  this requires a negative turbulent diffusivity $(\beta<-\eta)$, which is
  of course not a feature of daily-life turbulence but is possible
  under certain circumstances.  In standard turbulence conditions at
large $Rm$, $\beta\gg \eta>0$.}. This conclusion may seem puzzling
at first in view of Cowling's theorem. If we interpret large-scale
averages as azimuthal averages, how is it possible that
\equ{eq:MFequation}, which only involves axisymmetric quantities, has
growing solutions ? The key here is to realise that we have
massaged the original problem quite a bit to arrive at
this result. \Equ{eq:MFequation} is not the pristine
axisymmetric induction equation but a simpler 
\textit{model} equation. The exact evolution equation 
for $\meanvB$, \equ{eq:Bbar}, has a dynamo-driving 
term $\meanvEMF$, which is quadratic in non-axisymmetric
(or small-scale) fluctuations. In order to arrive at
\equ{eq:MFequation}, we have simply expressed this term using a
simple closure \equ{eq:meanfieldexpansion} motivated by the linear
nature of \equ{eq:Btilde}. In other words, the three-dimensional
nature of the dynamo is now hidden in the mean-field coefficients
$\alpha$ and $\beta$. We will demonstrate in \sect{MFcoeff} that 
these coefficients are indeed functions of the statistics of
small-scale fluctuations.

\subsubsection{$\alpha^2$, $\alpha\Omega$ and $\alpha^2\Omega$ dynamo solutions\label{alphaomega}}
\begin{figure}
\centering\includegraphics[width=\textwidth]{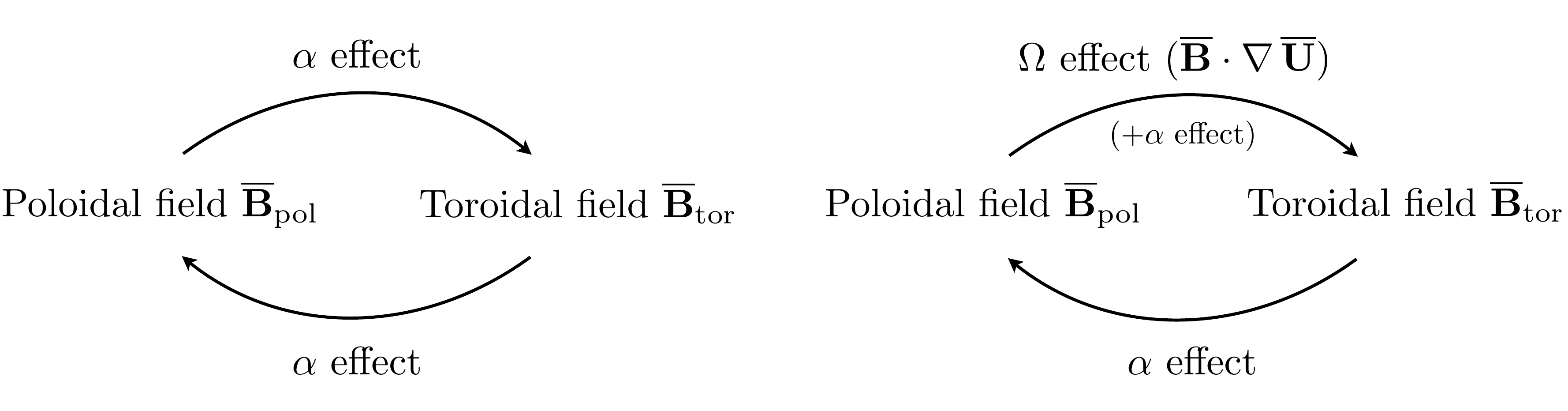}
\caption{The $\alpha^2$ (left), $\alpha\Omega$, and $\alpha^2\Omega$
  (right) mean-field dynamo loops.\label{figMFloops}}
\end{figure}

In a Cartesian domain with periodic boundary conditions, we may seek
simple solutions of \equ{eq:MFequation} in the form
\begin{equation}
\label{eq:planewave}
  \meanvB=\meanvB_\vec{k}\exp\left[s
    t+i\vec{k}\cdot\vec{x}\right] + \mathrm{c.c.}~,
\end{equation}
subject to $\vec{k}\cdot\meanvB_\vec{k}=0$.
We first consider the case with no mean flow or shear. Substituting
\equ{eq:planewave} into \equ{eq:MFequation} with $\meanvU=\vec{0}$,
and solving the corresponding eigenvalue problem, we find a branch of
growing dynamo modes with a purely real eigenvalue 
corresponding to a dynamo growth rate
\begin{equation}
  \label{eq:gammaalphasquare}
  \gamma=|\alpha| k- (\eta+\beta)\, k^2~.
\end{equation}
For $\beta>0$, the maximum growth rate and optimal wavenumber are
\begin{equation}
  \label{eq:optimalalphasquare}
  \gamma_\mathrm{max}=\f{\alpha^2}{4\,(\eta+\beta)},\quad  k_\mathrm{max}=\f{|\alpha|}{2\,(\eta+\beta)}
\end{equation}
(for consistency, we should
have $k_\mathrm{max}\ell_0<1$, but to check this we first need to
learn how to calculate $\alpha$ and $\beta$). The dynamo is
possible here because the $\alpha$ effect couples
the two independent components of $\meanvB$
in both ways through the curl operator (in particular, it also couples the
toroidal field component to the poloidal field component in
cylindrical or spherical geometry). For this reason, this mean-field
dynamo model is usually
called the $\alpha^2$ dynamo. It is this kind of coupling,
illustrated in \fig{figMFloops} (left), that drives the helical
large-scale kinematic dynamo effect in the simulations shown in
\fig{figMFP81large} and \fig{figbranden01large}. A discussion of
magnetic helicity and linkage dynamics in the $\alpha^2$ dynamo,
complementary to the Parker mechanism picture shown earlier in
\fig{figtwisting}, can be found in \cite{blackman14}.

What about the astrophysically relevant situation involving a
mean shear flow or differential rotation, on top of some smaller-scale
turbulence ? While this case is of course most directly connected to
problems in global spherical and cylindrical geometries, it it most
easily analysed in the Cartesian shearing sheet model
introduced in \sect{shearingsheet}. In the presence of linear shear,
$\meanvU=\vU_S=-Sx\,\vec{e}_y$, the mean-field \equ{eq:MFequation} reads
\begin{equation}
  \label{eq:alphaOmega}
  \dpart{\protect\meanvB}{t}=-SB_x\vec{e}_y+\alpha\curl{\meanvB}+(\eta+\beta)\Delta\meanvB~.
\end{equation}
The $\meanvU\cdot\grad{\,\meanvB}$ term
on the l.h.s. of \equ{eq:MFequation} vanishes for a purely azimuthal
mean flow because $\meanvB$ is axisymmetric\footnote{We
  would have to retain this term in order to describe the
  advection of the mean field in meridional planes by a 
  poloidal mean flow. This effect does not amplify the magnetic
  field on its own, but it can redistribute it spatially in a way that 
alters the nature of a large-scale dynamo in comparison to a fully
homogeneous process. Meridional circulation is for instance thought 
to be important for the global solar dynamo mechanism, see
e.g. \cite{charbonneau14}.} (independent of $y$ in our notation).
The couplings in \equ{eq:alphaOmega} allow different
forms of large-scale dynamo action illustrated in
\fig{figMFloops} (right). In the most standard case, commonly
referred to as the $\alpha\Omega$ dynamo, toroidal/azimuthal ($y$) field is
predominantly generated out of the poloidal field (of which $B_x$ is
one component) via the $\Omega$ effect (the first term on the
r.h.s.), while the $\alpha$ effect associated
with the small-scale helical turbulence turns the toroidal field back
into poloidal field. In $\alpha^2\Omega$ dynamos, the $\alpha$ effect
also significantly contributes to the conversion of \red{poloidal field
into toroidal field.}

Let us briefly look at axisymmetric mean-field solutions%
\footnote{Earlier in the text, we opted not to make a
  distinction between azimuthal and small-scale or ensemble averages.
  When this distinction is made, it is formally possible to seek
  non-axisymmetric mean-field solutions of \equ{eq:alphaOmega} with
  an azimuthal wave number $k_y$ such that $k_y\ell_0\ll 1$. The
  analysis of such solutions is a bit more involved due to the
  presence of the shear \citep{branden05}.}
 $\meanvB(x,z,t)$ of \equ{eq:alphaOmega} in the
 $\alpha\Omega$ case. We again seek plane-wave
solutions in the form of \equ{eq:planewave}, with a wavenumber
$\vec{k}=(k_x,0,k_z)$. In the $\alpha\Omega$ limit $S\gg \alpha k_z$,
the solution of the linear dispersion relation leads
to a branch of growing dynamo solutions if the so-called 
dynamo number $D\equiv\alpha S k_z/\left[2(\eta+\beta)^2k^4\right]$  is larger
than unity \citep{branden05}. The unstable eigenvalue 
in this case has an imaginary part, 
\begin{equation}
\mathcal{R}(s)\equiv\gamma=\f{1}{2}\left|S\alpha k_z\right|^{1/2}-(\eta+\beta) k^2,
\end{equation}
\begin{equation}
\mathcal{I}(s)\equiv\omega=\f{1}{2}\left|S\alpha k_z\right|^{1/2},
\end{equation}
i.e. the solutions take the form of growing oscillations. The
interplay between the $\alpha$ effect and the $\Omega$ effect
can therefore produce dynamo waves involving periodic
field reversals. This conclusion, first obtained by \cite{parker55},
was key to the realisation
that differentially rotating turbulent MHD systems can host
large-scale oscillatory dynamos, and sparked an enormous amount of
research on planetary, solar, stellar and galactic dynamo modelling.

\subsubsection{Calculation of mean-field coefficients: First Order Smoothing\label{MFcoeff}}
The mean-field formalism appears to be a very convenient
and intuitive framework to model large-scale turbulent dynamos and
to simplify their underlying nonlinear MHD complexity. However, we have so
far ignored several difficult key questions: under which
conditions is the mean-field ansatz justified ? And are there
systematic ways to derive mean-field coefficients such as $\alpha$ and
$\beta$ from first principles, given a velocity field with prescribed
statistical properties ?

The former question largely conditions the answer to the latter.
To illustrate the nature of the problem, recall that we have
carefully avoided to discuss how to handle the tricky term in
\equ{eq:Btilde} in our earlier discussion of the mean-field ansatz.
However, if we want to calculate $\alpha$, we need to do something
about it ! The radical closure option is to simply neglect this term.
This approximation is often called the First Order Smoothing
Approximation (FOSA), Second Order Correlation Approximation (SOCA),
Born approximation, or quasilinear approximation. It has the merit of
simplicity, but can only be rigorously justified in two limiting cases:
\smallskip
\begin{itemize}
\item small velocity correlation times (either small Strouhal
  numbers $St=\tau_c/\tau_{\mathrm{NL}}\ll 1$, or random wave fields)
\smallskip
\item low magnetic Reynolds numbers 
  $Rm=\tau_\eta/\tau_{\mathrm{NL}}\ll 1$~.
\end{itemize}
\smallskip
where $\tau_\mathrm{NL}=\ell_0/u_{\mathrm{rms}}$ and
$\tau_\eta=\ell_0^2/\eta$ in the two-scale model (assuming $Pm=O(1)$).
To see this, we borrow directly from \cite{moffatt78} and consider the ordering
of \equ{eq:Btilde},
\begin{equation}
\label{eq:Btildeasymp}
\underbrace{\dpart{\fluctvB}{t}}_{O(\fluctB_{\mathrm{rms}}/\tau_{c})}=
\underbrace{\curl{\left[\left(\fluctvU\times\meanvB\right)\right]}}_{O(\meanB/\tau_{\mathrm{NL}})}
+\underbrace{\curl{\left[\left(\fluctvU\times\fluctvB\right)-\left(\mean{\fluctvU\times\fluctvB}\right)\right]}}_{\mathrm{\small
    Tricky\, term}:\,O(\fluctB_{\mathrm{rms}}/\tau_{\mathrm{NL}})}
+\underbrace{\eta\Delta\fluctvB}_{O(\fluctB_{\mathrm{rms}}/\tau_\eta)}~.
\end{equation}
where $\meanvU=\vec{0}$ has been assumed for simplicity. 
The first important thing to notice here is that the typical rate of
change of $\fluctvB$ on the l.h.s. is ordered as the
inverse correlation time of the velocity field, because significant
field-amplification requires coherent flow stretching episodes whose
typical duration is $O(\tau_c)$. Another handwaving way to look at
this is to formally integrate \equ{eq:Btildeasymp} up to a time $t$,
and to approximate the integral of the tricky term by $\tau_c$ (the
times during which $\fluctvU$ remains self-correlated)
times the integrand. Either way, we see that the tricky term is
negligible compared to the l.h.s. in the limit $St\ll 1$. 
The problem, of course, is that $St$ is usually $O(1)$ for
standard fluid turbulence such as Kolmogorov turbulence.
In this case, the only limit in which the tricky term is 
formally negligible is the diffusion-dominated regime $Rm\ll
1$ in which the resistive term dominates. Unfortunately, this
regime is not very interesting from an astrophysical perspective
either.

\subsubsection{FOSA derivations of $\alpha$ and $\beta$ for
  homogeneous helical turbulence\label{FOSAderiv}}
The calculation of mean-field electrodynamics coefficients under FOSA
can be done with a variety of mathematical methods, most if not all of
which have already been skillfully laid out in textbooks and reviews
by the inventors and main practicioners of the theory
\citep[e.g.][]{moffatt78,krause80,branden05,raedler07}.
Many of these presentations are very detailed though, and can feel a
bit intimidating or overwhelming to newcomers in the field. The aim of
this paragraph is therefore to provide concise, low-algebra versions
of the calculations of the $\alpha$ and $\beta$ effects distilling the
essence of these methods.

In order to illustrate in the most simple possible way how this
kind of derivation works, we start with a physically intuitive
calculation in the $St\ll 1$, $Rm\gg 1$ regime. In this limit, both
the tricky term and the resistive term can
be neglected in \equ{eq:Btildeasymp}, but we cannot formally
neglect the first term on the r.h.s. of the equation, because there
is no particular reason to assume that $\fluctvB$ and
$\meanvB$ should be of the same
order. In fact, they are not and, as mentioned in \sect{MFansatz}, it
is precisely this term that makes a large-scale dynamo possible
in the first place. In order to calculate $\alpha$ and $\beta$, we
simply substitute the formal integral solution $\fluctvB(\vec{x},t)$ of
\equ{eq:Btildeasymp} into the expression of the mean EMF,
\begin{equation}
  \meanvEMF(\vec{x},t)\equiv\mean{\fluctvU(\vec{x},t)\times\fluctvB(\vec{x},t)}  = 
\mean{\fluctvU(t)\times\int_{t_0}^t\curl{\left[\fluctvU(\vec{x},t')\times\meanvB(\vec{x},t')\right]}\diff
  t'}~.\label{eq:EMFFOSA}
\end{equation}
In writing this expression, we have implicitly assumed that $t-t_0$
is much larger than the typical flow correlation and resistive
timescales, so  that the correlation
$\mean{\fluctvU(t)\times\fluctvB(t_0)}$
can be neglected, and the integral is understood to be independent of
the lower bound $t_0$. The spatial dependence of
$  \meanvEMF$ is to be understood as a large-scale
dependence on the scale of the mean field itself. The r.h.s. of \equ{eq:EMFFOSA} only
contains correlations that are quadratic in fluctuations, as a result
of the neglect of the tricky term in \equ{eq:Btildeasymp}, hence the
name Second Order Correlation Approximation (or First Order Smoothing
Approximation, if we take the equivalent view that it is the quadratic
term in \equ{eq:Btildeasymp} which has been neglected). 
Now, a bit of tensor algebra shows that \equ{eq:EMFFOSA} in the
isotropic case reduces to
\begin{equation}
\label{eq:EMFFOSAiso}
  \meanvEMF
(\vec{x},t) =  \int_0^t
\left[\hat{\alpha}(t-t')\meanvB(\vec{x},t')-\hat{\beta}(t-t')\curl{\meanvB}(\vec{x},t')\right]
\diff t'~,
\end{equation}
where 
\begin{equation}
  \hat{\alpha}=-\f{1}{3}\,\mean{\fluctvU(t)\cdot\fluctvomega(t')}~,\quad  \hat{\beta}=\f{1}{3}\,\mean{\fluctvU(t)\cdot\fluctvU(t')}~,
\end{equation}
and $\fluctvomega=\curl{\fluctvU}$.
The coefficients $\alpha$ and $\beta$ are uniform in this derivation
as a result of the assumption of statistical homogeneity.
Now, since the correlation time of the flow is assumed small compared
to its typical turnover time, we can approximate the integrals in
\equ{eq:EMFFOSAiso} by $\tau_c$ times the integrand at $t'=t$, i.e.
\begin{equation}
\label{eq:EMFFOSAresult}
  \meanvEMF = \alpha\,\meanvB-\beta\,\curl{\meanvB}
\end{equation}
with
\begin{equation}
  \label{eq:alphasmalltau}
\alpha= -\f{1}{3}\tau_c
\mean{\left(\fluctvU\cdot{\fluctvomega}\right)},\quad
\beta=\f{1}{3}\tau_c\mean{|\fluctvU|^2}~.
  \end{equation}
As expected by virtue of the FOSA, \equ{eq:EMFFOSAresult} has the form
of the isotropic version of the mean-field
expansion~(\ref{eq:meanfieldexpansion}) postulated in \sect{MFansatz},
but we have now also found explicit expressions for $\alpha$ and $\beta$.

The previous calculation relies on a description of the statistical
properties of the flow in configuration (correlation) space, and is
representative of the general
method used by Krause, Steenbeck and R\"adler in the 1960s to
develop a much broader theory of mean-field electrodynamical effects in
differentially rotating stratified flows \citep[their most important
papers, originally published in German, have been translated in
English by \cite{roberts71}, and form the core of the textbook
by][]{krause80}. However, mean-field coefficients can also
be derived by means of a spectral representation of the turbulent
dynamics. This formalism is
perhaps slightly less intuitive physically than the procedure
outlined above, but it makes the calculations a bit more compact
and straightforward in the general case, and is also frequently
encountered in the literature \citep[the reference textbook here
is][]{moffatt78}. In order to introduce
this alternative in a nutshell, we concentrate uniquely on the problem
of an $\alpha$ effect generated by helical homogeneous incompressible
turbulence acting upon a uniform mean magnetic field. However, we do
not \textit{a priori} restrict the calculation to one of the two
possible FOSA limits in this case and, in anticipation of a further
discussion of rotational effects in \sect{SCeffect}, we do not assume
isotropy from the start either \citep[a more detailed version of this
derivation, also including a determination of the $\boldsymbol{\beta}$
tensor for a non-uniform field, can be found
in][]{moffatt82}. Denoting the space-time Fourier transforms of $\fluctvU$ and
$\fluctvB$  over the small-scale, fast variables by
$\hat{\fluctvU}(\vec{k},\omega)$ and $\hat{\fluctvB}(\vec{k},\omega)$,
we first solve \equ{eq:Btildeasymp} in spectral space subject to FOSA as
\begin{equation}
\hat{\fluctvB}(\vec{k},\omega)=\f{i\left(\meanvB\cdot\vec{k}\right)}{\left(-i\omega+\eta
    k^2\right)}\hat{\fluctvU}(\vec{k},\omega)~.
\label{eq:Btildespec}
\end{equation}
Substituting this expression into the general expression for the mean EMF,
\begin{equation}
  \meanEMF_i=\int\!\!\int\!\!\int\!\!\int \varepsilon_{ilm}
   \mean{\hat{\fluctU}_l^*(\vec{k}',\omega')\hat{\fluctB}_m(\vec{k},\omega)}\mathrm{e}^{i[(\vec{k}-\vec{k}')\cdot\vec{x}-(\omega-\omega')t]}  \diff \vec{k}\,  \diff \omega\,\diff \vec{k}'\,  \diff \omega'~,
 \end{equation}
 and introducing the spectrum tensor of the turbulence $\boldsymbol{\Phi}_{lm}(\vec{k},\omega)$, defined by
 \begin{equation}
      \mean{\hat{\fluctU}_l^*(\vec{k},\omega)\hat{\fluctU}_m(\vec{k}',\omega')}=\boldsymbol{\Phi}_{lm}(\vec{k},\omega)\delta(\vec{k}-\vec{k}')\delta(\omega-\omega')~,
    \end{equation}
    we find after integration over $\vec{k}'$ and $\omega'$ that
    \begin{equation}
  \meanEMF_i=\left(\int\!\!\int\f{i\varepsilon_{ilm}k_j}{\left(-i\omega+\eta
    k^2\right)}\boldsymbol{\Phi}_{lm}(\vec{k},\omega)\,\diff \vec{k}\,
\diff \omega\right) \meanB_j~.
\label{eq:alphaspec}
\end{equation}
Recalling the mean-field ansatz for a uniform mean field,
$\meanEMF_i=\alpha_{ij}\meanB_j$, the term in
parenthesis is easily identified with $\alpha_{ij}$. In order to
simplify this expression further, we introduce the helicity spectrum
function
\begin{equation}
    H(\vec{k},\omega)=-i k_n\varepsilon_{nlm} \boldsymbol{\Phi}_{lm}(\vec{k},\omega)~,
  \end{equation}
and remark that 
  \begin{equation}
  \varepsilon_{ilm}
  \boldsymbol{\Phi}_{lm}(\vec{k},\omega)=\f{k_i}{k^2}H(\vec{k},\omega)
  \label{eq:helicityincompressible}
\end{equation}
for an incompressible flow (i.e., the l.h.s., being a cross-product of
two vectors $\hat{\fluctvU}(\vec{k},\omega)$ and
$\hat{\fluctvU}^*(\vec{k},\omega)$ both perpendicular to $\vec{k}$, is
oriented along $\vec{k}$). Substituting
\equ{eq:helicityincompressible} into \equ{eq:alphaspec} and using the
condition $H(-\vec{k},-\omega)=H(\vec{k},\omega)$ deriving from the
reality of $\fluctvU$ to eliminate the imaginary part of the
integral, we arrive at the result
\begin{equation}
  \alpha_{ij}=-\eta\int\!\!\int
  \f{k_ik_j}{\omega^2+\eta^2k^4}H(\vec{k},\omega)\,\diff\vec{k}\,\diff{\omega}
  \label{eq:alphaspecaniso}
\end{equation}
showing that the $\boldsymbol{\alpha}$ tensor is a weighted integral
of the helicity spectrum of the flow under generic anisotropic
conditions. In the isotropic case,
$H(\vec{k},\omega)=H(k,\omega)$, and $\alpha_{ij}=\alpha\,\delta_{ij}$
with
  \begin{equation}
\alpha=\f{\alpha_{ii}}{3}=-\f{\eta}{3}\int\!\!\int
\f{k^2}{\left(\omega^2+\eta^2k^4\right)}H(k,\omega)\,\diff\vec{k}\,\diff{\omega}~.
\label{eq:alphaisospec}
\end{equation}
One of the main advantages of the spectral formalism, clearly, is that it
makes it easy in principle to investigate either analytically or
numerically the dynamo properties of flows with different prescribed
spectral distributions, from simple ``monochromatic''  helical wave
fields to more complex random flows characterised by a broad range
of frequencies and wavenumbers. As stressed by \cite{moffatt78} though,
one must be wary of the application of the formalism at large $Rm$ if
the flow under consideration has some significant helicity and energy
at frequencies $\omega<u_0/\ell_0$, in which case the FOSA approximation
cannot be justified. On the other hand,
we can safely use \equ{eq:alphaisospec} to derive $\alpha$
analytically in the $St=O(1)$, $Rm\ll 1$ limit. In this case,
$\omega\ll \eta k^2$ and the weight function in the integral in
\equ{eq:alphaisospec} can be approximated by $1/(\eta^2k^2)$. 
Using the property
  \begin{equation}
  \mean{\left(\fluctvU\cdot{\fluctvomega}\right)}=\int\!\!\int
  H(\vec{k},\omega)\,\diff\vec{k}\,\diff{\omega}~,
\end{equation}
we obtain the low-$Rm$ result \citep{moffatt70}
\red{
  \begin{equation}
 \label{eq:alphalowRm}
\alpha=-\f{1}{3}\tau_{\eta}^{(H)}
\mean{\left(\fluctvU\cdot{\fluctvomega}\right)}~
\end{equation}
with
\begin{equation}
  \tau_{\eta}^{(H)}=\f{1}{\eta}\f{\displaystyle{\int\!\!\int 
    k^{-2}H(\vec{k},\omega)\,\diff\vec{k}\,\diff{\omega}}}{\displaystyle{%
    \int\!\!\int
  H(\vec{k},\omega)\,\diff\vec{k}\,\diff{\omega}}}~.
\end{equation}}
An extension of the calculation to a non-uniform mean field similarly
leads to
\red{
\begin{equation}
  \label{eq:betalowRm}
  \beta=\f{1}{3}\tau_{\eta}^{(E)}\mean{|\fluctvU|^2}~,
\end{equation}
with $\tau_{\eta}^{(E)}$ similarly defined as a weighted integral of the
kinetic energy spectrum $E(k,\omega)$.}

The previous results all lead to the comforting conclusion
that the $\alpha$ effect is directly proportional to the net kinetic
helicity of the flow, in line with Parker's original intuition of
large-scale dynamo action driven by cyclonic convection. The only
difference between \equ{eq:alphasmalltau} and \equ{eq:alphalowRm}
derived in the low-$St$ and low-$Rm$ FOSA limits respectively is the
characteristic time involved. In both cases, the results
translate that the twisting of magnetic-field lines
into Parker loops occurs on a short time compared to the
dynamical turnover time. The amount of coherent twisting by each
individual swirl is limited by magnetic diffusion in the $Rm\ll 1$
case, and by the fast decorrelation of the flow in the $St\ll 1$
case. The small but systematic effects
of these ``impulsive'' twists simply add up statistically.

The problem, however, gets significantly more complicated if we
consider the more realistic and astrophysically relevant regime
$St=O(1)$, $Rm\gg 1$ for which the field is essentially frozen-in.
For instance, when $\tau_c=O(\ell_0/u_0)$, a Parker loop can do a full
$360^\circ$ turn before the swirl decorrelates, leaving us with zero
net magnetic field in the direction perpendicular to the original field
orientation. While this scenario is extreme, the point is that the net
effect of a statistical ensemble of generic turbulent swirls with
$St=O(1)$ at large $Rm$ is much harder to assess than in the
calculations above due to cancellation effects. This kind of
complication notably creates some significant difficulties with the
determination of the kinematic value of $\alpha$ at large $Rm$ for
some families of chaotic flows with long correlation times
\citep*{courvoisier06}, although not
necessarily for generic isotropically-forced helical turbulence
with $St=O(1)$ \citep{sur08}. On top of all that, we will also
discover later in \sect{coexistence} that the neglect of the tricky
term is actually generically problematic at large $Rm$, even when
$St\ll 1$, as a result of the excitation of small-scale dynamo fields.

Note finally that the FOSA result~(\ref{eq:alphalowRm}), combined with
$\beta\ll \eta$, ensures that the theory is consistent with the
two-scale assumption in the regime of low $Rm$. In the $\alpha^2$
dynamo problem, for instance, the wavenumbers $k$ at which growth occurs
at low $Rm$ are of the order \red{$k\,\ell_0\sim \alpha\ell_0/\eta\sim Rm^2\ll
1$ (assuming $|\fluctvomega|_{\mathrm{rms}}\sim u_{\mathrm{rms}}/\ell_0$)}. In
the $St\ll 1$, $Rm\gg 1$ regime, on the other hand,
\equ{eq:alphasmalltau} applies, $\beta\gg\eta$, and
\red{$k\,\ell_0\sim\alpha\ell_0/\beta\sim
\mean{\left(\fluctvU\cdot{\fluctvomega}\right)}\ell_0/
\mean{\fluctvU^2}$. The theory is therefore formally only
self-consistent in this regime if the flow has small fractional
helicity, i.e. $|\fluctvomega|_{\mathrm{rms}}\ll u_\mathrm{rms}/\ell_0$.}

\subsubsection{Third-order-moment closures: EDQNM and $\tau$-approach*\label{MTAclosure}}
While the FOSA closure has been central to the historical development 
of mean-field electrodynamics, its very limited formal domain of
validity implies that its predictive power and practical applicability 
is also formally very limited. Other closure schemes, such as the
eddy-damped quasi-normal Markovian (EDQNM) \citep*{orszag70,pouquet76}
or minimal $\tau$ approximation (MTA) closures
\citep*{vainshtein83,kleeorin90,blackman02a} operating
at the next order in the hierarchy of moments, have been implemented
in the context of large-scale dynamo theory in order to attempt to
remedy this problem. To illustrate the idea behind these closures,
consider the time-derivative of the mean EMF,
\begin{equation}
\label{eq:meanEMF}
\dpart{\meanvEMF}{t}\equiv\mean{\fluctvU\times\dpart{\fluctvB}{t}}+\mean{\dpart{\fluctvU}{t}\times\fluctvB}~,
\end{equation}
and substitute $\dpartshort{\fluctvB}{t}$ by its expression in
\equ{eq:Btildeasymp}, and $\dpartshort{\fluctvU}{t}$ by its
expression given by the Navier-Stokes equation for
$\fluctvU$, assuming that the contribution of the Lorentz
force is of the form $\meanvB\cdot\grad{\fluctvB}$
\citep[this is formally only valid for small
$|\meanvB| \ll\fluctB_\mathrm{rms}$, see e.g. discussion in
][]{proctor03}. From this equation, it is easy to see that the tricky
term in \equ{eq:Btildeasymp} and the inertial term in the Navier-Stokes
equation introduce third-order correlations in the resulting evolution
equation for the second-order moment $\meanvEMF$.
In order to close this equation, one possibility is to model  the
effects of these triple correlations (fluxes) in terms of a simple
relaxation of the EMF, $-\meanvEMF/\tau$
($\tau$, the relaxation time, is sometimes assumed to be scale
dependent). This qualitatively amounts to modelling the effects of the
turbulent fluxes as a diffusion-like/damping effect.
The result of the simplest version of this kind of calculation after
applying this new closure ansatz is \citep[e.g.][]{blackman02a,
  branden05}:
\begin{equation}
\label{eq:MTA}
\dpart{\meanvEMF}{t}=\alpha'\,\meanvB-\beta'\,\curl{\meanvB}-
\frac{\meanvEMF}{\tau}~,
\end{equation}
where now
\begin{equation}
\label{eq:MTAalpha}
\alpha'=-\f{1}{3}\left[\,\mean{\fluctvU\cdot\fluctvomega}-\mean{(\curl{\fluctvB})\cdot\fluctvB}\,\right]~,\quad\beta'=\f{1}{3}\mean{|\fluctvU|^2}.
\end{equation}
One seeming advantage of this scheme, compared to FOSA, is
that $\tau$ is not formally required to be small in comparison to the
typical eddy turnover time. However, it is not a panacea either,
because there is no guarantee that the net effect of
triple correlations is to relax the mean EMF in the form prescribed by
an MTA-like closure. In fact, it is all but certain that this is not
strictly true for most turbulence problems, and that even the
qualitative validity of this kind of prescription is problem- and
regime-dependent. In the mean-field dynamo context, it has for instance
notably been found that analytical predictions that come out of
the MTA are not always in agreement with the FOSA predictions in the
(admittedly rather peculiar) regimes in which the latter is
formally valid \citep[e.g.][see \sect{SCeffect} below for a
discussion of a specific example]{raedler07}. On the other hand,
analyses of numerical simulations of turbulent transport and
large-scale dynamo problems suggest that an MTA closure
with $\tau/\tau_{\mathrm{NL}}=O(1)$ is not entirely
unfounded either in some common fluid turbulence regimes
\citep*{branden04,branden05b} (of course, that $\tau$ in this model
should be of the order of $\tau_{\mathrm{NL}}$ is expected from a
simple order of magnitude analysis of the different correlation
functions involved in the closure). Overall, it seems like we have no
choice at the moment but to accept the theoretical uncertainties and
confusion that come with this kind of problem. Closures are 
notoriously difficult in all areas of turbulence research
\citep{krommes02}, and dynamo theory is no exception to this.

Keeping these caveats in mind, it is nevertheless interesting to note 
that the MTA closure applied to the helical dynamo problem gives rise
in \equ{eq:MTAalpha} to a magnetic contribution to the $\alpha$ effect
 associated with small-scale helical magnetic
fluctuations interacting with the turbulent velocity field. The
questions of the exact meaning, validity and applicability of this
result are beyond the realm of pure kinematic theory, and will be
critically discussed in \sect{largescalesat} in the context of
saturation of large-scale dynamos.

\subsection{Mean-field effects in stratified, rotating, and shearing flows\label{otherstat}}
While it is eminently instructive and captures the essence of the
$\alpha$ effect, the problem of large-scale dynamo in a homogeneous,
isotropic helical flow  discussed so far in this
section is quite idealised, and does not capture the full physical
complexity of large-scale
dynamos in astrophysical, planetary and even experimental MHD
 flows involving a combination of rotation, shear, stratification
 and thermal physics. \red{Intuitively, we may for instance expect that
statistical magnetic-field generation effects and turbulent magnetic diffusion
  become anisotropic in the presence of a large-scale
  stratification, inhomogeneity, or rotation, or in the presence of
  a large-scale magnetic field in the dynamical regime of the dynamo}.
 This, in itself, gives rise to significant technical
 complications. The problem, however, is not just one of generalising
 some already familiar effects to less symmetric cases.
 Inhomogeneity, stratification, rotation, and shear all break
 some symmetry and can therefore be expected to give rise to a
 zoo of additional large-scale statistical effects already at the
 kinematic level.

  Before supercomputing became everyday routine, mean-field
  electrodynamics provided the only available, albeit not necessarily
  always very physically transparent, means to explore this seemingly
  outstanding statistical complexity in a somewhat systematic way.
  Despite all their limitations, theoretical
  calculations of this kind have therefore played an essential role in the
  development of large-scale astrophysical dynamo theory and have had
  a profound, long-lasting impact on how the community speaks and
  thinks of large-scale statistical effects. The classic reference on this
  subject is the book of \cite{krause80}. The aim of the next few
  paragraphs is to single out several such effects that appear to
  be most generically relevant to large-scale astrophysical, planetary
  or experimental dynamos, and to explain how they arise within the
  framework of mean-field electrodynamics.
  
\subsubsection{$\alpha$ effect in a stratified, rotating flow\label{alphastrat}}
As we discovered in \sect{LSsym}, an $\alpha$ effect is only
possible in flows that are not parity-invariant, a particular example of
which is helically-forced, pseudo-isotropic homogeneous turbulence. The kinematic
derivations presented in \sect{FOSAderiv} showed that $\alpha$ in this
problem is directly proportional (in the FOSA regimes) to the
net average kinetic helicity of the flow. Here, we will see that the $\alpha$
effect is also non-zero in a rotating, stratified flow involving a
gradient of density and/or kinetic energy, such as stratified rotating
thermal convection typical of many stellar interiors (see
e.g. \cite{branden05} for a similar discussion). Beyond its obvious
astrophysical relevance, this particular problem provides a good
illustration of how one can use symmetry arguments to simplify
matters. Zooming in on a patch of stratified, rotating
turbulence, we can locally identify two preferred directions in the
system, that of gravity denoted by the unit vector $\hat{\vec{g}}=\vec{g}/g$,
and that of rotation, denoted by the unit vector $\hat{\vOmega}=\vOmega/\Omega$.
The pseudo-tensor $\boldsymbol{\alpha}$ can only be
constructed from these two vectors in the following way
(to lowest order in $g$ and $\Omega$):
\begin{equation}
\alpha_{ij}=\alpha_0\,
\hat{\vec{g}}\cdot\hat{\vOmega}\,\delta_{ij}+
\alpha_1\hat{g}_j\hat{\Omega}_i+\alpha_2\hat{g}_i\hat{\Omega}_j~.
\end{equation}
Of course, we still have to calculate the three mean-field
coefficients. This is difficult to achieve in the general case,
but explicit expressions can be obtained for a few
analytically-prescribed flows. One of the most well-known results
in this context is the expression
\begin{equation}
  \alpha_0\,\hat{\vec{g}}\cdot\hat{\vOmega}=-\f{16}{15}\tau_c^2\mean{|\fluctvU|^2}\vOmega\cdot\grad{\,\ln\left(\rho\,
      \sqrt{\mean{|\fluctvU|^2}}\right)}~,\quad  \alpha_1=\alpha_2=-\f{\alpha_0}{4}~
  \label{eq:alphaconv}
\end{equation}
derived in the $St\ll 1$ FOSA regime for an anelastic flow model
(characterised by $\div{~\left(\rho\fluctvU\right)}=0$)
encapsulating the effects of stratification through a simple
exponential vertical-dependence parametrisation of the background
density and turbulence intensity
\citep{krause67,steenbeck69}. \Equ{eq:alphaconv} obviously provides
a more explicit mathematical validation of Parker's idea that rotating convection
can generate an $\alpha$ effect than \equ{eq:alphasmalltau}, however it
also suggests that some stratification along the rotation axis, not
just rotation, is actually required in a rotating flow to obtain a
non-zero $\boldsymbol{\alpha}$ tensor. The physical reason
underlying this result was identified by \cite{steenbeck66}.
In a stratified environment, rising fluid expands horizontally, while
sinking fluid is compressed. Thereby, and upon the action of the Coriolis
force, upflows in the northern hemisphere are made to rotate clockwise,
while downflows are made to rotate counter-clockwise. In both cases,
the flow acquires negative helicity. The effect is opposite in the
southern hemisphere. The net result, expressed by \equ{eq:alphaconv},
is an antisymmetric distribution of $\alpha$ with respect to the
equator. This argument and \equ{eq:alphaconv} also predict
that $\alpha_{\varphi\varphi}$ (or $\alpha_{yy}$ in the Cartesian model
with a vertical rotation axis), the key coefficient that couples back the
toroidal field to the poloidal field in an $\alpha\Omega$ dynamo, is
positive in the northern hemisphere. We will discover later in
\sect{numexplor} that this result is actually problematic in the solar
dynamo context.

The stratification effect described above formally vanishes in
the incompressible Boussinesq regime of convection rotating about a
vertical axis, because in this limit all motions are assumed to take
place at a vertical scale much smaller than the typical density
scale-height. Note however that the presence of vertical boundaries in
the system also generates converging and diverging motions that get
acted upon by the Coriolis force. It is actually this type of
horizontal motions, not expansions or compressions in a  stratified
atmosphere, that Parker apparently had in mind when he
introduced the idea of an  $\alpha$ effect generated by cyclonic
convection (he used the words ``influxes'' and ``effluxes''). In his
 work, Parker focused on helical motions of a given sign,
but in Boussinesq convection the dynamics of overturning rotating
convective motions actually leads to the generation of flow helicity
of opposite signs at the top and bottom. In plane-parallel Boussinesq
convection between two plates and rotating about the vertical axis
(the rotating Rayleigh-B\'enard problem) in particular, the existence of an
``up-down'' dynamical mirror symmetry with respect to the mid-plane of
the convection layer \citep{chandra61} implies that the profile of
(horizontally-averaged) flow helicity induced by the Coriolis force
is antisymmetric in $z$. Accordingly, the volume-averaged flow
helicity of the system is zero in this case%
\footnote{This, however, does not formally rule out the existence
in rotating Rayleigh-B\'enard convection of helical large-scale
kinematic dynamo eigenmodes consistent with the vertical boundary
conditions and with the symmetry of the vertical profile of
$\alpha(z)$, see e.g. \cite{soward74} and \cite{hughes08}.}.
In the density-stratified case, this up-down
symmetry is broken
\citep[e.g.][]{graham75,gough76,graham78,massaguer80} and the vertical
profile of horizontally-averaged helicity has no particular symmetry 
\citep{kapyla09}. The volume-averaged helicity does not therefore in
general vanish in this case, illustrating further that it is
stratification effects that generate a helicity imbalance in this
problem.

  \subsubsection{Turbulent pumping*\label{pumping}}
  In an inhomogeneous fluid flow, the statistics of the velocity
  field depend on the coordinate along the inhomogeneous direction
  $\hat{\vec{g}}$. If we thread such a flow with a
  large-scale magnetic field, then the latter will be
  expelled from the regions of higher turbulence intensity to the
  regions of lower turbulent intensity. In a stratified turbulent
  fluid layer in particular, $u_{\mathrm{rms}}$ usually decreases
  along $\vec{g}$ as density increases, and we therefore expect
  the field to be brought downwards. This is usually referred to as
  diamagnetic pumping \citep{zeldovich56,raedler68,moffatt83}. While
  this effect is not inducing magnetic field on its own, it can
  contribute to the large-scale transport of the field, and may
  notably lead to its accumulation deep into stellar interiors.
 
  From a mean-field theory perspective, inhomogeneity implies
  the presence of a non-vanishing
  $\boldsymbol{\gamma}\times\meanvB$ term in
  \equ{eq:MFexpansiongeneral}, as we now have a preferred
  direction to construct a true
  vector $\boldsymbol{\gamma}=\gamma\,\vec{g}~$. Remark that
$\boldsymbol{\gamma}\times\meanvB$ has the same
 form as $\meanvU\times\meanvB$, so that 
 $\boldsymbol{\gamma}$ is easily interpreted as an effective
 large-scale velocity advecting the field. A detailed FOSA
 derivation in the $St\ll 1$ limit, for a simple flow model
 encapsulating the effects of inhomogeneity through an exponential
 dependence of the turbulence intensity, gives \citep{krause67} 
  \begin{equation}
  \boldsymbol{\gamma}=-\f{1}{6}\tau_c\grad{\mean{|\fluctvU|^2}}~,
\end{equation}
confirming the diamagnetic character of the effect. Note finally that
the further presence of rotation or large-scale flows in the problem
raises additional contributions to $\boldsymbol{\gamma}$, not all of
which are oriented along $\vec{g}$ \citep[see e.g.][]{raedler06}. In
spherical geometry, this notably makes for the possibility of
large-scale pumping along the azimuthal direction.

\subsubsection{R\"adler and shear-current effects*\label{SCeffect}}
  Additional statistical effects distinct from the standard
  $\alpha$ effect formally arise if the turbulence is rendered
  anisotropic by the presence of large-scale rotation or shear, even
  in the absence of stratification. These effects are not
  necessarily very intuitive physically but their mathematical form
  can be captured easily within the framework of mean-field
  electrodynamics using some of the symmetry arguments introduced in
  \sect{LSsym}. Namely, in the presence of rotation,
  we can now form a non-vanishing  
  $\boldsymbol{\delta}\times\left(\curl{\meanvB}\right)$
  term in the expression~(\ref{eq:MFexpansiongeneral}) for the mean EMF by making
  $\boldsymbol{\delta}$ proportional to the rotation vector,
  $\boldsymbol{\delta}\equiv\delta_\Omega\vOmega$.
  The associated statistical effect is generally referred to as
  the R\"adler effect after its original derivation by
  \cite{raedler69a,raedler69b} or, more explicitly,
  as the ``$\vOmega\times\meanvJ$'' effect.
  Similarly, we may expect that $\boldsymbol{\delta}$ is formally
  non-vanishing if the turbulence is sheared by a large-scale flow
  $\meanvU$ with associated vorticity
  $\meanvW=\curl{\meanvU}$, in which case
  $\boldsymbol{\delta}\equiv\delta_W\meanvW$. This
  effect is generally referred to as the shear-current effect,
  or ``$\meanvW\times\meanvJ$'' effect
  \citep*{rogachevskii03}.
  
  The R\"adler effect has a subtle connection to parity
  invariance and flow helicity. To see this, let us consider
  a flow whose statistics are rendered anisotropic (more precisely
  axisymmetric) by the presence of rotation. In particular, the energy
  and helicity spectra are such that
  $E(\vec{k},\omega)=E(k,\mu,\omega)$ and
  $H(\vec{k},\omega)=H(k,\mu,\omega)$, where
  $\mu=\hat{\vec{k}}\cdot\hat{\vOmega}$. A spectral FOSA derivation
  similar to that outlined in \sect{FOSAderiv} shows that \citep{moffatt82}
  \begin{equation}
    \delta_\Omega\Omega=\f{2\eta^2}{5}\int\!\!\int \f{k^3\mu\,\omega H(k,\mu,\omega)}{\left(\omega^2+\eta^2k^4\right)^2}\,\diff\vec{k}\,\diff{\omega}~,
    \label{eq:deltaomegaaxi}
  \end{equation}
  and
  \begin{equation}
    \alpha_{ij}=\alpha\,\delta_{ij}+\alpha_1\left(\delta_{ij}-3\
    \hat{\Omega}_i    \hat{\Omega}_j\right)~,
        \label{eq:alphatensaxi}
      \end{equation}
      with 
      \begin{equation}
\alpha=-\f{\eta}{3}\int\!\!\int\f{k^2}{\left(\omega^2+\eta^2k^4\right)}H(k,\mu,\omega)\,\diff\vec{k}\,\diff{\omega}
        \label{eq:alphaaxi}
      \end{equation}
      and
            \begin{equation}
\alpha_1=\f{\eta}{6}\int\!\!\int\f{k^2\left(3\,\mu^2-1\right)}{\left(\omega^2+\eta^2k^4\right)}H(k,\mu,\omega)\,\diff\vec{k}\,\diff{\omega}~.
        \label{eq:alpha1axi}
      \end{equation}
    These expressions show that both $\delta_\Omega$ and $\alpha_{ij}$
    are weighted integrals of the helicity spectrum of the flow, and
    therefore vanish if the flow has no helicity at all.
    However, notice that the weight function in the $\delta_\Omega$ integral
    is odd in $\mu$, and even in the $\alpha_{ij}$ integrals.
    This implies that $\delta_\Omega$, unlike the
    $\alpha$ effect, can be non-zero even if there is an equal
    statistical amount of right- and left-handed velocity fluctuations,
    and the flow as a whole has zero net-helicity. In particular,
    $\delta_\Omega\neq 0$ and $\boldsymbol{\alpha}=\tens{0}$ if
    $H(k,-\mu,\omega)=-H(k,\mu,\omega)$. This situation corresponds
    to a flow with an up-down  symmetry with respect to planes
    perpendicular to the rotation axis, such as the incompressible
    Rayleigh-B\'enard convection rotating about the vertical
    axis discussed in \sect{alphastrat}. A corollary of
    this result is that the R\"adler effect is formally
    present in unstratified rotating flows. The derivation of
    \cite{moffatt82} also makes it clear that the effect survives
    even if the forcing of the flow is itself non-helical, and
    all the flow helicity is induced by the effects of the Coriolis
    force. In other words, we expect the R\"adler effect to
    be present even in incompressible, homogeneous turbulence
    forced isotropically and non-helically, as long as the system
    rotates. In a generic stratified, rotating, turbulent
    system, both $\alpha$ and R\"adler effects are expected to
    coexist, although it is sometimes argued that the latter
    should be smaller due to the presence of an extra
    slow spatial derivative \red{($\curl{\meanvB}$)}
    in the expression of its EMF. Analyses of
    simulations of stratified rotating convection in which both effects
    are present suggest that the R\"adler effect is smaller than the
    $\alpha$ effect, but not altogether negligible, with
    $\delta_\Omega$ being comparable in magnitude to the isotropic
    turbulent diffusivity coefficient $\beta$ \citep{kapyla09}.

  To understand how the R\"adler and shear-current effects may affect
  large-scale dynamos in rotating shear flows, let us
  have a look at the mean-field dynamo problem for small-scale
  homogeneous turbulence forced isotropically and non-helically in the
  simplest possible unstratified, rotating shearing sheet configuration,
  $\vOmega=\Omega\,\vec{e}_z$, $\vU_S=-Sx\,\vec{e}_y$,
  for which $\meanvW=-S\,\vec{e}_z$ and the only non-zero
  components of the deformation tensor are
  $\mean{D}_{xy}=\mean{D}_{yx}=-S/2$.
  Under these assumptions, there are no mean $\boldsymbol{\alpha}$ and
  $\boldsymbol{\gamma}$ effects and the kinematic evolution equations
  for the $x$ and $y$ components of a $z$-dependent mean magnetic field
   (defined as the average over $x$ and $y$ of the total magnetic
   field) can be cast in the simple form 
      \begin{eqnarray}
      \dpart{\meanB_x}{t}& = &-\eta_{yx}\ddpart{\meanB_y}{z}+(\eta+\eta_{yy})\ddpart{\meanB_x}{z}~,\label{Bxoffdiag}\\
      \dpart{\meanB_y}{t} & = & -S B_x-\eta_{xy}\ddpart{\meanB_x}{z}+(\eta+\eta_{xx})\ddpart{\meanB_y}{z}~\label{Byoffdiag}~,
      \end{eqnarray}
      where we have introduced a contracted generalised anisotropic
      turbulent diffusion tensor $\eta_{ij}$ appropriate to the
      configuration of the problem, namely
      \begin{equation}
        \label{eq:defetaij}
        b_{ijz}=\eta_{il}\varepsilon_{jzl}~.
      \end{equation}
      Using \equs{eq:defalphabeta}{eq:defkappa}, it can be shown that
      \begin{equation}
\eta_{xx}=\eta_{yy}=\beta~,
\end{equation}
\begin{equation}
  \eta_{yx}= \Omega\,\left(\delta_\Omega-\f{\kappa_{\Omega}}{2}\right)-S\,\left[\delta_W-\f{1}{2}\left(\kappa_W-\beta_D+\kappa_D\right)\right]~,
\end{equation}
\begin{equation}
  \eta_{xy} =
  -\Omega\,\left(\delta_\Omega-\f{\kappa_{\Omega}}{2}\right)+S\,\left[\delta_W-\f{1}{2}\left(\kappa_W+\beta_D-\kappa_D\right)\right]~,
  \end{equation}
where $\beta$ is the usual isotropic turbulent diffusion coefficient,
$\beta_D$ is an anisotropic contribution to the $\boldsymbol{\beta}$
tensor arising from the presence of the large-scale strain
$\mean{\tens{D}}$ associated with the shear flow, and $\kappa_\Omega$,
$\kappa_W$ and $\kappa_D$ are contributions to the
mean-field $\boldsymbol{\kappa}$ tensor arising (similarly to
$\delta_\Omega$ and $\delta_W$) from the presence of  rotation,
large-scale vorticity, and strain associated with the shear flow
\citep[for a detailed derivation, see][]{raedler06,squire15PRE}.

We see that all these effects amount to a special kind of off-diagonal
diffusion that couples the components of the magnetic field
perpendicular to the new special direction introduced in the
problem. Can these couplings, most importantly that associated with
the $\eta_{yx}$ coefficient that couples back the toroidal field to
the poloidal field, drive a dynamo of their own ? To discover this, we
can derive from \equs{Bxoffdiag}{Byoffdiag} the complex growth rate of a
horizontal mean-field mode with a simple $\exp (st+i k_zz)$ dependence,
\begin{equation}
  s=k_z\sqrt{\eta_{yx}\left(-S+\eta_{xy}k_z^2\right)}-(\eta+\beta) k_z^2~.
  \end{equation}
  Further assuming $|S|\gg \eta_{xy} k_z^2$, similarly to what
  we did when we derived the $\alpha\Omega$ dispersion relation in
  \sect{alphaomega}, we find that a necessary condition for dynamo
  growth is
  \begin{equation}
    S\eta_{yx}<0~,
    \label{radlersccondition}
  \end{equation}
  i.e. $S$ and $\eta_{yx}$ must have opposite sign.
  If the dynamo is realisable, the maximum growth rate and optimal
  wavenumber are
  \begin{equation}
    \label{eq:optimalraedlerSC}
    \gamma_{\mathrm{max}}=\f{|S\eta_{yx}|}{4\,(\eta+\beta)}~,\quad
      k_{z,\mathrm{max}}=\f{\left(S\eta_{yx}\right)^{1/2}}{2\,(\eta+\beta)}~.
 \end{equation}
  But under which  circumstance(s), if any, is the
  condition~(\ref{radlersccondition}) satisfied ? 
  To answer this question, we need to calculate $\eta_{yx}$ using a closure
  assumption such as the FOSA or MTA. At this point, things become tricky.
  Indeed, FOSA calculations \citep{krause80,moffatt82} show that the
  R\"adler effect alone can promote a dynamo ($\eta_{yx}S<0$) when
  the rotation is anticyclonic (corresponding to positive $S/\Omega>0$
  with our convention), but not the shear-current effect 
  \citep[$S\eta_{yx}\geq 0$,][]{raedler06,ruediger06,sridhar09,sridhar10,squire15PRE}.
  Calculations based on the MTA closure agree with FOSA calculations
  as far as the R\"adler effect is concerned, but find dynamo growth
  ($S\eta_{yx}<0$) for the shear-current effect
  \citep{rogachevskii03,rogachevskii04} ! Which one, if any,
  is right in practice ? We know that the results derived with FOSA at
  least hold rigorously if $Rm\ll 1$ or $St\ll 1$, which establishes
  the absence of a shear-current-driven dynamo  and the existence of
  a R\"adler-effect-driven dynamo in these limits (at least for the
  kind of sheared or rotating random flows considered in the FOSA
  derivations). However, there is no guarantee that these conclusions
  extend to standard MHD turbulence regimes. As discussed in
  \sect{MTAclosure}, the MTA is an empirical closure model that
  seems to be in reasonable agreement with numerical results for
  a few simple turbulent-transport problems with $St=O(1)$, but
  it cannot be rigorously shown to be valid in any particular
  regime either, leaving us in the dark for the time being. 

  Looking at the bigger picture, we see that the previous discussion
  of a shear-current-effect dynamo (or lack thereof) raises the more
  general, very intriguing question of the possibility of large-scale
  dynamo excitation in non-rotating sheared turbulence with zero mean
  helicity. This problem, including the sign of $S\eta_{yx}$ in
  non-rotating sheared turbulence, will be discussed much more
  extensively in \sect{sheardynamo} in the light of several recent
  numerical developments.
 
\subsection{Difficulties with mean-field theory at large $Rm$\label{coexistence}}
\subsubsection{The overwhelming growth of small-scale dynamo fields}
In the previous paragraphs, we discussed the very limited formal
asymptotic mathematical range of parameters under which kinematic
mean-field electrodynamics can be rigorously derived, but there is
actually an even bigger potential \textit{physical} problem looming
over the theory at large $Rm$. Indeed, we have learned in
\sect{smallscale} that homogeneous, isotropic turbulence independently
generates small-scale magnetic fluctuations $\fluctvB$ through a
small-scale dynamo effect, provided that $Rm$ exceeds a value
$Rm_{c,\mathrm{ssd}}$ of a few tens to a few hundreds, depending on
$Pm$ (\fig{figgrowthratePm}). This dynamo produces
equipartition field strengths over a dynamical (turbulent) turnover
timescale, which is usually much smaller than the rotation or
shearing timescales over which kinematic mean-field dynamo modes
grow. This strongly suggests that the derivation of a meaningful
solution of the generic problem of large-scale magnetic-field growth
at large $Rm$ should start from a state of saturated small-scale MHD
turbulence characterised by $\mean{|\fluctvB|^2}\gg|\meanvB|^2$
and $\mean{\rho|\fluctvU|^2}\sim\mean{|\fluctvB|^2}$,  rather
than from a state of hydrodynamic turbulence. This was 
recognised as a major issue for large-scale astrophysical dynamo
theory after the work of \cite{kulsrud92} described in
\sect{largepmselection} \citep[see also the review by][]{kulsrud97},
although obviously the tricky questions of the interactions of
small-scale MHD turbulence and large-scale dynamos already pervaded
several earlier calculations, including the \cite{pouquet76} paper
mentioned earlier (see also \cite{ponty11} for a specific numerical
example of a classic large-scale helical dynamo being overwhelmed by
a small-scale dynamo as $Rm$ increases).

We are unfortunately not in a very good place to address this
problem at this stage of exposition of the theory.
A key observation is that small-scale dynamo modes were purely and simply
discarded in the FOSA treatment as a result of the neglect of the
tricky term in \equ{eq:Btilde}. Mathematically, we can think of these
modes as fast-growing homogeneous solutions of \equ{eq:linear},
which have nothing to do with $\meanvB$. In other words,
our earlier assumption that magnetic fluctuations $\fluctvB$
are small and uniquely the product of the stretching and tangling of
the mean field $\meanvB$ is incorrect at large $Rm$. This in turn
begs the question of the interpretation and practical applicability 
of the linear mean-field ansatz~(\ref{eq:meanfieldexpansion}) in this 
regime \citep{cattaneo09b,hughes10,cameron16}. In order to make
progress on a unified theory of large- and small-scale
dynamos, our first priority should therefore be to derive
a linear theory that at the very minimum accommodates both types 
of dynamos. Such a theory is  available in the form of a helical
extension of the Kazantsev model discussed in \sect{kazantsev}.

\subsubsection{Kazantsev model for helical turbulence*\label{kazantsevhelical}}
A generalisation of the Kazantsev model to helical flows was first
derived by \cite{vainshtein70} shortly after the introduction of
mean-field electrodynamics. Vainshtein used a Fourier-space
representation \citep[see  also][]{kulsrud92,berger95}, but in the
following we will stay in the correlation-vector space as a matter
of continuity with \sect{kazantsev}. Helicity can be accommodated in
the model by adding a new term to the correlation tensor of the
flow, 
\begin{equation}
  \label{eq:kappaijhelical}
  \kappa^{ij}(\vec{r})=\kappa_N(r)\left(\delta^{ij}-\frac{r^ir^j}{r^2}\right)+\kappa_L(r)\f{r^ir^j}{r^2}+g(r)\varepsilon^{ijk}r^k~,
\end{equation}
where $g$ is a scalar function that vanishes for parity-invariant
flows. A corresponding helical term is added to the magnetic
correlation tensor,
\begin{equation}
  \label{eq:Hijhelical}
H^{ij}(\vec{r},t)=H_N(r,t)\left(\delta^{ij}-\frac{r^ir^j}{r^2}\right)+H_L(r,t)\frac{r^ir^j}{r^2}+K(r)\varepsilon^{ijk}r^k~.
\end{equation}
Proceeding as in \sect{kazantsev} to close the problem, we obtain 
a system of two coupled equations  for $H_L(r)$ and $K(r)$
\citep*{vainshtein86,subramanian99,boldyrev05},
\begin{eqnarray}
\label{eq:HLequationhelical}
\dpart{H_L}{t} & = & \frac{1}{r^4}\dpart{}{r}\left(r^4\kappa\dpart{H_L}{r}\right)+G H_L-4hK~,
\\
\label{eq:Kequationhelical}
\dpart{K}{t} & = &
\f{1}{r^4}\dpart{}{r}\left[r^4\dpart{}{r}\left(\kappa K+h
    H_L\right)\right]~,
\end{eqnarray}
where $\kappa(r)$ is given by \equ{eq:kappaequation},
$h(r)=g(0)-g(r)$, and $G(r)=\kappa''+4\kappa'/r$. When $g=0$ (no
helicity in the flow), $H_L$ decouples from $K$,
\equ{eq:HLequationhelical} reduces to \equ{eq:HLequation}, and
\equ{eq:Kequationhelical} for the evolution of the magnetic helicity
correlator $K$ reduces to a diffusion equation with no source term. 
From there, it can be shown
\citep{vainshtein70,subramanian99,boldyrev01,boldyrev05}
that taking the $r\rightarrow\infty$ limit of
this problem leads to a mean-field $\alpha^2$ equation reminiscent
of \equ{eq:MFequation},
\begin{equation}
\label{eq:MFequationkazantsev}
\dpart{\left<\vB\right>}{t}=\alpha\curl{\left<\vB\right>}+(\eta+\beta)\Delta\left<\vB\right>~,
\end{equation}
where $\alpha\equiv g(0)$, $\beta\equiv \kappa_L(0)/2$, and the
mean field here is to be interpreted as an average of the full magnetic 
field over the Kraichnan velocity-field ensemble. Because the velocity
is correlated on scales $\ell_0$, this averaging procedure is
equivalent to a spatial average over scales much larger than
$\ell_0$. 

\begin{figure}
\centering\includegraphics[width=\textwidth]{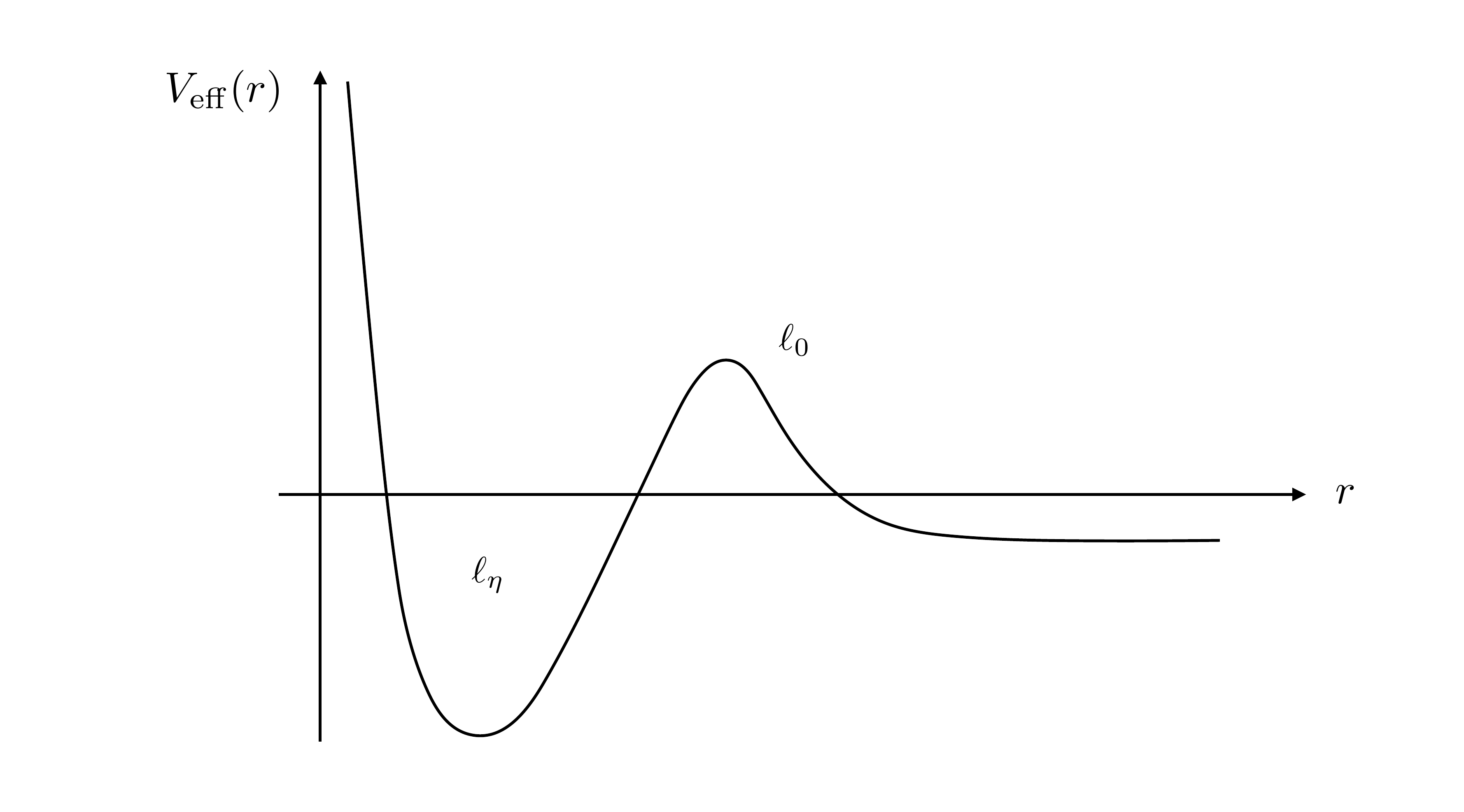}
\caption{Helical Kazantsev potential as a function of correlation length
  $r$. \label{figkazantsevpotentialhelical}}
\end{figure}

This seems promising, but let us now look at the full solutions of
\equs{eq:HLequationhelical}{eq:Kequationhelical} to gain a better
understanding of the situation. Using the transformation
\begin{equation}
H_L=\f{\sqrt{2}}{r^2}W_h~,\quad
K=-\f{\sqrt{2}}{r^4}\dpart{}{r}\left(r^2W_k\right)~,\quad
\vec{W}=\left(\begin{array}{c}W_h \\ W_k\end{array}\right)~,
\end{equation}
\cite{boldyrev05}
showed that \equs{eq:HLequationhelical}{eq:Kequationhelical} can be
cast into a self-adjoint spinorial form,
\begin{equation}
\dpart{\vec{W}}{t}=-\tilde{\mathsf{R}}^T\tilde{\mathsf{J}}\tilde{\mathsf{R}}\vec{W}~,
\end{equation}
where 
\begin{equation}
\tilde{\mathsf{R}}=\left(
\begin{array}{cc}
\sqrt{2}/r & 0 \\
 0 & \displaystyle{-\frac{1}{r^2}\dpart{}{r}r^2}
\end{array}
\right)~,
\end{equation}
\begin{equation}
\tilde{\mathsf{J}}=\left(
\begin{array}{cc}
\hat{E} & C \\
 C & B
\end{array}
\right)~,
\end{equation}
\begin{equation}
\hat{E}=-\f{1}{2}r\dpart{}{r}B\dpart{}{r}r+\f{1}{\sqrt2}(A-rA')~,
\end{equation}
and
\begin{equation}
A(r)=\sqrt{2}\left[2\eta+\kappa_N(0)-\kappa_N(r)\right]~,
\end{equation}
\begin{equation}
B(r)=\sqrt{2}\left[2\eta+\kappa_L(0)-\kappa_L(r)\right]~,
\end{equation}
\begin{equation}
C(r)=\sqrt{2}\left[g(0)-g(r)\right]r~.
\end{equation}
Therefore, the generalised helical case too can be diagonalised and 
has orthogonal eigenfunctions. It turns out that there are now two
kinds of growing eigenmodes, discrete bound modes (``trapped
particles'' in the quantum-like description) with growth rates
$\gamma_n$, and a continuous spectrum of free modes (``travelling
particles'') with growth rates $\gamma_\mathrm{free}$. The growth
rates are such that $\gamma_n>\gamma_0>\gamma_\mathrm{free}>0$, where
\begin{equation}
\gamma_0=\f{g^2(0)}{2\eta+\kappa_L(0)}=\f{2\alpha^2}{4(\eta+\beta)}~,
\end{equation}
is twice the maximum standard mean-field $\alpha^2$ dynamo growth rate
for the magnetic field \citep[see e.g.][remember that we are looking
at the growth rates of the quadratic magnetic correlator here]{malyshkin07}.
In other words, not only do we have trapped growing modes reminiscent
of the small-scale dynamo modes derived in \sect{kazantsev}, there are
now also free ``large-scale'' modes asymptotic to helical mean-field
dynamo modes in the limit $r\rightarrow \infty$.  Of course, this
convergence is not totally surprising considering that the $\alpha^2$
mean-field theory can be derived rigorously in the asymptotic
two-scale approach in the limit of short correlation times. The family
of free large-scale growing modes can be envisioned in the quantum-like
description as caused by a helicity-induced modification of the potential
at large scales. When $r/\ell_0\gg 1$, the effective helical potential is
\begin{equation}
V_\mathrm{eff}(r)\simeq\f{2}{r^2}-\f{\alpha^2}{\left(\eta+\beta\right)^2}
\end{equation}
and therefore tends to a strictly negative constant value as
$r\rightarrow \infty$, allowing large-scale eigenfunctions with
negative energy (positive growth rates). The shape of the full
potential is shown in \fig{figkazantsevpotentialhelical}, 
to be compared with \fig{figkazantsevpotential} for the non-helical case.

We see that the smaller the helicity of the flow (as represented by
$\alpha$), the smaller the growth rate of the free large-scale modes
must be, due to the  constraint $\gamma_0>\gamma_\mathrm{free}$. The
growth rate of the bound modes, on the other hand, is not bounded from
above by the amount of helicity in the flow. In all cases, the model
predicts that the bound modes grow faster than the free modes, and are
therefore expected to dynamically saturate the turbulence before the
free modes saturate.  

\begin{figure}
\centering\includegraphics[width=0.499\textwidth]{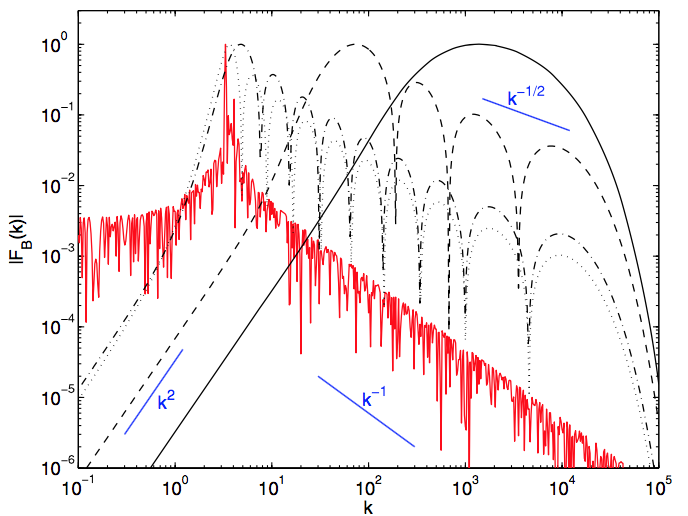}\centering\includegraphics[width=0.493\textwidth]{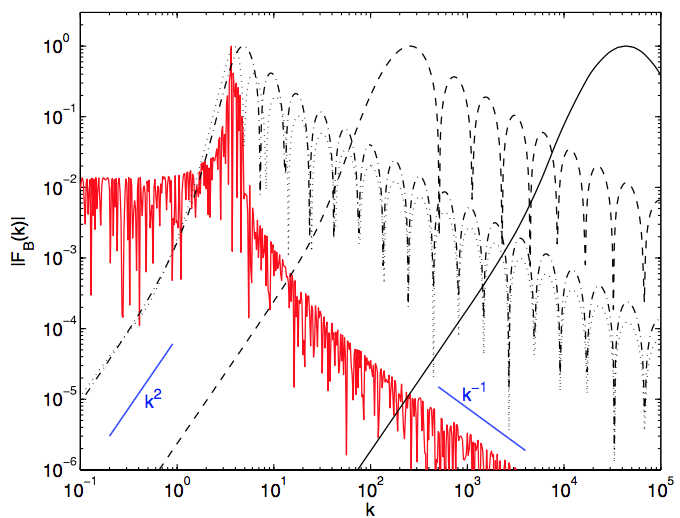}
\caption{Fourier spectra of selected dynamo eigenfunctions in
  the helical Kazantsev model (black thin lines: bound modes, red
  spiky solid line: most unstable unbound mode). Left: $Pm=150$
  case. Right: $Pm=6.7\times 10^{-4}$ case. The kinetic
  helicity of the flow is maximal in both cases \citep[adapted
  from][]{malyshkin09}. \label{figkazantseveigenfunctionshelical}}
\end{figure}

Another major conclusion of this analysis, however, is that the
kinematic helical dynamo problem at large $Rm$ is not reducible
to a simple dichotomy between fast, small-scale dynamo modes and slow,
large-scale dynamo ones. Solving
\equs{eq:HLequationhelical}{eq:Kequationhelical}
numerically for a helical velocity field with a Kolmogorov spectrum,
\cite{malyshkin07,malyshkin09} found that the faster-growing bound
modes that reduce to ``small-scale dynamo'' modes when $h=0$
in \equ{eq:Kequationhelical} themselves
develop correlations on scales $r>\ell_0$ in the presence of helicity,
and are therefore expected to contribute to the growth of the large-scale
magnetic field. A comparison between some of the bound eigenfunctions
and the fastest-growing unbound eigenfunction is shown in
\fig{figkazantseveigenfunctionshelical} for two different regimes.
The results show that the fastest kinematic modes in helical
turbulence at large $Rm$ are not pure mean-field or
small-scale dynamo modes, but hybrid modes energetically
dominated by small-scale fields, with a significant 
large-scale magnetic tail.

\subsubsection{$Pm$-dependence of kinematic helical dynamos}
It is finally worth pointing out in relation to our discussion in
\sect{lowPmsection} of small-scale dynamos at low $Pm$ that the
helical Kazantsev model predicts that the presence of flow helicity at the
resistive scale $\ell_\eta\sim Rm^{-3/4}$ has the effect of decreasing
$Rm_c$ for bound modes at low $Pm$ in comparison to the non-helical
case, down to a value comparable to or even smaller than in the
large-$Pm$ case \citep{malyshkin10}. In other words, helicity/rotation
facilitates the  low-$Pm$ growth of modes identified as small-scale
dynamo modes in the non-helical case. While this problem deserves further numerical
scrutiny, a rotationally-induced decrease of $Rm_c$ has for instance been
reported in simulations of dynamos driven by turbulent convection at
$Pm=1$ \citep{favier12}, and in simulations of kinematic dynamos
driven by rotating 2.5D flows, but forced non-helically
\citep*{seshasayanan17}. More generally, numerical simulations suggest
that large-scale dynamos in rotating/helical turbulent flows are much
less dependent on $Pm$ than small-scale ones \citep[see
e.g.][]{mininni07,branden09c}. A handwaving physical explanation for
this is that the large-scale field always feels the whole sea of
small-scale turbulent velocity and magnetic fluctuations. Pure
``non-helical'' small-scale dynamo fields, on the other hand, appear
to be much more dependent on the details of the flow and dissipation.

\subsection{Dynamical theory\label{largescalesat}}
Having completed our tour of linear theory, we are now in a slightly
better position to discuss the dynamical evolution and saturation of
large-scale dynamos, in particular helical dynamos driven by an
$\alpha$ effect. As with small-scale dynamos, we would
like to understand by which dynamical mechanisms saturation occurs,
at which strength the magnetic field saturates (especially its
large-scale component), and on which timescale this
happens. Due notably to the problems described in \sect{coexistence}, 
formulating a consistent mathematical theory of nonlinear large-scale
dynamos at large $Rm$ has kept a lot of theoreticians busy for many
years, and still remains one of the most formidable problems in the
field today.

\subsubsection{Phenomenological considerations}
The simplest possible outcome of large-scale dynamo saturation, which
is also arguably the most desirable one to explain the characteristics
of magnetic fields in astrophysical systems such as our Sun or galaxy,
is that the large-scale magnetic field reaches equipartition with the
underlying turbulence,
\begin{equation}
\label{eq:equipartLS}
|\meanvB|^2\sim \mean{\rho|\fluctvU|^2}\equiv B_\mathrm{eq}^2~.
\end{equation}
Whether and how such a state can be achieved, however, is quite an
enigma. Note in particular that the two dynamical fields involved in
\equ{eq:equipartLS}, $\meanvB$ and $\fluctvU$, are
at very different scales. How can we get these two fields to
communicate and equilibrate in the dynamical regime, considering that
the part of the Lorentz force  $\meanvJ\times{\meanvB}$ due solely to 
$\meanvB$ acts on a much
larger scale than that of the dynamo-driving flow ? Independently of
whether \equ{eq:equipartLS} holds, it seems clear that the saturation
process must involve dynamical interactions between velocity
fluctuations $\fluctvU$ and  magnetic field fluctuations
$\fluctvB$ at scales similar to or below $\ell_0$, the forcing
scale of the turbulence. Besides, for the helical large-scale dynamo
problem, we have actually found in \sect{kazantsevhelical} that
small-scale magnetic fluctuations are in all likelihood already much
more energetic than the large-scale field in the kinematic stage of
the dynamo, and should therefore reach equipartition with the
flow well before the large-scale field does. Once the Lorentz
force associated with the small-scale fluctuations starts to affect
the flow, it is not obvious that $\meanvB$ itself can keep
growing up to equipartition on dynamically relevant timescales. 
This problem should arise even in the absence of fast-growing
small-scale dynamo modes. Indeed, according to
\fig{figkazantseveigenfunctionshelical}, even unbound mean-field
modes are characterised by $\mean{|\fluctvB|^2}\gg|\meanvB|^2$.

A classic argument illustrating the nature of the problem in a rather
dramatic way is as follows. Consider a hydrodynamic velocity
fluctuation $\fluctU_0$ at scale $\ell_0$, with a typical shearing rate
$\omega_0=\tau_{\mathrm{NL}}^{-1}\sim u_0/\ell_0$, threaded
by a weak large-scale field $\meanvB$ well below
equipartition. In the initial stages of the evolution, the 
stretching of field lines will induce magnetic fluctuations
$|\fluctvB|(t)\sim |\meanvB|\, \omega_0 t$ with increasingly
smaller-scale gradients characterised by a typical
wavenumber $k(t)\sim k_0\,\omega_0 t$. These increasingly thinner
magnetic structures will hit the resistive scale at a time $t_\mathrm{res}$
defined by $\eta k^2(t_\mathrm{res})\sim \omega_0$, at which point the typical
energy of the fluctuations will be of the order $\mean{|\fluctvB|^2}\sim
Rm^p|\meanvB|^2$, with $p=O(1)$ (the exact exponent of this
relation, usually referred to as the Zel'dovich relation,
\red{formally} depends on the number of dimensions and type of
magnetic structures, see \cite*{zeldovich56,moffatt78,zeldovich83},
\red{but appears to be close to unity in practice}). At
large $Rm$ typical of astrophysical regimes, this formula predicts
$\mean{|\fluctvB|^2}\gg|\meanvB|^2$, suggesting that dynamical effects
become significant when $\mean{|\fluctvB|^2} \sim B_\mathrm{eq}^2$
and $|\meanvB|^2\sim B_\mathrm{eq}^2/Rm^p\ll
B_\mathrm{eq}^2$ \citep{vainshtein91,cattaneo91,vainshtein92}.  
If this line of reasoning is correct, it would suggest that
large-scale astrophysical dynamos first start to saturate through the
dynamical feedback of small-scale fluctuations at dramatically
(asymptotically) low level of large-scale fields, and may therefore
have a very hard time powering the near-equipartition large-scale
magnetic fields that we observe in the Universe \red{on timescales shorter
than the \red{cosmological Hubble time}}.  Such a strong potential
dependence of large-scale helical dynamos  on $Rm$ in the nonlinear
regime is commonly referred to as \textit{catastrophic quenching}.

It is important to stress that this particular argument is not
universally accepted and does not constitute a proof that 
catastrophic quenching occurs, notably because it is based on
a description of a transient evolution of magnetic fluctuations,
starting from a large-scale field, rather than on a statistically
steady 3D dynamo eigenfunction such as shown in
\fig{figkazantseveigenfunctionshelical} (see e.g. \cite{blackman05}
for a critical discussion of the applicability of Zel'dovich relations in this
context). Nevertheless, it provides a clear illustration of 
the broader problem that the generation of dynamically strong,
predominantly small-scale fields can pose to large-scale magnetic
field growth at large $Rm$, and also suggests that the microphysics
of dissipation at the resistive scale can play a subtle but important
role in this problem.

\subsubsection{Numerical results\label{numobs}}
Let us now discover what our trusty brute-force simulations of
large-scale helical dynamos driven by pseudo-isotropic turbulence in a
periodic box have to say about this
problem. First, numerical results confirm that the kinematic dynamo
stage is dominated by small-scale magnetic fluctuations, and that
saturation at small-scales does indeed occur well before the
large-scale component reaches equipartition with the flow. This
appears to be true at $Rm$ both smaller \citep[e.g.][]{branden01} and
larger \citep*[e.g.][]{mininni05,subramanian14,bhat16a}
than the critical $Rm$ for small-scale dynamo action
\red{(\cite*{bhat19} have recently been argued that a secondary,
  transient, ``quasi-linear'' growth phase of the large-scale mean
  field may occur once the small-scale dynamo saturates)}. Second,
all simulations of this kind to date are ultimately plagued
by catastrophic quenching, albeit not quite of the form discussed
above. In the absence of shear ($\alpha^2$ case), the large-scale
field does ultimately saturate
at significant dynamical levels actually exceeding equipartition 
(as noted earlier, the dynamo in this configuration generates a
large-scale force-free field, i.e. the mean-field itself has no
back-reaction on the flow), but it only does so on a timescale of
the order of the prohibitively large resistive time at the scale of 
the mean field, not on a dynamical timescale (see e.g. Fig.~8.6 of
\cite{branden05} derived from the \cite{branden01} simulation
set). A ``catastrophic'' timescale of saturation of the
dynamo is also manifest in simulations of helical dynamos in the
presence of large-scale shear ($\alpha\Omega$ case) at $Rm=O(100)$
\citep*{branden01b}. 

\begin{figure}
\includegraphics[width=\textwidth]{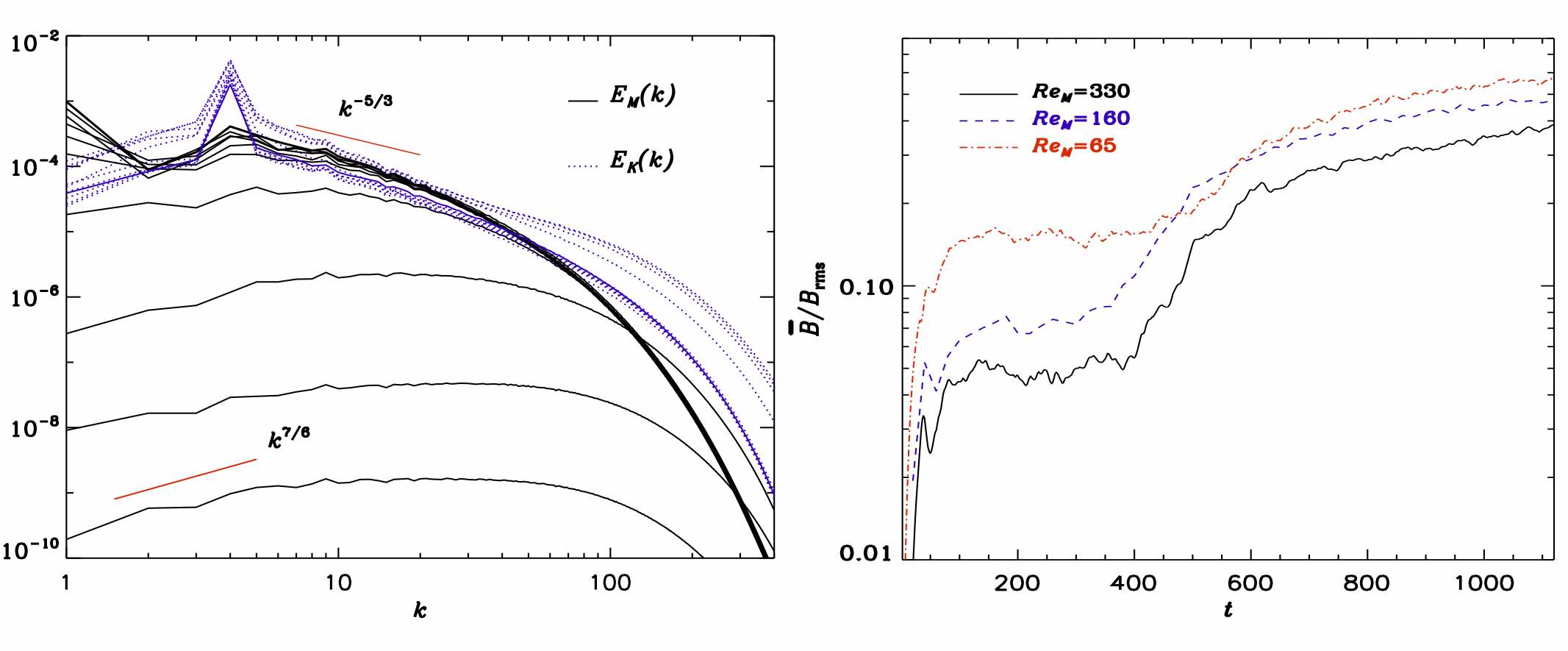}
  \caption{Left: Magnetic (full black lines) and kinetic energy 
    (dotted blue lines) spectra in a simulation of large-scale
helical dynamo action at $Pm=0.1$ and $Rm\equiv
u_\mathrm{rms}\ell_0/(2\pi\eta)=330$. Each spectrum corresponds 
to a time separation of a hundred turbulent turnover times. Right:
time-evolution of $|\meanvB|/B_\mathrm{rms}$ in three simulations
with $Pm=0.1$ and different $Rm$ \citep[adapted from][]{bhat16a}.}
  \label{figBhatspectra}
\end{figure}

To illustrate these results, we show in \fig{figBhatspectra} the
time-evolution of the magnetic energy spectrum in one of the
highest-resolution helical dynamo simulations to date with $Pm=0.1$
and $Rm=330$ \citep{bhat16a}, and the evolution of the ratio
$|\meanvB|/B_\mathrm{rms}$ in three simulations with the
same $Pm$ but different $Rm$, taken from the same study
(the turbulence is forced at $k_0=4 k_L$ in these simulations). 
The spectrum of the kinematic eigenfunction clearly peaks at small 
scales $k/k_L\sim 30$ comparable to the resistive scale, and at first the
whole eigenfunction grows exponentially as expected in a linear regime,
until saturation occurs around $t\sim 400\,
(k_0u_\mathrm{rms})^{-1}$. The larger-scale components of the field
continue to grow nonlinearly after that, with the peak of the
spectrum progressively shifting to larger scales. The energy of the
``mean-field'' (defined as the energy in the $k/k_L=1-2$ modes in the
simulation) grows very slowly in the nonlinear regime, and only
becomes comparable to that of the total saturated
r.m.s. magnetic energy after $10^3$ turnover times ($B_\mathrm{rms}$
is itself only of the order $10\%$ of $B_\mathrm{eq}$ in this particular
low-$Pm$ simulation). Note also that the mean field never reaches
saturation in these already very long simulations, and that its
typical time of nonlinear evolution seemingly increases with $Rm$.
These results are therefore strongly suggestive of a catastrophic
resistive scaling of the timescale of evolution of the dynamo in the
nonlinear regime.

\subsubsection{Magnetic helicity perspective on helical dynamo quenching\label{quenchingmodels}}
Is it possible to make sense of these results ? Let us start
from the basics and ask what kind of dynamical effects may be
important in the problem. Leaving aside the hard question of the
interplay between small and large-scale dynamos for a moment, we have
actually already identified a possible channel of back-reaction of the
magnetic field on helical motions in our discussion in
\sect{writhetwist} of magnetic helicity dynamics in Parker's
mechanism, and that is through the magnetic tension associated with
the small-scale curvature of the twisted field. This somewhat
intuitive idea was first translated into a mathematical model in a
paper by \cite{pouquet76}, who used an EDQNM closure to derive the
expression
\begin{equation}
\label{eq:MTAalphafinal}
\alpha=-\f{1}{3}\tau\left[\,\mean{\fluctvU\cdot\fluctvomega}-\mean{(\curl{\fluctvB})\cdot\fluctvB}\,\right]~.
\end{equation}
This result can also be derived from the MTA closure 
\equs{eq:MTA}{eq:MTAalpha} presented in \sect{MTAclosure}, assuming 
a steady state for the EMF. The second term on the r.h.s. of this
equation can be traced back to the effect of the tension force term
$\meanvB\cdot\grad{\fluctvB}$ on $\fluctvU$,
and does therefore capture a back-reaction of a helical, twisted
magnetic field on the flow. An easy way to see this is by noticing
that $\alpha$ in \equ{eq:MTAalphafinal} vanishes for torsional
Alfv\'en waves. As will be discussed shortly, there are many subtleties
and caveats attached to the interpretation of \equ{eq:MTAalphafinal},
but let us also temporarily ignore them and simply acknowledge for the time 
being that this result gives us an incentive to look at the
problem of saturation from a magnetic helicity dynamics perspective.

\paragraph{\textit{Magnetic helicity conservation in the two-scale approach}.}
Arguably, the conservation \equ{eq:helicitynonideal} for the
total magnetic helicity is a particularly attractive feature in this
context and seemingly represents a good starting point to develop some
multiscale theory, as it provides a direct coupling between the small
and large-scale components of the field. In order to illustrate this, we
will again use the simple two-scale decomposition. Manipulating either
the small and large-scale components of the induction equation or
\equ{eq:helicitynonideal} directly,  and using
$\vEMF\cdot\vB=0$  (by definition of
$\vEMF$), it is straightforward to show that
\begin{equation}
\label{eq:lshelicity}
\dpart{}{t}(\meanvA\cdot\meanvB)+\div{\mean{\vec{F}}_{\mathcal{H}_{m,\mathrm{mean}}}}=2\,\meanvEMF\cdot\meanvB-2\eta\left(\curl{\meanvB}\right)\cdot\meanvB~,
\end{equation}
\begin{equation}
\label{eq:sshelicity}
\dpart{}{t}(\mean{\fluctvA\cdot\fluctvB})+\div{\mean{\vec{F}}_{\mathcal{H}_{m,\mathrm{fluct}}}}=-2\,\meanvEMF\cdot\meanvB-2\eta\,\mean{(\curl{\fluctvB})\cdot\fluctvB}~,
\end{equation}
where we have introduced the mean and fluctuating helicity fluxes
\begin{equation}
  \mean{\vec{F}}_{\mathcal{H}_{m,\mathrm{mean}}}=c\left(\mean{\varphi}\meanvB+\meanvE\times\meanvA\right)~,
\end{equation}
\begin{equation}
  \mean{\vec{F}}_{\mathcal{H}_{m,\mathrm{fluct}}}=c\left(\mean{\fluct{\varphi}\fluctvB}+\mean{\fluctvE\times\fluctvA}\right)=\mean{\vec{F}}_{\mathcal{H}_{m}}-\mean{\vec{F}}_{\mathcal{H}_{m,\mathrm{mean}}}~,
\end{equation}
the total magnetic-helicity flux $\vec{F}_{\mathcal{H}_{m}}$ is given by
\equ{eq:helicityflux}, $\fluctvA$ is the fluctuating small-scale
magnetic vector potential, and $\fluctvE$ in the context of this
equation is the small-scale fluctuating electric field (not a unit
vector). Averaging over a periodic domain, or over a domain bounded by
perfectly conducting boundaries, the surface integrals
associated with the flux terms vanish, leading to
\begin{equation}
\label{eq:lshelicityav}
\deriv{}{t}\left<\meanvA\cdot\meanvB\right>_V=2\left<\meanvEMF\cdot\meanvB\right>_V-2\eta\left<\left(\curl{\meanvB}\right)\cdot\meanvB\right>_V~,
\end{equation}
\begin{equation}
\label{eq:sshelicityav}
\deriv{}{t}\left<\mean{\fluctvA\cdot\fluctvB}\right>_V=-2\left<\meanvEMF\cdot\meanvB\right>_V-2\eta\left<\mean{(\curl{\fluctvB})\cdot\fluctvB}\right>_V~,
\end{equation}
where $\left<\cdot\right>_V$ denotes a volume average. 
Let us stress at this stage that the equations above
are independent of any dynamical closure 
for $\meanvEMF$, such as \equ{eq:MTAalphafinal}.
Besides, all these equations can be derived from the induction
equation alone and therefore provide no information of their own
regarding the dynamical effects of the Lorentz force. Finally, note
that the production terms of large and small-scale  helicities, $\pm
2\,\meanvEMF\cdot\meanvB$, are equal
in magnitude but opposite in sign. This is the mathematical
translation of the conservation of the total magnetic helicity, and 
of our earlier observation in \sect{writhetwist} that
large-scale helical dynamos generate magnetic helicity of one
sign at large scales, and the same amount of helicity of opposite 
sign at small scales. 

It is clear at this stage that we need a closure
prescription for $\meanvEMF$ of the kind
provided by \equ{eq:MTAalphafinal} in order to solve for the full
dynamical evolution of the system in a way that consistently factors 
in the effects of the Lorentz force. Before we go  down this path,
however, let us see whether we can learn something 
about the long-time evolution of the system from
\equs{eq:lshelicityav}{eq:sshelicityav} alone. To this end,
imagine a situation in which $\mean{|\fluctvB|^2}>|\meanvB|^2$
in the kinematic regime, so that the small-scale fluctuating field component
attains dynamical levels first. We then assume that the small-scale
field reaches a statistical steady state with $\mean{|\fluctvB|^2}$
of the order of $B_\mathrm{eq}^2$ at a time $t_\mathrm{sat,fluct}$,
and that this also corresponds to a steady
state for the small-scale magnetic helicity. How does the large-scale
field evolve for $t>t_\mathrm{sat,fluct}$ ? In this  regime, the
time-derivative in \equ{eq:sshelicityav} becomes negligible, leaving us with
\begin{equation}
\label{EMFjdissip}
\left<\meanvEMF\cdot\meanvB\right>_V=-\eta\left<\mean{(\curl{\fluctvB})\cdot\fluctvB}\right>_V~.
\end{equation}
This simply tells us that $\fluctvU$ and $\fluctvB$
(and therefore $\meanvEMF$) have dynamically co-evolved in
such a way that a balance between the production and
dissipation of small-scale magnetic helicity is established. But we
also know that the small-scale helicity production term is equal in
magnitude to that of large-scale magnetic helicity as a result of total helicity
conservation. \Equ{EMFjdissip} therefore suggests that the growth of
the large-scale field in this ``partially-saturated'' regime is tied
to the microscopic ohmic dissipation of the (steady) small-scale
magnetic helicity \citep{gruzinov94}.

In the simplest $\alpha^2$ dynamo case, the kinematic mean-field 
mode is an eigenvector of the curl operator, so that 
\begin{equation}
\label{lshelicitycoulomb}
(\curl{\meanvB})\cdot{\meanvB}=k_L^2\meanvA\cdot\meanvB=\mp
k_L|\meanvB|^2
\end{equation}
 in the Coulomb gauge, $\div{\vA}=0$ (here we
have assumed for simplicity that the large-scale field has a scale
comparable to the system size $L\equiv 1/k_L$). Similarly,
\begin{equation}
\label{sshelicitycoulomb}
(\curl{\fluctvB})\cdot{\fluctvB}=k_0^2\,\fluctvA\cdot\fluctvB=\pm
k_0|\fluctvB|^2~.
\end{equation}
Using these results and substituting
\equ{EMFjdissip} in \equ{eq:lshelicityav}, we then find that 
\begin{equation}
\deriv{}{t}\left< |\meanvB|^2\right>_V\simeq
2\eta k_0k_LB_\mathrm{eq}^2-2\eta k_L^2\left< |\meanvB|^2 \right>_V.
\end{equation}
The solution of this equation is
\begin{equation}
  \label{eq:slowsat}
  \left< |\meanvB|^2\right>_V\simeq\f{k_0}{k_L}B_\mathrm{eq}^2\left[1-\mathrm{e}^{-2\eta k_L^2(t-t_\mathrm{sat,fluct})}\right]~,
\end{equation}
i.e. the mean-field saturates at a super-equipartition value, but
only on the catastrophically slow mean-field resistive timescale
$t_\mathrm{sat,mean}=t_\mathrm{\eta,mean}=(\eta k_L^2)^{-1}$. In the
transient phase $(t-t_\mathrm{sat,fluct})/t_\mathrm{\eta,mean}\ll 1$,
the mean field in this model grows linearly with time,
\begin{equation}
  \label{eq:slowsattransient}
  \left< |\meanvB|^2\right>_V\simeq
\left< |\meanvB|^2\right>_V (t=t_\mathrm{sat,fluct})+
B_\mathrm{eq}^2\gamma_\mathrm{\mathrm{sat}}(t-t_\mathrm{sat,fluct})~,
\end{equation}
where $\gamma_{\mathrm{sat}}\equiv 2\eta k_0 k_L$, so that the
timescale to reach equipartition can actually be much shorter 
than $t_\mathrm{\eta,mean}$ for significant scale separations. 
Interestingly, this model, originally derived by \cite{branden01}
(see \cite{branden01b} for a similar analysis of the
$\alpha\Omega$ dynamo), matches quite well simulation results of the
nonlinear evolution of helical dynamos in periodic boxes for $Rm$
below or comparable to $Rm_{c,\mathrm{ssd}}$. The result does not 
appear to be explicitly dependent on a closure as we usually
envision them, but note that it is nevertheless heavily dependent on
the assumption of a preliminary saturation of the small-scale field.
This assumption is not unreasonable at all, but it is clearly
additional information on the dynamics (as opposed to the
kinematics) of the field that we inject by hand in the induction
equation to understand the evolution of the system in the nonlinear
regime. Overall though, it seems that we can already learn something on
saturation without having to bother too much about the tricky details
of small-scale dynamical closures such as that suggested by
\equ{eq:MTAalphafinal}.

\paragraph{\textit{$\alpha$-quenching models}.}
As mentioned earlier, solving the full dynamical time-evolution requires
closing \equs{eq:lshelicityav}{eq:sshelicityav}, and it is very
tempting to make use of the mean-field ansatz with $\alpha$ given by
\equ{eq:MTAalphafinal} to do this. This closure has some seemingly nice
features: first, it suggests that saturation starts to take place when
the small-scale field reaches equipartition, which is consistent with
our earlier arguments and numerical results. Second, it seems to
capture the effects of the magnetic tension associated with the
small-scale helical twisting of field lines. Finally, it directly
involves a small-scale magnetic current helicity, raising the
prospect of a relatively straightforward mathematical closure of
\equs{eq:lshelicityav}{eq:sshelicityav} when used in combination 
with \equs{lshelicitycoulomb}{sshelicitycoulomb}.
As emphasised by several authors though, one has to be very careful
with the interpretation and definition of $\fluctvB$ in
\equ{eq:MTAalphafinal} in the context of the dynamo problem
\citep[e.g.][]{blackman99,blackman00,proctor03}. The reason is that
in the original study of \cite{pouquet76}, $\fluctvB$  is
not tied to the large-scale field $\meanvB$  through
\equ{eq:linear}, but stands for the magnetic component of some
underlying small-scale background turbulent helical magnetic field
predating a large-scale dynamo. In \sect{MTAclosure}, the problem
was also linearised around such a state to derive
\equ{eq:MTAalpha}. In other words, the magnetic current helicity term
in these derivations cannot be easily formally associated with a dynamical
quenching of the $\alpha$ effect resulting from the exponential growth
of the small-scale helical component of the dynamo field.

Overall, we should be mindful that \equ{eq:MTAalphafinal} probably
at best only qualitatively captures some, but not all the possibly
relevant dynamical effects that affect the large-scale growth of
helical dynamos. It is clear from the algebra that a magnetic current
helicity term associated with small-scale magnetic fluctuations 
specifically generated by the helical tangling and stretching of a
growing large-scale dynamo field $\meanvB$ should also
contribute to $\alpha$ and reduce the kinematic helical dynamo effect
when these fluctuations reach equipartition with the flow, but there
is no guarantee that this effect is either dominant or essential to
saturation. Remember for instance that we have so far swept all the
small-scale dynamo fields present at large $Rm$ under the carpet,
and those are not obviously accounted for in the closure procedures
leading to \equ{eq:MTAalphafinal}. 

Keeping all these caveats in mind, let us nevertheless quickly review
some simple flavours of so-called $\alpha$-quenching models that 
result from using \equ{eq:MTAalphafinal} as a closure. Roughly
speaking, static quenching models
\citep*[e.g.][]{gruzinov94,bhattacharjee95,vainshtein98,field99}
are relevant to the partially-saturated regime in which the small-scale
field has already saturated, and essentially provide closed
expressions for $\alpha(Rm,|\meanvB|)$ that only
depend on time through the time-dependence of $|\meanvB|$
itself. For instance, the expression
\begin{equation}
\alpha(|\meanvB|)=\f{\alpha_\mathrm{kin}}{1+
  \left(|\meanvB|/B_\mathrm{eq}\right)^2}~,
\label{eq:alphanotquenched}
\end{equation}
describes a regular ``quasi-linear'' quenching of the dynamo
as the large-scale field grows towards equipartition \red{(see
  e.g. \cite*{gilbert90,fauve03} for derivations of expressions of this kind
  for a variety of flows, using weakly nonlinear asymptotic multiscale
  analyses close to the dynamo threshold).} On the other hand, the formula
\begin{equation}
\alpha(|\meanvB|)=\f{\alpha_\mathrm{kin}}{1+Rm
  \left(|\meanvB|/B_\mathrm{eq}\right)^2}~,
\label{eq:alphaquenched}
\end{equation}
describes a form of catastrophic quenching. Here $\alpha_\mathrm{kin}$
refers to the (unquenched) value of $\alpha$ in the kinematic regime.
The catastrophic quenching expression~(\ref{eq:alphaquenched})
actually follows directly from combining \equ{eq:MTAalphafinal} and
\equ{EMFjdissip} \citep[][]{gruzinov94}. A castastrophic dependence of
the $\alpha$ effect on $Rm$ of this kind was notably
observed in the first numerical study of $\alpha$-quenching
\citep{cattaneo96b}. In this experiment, a uniform, externally imposed
mean field was used to probe the EMF response to the
introduction of a large-scale field of an
underlying, fully developed turbulent helical MHD turbulence in a
periodic domain. This set-up, while not a full-on helical dynamo
experiment, has
the advantage of keeping things as close as possible to the conditions 
of the derivation of \equ{eq:MTAalphafinal}. This, in hindsight, probably
made the attainment of the result~(\ref{eq:alphaquenched}) quite
inevitable in these simulations, considering that \equ{EMFjdissip} 
is also necessarily satisfied if the measurements of the EMF
are carried out in a steady state. The possible limitations and flaws
of static quenching models and of the aforementioned numerical
experiments exhibiting catastrophic quenching have been debated at
length in the literature
\citep[e.g.][]{field99,field02,proctor03,branden05,hughes18}.
In the end, \equ{eq:alphaquenched} appears to be no more than a simple
translation in mean-field terms of the earlier finding
that the EMF along the mean field  (expressed as a difference between
the small-scale kinetic and current helicity in the MTA closure framework)
 in the partially-saturated regime must have dynamically decreased
down to a level at which it matches the resistive dissipation of helicity.

\Equ{eq:alphaquenched} should be regarded with utmost caution, 
as it is not generically valid and should not be blindly applied or
trusted in all possible situations. First and foremost, remember that
it is the result of using both the mean-field ansatz and a dynamical
closure formula whose interpretation remains
disturbingly fragile, and whose domain of applicability is clearly
limited. Second, \equ{eq:alphaquenched} was derived in a regime
in which the small-scale field has already fully saturated. While the
first limitations are very hard to overcome, the second can be easily
circumvented. Using \equ{eq:MTAalphafinal} in combination with
\equs{lshelicitycoulomb}{sshelicitycoulomb}, it is easy to see that
\equs{eq:lshelicityav}{eq:sshelicityav} describing the time-dependent
dynamics of magnetic helicity can be turned into a relatively simple
nonlinear dynamical quenching model in which $\alpha$ itself becomes
time-dependent \citep[e.g.][]{kleeorin82,zeldovich83,field02}. These
 models predict for instance that the catastrophic quenching of the
$\alpha^2$ dynamo is not engaged when $|\meanvB|^2\sim
B_\mathrm{eq}^2/Rm$, as \equ{eq:alphaquenched} would suggest, but when
$|\meanvB|^2\sim (k_L/k_0) B_\mathrm{eq}^2$, which is admittedly
a bit less catastrophic \citep{field02}. Overall, however, the
long-time asymptotics of two-scale dynamical quenching models
of large-scale helical dynamos in homogeneous periodic domains 
reduce to the \cite{branden01} model described earlier, and therefore
also predict that such dynamos are ultimately resistively limited. The
main difference with static models is that they provide a closed 
description of the dynamical transition into  resistive saturation.

\paragraph{\textit{Is there a way out of catastrophic quenching ?}}
Our discussion so far suggests that the catastrophic quenching of
large-scale helical dynamo fields observed in numerical
simulations in periodic domains is an inevitable outcome of the
back-reaction of small-scale fields on the flow. However, all these
models and numerical simulations are limited in some way, either
because they rely on over-simplifying assumptions, are too idealised,
or are not asymptotic in $Rm$. Their conclusions should therefore
probably not be taken as the final word
on catastrophic quenching, and we may ask what kind of physics
overlooked until this point could potentially improve the efficiency
of such dynamos. 

One possibility would be for the dissipation of magnetic
helicity to behave in unforeseen ways at very large $Rm$. In the
two-scale model of the saturated $\alpha^2$ dynamo, the r.h.s. of
\equ{EMFjdissip} is directly proportional to $\eta$ (assuming $k_0$ is
independent of $\eta$ and $\mean{|\fluctvB|^2}\sim
B_\mathrm{eq}^2$) and therefore
quickly goes to zero as $\eta\rightarrow 0$  ($Rm \rightarrow
\infty$). Accordingly, so must
the large-scale saturated EMF, resulting in the resistively limited
growth of the large-scale field. But a different outcome may be
possible if the actual dissipation of magnetic helicity in the full
MHD  problem remains finite or asymptotes to zero as a relatively
small power of $\eta$ as $\eta\rightarrow 0$ (definitely $<1$, for the
whole dynamo process to  become astrophysically relevant).
It is often argued, though, that the dissipation of magnetic helicity
is generically less efficient than that of magnetic energy for
physically reasonable magnetic energy and helicity spectra, with the
implication that the resistively limited growth of the dynamo
is quite inevitable \citep[e.g.][]{branden02,branden05,blackman15}. A
common argument made in support of this claim is that if the
volume-averaged small-scale magnetic energy dissipation
$\left<D_{\eta,\mathrm{fluct}}\right>_V=\eta\left<|\grad{}
\fluctvB|^2\right>_V$ in the saturated state is finite 
and independent of $\eta$ as $\eta\rightarrow 0$, then $\ell_\eta\sim
\eta^{1/2}$ and
$\eta\left<(\curl{\fluctvB})\times\fluctvB\right>_V\sim
\eta^{1/2}$ should still asymptote to zero as $\eta\rightarrow 0$,
just not quite as drastically as in the two-scale model
\citep{branden01,branden02}.

Another possible way out of catastrophic quenching put forward 
by \cite{blackman00} \citep[see also][]{kleeorin00,ji02,branden02} may
be through the removal of small-scale magnetic helicity from the
system via magnetic-helicity flux losses through its boundaries. 
To understand this proposition, let us again consider \equ{eq:sshelicity},
but this time without assuming any spatial periodicity of perfectly
conducting boundaries. The spatial average of the dynamical evolution
equation for the small-scale magnetic helicity now reads
\begin{equation}
\label{eq:sshelicityavflux}
\deriv{}{t}\left<\mean{\fluctvA\cdot\fluctvB}\right>_V=-\f{1}{V}\int_{\partial
V}{\mean{\vec{F}}_{\mathcal{H}_{m,\mathrm{fluct}}}}\cdot\diff\vec{S}-2\left<\meanvEMF\cdot\meanvB\right>_V-2\eta\left<\mean{(\curl{\fluctvB})\cdot\fluctvB}\right>_V~.
\end{equation}
The main idea is that a dynamically significant large-scale EMF may
survive in the regime of small-scale saturation if a dominant balance 
\begin{equation}
\label{helicityfluxbalance}
\left<\meanvEMF\cdot\meanvB\right>_V\sim
-\f{1}{2V}\int_{\partial V}{\mean{\vec{F}}_{\mathcal{H}_{m,\mathrm{fluct}}}}\cdot\diff\vec{S}
\end{equation}
between the EMF term and a non-resistive boundary flux term is
established, rather than \equ{EMFjdissip} \citep[the full time-dependent
\equ{eq:sshelicityavflux} can also be used  as a basis for dynamical
quenching models, see e.g. discussion in][]{branden18}. There
have been a variety of phenomenological proposals as to how the
removal of small-scale magnetic helicity may occur in astrophysical
conditions, notably with the help of differential rotation and/or
large-scale winds and magnetic coronae
\citep[e.g.][]{vishniac01,blackman03,branden05}. Numerical efforts to
make large-scale helical dynamos work in the presence of open
boundaries remain work in progress and have so far only given mixed
results. Non-zero helicity fluxes
\citep*{branden01c,kapyla10,hubbard11,hubbard12,delsordo13} have been
measured in simulations, but they do not seem to reach the
balance~(\ref{helicityfluxbalance})
in the range of $Rm$ investigated so far, and do not unambiguously
produce fast large-scale dynamo fields on dynamical timescales. Some
recent simulations even suggest that helicity flux losses
preferentially occur at large scales \citep{branden19}. For a more
detailed presentation of this line of
research, we refer to the recent review of \cite{branden18}, in which
it is notably conjectured that helicity fluxes should become dominant
dynamically at asymptotically large $Rm$ (see for instance
\cite{delsordo13} for numerical results up to $Rm=O(10^3)$ suggestive
of this trend).

Whether any of the previous arguments survives the test of numerical
simulations at asympotically large $Rm$ and $Re$ remains to be
found. The discussion above, at the very least, suggests that
significant further progress on the question may require a much deeper
understanding of the dynamics of magnetic helicity dissipation,
reconnection and magnetic relaxation in high-$Rm$ nonlinear MHD than
we currently have, and this problem also appears to be very difficult
to solve in a general way on its own (\cite{moffatt15,moffatt16}, see
also our ealier discussion in \sect{reconnect}). 
Let us finally point out that the whole magnetic helicity dynamics
approach to saturation is not universally accepted, notably because
magnetic helicity conservation itself is a by-product of the ideal
induction equation, and because non-helical
small-scale dynamo fields may also play a role in saturation. Two very
different takes on this problem can be found in the recent reviews of
\cite{branden18} and \cite{hughes18} published in this journal. A
selection of analytical and numerical results offering different
perspectives on saturation at large $Rm$ is presented in the next
paragraph, and in \sect{largeRmfrontier}.

\subsubsection{Dynamical saturation in the helical Kazantsev model*}
In the previous discussion of dynamical quenching, we did not really
pay attention to the origin of the small-scale fields saturating the
large-scale dynamo. Some of the nonlinear numerical simulations of the
$\alpha^2$ dynamo we mentioned had $Rm$ smaller than
$Rm_{c,\mathrm{ssd}}$, while others had $Rm$ larger than
that. Catastrophic quenching appeared to be the norm for all
larger-scale helical dynamo simulations in periodic domains, up to
$Rm=O(10^3)$. Overall, it is therefore not clear at this stage whether
small-scale dynamo fields have a specific dynamical impact on the
evolution of large-scale fields at large $Rm$.

Perhaps not entirely surprisingly considering its complexity, 
analytical results on this problem are scarce. An instructive
calculation possibly linking the dynamical evolution of the large-scale 
field to the saturation of small-scale dynamos can be found in the
work of \cite{boldyrev01}, who considered the extension to the 
helical case of the derivation of the small-scale dynamo magnetic
p.d.f. discussed in \sect{pdfformalism}. In the derivation of 
the non-helical small-scale dynamo problem, there was no helical
contribution to
the velocity correlation function, and we could also invoke the
statistical isotropy of the magnetic field to simplify the problem
into a 1D Fokker-Planck evolution
\equ{eq:fokkerpdfiso} for the p.d.f. of the magnetic-field strength,
from which the log-normal statistics of the field strength were
inferred. In the helical case, however, this isotropy is broken by
the growth of the mean field $\left<\vB\right>(x,t)$, and it
becomes necessary to solve a more general Fokker-Planck equation for
the full field p.d.f.
\begin{equation}
  \label{eq:Pb}
  P[\vB](t)=P_B[B](t)G[\hatvB](x,t)~,
\end{equation}
including the p.d.f. of magnetic-field orientations
$G[\hatvB](x,t)$. In the non-resistive, large-$Pm$ limit
in which the flow correlator can be expanded according to
\equ{eq:kappaexpand}, it can be shown using the characteristic function
technique introduced in \sect{pdfformalism} that
\begin{equation}
  \label{eq:Gb}
  \dpart{G}{t}=\f{\kappa_0}{2}\Delta G+ \left[\f{\kappa_2}{2}(\delta^{ik}-\hatB^i\hatB^k)\ddpartmixed{}{\hatB^i}{\hatB^k}-\kappa_2\hatB^i\dpart{}{\hatB^i}+g\varepsilon^{ikl}\hatB^i\dpart{}{\hatB^k}\nabla_l\right]G
\end{equation}
in 3D, in the presence of flow helicity\footnote{There is a factor 1/2
  difference between this expression and that given by
  \cite{boldyrev01} due to his different treatment of the integration
  of the $\delta(t-t')$ function (Stanislav Boldyrev, private
  communication).}. \red{We seek solutions of this equation in the
  form of a series in powers of $\hatB^i$, with terms of increasing
  order describing the angle distribution of the magnetic field on an
  increasingly fine level. \Equ{eq:Gb} then suggests that we write
\begin{equation}
  \label{eq:Gbsol}
  G[\hatvB](x,t)=
  1+\hatB^i\mean{B^i}(x,t)\mathrm{e}^{-\kappa_2 t}+O(\hatvB\hatvB)~\mathrm{terms}
\end{equation}
with the higher order terms decaying faster than the first order
term. Keeping only the latter is sufficient to study the
behaviour of $\mean{B^i}$. Substituting \equ{eq:Gbsol} in \equ{eq:Gb},
we then simply recover the evolution equation for the large-scale
field in mean-field dynamo theory}
\begin{equation}
\label{eq:MFequationkazantsevagain}
\dpart{\meanvB}{t}=\f{\kappa_0}{2}\Delta
\meanvB+g\curl{\meanvB}~.
\end{equation}
This result is of course not very surprising, considering that the
calculation is essentially another way to derive the helical Kazantsev
model discussed in \sect{kazantsevhelical}. However, it
brings the growth of the mean field under a slightly
different light. \Equ{eq:Gbsol} shows
that the anisotropic part of the magnetic-field distribution, to which the
mean field is directly tied, decays over time due to the isotropisation
of the field by the turbulence. Taken at face value, this would
suggest that the mean field should decay. However, in the kinematic
regime, the average field strength $\left<B\right>$ still grows
exponentially according to \equ{eq:magneticmomentskazantsev}, at a rate
$\kappa_2$ that compensates exactly the decay rate of the anisotropic
part of $G$. Integrating the first moment of the full probability distribution
function~(\ref{eq:Pb}) over all possible magnetic-field strengths and
orientations, we then find that the actual mean field $\left<\vB\right>$
(now understood as an average over the Kraichnan ensemble)
is simply proportional to $\meanvB$, whose evolution is 
governed by \equ{eq:MFequationkazantsevagain}. And, of course, we know
that this equation has unstable $\alpha^2$ solutions.
 
The interesting aspect of this calculation in comparison
to \sect{kazantsevhelical} becomes clear if we consider
a regime in which the small-scale field is saturated. We already
briefly discussed in \sect{nlkazantsev} different ways in which the
Kazantsev model may  be ``patched'' to describe such a regime, but 
for the purpose of this discussion it is enough to assume 
that the moments of the magnetic-field strength stop growing 
when the small-scale field saturates. In this situation, the
relaxation of the  p.d.f. of magnetic orientations is not balanced
anymore by the growth of magnetic-field strength, and it cannot
be balanced by the growth of $\meanvB$ through
\equ{eq:MFequationkazantsevagain} either as a result of a flow
helicity realisability condition $g^2<(5/8)\kappa_0\kappa_2$. The
calculation therefore predicts a quenching of the growth of the
large-scale field as a result of the saturation of the magnetic-field
strength. \cite{boldyrev01} argues that this quenching should
be independent of $Rm$ (non-catastrophic) if the saturation of the
field strength occurs while the small-scale field is still in the
diffusion-free regime described in \sect{largepmselection}. This
conclusion only reinforces the feeling that the question of quenching
is intimately related to the physics of magnetic dissipation.

This theory of saturation, unlike that presented in
\sect{quenchingmodels}, does not explicitly invoke helicity
conservation. It is not  \textit{a priori} obvious whether and
how the two approaches can be bridged, although of course one of the
(kinematic) helical Kazantsev model equations is the evolution
\equ{eq:Kequationhelical} for the magnetic helicity
correlator. Overall, this discussion and \sect{kazantsevhelical}
nevertheless make it clear that the helical Kazantsev model represents
one of the most promising (and well-posed) analytical frameworks for
future research on nonlinear large-scale helical dynamos at large
$Rm$.

\subsubsection{Quenching of turbulent diffusion}
In the previous subsections, we discussed the phenomenology
of large-scale helical mean-field generation in a turbulent MHD fluid,
and derived this effect mathematically rigorously within the limits of
the FOSA in the form of an $\alpha$ dynamo coupling. We then discovered
the many uncertainties surrounding the question of how the $\alpha$
effect changes as magnetic fields of dynamical strength
develop. Furthermore, in the
process of our simple kinematic mean-field derivation, we found another
effect, the $\beta$ effect, which has a clear interpretation as
a turbulent magnetic diffusion. The question therefore naturally
also arises as to how this effect changes as dynamical saturation
takes place. Considering that turbulent magnetic diffusion
also significantly affects the growth of large-scale magnetic fields
(just ask your favourite colleague doing liquid-metal dynamo
experiments at very large $Re$ for confirmation !), it is important to
at least mention  this question here. Its resolution, however,
unsurprisingly turns out to be just as difficult and technical as that
of the saturation of the $\alpha$ effect and, in order to keep the
matters at a simple introductory level, we will therefore not dive
much into it here. A thorough presentation of the fundamental
theoretical aspects of the problem (and of some of the main arguments
stirring the debate in the community) can be found in \cite{diamond05}.

In a nutshell, turbulent diffusion is strongly quenched in two
dimensions (with an in-plane magnetic field). This result can be
simply traced back to the conservation of the out-of-plane mean-square
magnetic vector potential \citep[e.g.][]{cattaneo91,vainshtein92} in
2D. The three-dimensional case, however, turns out to be significantly more
complex due to the extra freedom of motion that tangled magnetic-field
lines enjoy in 3D. Analytical calculations similar to that of
\cite{gruzinov94} leading to the catastrophic $\alpha$-quenching
\equ{eq:alphaquenched} for instance suggest that there is no quenching
of the kinematic turbulent diffusion at all in three dimensions (see
also \cite{avinash91} for a derivation in the FOSA limit of the
vanishing of a ``magnetic'' $\beta$ effect associated with
magnetically-driven fluctuations), while other analytical studies
predict a full range of possible outcomes,
including regular \citep{kitchatinov94,rogachevskii00} and
catastrophic quenching \citep{vainshtein92}. In practice, numerical
experiments, in their typical way of not revealing anything
particularly definitive or extreme, suggest that a significant,
albeit not necessarily catastrophic quenching of turbulent magnetic
diffusion occurs in dynamical regimes
\citep{branden01,blackman02b,yousef03b,branden08b,kapyla09,gressel13,karak14,simard16}.
As with $\alpha$ quenching though, it should be kept in mind that
no such simulation is truly asymptotic in $Rm$ and that the effects of
a small-scale dynamo on the results are still not well understood.
Also, some particular kinds of numerical experiments which
at first glance appeared well-suited to study investigate dynamical
quenching in a simple, systematic way, have turned out to be far from
ideal in this respect in practice. This is for instance the case of
our good old simulations of the $\alpha^2$ turbulent dynamo which, besides
having the conservation of magnetic helicity hard-wired into them,
tend to generate large-scale force-free Beltrami fields, resulting in
a degeneracy between the $\alpha$ and $\beta$ effects \citep[see
e.g.][]{blackman02b,branden08b}.

\subsection{Overview of mean-field dynamo theory applications\label{appli}}
Having discussed some of the most fundamental aspects of linear and
nonlinear large-scale dynamos, we will now go through an overview of the most
common applications of their popular mean-field theory. These are essentially
of two kinds: as low-dimensional nonlinear mathematical models aiming
at describing in relatively simple effective ways the complex
nonlinear behaviour of large-scale astrophysical and planetary MHD
dynamos, and as a numerical analysis tool to fit and extract effective
dynamical information from multidimensional nonlinear MHD simulations.

\subsubsection{Low-dimensional nonlinear mean-field models\label{lowdim}}
Low-dimensional dynamical systems derived from simple dynamical
and symmetry considerations are well-suited to study
bifurcations, weakly-nonlinear solutions and chaotic behaviour
in fluid dynamos that operate in the vicinity of their critical $Rm$,
and are for instance commonly used to shed some light on
the dynamics and reversals of system-scale magnetic fields
in experimental dynamos close to threshold
\citep{ravelet08,berhanu09,petrelis08,petrelis10}.
But what about hugely supercritical systems, such
as those encountered in astrophysics ? In this context, it
has long been argued on the basis of generic dynamical and symmetry arguments
that a low-dimensional, weakly-nonlinear approach still makes sense,
provided that we think of the various transport and coupling
coefficients involved in the equations, such as the viscosity or
magnetic diffusivity, as being dominated by turbulent
processes rather than microscopic processes.

Mean-field electrodynamics (or rather, mean-field magnetohydrodynamics
in the nonlinear regime) can be viewed as a particular application
of these principles in the dynamo context, and has therefore long
been used to devise low-dimensional dynamical representations of
large-scale astrophysical and planetary magnetism at
highly-supercritical $Rm$ (and large $Re$), see
e.g. \cite{meinel90,stefani06a}. At their core,
astrophysical mean-field dynamo models are relatively simple systems
of nonlinear partial differential equations governing the evolution of
large-scale magnetic and velocity fields coupled through mean-field
coefficients, such as in \equ{alphaomega}, and nonlinear terms.
Different large-scale dynamical transport processes, such as the
advection of the large-scale field by meridional circulations, or its
rise under the effect of magnetic buoyancy, can be included with varying
degrees of mathematical rigour on the basis of phenomenological
considerations and observational incentives. Some nonlinear terms usually
included in these models,  such as the effect of the large-scale
Lorentz force $\meanvJ\times\meanvB$ on mean
flows for instance, correspond to pristine dynamical nonlinearities 
that may also contribute to the equilibration of large-scale dynamos 
\citep[e.g.][]{malkus75}. Other nonlinearities stem
from introducing more or less empirical magnetic-field dependences in
the mean-field coefficients, such as the dynamical quenching effects
discussed in \sect{largescalesat}.

Many variations and phenomenological prescriptions are possible within
this framework. To illustrate in the simplest possible way this popular
modelling approach and how it makes use of the theory discussed so
far, we will consider the following system of equations, which aims at
describing axisymmetric mean-field
dynamo action in the convection zone of a differentially rotating star
such as the Sun through a minimal nonlinear extension of the local
Cartesian $\alpha\Omega$ dynamo model introduced in \sect{alphaomega}
(\cite*{jones83,weiss84,jennings91}, \red{see \cite{fauve07} for a
similar discussion}):
\begin{equation}
  \label{eq:jennings1}
  \dpart{\mean{A}}{t}=\f{C_\alpha\cos z}{1+\tau  \meanB^2}\meanB+\ddpart{\mean{A}}{z}~,
\end{equation}
\begin{equation}
  \label{eq:jennings2}
  \dpart{\meanB}{t}=\f{C_S\sin z}{1+\kappa
    \meanB^2}\dpart{\mean{A}}{z}+\ddpart{\meanB}{z}-\lambda \meanB^3~,
\end{equation}
\begin{equation}
\label{eq:jennings3}
\mean{A}=\meanB=0\quad\mathrm{at}\quad z=0,\pi~.
\end{equation}
Distances and times in these equations are expressed in units of
stellar radius $R$ and turbulent magnetic diffusion time
$R^2/\beta$ respectively (it is assumed that $\beta\gg
\eta$), $x$ corresponds to the radial
direction in spherical geometry, $y$ to the
azimuthal direction, $z$ to the colatitude\footnote{For the sake of
notation consistency, we have used the same coordinate system
as in \sect{alphaomega} in \equs{eq:jennings1}{eq:jennings3}. These
coordinates differ from those used by \cite{weiss84} and
\cite{jennings91}, for whom $x$ refers to the colatitude
and $z$ to the radial direction. The explicit presence of $D$ 
in their equations instead of $C_\alpha$ and $C_\Omega$ here
stems from a rescaling of $\meanB$ relative to $\mean{A}$.}. All quantities are
assumed to depend on time and $z$ only,  and $\meanvB(z,t)\equiv
(-\partial \mean{A}/\partial z,\meanB,0)$ is expressed in
equipartition units ($\mean{A}$ is the poloidal flux
function). This particular model also assumes that $\alpha\equiv
\alpha_0\cos z$ in the kinematic regime is antisymmetric with respect
to the equator, as expected in the solar convection zone
  and consistent with \equ{eq:alphaconv}, while the background
shear flow $\meanvU=-S(z)x\,\vec{e}_y$ associated with the
differential rotation ($S\equiv -r d\Omega/dr$) in that same regime
vanishes at the poles, $S\equiv S_0\sin z$.  The control parameter of
the linear dynamo instability is again the dynamo number, defined here
as $D=C_\alpha C_S=\alpha_0 S_0 R^3/\beta^2$, where
$C_\alpha=\alpha_0R/\beta$ and $C_S=S_0 R/\beta$.
The model also includes non-catastrophic static dynamical quenching
of both $\alpha$ and $\Omega$ effects in the presence of a dominant
azimuthal field $B$ \citep{stix72,kleeorin81}, of the form given in
\equ{eq:alphanotquenched}, and an empirical
cubic nonlinearity mimicking losses of magnetic flux through magnetic
buoyancy. These effects are parametrised by three free parameters
$\tau$, $\kappa$ and $\lambda$ respectively. 

\begin{figure}
\centering\includegraphics[height=0.35\textheight]{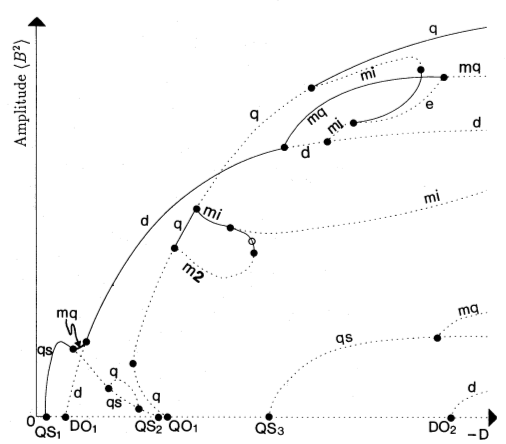}
\caption{Example bifurcation diagram of the nonlinear $\alpha\Omega$ stellar mean-field 
  dynamo \equs{eq:jennings1}{eq:jennings3} as a function of $-D$,
  computed for $\kappa=\lambda$ and $\tau=0$ \citep[adapted
  from][]{jennings91}.\label{figjennings91}}
\end{figure}

As simple as it looks, this model already exhibits some interesting
dynamical phenomena, including some symmetry breaking from pure
oscillatory dipole and quadrupole solutions. An example bifurcation
diagram of this system as a function of $D$ is shown in
\fig{figjennings91}. This diagram, \red{which describes changes in
  the amplitude of nonlinear solutions and their branching into new
  solutions as the control parameter of the system is increased}, is
typical of the solutions
of many astrophysical mean-field dynamo models. It is of course
possible, and in fact very common in practical applications such as
solar dynamo cycle prediction, to devise similar models in 2D or even
3D (if non-axisymmetric solutions are sought), and in more
astrophysically realistic cylindrical or spherical geometries \citep[see
e.g.][]{jouve08}. Obviously, the results of any such model depends on
the values of its free parameters, on the mean-field  and nonlinear terms
included, and on their particular mathematical form. Larger MHD 
dynamical systems including nonlinear evolution equations for
the large-scale velocity field $\meanvU$ and
more advanced parametrisations such as time-delays, spatial
non-localities or localised couplings usually exhibit an even 
larger dynamical complexity and chaotic behaviour including
dynamo-cycle modulations \citep[e.g.][see 
\cite{weiss05} for a detailed discussion and further
references]{ruzmaikin81b,weiss84,jones85,tobias96}. Models of this
kind notably include the popular interface ``flux-transport'' models
of the solar dynamo inspired by the early work of \cite{babcock61} and
\cite{leighton69}, see also \cite{parker93} and reviews by
\cite{charbonneau10,charbonneau14}.

With time, applied astrophysical dynamo modelling of this kind has
turned into an industry. The popularity of these models is of course
related to their minimal formal
mathematical complexity, to their dynamical phenomenological simplicity,
and to the practical convenience and flexibility that low-dimensional
dynamical systems with theoretically unconstrained parameters
offer. Historically, this approach has been instrumental in providing
fundamental qualitative insights into the nonlinear dynamical essence of
large-scale astrophysical and planetary magnetism at a time when
three-dimensional MHD simulations were not available
\citep[for a perspective on this school of thinking,
see again][]{weiss05}. On the other hand, the impossibility to derive any
such model rigorously from the pristine MHD equations under realistic
assumptions, their degeneracies, and the arbitrary amounts of fine-tuning and
\textit{ad hoc} refinement that they are amenable to are considered
by many, this author included, as severe structural theoretical
weaknesses. It notably requires a big leap of faith to believe that such
models can lead to reliable quantitative predictions on the magnetism
and dynamics of just about any known astrophysical system. To be fair
to this approach, the intrinsically chaotic nature of most known dynamos
implies that prediction using any kind of nonlinear model, not just
mean-field ones, is a particularly tricky business. A
critical discussion and illustration of these problems
can notably be found in two papers by \cite*{tobias06} and
\cite*{bushby07}. \cite{charbonneau14} provides a good overview
of the various possible applications of mean-field modelling in the
solar dynamo context, ranging from cycle-prediction activities to the
phenomenological interpretation of global simulations discussed
in \sect{globalsim} below. 

\subsubsection{Mean-field electrodynamics as a numerical analysis tool\label{numtool}}
Once an essentially theoretical and observational interpretation tool,
mean-field electrodynamics has gradually morphed
into a practical simulation analysis technique and computer-assisted
astrophysical modelling tool in the supercomputing era. A systematic
reduction of the challenging dynamical complexity of high-resolution 3D
simulations into lower-dimensional mean-field models is of course
appealing, and is now indeed routinely performed using tools that measure
\textit{in situ} (in the simulation data) the local EMF responses of
the flow to the introduction of large-scale neutral test fields (the
so-called test-field method of \cite{schrinner05}), or the actual
$\meanvEMF(\meanvB)$ relationship in the simulation
\citep{branden02b,racine11,tobias13b,charbonneau14,squire16JPP}, and
project the results onto a tensorial mean-field relationship
\citep[or a generalised convoluted version of it that factors in
time-delays and spatial non-locality,][]{branden18}.
\red{The test-field method, notably, solves \equ{eq:Btilde} for the
  fluctuations without any approximation, making it possible to
  measure the exact total EMF acting on the mean field even in the
  presence of small-scale magnetic fluctuations induced by a
  small-scale dynamo. These various approaches} are considered by many
as a pragmatic, convenient way of simplifying the overall dynamical
picture and modelling of otherwise theoretically intractable
turbulent dynamos problems \citep{branden10}, in that it provides
the convenience and comfort of a data-driven interpretation of
complex simulation results in relatively simple dynamical terms
(e. g. ``this simulated dynamo behaves as an $\alpha\Omega$ dynamo'').
It is for instance widely exploited to  distill effective
low-dimensional nonlinear mean-field dynamo models from simulations of
large-scale planetary and stellar dynamos driven by rotating
convection \citep{schrinner07,charbonneau10,schrinner11,schrinner11b,schrinner12,miesch12,charbonneau14,brun17}
and accretion-disc dynamos, 
\citep{branden95,gressel10,blackman12,gressel15,branden18}, all of
which will be further discussed in \sect{complexLS}.

While it is undoubtedly
practical and may be sufficient to understand some important dynamical
aspects of natural dynamos at the phenomenological level, it is
important to stress that this reductionist approach does not in itself
vindicate the existing mean-field theory, and should not distract us
from seeking a better, self-consistent theory of large-scale turbulent
dynamos. In particular, this kind of analysis cannot easily escape its
main criticism, that mean-field electrodynamics cannot be generically
rigorously justified for nonlinear MHD at large $Rm$ in the current
state of our understanding. Finally, it should be pointed out that
choosing to view all the dynamics through the mean-field prism by
decomposing the effective dynamics into many seemingly independent
statistical mean-field effects (related to shear, rotation, helicity,
etc.) creates a significant risk of disconnection between the analysis
and the underlying three-dimensional nonlinear physical dynamics,
which often involves several physical effects working together. A
clear illustration of this issue will be provided in our discussion of
accretion-disc dynamos and the magnetorotational instability in \sect{subtostat}.

\section{The diverse, challenging complexity of large-scale dynamos\label{complexLS}}
The observation that large-scale dynamos in nature 
are almost invariably associated with small-scale turbulence, rotation
and shear has been one of the main drivers of the development of a
seemingly universal statistical theoretical framework, the mean-field
electrodynamics theory presented in \sect{largescale}, at the centre
of which lies the $\alpha$ effect and the paradigm of helical
$\alpha^2 $ and $\alpha\Omega$ dynamos. Astrophysical and planetary
fluid flows, however, occur in diverse geometries, in diverse thermal, rotation,
Reynolds and Prandtl number regimes (\fig{figPmlandscape}), and
can accordingly be excited by very different means and
instabilities in principle. We should therefore perhaps expect that
all large-scale dynamos do not operate in the same way, and
that their physical and dynamical peculiarities
may not necessarily be easy or even possible to capture with a
single universal formalism, not least one that is not formally valid
in commonly encountered turbulent MHD regimes.

Whether or not we have confidence in the idea that actual large-scale
dynamos at large $Rm$ can be described by a low-dimensional
statistical theory with some degree of universality (and the present
author does so to some extent), it is fundamental that we
keep exploring in parallel their full three-dimensional complexity without
any theoretical preconception. Only by doing this can we hope to gain
an in-depth understanding of the diverse dynamics at work in these systems, 
to relate them to descriptive mathematical theories in physically
meaningful ways, and to identify and fix the possible flaws
of existing theoretical paradigms. Over the last twenty years or so,
supercomputing has made it possible to start investigating many
aspects of large-scale dynamos from a variety of new angles, ranging
from the exploration of asymptotic regimes of astrophysical and planetary
dynamos using sheer numerical power, to numerical experiments focused
on some very fundamental dynamical questions pertaining to the general
instability and statistical problems. Very often, these efforts 
lead to new insights or results that do not  seem to fit
easily with the existing theory, or require that we use, extend,
revise or reinterpret it in previously unforeseen ways.

The aim of this section is to provide an overview of the challenging
dynamical and physical complexity that is continously emerging from a
combination of old and new theoretical ideas, numerical simulations,
and in a few cases also observations of a variety of large-scale
dynamos driven by rotating convection, sheared turbulence, or even
MHD instabilities such as the magnetorotational instability.
As with many other aspects of this review, it is however almost
impossible (and counterproductive pedagogically) to aim
for exhaustivity on such a vast subject. The
next paragraphs therefore concentrate on a selection of problems
that the present author feels at least moderately qualified to write
about, and believes are quite representative of the exceptional dynamical
diversity and outstanding theoretical challenges that large-scale
natural dynamos currently present us with.

\subsection{Dynamos driven by rotating convection: the solar and geo- dynamos\label{numexplor}}
The solar and geo- dynamos are, for human beings, the epitomes of large-scale
MHD dynamos in the Universe, and are undoubtedly the most thoroughly
studied and best-documented natural processes of this kind both
observationally and numerically. These two dynamos belong to the
same family of low-$Pm$ large-scale stellar and planetary MHD dynamos
driven by helical thermal convection in a differentially rotating spherical
shell, however they are very different from each other in many
respects, and their study is strongly illustrative of the generic
difficulties that we have to face as we attempt to understand almost
any laboratory or natural dynamo. Although we are going to discuss these
two problems in some detail, our main objective in this paragraph is not
to provide a specialised review of any of them either,
but rather to highlight some important trends and phenomenological
aspects of these problems that relate to what we have discussed so
far, and to the broader study of large-scale dynamos. Readers
interested in specialised reviews are encouraged to consult
\cite{branden05,charbonneau10,miesch12,charbonneau14,brun17,branden18}
on solar and stellar dynamos and \cite{christensen10,jones11,roberts13}
on geo- and planetary dynamos.

\subsubsection{A closer look at the dynamo regimes of the Sun and the Earth}
Stellar and planetary dynamos driven by rotating convection, of which
the solar and geo- dynamos are the most nearby instances,
share many important dynamical features: they excite a broad
spectrum of magnetic fluctuations due to the large $Rm$
involved, sustain a significant large-scale field component, 
and in some (but not all) cases display some dynamical variability,
including large-scale field reversals. The liquid-metal alloy in the
Earth's core and the hydrogen gas in the Sun are both low-$Pm$
MHD fluids, meaning that turbulence in these systems
extends down to scales well below the magnetic dissipation scale. The
dynamics of rotating stellar and planetary interiors is also generically
characterised by extremely low Ekman numbers $E=\nu/(\Omega L^2)$.
In both the Earth's core and the solar convection zone,
$E=O\left(10^{-15}\right)$.

The Sun and the Earth, however, are not in the same rotation regime.
The ratio between inertial and Coriolis forces at the turbulence forcing
scale, the Rossby number $Ro=u_0/(\Omega \ell_0)$, is $O(10^{-6})$ in the
Earth's core and $O(0.1)$ in the solar convection zone. In other
words, flows in the Earth's core down to very small scales 
are very strongly affected by rotation. In comparison, solar
surface convection at scales smaller than that of the rather
large-scale supergranulation flows \cite[30~000~km, see][]{rincon18}
is essentially unaffected by rotation. Rapidly-rotating convection in
the Earth's core is generally thought to be statistically organised into columns
aligned with the rotation axis (at least if we think of the
hydrodynamic regime rather than the dynamo-saturated regime), a
natural consequence of the Taylor-Proudman theorem
\citep{greenspan68}. In contrast, we know from helioseismic inversions
that the solar differential rotation in the bulk of the convection
zone is not organised along axial cylinders, but rather in a
spoke-like pattern (see e.g. \cite{sekii03} for a review). This
difference is not only a matter of inertial effects, though, as
thermal winds driven by latitudinal entropy gradients are
thought to be a major determinant of the Sun's internal rotation
profile \citep{miesch05,miesch06,balbus09}.

A second major difference
between these two systems is the level at
which their dynamo fields saturate. The magnetic energy in the
Earth's core is estimated to be four orders of magnitude larger than
the kinetic energy \citep[e.g. \cite{gillet10,gillet11}, see also
remarks in][]{aubert17}. Measurements of
dynamical kG-strength magnetic fields at the solar surface
\citep{solanki06} and the detection of large-scale
torsional oscillations \citep[e.g.][]{vorontsov02} point to a solar
dynamo magnetic field in rough equipartition with flows
at both large and small scales.

Altogether, these key observations suggest that the solar and geo-
dynamos do not operate in the same dynamical regime, despite both
being low-$Pm$ and convection-driven. The strong
rotational effects in the Earth's liquid convective core and its
strong magnetic field point to a dynamical balance 
between Magnetic, Archimedean and Coriolis forces (the so-called MAC or
magnetostrophic balance), with subdominant turbulent inertial
and viscous effects. This notably leads to the prediction that the
asymptotic state in which the nonlinear geodynamo resides
should be essentially characterised by a vanishing of the azimuthal
Lorentz force averaged over axial cylinders, the so-called Taylor
constraint \citep[][not to be confused with the Taylor-Proudman
theorem]{taylor63}. The weaker magnetic-field and rotational influence
in the Sun, on the other hand, suggest that the dominant dynamical
balance of the solar dynamo involves turbulent inertia, with the
Coriolis force having a real, but more
subtle influence on the dynamo. Accordingly, there
is no equivalent to the Taylor constraint in solar dynamo theory. 
This also seemingly implies that magnetic saturation proceeds
differently in the two systems. The dynamical feedback of the
magnetic field on the turbulence is thought to be real but
energetically subdominant in the strong-field magnetostrophic
dynamical regime 
\citep{roberts88,roberts92,davidson13b,dormy16,hughes16}
but it is usually deemed essential to the saturation of the solar
dynamo. In this respect, our earlier theoretical discussion of dynamical
saturation in \sect{largescalesat} appears to be  more directly
relevant to the solar dynamo than to the geodynamo.

\begin{figure}
\includegraphics[width=0.99\textwidth]{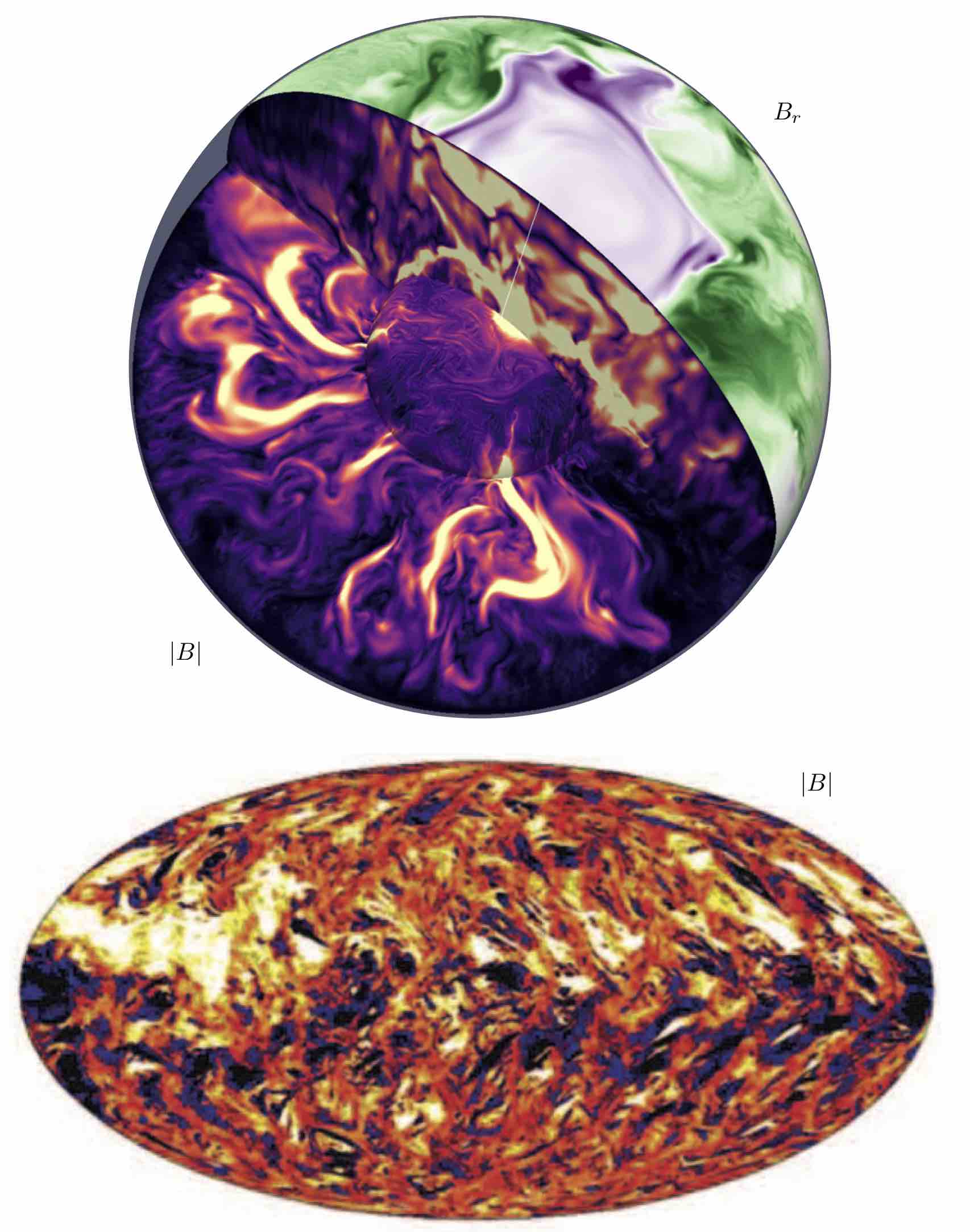}
\caption{Two very different simulations of large-scale dynamos
  driven by rotating turbulent convection at large $Rm$. Top
 \citep[adapted from][]{schaeffer17}:
  rendering of the magnetic-field strength, and radial magnetic field
  at the outer boundary, in a nonlinear geodynamo simulation at
  $Rm=514$, $Pm=0.1$, $Ro=2.7\times 10^{-3}$ and $E=10^{-7}$ (in the
  liquid iron Earth's core, $Rm=O(10^3)$, $Pm=O(10^{-6})$,
  $Ro=O(10^{-6})$ and $E=O(10^{-15})$).
  Bottom \citep[adapted from][]{hotta16}: horizontal projection of
  the magnetic field-strength in high-resolution simulations of the
  solar dynamo with implicit numerical diffusion, at an estimated
  $Pm=O(1)$, $Rm \simeq 2000$, $Ro=10^{-1}-1$, and $E\simeq 10^{-5}$
  (in the strongly stratified gaseous hydrogen solar convection zone,
  $Pm=10^{-6}-10^{-2}$, $Rm=10^6-10^{10}$, $Ro=O(10^{-1})$ and
  $E=O(10^{-15})$). In both cases, $Rm$, $Ro$ and $E$ are defined on
  the thickness of the convective layer and typical (r.m.s. or mixing
  length) convective flow velocity.\label{figgeosolardynamo}}
\end{figure}

\Fig{figgeosolardynamo} shows snapshots of the
magnetic field in two of the highest-resolution geo- and solar
dynamo simulations to date. The magnetic field in the geodynamo simulation
notably appears to have a much stronger latitudinal dependence and
\red{geometric connection to the cylinder tangent to the inner core
and oriented along the rotation axis (the so-called tangent cylinder)}
than that in the solar dynamo simulation, although there is also some
clear latitudinal dependence in the latter. The solar dynamo simulation is also
characterised by much more statistically homogeneous, smaller-scale
magnetic fluctuations in strong dynamical interaction with the turbulence.

\subsubsection{Global simulations of dynamos driven by rotating convection\label{globalsim}}
MHD simulations in spherical geometry
such as those shown in \fig{figgeosolardynamo}
have become a valuable tool to probe the rather extraordinary
geometric and thermodynamic complexity of stellar and planetary
dynamos. Their main strength is their capacity to factor in many 
potentially relevant dynamical phenomena (e. g. rotating convection,
magnetic buoyancy, thermal winds, meridional circulations and other
large-scale flows, strong shear layers) as well
as important geometric constraints (e. g. tangent cylinders), whose
individual or combined effects are harder or even sometimes impossible
to foresee in more idealised local Cartesian settings, and are
difficult to capture with statistical theories. A selection of work
reflecting the historical evolution and increasing massive
popularity of global simulations or large-scale dynamos driven by
convection in rotating spherical shells includes
\cite*{zhang89,glatzmaier95,kuang97,christensen99,christensen06,kutzner02,kageyama08,takahashi08,sakuraba09,soderlund12,dormy16,sheyko16,yadav16,aubert17,schaeffer17,sheyko18}
in the geo- and planetary dynamos contexts, and
\cite*{gilman81,gilman83,valdettaro91,brun04,dobler06,browning06,browning08,ghizaru10,brown11,kapyla12,nelson13,fan14,augustson15,yadav15,yadav16b,hotta16,strugarek17,warnecke18}
in the solar and stellar dynamo contexts.

\paragraph{\textit{Large-scale fields.}}
Similarly to the real systems that they try to emulate, many
high-resolution global simulations of convection-driven dynamos
now show organised, time-dependent (and sometimes reversing)
dynamically-strong large-scale fields emerging from a much more
disordered field component (see e.g. \cite{stefani06b,wicht09,amit10,petrelis10}
for descriptions of possible dynamo-reversal mechanisms in the geodynamo context,
and \cite{charbonneau14} for a theoretical perspective on solar dynamo
reversals). It remains unclear how these findings can be
articulated with the results of the homogeneous simulations
in periodic Cartesian domains \red{presented in \sect{numobs}.}
One possibility is that helicity fluxes discussed at the end of
\sect{quenchingmodels} are indeed important in global simulations with
open magnetic boundary conditions and somehow alleviate catastrophic
quenching. Another possibility is that the growth, dynamical
saturation and evolution timescales of the large-scale dynamo modes
observed in global simulations are indeed subtly tied to
the microscopic diffusion time at system scales, but that
much higher $Rm$ must be achieved in simulations to reveal the full
extent of catastrophic quenching. Some global simulation results
at relatively low $Rm$ (of a few tens) are indicative of a strong
quenching \citep{schrinner12,simard16}. \red{Interestingly, a breadth
of recent high-resolution local simulations also now
suggest that cyclic large-scale dynamo fields can be generated by
compressible rotating convection in Cartesian geometry, albeit with a
period seemingly controlled by the resistive time at the system scale
\citep{bushby18}. Some authors have argued that the solar cycle
itself may be close to resistively
limited \citep[see for instance][Sect.~9.5]{branden05}. Note finally
that the quenching issue may be of lesser importance in the context of
the geodynamo, as the latter is characterised by a smaller $Rm$.}

\begin{figure}
\includegraphics[width=\textwidth]{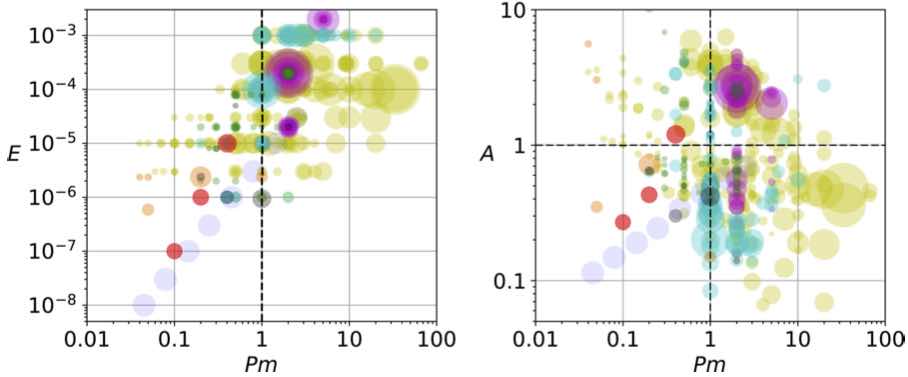}
\caption{The quest for asymptotic geodynamo regimes \citep[adapted
  from][]{schaeffer17}. Each circle represents a published direct or
  large-eddy numerical simulation, and the area of the discs represent
  $Rm$ (again defined on the thickness of the convective layer and
  r.m.s. flow velocity). The pale-blue discs down to $E=10^{-8}$ at
  $Pm<1$ and $Rm=O(10^3)$ \red{correspond to the LES simulations of
  \cite{aubert17}, the red ones to the DNS simulations of
  \cite{schaeffer17}}. $A^2$ is the ratio
  between kinetic and magnetic energy. Many simulations have
  super-equipartition saturated fields, although such states appear
  more difficult to achieve \red{computationally} at low $Pm$ ($Rm=O(10^3)$,
  $Pm=O(10^{-6})$, $E=O(10^{-15})$, and $A=O(10^{-2})$ in the liquid
  iron core of the Earth).\label{figschaefferparam}}
\end{figure}

\paragraph{\textit{Parametric explorations.}}
Unfortunately, global simulations require even
more important sacrifices than local Cartesian simulations in terms of
scale separation in order to encompass global system scales, turbulent
forcing scales, inertial scales and bulk and boundary layer dissipation
scales. The numbers given in the caption of \fig{figgeosolardynamo}
show that there remains a wide parameter gap between even the most massive
simulations to date and the Sun or the Earth. A recurring question
with any new generation of simulations is therefore to what extent
their results are representative of the asymptotic dynamical regimes in which the
systems that they try to emulate operate. A long-time strategy of the
geodynamo community to address this question has been to carry out a
methodical parametric numerical exploration of dynamical force
balances, scaling laws, and nonlinear dynamo states
along parameter paths consistent with the natural ordering of scales
in the problem \citep[][]{christensen99,olson06,christensen06,takahashi08,soderlund12,schrinner12,
  stelzer13,dormy16,sheyko16,yadav16,aubert17,schaeffer17,sheyko18}.
This approach, illustrated in \fig{figschaefferparam},
is now also gaining traction in astrophysics, notably on the problem
of dynamos in rapidly-rotating fully convective stars
\citep{christensen09,morin11,gastine12,gastine13,raynaud15} or
convective stellar cores \citep{augustson16}\red{, see also
  \cite{augustson19} for a recent discussion and meta-analysis
  of several sets of convective dynamo simulations}.

\paragraph{\textit{Geodynamo simulations}.}
Much analytical and numerical work has been done to understand the
roles that the many kinds of flows relevant to fast-rotating planetary
cores \citep{fearn98} may have on their internal dynamos. These
include fast-rotating thermal convection and vortices
\citep{childress72,busse75,busse76,roberts78,roberts92,kageyama97,sarson98,olson99,stellmach04,sreenivasan11,jones11,guervilly15,calkins16},
helical waves \citep{braginskii64,braginskii64b,moffatt70b,schaeffer06,davidson14,davidson18}
Ekman boundary layers \citep{ponty01,schaeffer06}, and zonal flows
and internal shear layers \citep{simitev09,sheyko16} (this list of
references is not meant to be exhaustive but to provide relevant entry
points in the literature). Various analyses have \red{recently} shown
that some aspects of the dynamics of geodynamo simulations conducted
even just a few years ago down to $E=O(10^{-5})$, $Pm=O(1)$ and
$Rm=O(100)$ were still significantly affected by viscous effects
\red{\citep[\cite{soderlund12,king13,davidson13b,oruba14,dormy16}, see
  also see also theoretical arguments in][]{petrelis01,fauve07}}.
The weak-field (equipartition or sub-equipartition) dynamo
branch solutions obtained in these ``classical'' simulations have been
interpreted as the product of an $\alpha^2$-type Parker mechanism
driven by helical convection columns involving an axial flow component
generated through the viscous coupling of the bulk convection with the
boundary layer \citep{roberts13}. The latest generation of
high-resolution simulations of the geodynamo extending down to
$E=O(10^{-7})$, $Pm=0.05$ and $Rm=O(10^3)$, on the other hand, now
seems on the verge of convergence towards asymptotic strong-field
magnetostrophic dynamo states \citep{yadav16,aubert17,schaeffer17,sheyko18}.
These simulations paint a very complex, inhomogeneous, and multiscale
dynamical picture that is not easy to reconcile with a simple
statistical theoretical description \citep[although interestingly
it appears to be possible to construct magnetostrophic mean-field
dynamo solutions, see][]{wu15,roberts18}. The highly non-perturbative
effects of the super-equipartition magnetic field on the dynamics of
the fluid are a major theoretical complication, with the detailed
force balance and flow properties depending both on the region
(i.e. inside or outside the tangent cylinder, or in the boundary
layers) and scale considered \citep{aubert17}.

\paragraph{\textit{Solar dynamo simulations}.}
As complex as the geodynamo is to simulate realistically and to
interpret theoretically, modelling the solar dynamo appears to
be even more of a quagmire, so much so that a credible ``numerical''
dynamo solution (in the sense of being both in a credible parameter
regime with the right scale-orderings, and showing convincing
dynamical similarities with the
observational characteristics of the solar cycle) still does not
appear to be in sight at the moment. One of the main difficulties is
that the solar dynamo involves many different processes that all
appear to be of the same order. As explained earlier, the rotation
regime of the Sun is not quite as asymptotic as that of the Earth's
core, and as a result turbulent inertial effects are likely to play a
much more important dynamical role in the problem of its magnetic-field
sustainment, as indicated by the near-equipartition levels of
saturation. Turbulent transport of angular momentum through Reynolds
stresses is also thought to significantly contribute to the
maintenance of the Sun's differential rotation, whose exact
distribution has a strong impact on the generation of toroidal
fields. The solar differential rotation profile is itself notoriously
difficult to reproduce even in hydrodynamic numerical simulations,
with the effect of thermal winds driven by dynamically established
latitudinal entropy gradients being another key factor to consider in
this context.

The thermodynamics and internal structure of the Sun adds yet another
layer of complexity to the problem: the strong density stratification
of the bulk of the solar convection zone, for instance, is fertile
to many additional MHD phenomena relevant to the dynamo,
including magnetic-buoyancy instabilities (more on this in
\sect{subcritical}) and the turbulent pumping of magnetic
fields (see \sect{pumping} and
\cite{nordlund92,branden96,tobias98,tobias01,browning06} for
simulations). The inner radiative zone of the Sun and its outer convection
zone are also coupled through a thermodynamically and dynamically
complex shear layer, the tachocline \citep*{stixbook,hughes12}. From a
dynamo theory perspective, this region is a mixed blessing. On the one
hand, it is an obvious locus for the generation of strong toroidal
magnetic fields through the $\Omega$ effect \citep{parker93},
potentially providing us with half of a solar dynamo mechanism without
having to think too much. On the other hand, this layer turns out to
be extremely difficult to model consistently both analytically and
numerically, and it creates all sorts of dynamical complications
ranging from being prone to its own large-scale MHD instabilities
(more on this in \sect{otherinstab}) to being a complex processing and
storage unit of angular momentum and thermal, potential, kinetic and
magnetic energy. In other words, the lower boundary condition in the
solar dynamo problem is arguably much more tricky and dynamically
active than in the geodynamo problem.

One of the major problems of early global simulations of
the solar dynamo is that they showed very few signs of strong
large-scale field organisation, and instead essentially looked like
standard Cartesian simulations of turbulent small-scale
fluctuation dynamos (\sect{smallscale}) mapped on a sphere
\citep[e.g.][]{brun04,dobler06}. The  consistent
obtention of more organised large-scale and often cyclic
fields in simulations of increasing structural and dynamical
complexity over the last ten years
\citep[e.g.][]{ghizaru10,brown10,brown11,kapyla12,nelson13,fan14,%
  augustson15,hotta16,strugarek17,warnecke18,strugarek18}
therefore marks an important milestone in the field.
While the $\alpha^{(2)}\Omega$ dynamo paradigm remains a pillar
of the phenomenological interpretation of many such simulations
\citep{racine11,charbonneau14,brun17,branden18}, many important
dynamical effects such as small-scale dynamo activity, tachoclinic
dynamics, turbulent pumping, magnetic buoyancy, and large-scale
transport of magnetic fields by meridional circulations appear to be
significant in many instances. All these effects work together in the
simulations to produce (or impede) the generation of large-scale field
and are extremely difficult to disentangle in practice, making it very
hard to assess with confidence what exactly drives and controls the
solutions (not even mentioning the actual solar dynamo). Note that
large-scale effects distinct from an $\alpha$ effect excited by
rotating stratified convection must be important in the
solar dynamo problem, because the theoretical prediction (also
observed in simulations) of a positive $\alpha$ effect in the northern
hemisphere (\sect{alphastrat}),
combined with helioseismic measurements of the differential rotation
profile in the Sun, leads to the prediction of a polewards migration
of pure $\alpha\Omega$ dynamo waves according to the so-called
\cite{yoshimura75} rule, opposite to the actual record
of sunspot patterns.

Another significant complication to the analysis and comparison
of the current generation of numerical solutions is the use of
(different) subgrid scale models. For instance, all other things being
close, different codes tend to produce cyclic dynamo solutions with
significantly different cycle periods, and these differences have been
attributed to the details of the numerical implementation
\citep{charbonneau14}. This observation once again
raises the problem of the subtle but seemingly critical role of
small-scale dynamics and dissipative processes in controlling
large-scale dynamos.

\subsection{Large-scale shear dynamos driven by turbulence with zero net
  helicity\label{sheardynamo}}
\subsubsection{Numerical simulations}
A very different, but equally interesting development in
large-scale dynamo theory and modelling over the past ten years
has been the explicit numerical demonstration (by several independent
groups using different methods and studying different flows) that
small-scale turbulence with \textit{zero net helicity}, but embedded
in a large-scale shear flow, can drive a large-scale dynamo
\citep{yousef08a,branden08,kapyla08,hughes09,singh15}. In the simplest
possible Cartesian case, a shearing-box extension of the
\cite{meneguzzi81} set-up, turbulence is forced at a small-scale
$\ell_0$ by an external, $\delta$-correlated-in-time non-helical body
force, and is embedded in a linear shear flow $\vU_S=-Sx\,\vec{e}_y$.
This results in the generation of a large-scale horizontal magnetic field
$\left[\meanB_x(z,t),\meanB_y(z,t)\right]$ with a typical wavenumber $k_z\ell_0\ll 1$ (the
overline here denotes an average in the $xy$-plane). Magnetic field
generation of this kind appears to be possible both with and without
overall rotation $\vOmega=\Omega\,\vec{e}_z$, and
survives even in the presence of a small-scale dynamo at
$Rm>Rm_{c,\mathrm{ssd}}$ \citep{yousef08b,squire15PRL,squire16JPP}, see
\fig{figsquireSD}. This mechanism, now commonly referred to as the
shear dynamo, can also produce large-scale field reversals typical of
dynamo cycles \citep*{teed17}.

\subsubsection{Shear dynamo driven by a kinematic stochastic $\alpha$ effect}
\begin{figure}
\includegraphics[width=\textwidth]{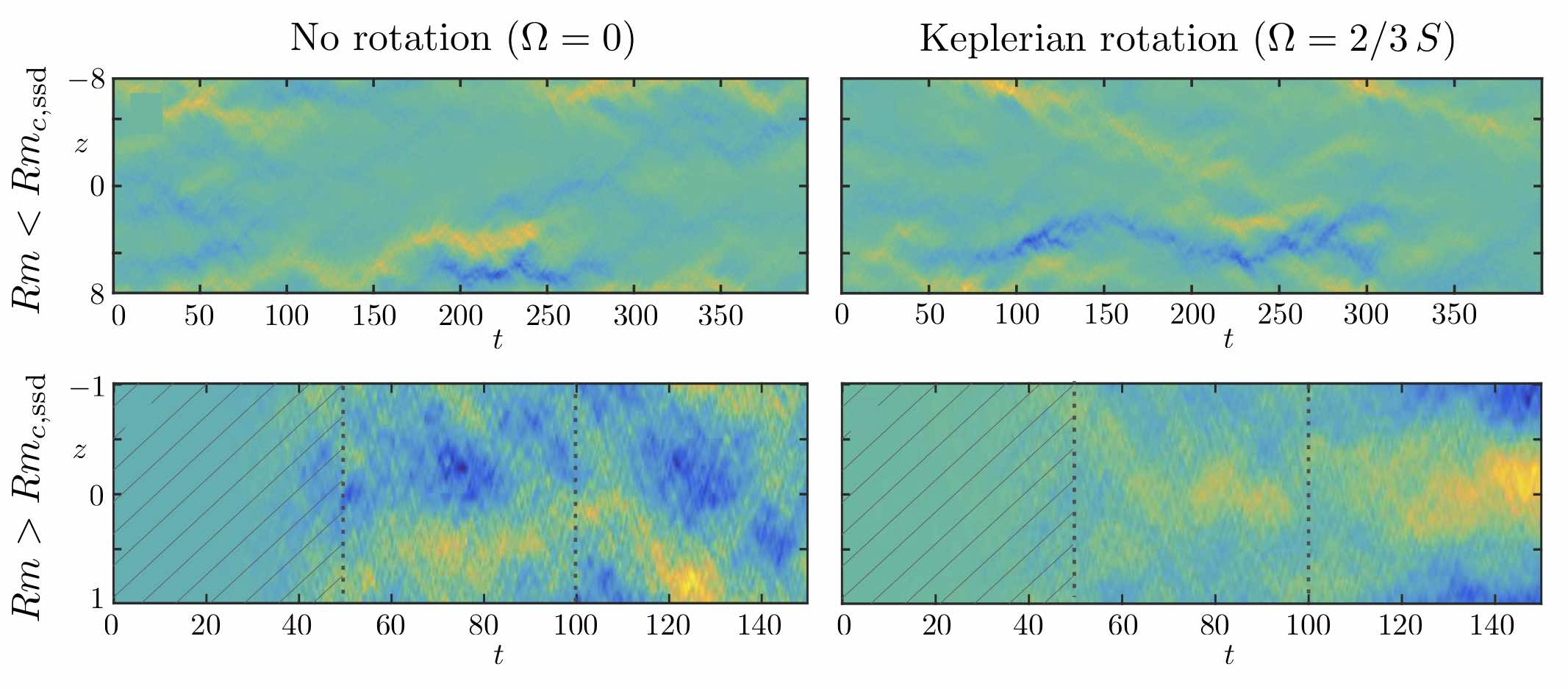}
\caption{Space-time diagrams showing the evolution of a large-scale
dynamo magnetic field $\meanB_y(z,t)$  in local Cartesian shearing box
numerical simulations of dynamo action in the presence of forced,
 \textit{non-helical} small-scale turbulence, shear
 $\vU_S=-Sx\,\vec{e}_y$, with or without global rotation
 $\vOmega=\Omega\,\vec{e}_z$. Top row: simulations
 at $Rm$ lower than $Rm_{c,\mathrm{ssd}}$ for small-scale dynamo action
   \citep[adapted from][]{squire15ApJ}. Bottom row: simulations at $Rm$
   larger than $Rm_{c,\mathrm{ssd}}$. Hatches indicate the phase of
   small-scale dynamo growth \citep[adapted
   from][]{squire15PRL,squire16JPP}. Left: simulations with shear, but
   no rotation. Right: simulations with
   Keplerian rotation, $\Omega=(2/3)\,S$. The box size $(L_x,L_y,L_z)$
   is $(1,1,16)$ in the top-row calculations with no small-scale
   dynamo, and $(1,4,2)$ in the more computationally demanding
   bottom-row calculations with a small-scale dynamo.
   All four simulations are in a regime of $Re$ and $Rm$ that is
   linearly and nonlinearly stable to hydrodynamic shear instabilities
   and MHD instabilities such as the magnetorotational instability. In
   all cases, the turbulence is
   forced at a scale $\ell_0=L_x/3$ much smaller than that at which the
   large-scale dynamo fields develop, and has a turnover rate
   $u_{\mathrm{rms}}/\ell_0$ comparable to the shearing rate
   $S$.\label{figsquireSD}}
\end{figure}

\paragraph{\textit{Non-rotating case.}}
   That non-rotating turbulence with zero net helicity can drive a
   large-scale dynamo may seem quite surprising at first in the light
   of the discussion in \sect{LSsym} of the seemingly important role of
   reflection symmetry in the problem, and it has indeed proven quite
   difficult to make sense of these results. In view of its potential
   broad range of application in astrophysical dynamo theory and of
   the interest that it has drawn in recent years, this problem is
   definitely worthy of a brief review here. It is however
   unfortunately also among the most technical matters in dynamo
   theory, and will therefore only be treated at a rather superficial
   technical level. Readers are referred to the work
   of \cite*{heinemann11}, \cite{sridhar14} and \cite{squire15ApJ} for
   more thorough and quantitative, yet very clear presentations of the
   many analytical facets of the problem.

   One of the first effects put forward as an explanation
   of the results of simulations of the non-helical shear dynamo has
   been the kinematic shear-current effect introduced in
   \sect{SCeffect}. In fact, the original analytical study of this
   effect provided one of the main motivations for the first numerical
   simulations of the shear dynamo by \cite{yousef08a}. Its relevance
   as a dynamo-driving mechanism, however, remained quite
   controversial for a few years, as different closure
   calculations appeared to lead to conflicting conclusions
   (see discussion in \sect{SCeffect}).
   Measurements of mean-field responses (using the test-field method
   discussed in \sect{numtool}) in several independent numerical
   simulations of non-rotating, sheared, non-helical turbulence with
   $St=O(1)$, now seem to have confirmed that the contribution of the
  shear-current effect to $\eta_{yx}$ has the wrong sign for a
  large-scale dynamo, both in the FOSA regime of low $Rm$ (as it
  should be) and for moderate $Rm$ up to $O(100)$
  \citep{branden08,squire15ApJ,singh15}.

   A further indication that the shear dynamo is
   not driven by the shear-current effect, at least in the
   non-rotating case, is that the phase of the large-scale field
   $\meanvB$ in the simulations evolves randomly in
   time. This can be seen for instance in the time-distance
   representations of non-rotating shear dynamo simulations shown in
   \fig{figsquireSD} (left). Accordingly, if we were to perform a
   statistical ensemble average (denoted by $\left<\cdot\right>$) over
   independent realisations of the forcing (or over long-enough times),
   we would find that $\left<|\meanvB|^2\right>$ grows, but not
   $\left<\meanvB\right>$. If the dynamo was driven by a
   systematic, coherent statistical effect such as the shear-current
   effect, on the other hand, we would expect this phase to follow a
   deterministic pattern and $\left<\meanvB\right>$ to grow, in
   accordance with \equs{Bxoffdiag}{Byoffdiag}. In that sense, the shear
   dynamo can be said to be ``incoherent''
   \citep{heinemann11,squire15ApJ}. This peculiarity is now considered
   as one of the key features of the shear dynamo, but was not
   immediately recognised at the time of its first numerical simulations.

   An alternative possibility that may be squared more easily
   with this numerical observation is that the non-rotating shear
   dynamo below the small-scale dynamo threshold $Rm_{c,\mathrm{ssd}}$
   is driven by a stochastic $\alpha$ effect associated with fluctuations
   of kinetic helicity with zero mean, assisted by the $\Omega$
   effect. The general idea that helicity fluctuations with  zero mean
   may be capable of driving a large-scale dynamo actually also largely
   predates the numerical  discovery of the shear dynamo, and
   has evolved along several distinct lines through the history
   of dynamo research. Its origins can be traced back to the work of
   \cite{kraichnan76}, who inferred from a simple stochastic model of
   short-correlated\footnote{``Short-correlated'' here must be
     understood in a restricted sense. Namely,
     $\alpha$ is allowed to fluctuate on a timescale $\tau_\alpha$
     much longer than the typical correlation time of the small-scale
     flow (otherwise there is no scale separation), but much shorter
     than $\ell_\alpha/\alpha_\mathrm{rms}$, where
     $\alpha_\mathrm{rms}$ is the typical r.m.s. value of the
     fluctuating $\alpha$ and  $\ell_\alpha$ is a typical (large) spatial
     scale of the fluctuations of $\alpha$. The latter approximation
     is necessary to solve the problem with quasilinear theory.}
   fluctuations of the diagonal terms of the
   $\boldsymbol{\alpha}$ tensor that the large-scale field in the
   presence of helicity fluctuations could be prone to an
   instability of negative-eddy-diffusivity type.
   \cite{moffatt78} further showed that this effect is supplemented by
   an effective drift of the magnetic field when the fluctuations of
   $\alpha$ are also allowed to fluctuate in space
   (for zero $\alpha$-correlation-time $\tau_\alpha$, this effect does
   not induce any field). The full evolution equation for
   the magnetic field $\left<\meanvB\right>$
   ``super-ensemble''-averaged over both statistical
   realisations of $\alpha$ and small-scales is
   \begin{equation}
     \dpart{\left<\meanvB\right>}{t}=\curl{\left(\meanvU_M\times\left<\meanvB\right>\right)}+\left(\eta+\beta+\eta_K\right)\Delta\left<\meanvB\right>~,
     \label{KMequation}
     \end{equation}
     where
     \begin{equation}
       \eta_K=-\int_0^\infty\diff         \tau\left<
\alpha(\vec{x},\tau)\alpha(\vec{x},0)\right>\quad\mathrm{and}\quad\meanvU_M=\int_0^\infty\diff
       \tau\left<\alpha(\vec{x},\tau)\grad{\alpha(\vec{x},0)}\right>
       \label{KMcoeff}
     \end{equation}
     are called the Kraichnan diffusivity and Moffatt drift. The
     spatial and time dependence of $\alpha$ here must be understood as
     a slow dependence on scales much larger than that of the
     turbulence itself. Note finally that possible off-diagonal
     contributions to the fluctuating $\alpha$
     tensor are discarded in this derivation and that
      $\left<\boldsymbol{\alpha}\right>=\tens{0}$ is assumed. A significant issue with the
      mean-field \equ{KMequation} is that it promotes growth at the smallest
     scales available (as the instability growth rate is proportional
     to $k^2$), thereby compromising the scale-separation between the
     mean field and fluctuations on which the theory is constructed. 
     The Kraichnan-Moffatt (KM) mechanism, at least in its basic form,
     is therefore in all likelihood  not what drives the shear dynamo
     observed in simulations either. The
      latter does not  appear to be of negative-eddy-diffusivity type
   (the field does not grow at the
   smallest-scales of the simulations at all), is not excited in
   the absence of shear, and as we mentioned earlier does not
   generate a coherent mean field
   $\left<\meanvB\right>$. Besides, the KM mechanism
   requires strong fluctuations of the diagonal elements of
   $\boldsymbol{\alpha}$ in order for their r.m.s. effect to overcome
   the regular positive turbulent diffusivity $\beta$, and it has
   recently been pointed out that,
   even if this condition holds, an instability is only possible if
   the fluctuations of the off-diagonal elements of $\boldsymbol{\alpha}$ are
   comparatively much smaller \citep{squire15ApJ}. While such conditions
   may not be impossible to achieve in a non-shearing
   system, they do not appear to be typical of the sheared turbulence
   forced in the simulations of the shear dynamo. A generalised in-depth
   treatment of the KM mechanism in a variety of
   regimes, including the interesting case of small but non-zero
   $S\tau_\alpha$ in shearing regimes, can be found in the work of
   \cite{sridhar14}.

   More general manifestations of a stochastic $\alpha$ effect have
   been explored in the context of shearing systems such as galaxies
   or accretion discs. However, most studies of this problem have been conducted
   at a semi-phenomenological level within the framework of
   mean-field electrodynamics, usually by plugging ``by hand'' a
   stochastic model for $\alpha$ into $\alpha\Omega$ dynamo models
   \citep{sokolov97,vishniac97,silantev00,fedotov06,proctor07,kleeorin08,rogachevskii08,richardson12}.
   While such models generically produce growing dynamo solutions,
   their detailed results and conclusions appear to be strongly
   dependent on an \textit{ad hoc} prescription for the stochastic
   $\alpha$ model and on the procedure used to find a closed
   expression for the super-ensemble-averaged correlator
   $\left<{\boldsymbol{\alpha}\cdot \meanvB}\right>$ appearing in the derivation.
   A rigorous first-principles calculation demonstrating
   (in some analytically tractable limit) the existence of an incoherent
   shear dynamo driven by fluctuations of helicity in the simplest
   case of sheared, forced non-helical turbulence only appeared a few
   years ago in the  work of \cite{heinemann11} \citep[see
   also][]{mcwilliams12,mitra12,sridhar14}.
   Although it is perturbative at its core too, this calculation is
   technically more involved than the FOSA derivation of the standard
   $\alpha$ effect presented in \sect{MFcoeff}. We will therefore only
   explain its key features here. The main  interest of the derivation
   lies in that it rigorously captures the effect of the shear
   on the turbulent motions underlying helicity fluctuations. To
   achieve this, the authors introduce a   representation of the
   three-dimensional forced turbulent velocity field as an ensemble
   of plane ``shearing waves'' labelled by Lagrangian wavevectors
   $\vec{k}_0$ (we already encountered this mathematical
   representation in \sect{zeldo}). Under the effect of large-scale
   shearing, the Eulerian wavevectors of plane waves 
    evolve according to\footnote{We once again define
     $\vU_S=-Sx\,\vec{e}_y$ for internal consistency.
     \cite{heinemann11} use the opposite sign convention.}
   \begin{equation}
     \vec{k}(t)=\vec{k}_0+Sk_yt\,\vec{e}_x~.
   \end{equation}
   Using this representation, a closed evolution equation for the
   the Fourier-transformed mean-field $\meanvB(k_z,t)$ covariance
   four-vector, averaged over a statistical ensemble of
   realisations of a body force driving the turbulence, can be derived
   in the perturbative limits of low $Rm$ and  $S\tau_c\ll 1$
   (the ensemble average is again denoted by $\left<\cdot\right>$ below,
   while the overline denotes spatial averages in the $xy$-plane).
   Crucially, this equation couples the
       $\left<\meanB_x(k_z,t)\meanB_x(k_z,t)^*\right>$ 
   component of this vector to its
   $\left<\meanB_y(k_z,t)\meanB_y(k_z,t)^*\right>$ component
   through  an off-diagonal term proportional to the coefficient
   \begin{equation}
     D_{12}(t)=2 \int_0^\infty\diff
     t'\left<\alpha_{yy}(t)\alpha_{yy}(t-t')\right>= 4\sum_{\vec{k}_0}k_y^4\f{\left<|\Phi_{\vec{k}_0}(t)|^2\right>\left<|u_{z{\vec{k}_0}}(t)|^2\right>}{\nu\eta^2k^6(t)}~,
     \label{eqD12}
   \end{equation}
    where $\Phi_{\vec{k}_0}$ is the stream function associated with
    the horizontal velocity component of a shearing wave with Lagrangian
    wavevector $\vec{k}_0$. The usual $\Omega$ effect ensures the
    amplification of an azimuthal ($y$) mean-field component from the
    poloidal ($x$) component. Note that $D_{12}$ is always positive,
    which appears to be a necessary condition for the dynamo.
    That \equ{eqD12} involves the product between the
    energies of the horizontal vortical and  vertical components of the velocity
    field of the ensemble of shearing  waves makes it explicit that this dynamo
    owes its existence to the r.m.s. fluctuations of the
    kinetic helicity of the sheared turbulent velocity field
    (the net average helicity
    $k^2(t)\left<u_{z{\vec{k}_0}}(t)\Phi_{\vec{k}_0}(t)^*\right>$ for
    each shearing wavenumber
    is taken to be zero in the derivation, so that there is no
    systematic coherent $\alpha$ effect).  Despite being formally only
    valid at low $Rm$ and perturbative in the shear parameter, the
    \citeauthor{heinemann11}
    theory appears to predict a maximal dynamo growth
   rate $\gamma\sim |S|$ at a wavenumber $k_z\sim|S|^{1/2}$ consistent
   with the numerical results of \cite{yousef08a} obtained in the regime 
   $S\ell_0/u_{\mathrm{rms}}=O(1)$ and $Rm>1$.
   In view of our earlier remarks on the evolution of the phase of
   the mean field in the shear dynamo, it is also important to
   emphasise that the product of this derivation is a dynamo equation
   for the super-ensemble-averaged mean energy of the magnetic field
   in the $xy$-plane,    $\left<|\meanvB|^2\right>$, not for the
   super-ensemble-averaged mean field
   $\left<\meanvB\right>$ itself.  The latter does not
   grow in the theory, to lowest-order in the expansion parameters.

  That the non-rotating shear dynamo observed in simulations is
  primarily driven by an incoherent $\alpha$ effect as captured by
  the theory of \cite{heinemann11} is now further supported by
  parametric numerical explorations showing that the growth rate of the
  instability decreases if the horizontal dimensions $L_x$ and $L_y$
  of the numerical domain are increased
  while keeping the scale $\ell_0$ of the turbulent forcing
  fixed \citep{squire15ApJ}. This behaviour is exactly what one expects
  from a purely random effect by averaging over a larger number of
  fluctuations, and implies that such a dynamo formally disappears in a
  system of infinite size, contrary to a coherent dynamo relying on a
  systematic $\alpha$ effect in a helical flow. This is not a
  fundamentally existential problem for the shear dynamo though, as for all
  practical purposes we are only interested in single realisations of
  dynamos in finite systems. However, it would seem to imply that this dynamo
  effect could be much weaker than other possible effects in systems
  with very large-scale separations.

\paragraph{\textit{Rotating case.}}
Another interesting question seemingly relevant to dynamos in many
differentially rotating astrophysical bodies is what happens when we
add global rotation to the shear dynamo. The anticyclonic
Keplerian rotation regime characterised by $\Omega=(2/3)S$ in the
local shearing sheet approximation has been the most studied in this
context due to its relevance to accretion-disc dynamics. In this
regime of rotation and in the presence of a magnetic field, it is
well-known that an MHD fluid can become unstable to the
magnetorotational instability
\citep[MRI,][]{velikhov59,chandra60,balbus91} at wavenumbers such that
$\vec{k}\cdot\vU_A\sim \Omega$, \red{where $\vU_A$ is the Alfv\'en velocity
(\ref{eq:alfvenspeed})}, provided that $Rm$ is larger than
some critical $Rm$ whose exact value depends on the
magnetic-field configuration (and $Pm$). This is important because
the MRI can drive its own MHD turbulence, and can even sustain the
very magnetic field that mediates it through a nonlinear dynamo
process. We will study the latter problem in detail in
\sect{subcritical}, but for the
time being we will keep $Rm$ below the threshold of any kind of
manifestation of the MRI in order to study possible large-scale dynamo
effects excited solely by small-scale turbulence driven by a non-helical forcing.
As shown in \fig{figsquireSD} (right) then, the shear dynamo
studied earlier appears to survive in the Keplerian regime in a
range of $Rm$ where the MRI is not excited, which suggests that this
mechanism may be relevant to the excitation of a dynamo field in
accretion discs independently of an MRI-driven dynamo. This
  however would require that a robust turbulence-driving mechanism
  unrelated to the MRI, such as a hydrodynamic instability, is
  excited in Keplerian discs. Such a mechanism, if any, remains
  elusive \citep{fromang17}.

  It has recently been observed that the nature of the shear dynamo in the
Keplerian regime may be quite different from that in the non-rotating
regime though \citep{squire15ApJ}. An incoherent stochastic $\alpha$
effect is still present in the Keplerian case, however it appears to be
supplemented, and overwhelmed, by a more coherent statistical effect
reminiscent of the $\vOmega\times\meanvJ$ R\"adler
effect discussed in \sect{SCeffect}. This conclusion comes from the
observation that the coherent $\eta_{yx}$ coefficient measured in the
Keplerian simulations, unlike in the non-rotating simulations, is of
opposite sign to that of the shear $S$. This result is also consistent with the
observation made in \sect{SCeffect} that a R\"adler effect-driven
dynamo is only possible in anticyclonic regimes. Finally, \cite{squire15ApJ}
also argue that the phase of the mean field in Keplerian simulations is
more stable than in non-rotating simulations, indicating that the
dynamo is driven by a more coherent statistical effect in the former
case.

\subsubsection{Shear dynamo driven by nonlinear MHD fluctuations*\label{MHDsheardynamo}}
The previous results clearly show that kinematic large-scale
dynamos can be excited by a much wider variety of statistical effects than
the standard $\alpha$ effect on which much of our interpretation of so
many dynamos has historically relied. Accordingly, the shear dynamo is
now considered by many in the community as a promising way to circumvent
(some would say to avoid having to deal with) helical dynamo quenching
(\sect{largescalesat}). However, there is currently no guarantee either
that the shear dynamo is not itself affected by some strong form of dynamical
quenching at large $Rm$. Simulations of the shear dynamo problem above
the critical $Rm_{c,\mathrm{ssd}}$ for the small-scale dynamo (bottom row
of \fig{figsquireSD}) show that a large-scale dynamo can
coexist with a small-scale dynamo in this problem
\citep{yousef08b,squire15PRL,squire16JPP}, but this in itself does not
prove that large-scale growth is not quenched in some way in this
regime.

An interesting recent twist on this problem has been
the realisation that the presence of dynamical small-scale
magnetic fluctuations (characterised by
$\fluctB_\mathrm{rms}\sim\fluctU_\mathrm{rms}$), driven either by
the self-consistent saturation of a small-scale dynamo or by an
independent magnetic-forcing process (such as an antenna-driving or
MHD instability), may actually \textit{promote} rather than quench the
growth of the large-scale field in a large-scale turbulent shear flow
\citep{squire15PRE,squire15PRL,squire15ApJ,squire16JPP}. This effect can be
interpreted within the mean-field framework as a magnetic version of
the shear-current effect discussed in \sect{SCeffect}, in the sense
that the interaction between the shear and dynamical small-scale MHD
fluctuations appears to generate an off-diagonal turbulent
diffusivity $\eta_{yx}$ on its own \citep[a tentative physical
explanation of how this effect originates can be found
in][]{squire16JPP}. The claim,  backed up by both test-field
measurements of mean-field coefficients in numerical simulations
\citep{squire15PRL,squire15ApJ,squire16JPP} and analytical calculations
based on dynamical generalisations of the FOSA and MTA closures
 \citep{singh11,squire15PRE,rogachevskii04}, is that this magnetically-driven
 $\eta_{yx}$ coupling, unlike its kinematic shear-current effect analogue, has
 the right sign to excite a large-scale dynamo, and is actually
 strong-enough
 to overwhelm the latter in a state of saturated small-scale
 MHD turbulence. Accordingly, the ensuing shear dynamo is of the general
 type discussed in \sect{SCeffect}, and amplifies a coherent large-scale
 mean field according to \equs{Bxoffdiag}{Byoffdiag}. It is therefore
 distinct from the stochastic-$\alpha$ shear dynamo discussed in the
 previous paragraph. Both effects are expected to be present in
 non-helical sheared MHD turbulence at large $Rm$
 \citep{squire15PRL,squire16JPP}. Considering the very different intrinsic
 characteristics of the two mechanisms, which one dominates in
 practice is likely to depend on the specificities of the problem at
 hand. \red{Note finally that research on large-scale dynamos in non-rotating
 shear flows is currently rather active, with new potential mechanisms
 still being regularly proposed \citep[e.g.][]{ebrahimi19}.}

\subsection{Subcritical dynamos driven by MHD instabilities in shear flows\label{subcritical}}
A common requirement of all dynamo mechanisms encountered
so far is the presence of a three-dimensional flow \red{(or at least
2.5D, see \sect{cowling})}, possibly turbulent, predating (and causing)
the amplification of the magnetic field. There are however many
systems in which turbulence is itself driven by an MHD instability
whose excitation requires the presence of a large-scale magnetic field
or electrical current. A
particularly well-known example is angular-momentum transporting
turbulence driven by the MRI in a differentially rotating Keplerian
accretion disc, but other systems such as the solar tachocline or
radiative stellar interiors may also fall into
this category. But how does the instability-mediating magnetic field
itself originate in such systems ? If (for possibly purely theoretical
academic purposes) we make the hypothesis
that this magnetic field  does not have an extrinsic origin (which
it may well have in most astrophysical problems) and is internally
generated by a dynamo mechanism, we are confronted with a classic
chicken and egg problem: a flow is needed to induce magnetic
field, but the existence of this very flow rests on the presence of
the magnetic field~! We see that such MHD instability-driven dynamos,
if they exist, must be excited
\textit{nonlinearly}. We will discuss later in this section the
possibility that the dynamics of the large-scale field generated by
these dynamos may, at large $Rm$, be describable in terms of
large-scale statistical effects introduced earlier.
The intrinsic nonlinearity and peculiar conditions of their
excitation, however, warrants that we first approach them from a
distinct physical and dynamical perspective.

\subsubsection{Numerical evidence, and a few puzzling observations\label{instabdynamoevidence}}
In the years following the foundational work of \cite{balbus91} on the
MRI in accretion discs, numerical simulations of MHD in Keplerian
differential rotation flow regimes typical of accretion discs
blossomed. One of the important questions at the time was to establish
whether the MRI was a generic MHD-turbulence-inducing mechanism,
i.e. whether any kind of magnetic-field configuration would be
unstable, in 2D and in 3D. Accordingly, many different geometries and
magnetic configurations were tested in simulations, and among
them was a 3D Keplerian flow configuration with an initial
zero-net-flux magnetic field. The motivation for looking at this particular
configuration was to test whether MRI-driven turbulence could itself
sustain the magnetic field mediating the MRI in the first place. The numerical
answer to this question was found to be positive \citep{hawley96}, and
came with the following observation: \textit{``the use
of a kinematic dynamo model is inappropriate for an accretion disc ;
it assumes a preexisting turbulent state, the statistical independence
of the magnetic and velocity fluctuations, and neglects the magnetic
forces, all of which are inconsistent with our results. The turbulence
is driven by the very forces the kinematic dynamo excludes from the
outset''}. To prove the point, they artificially removed the Lorentz
force from the equations solved by their code, and found no sustained
MHD turbulence in that case. 

\begin{figure}
\centering\includegraphics[width=0.7\textwidth]{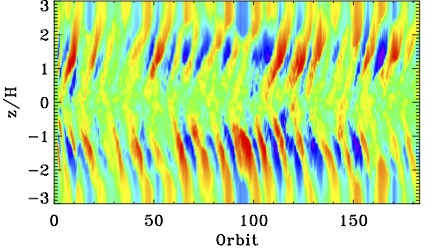}
\caption{Space-time diagram showing ``butterfly'' reversals of the
  large-scale, horizontally averaged azimuthal magnetic field
  $\overline{B}_y(z,t)$ in a local, zero net-flux simulation of MHD turbulence in
  Keplerian shear flow stratified in density along the $z$
  direction \citep*[adapted from][]{davis10}.\label{figdavis10}}
\end{figure}

Interestingly, it was also found in a parallel study by
\cite{branden95} that dynamo action in Keplerian flow
stratified along its rotation axis $\vec{e}_z$ can take the form of
large-scale magnetic cycles with reversing
polarities, a feature usually considered as a hallmark of time-dependent,
mean-field helical large-scale dynamos in differentially rotating
flows. Leaving aside some differences between the
different configurations used, let us simply mention that both numerical
observations, namely the possibility of \textit{nonlinear}
MRI dynamo action mediated by the Lorentz force, and statistical
dynamo cycles seemingly reminiscent of fundamentally
\textit{kinematic} mean-field $\alpha\Omega$ dynamos, were essentially
confirmed by subsequent independent numerical studies
\citep*[e.g.][]{stone96,fleming00,fromang07a,lesur08,davis10,gressel10,kapyla11,simon12}. 
A clear example of cyclic behaviour in stratified MRI dynamo
simulations is shown in \fig{figdavis10}.

In a seemingly disconnected line of
astrophysical fluids dynamics research, another fully nonlinear dynamo
relying on shear and a different MHD instability, the magnetic buoyancy
instability\footnote{Often referred to as the Parker instability, in
  reference to \citeauthor{parker55b}'s (\citeyear{parker55b}) work on
  sunspot formation contemporary with his pioneering dynamo work.}, was
found in local numerical simulations aiming at understanding the
large-scale magnetic dynamics of non-rotating stably-stratified shear layers
\citep*{cline03}. Chaotically modulated cyclic solutions were found in
that system too, despite there being no underlying small-scale
turbulence.

\begin{figure}
\centering\includegraphics[width=0.7\textwidth]{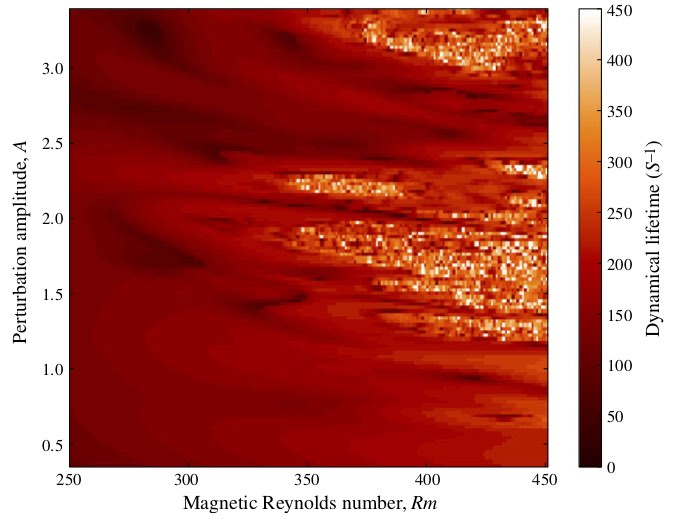}
\caption{Example of transition map for the MRI dynamo, constructed
  from the lifetimes of the dynamics measured in several thousands 
  of numerical simulations with different $Rm$ and initial
  perturbation amplitudes $A$, at fixed $Re=70$
  \citep[adapted from][]{riols13}.\label{figriols13}}
\end{figure}

Several of these numerical studies also come with a string of
even more puzzling observations. First, the dynamics in the
transitional range of $Rm$ investigated in MRI dynamo simulations has
a finite lifetime, but the typical statistical dynamical lifetime
averaged over many simulations grows exponentially with $Rm$
\citep*{rempel10}.  Second, the excitation of instability-driven
dynamos depends a lot, and in a very non-trivial way, on the strength
and form of the initial condition. The critical $Rm$ for the
transition is itself initial-condition-dependent. Finally, the
typical initial-perturbation strength required to excite the dynamo
seems to decrease with increasing $Rm$ on average \citep[see
e.g.][]{cline03}. Most of
these observations are conveniently captured in \fig{figriols13}, which
shows a typical map of the lifetimes of the dynamics measured using many
numerical simulations of the Keplerian MRI dynamo problem. Each point
in this map corresponds to a different simulation with a given $Rm$
and amplitude $A$ of the initial perturbation (all other things being
kept fixed, including the spatial form of the initial condition). A
striking feature of this kind of map is the intricate, fractal-like
geometric structure of the dynamo transition boundary in the phase
space of the system.

\subsubsection{Self-sustaining nonlinear processes\label{SSP}}
The transitional behaviour described above is definitely not that
expected from supercritical kinematic dynamo instabilities such as
the small-scale or large-scale dynamo instabilities described in the
previous sections. It took about ten years after the first MRI dynamo
simulations to realise that this kind of dynamo transition actually
had a lot in common with a classic, structurally nonlinear
hydrodynamic stability problem, the transition to turbulence of
non-rotating shear flows, and to start looking at the knowledge
accumulated on this problem to understand how instability-driven
dynamos are excited.

\begin{figure}
\includegraphics[width=\textwidth]{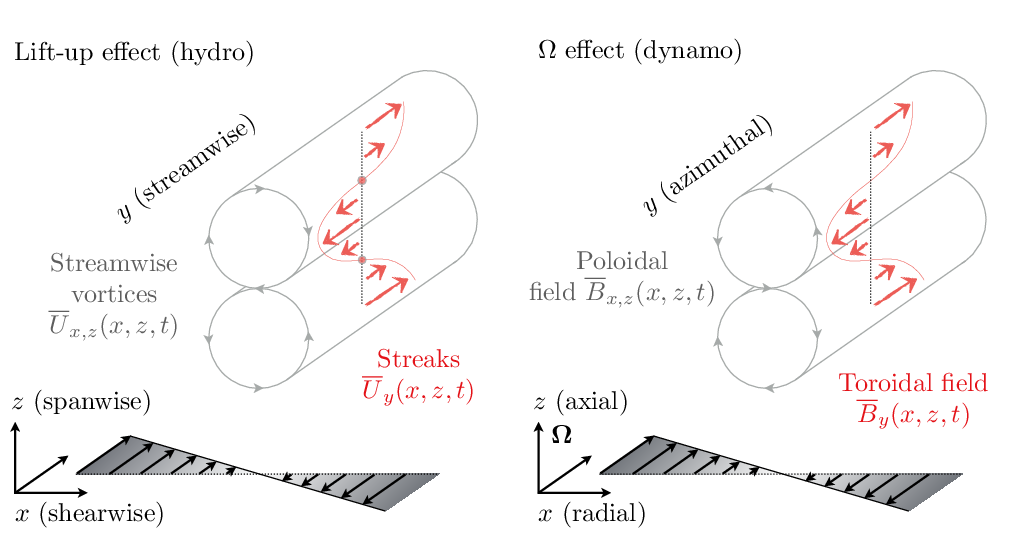}
\caption{Comparison of the first stage of the hydrodynamic shear flow and
  instability-driven dynamo SSPs in the shearing sheet geometry.
  Left: formation of streamwise-independent streaks of
  streamwise velocity $\meanU_y(x,z,t)$ (red arrows) by the
  lift-up effect in the hydro SSP. The full red circles show the
  locations of two inflexion points of the streaky flow profile along
  $z$. Right: amplification of azimuthal/toroidal magnetic field
  $\meanB_y(x,z,t)$ by the $\Omega$ effect in the dynamo SSP (red
  arrows). \label{figSSPsketch}}
\end{figure}

Hydrodynamic flows such as pipe Poiseuille flow and plane Couette flow
are mathematically linearly stable
but have nevertheless long been
known to transition to (hydrodynamic) turbulence at large enough,
finite kinetic Reynolds
number $Re$ \citep{reynolds83}. The subcritical, nonlinear nature of
their transition is established by the observation that it requires
finite-amplitude perturbations whose critical amplitude
decreases with $Re$ in a statistical sense \citep*[e.g.][]{darbyshire95,dauchot95,hof03}. 
Besides, similarly to the MRI dynamo, the typical lifetime of
turbulent dynamics in Poiseuille flow is finite, but increases
exponentially with $Re$ \citep{hof06}. Perhaps not suprisingly
considering its dynamical complexity, this nonlinear hydrodynamic
transition remained very poorly understood for more than a century
despite its relevance to many applied fluid dynamics problems, and it was
not until the early 1990s that a consistent phenomenological picture
started to emerge \citep[see reviews
by][]{eckhardt07,eckhardt09,mullin11,eckhardt18}.
An important milestone on this problem, which as we are about to
discover is also strongly relevant to instability-driven dynamos, was
the numerical discovery and subsequent
characterisation of a three-stage dynamical regeneration cycle, now
commonly referred to as the self-sustaining process \citep*[SSP,
see][]{hamilton95,waleffe95a,waleffe97}. In order to understand the
essence of this process, we consider again a linear shear flow in the
Cartesian shearing sheet model introduced in \sect{shearingsheet}, and
imagine that we introduce (by hand) two weak, counter-rotating,
streamwise-independent vortices (or rolls)
oriented along the streamwise direction $y$, and stacked along the $z$
direction (\fig{figSSPsketch} left). In the first stage of the SSP,
these two vortices advect shearing fluid with positive relative
streamwise velocity from $x<0$ to regions of lower streamwise velocity
at larger $x$, and fluid with negative relative streamwise velocity from
$x>0$ to regions of higher streamwise velocity at smaller $x$. This
simple linear mechanism, often referred to as the lift-up effect
\citep{ellingsen75,landahl80}, generates a $z$-modulated shear flow
$\meanU_y(x,z,t)$ (``streaks'') that grows linearly in time
in the absence of viscous effects and any other dynamics. This streaky
flow, unlike the original linear shear flow, has inflexion points, and
is therefore linearly unstable to infinitesimal streamwise-dependent
perturbations $\fluctvU(x,y,z,t)$. The growth of these
perturbations is the second step of the SSP. Finally, when these
perturbations reach finite amplitudes, the streaks break down and a
streamwise-independent Reynolds stress
$\mean{\fluctvU\fluctvU}$ is generated. Under
generic conditions, it turns out that  this stress can support the
streamwise vortices introduced initially against viscous decay, 
so that the process as a whole is self-sustaining. 

This description offers an important phenomenological clue about the
dynamical similarities of the non-rotating hydrodynamic shear
flow and instability-driven dynamo problems. Indeed, we have found that
the sustainment of streamwise-independent flow vortices in the
hydrodynamic problem requires streamwise-dependent fluctuations, just
like the sustainment of an axisymmetric poloidal magnetic field requires
non-axisymmetric fluctuations in the dynamo problem (\sect{cowling}).
This connexion can easily be understood mathematically if we recall
that the vorticity equation in hydrodynamics and the induction equation in MHD
have the exact same form. If we take the streamwise average of
the hydrodynamic vorticity equation, or the axisymmetric average of
the induction equation as we did in our discussion of Cowling's
theorem, we find in both cases that sustaining the
streamwise-independent (or axisymmetric) part of the dynamics
requires an ``inductive'' cross-correlation of
streamwise-dependent/axisymmetric fluctuations (in the hydrodynamic
problem, this inductive process is vortex stretching).  Note also that
non-axisymmetric fluctuations are not prescribed externally in either
of these problems, but must somehow be consistently produced
by an instability.

\begin{figure}
\includegraphics[width=\textwidth]{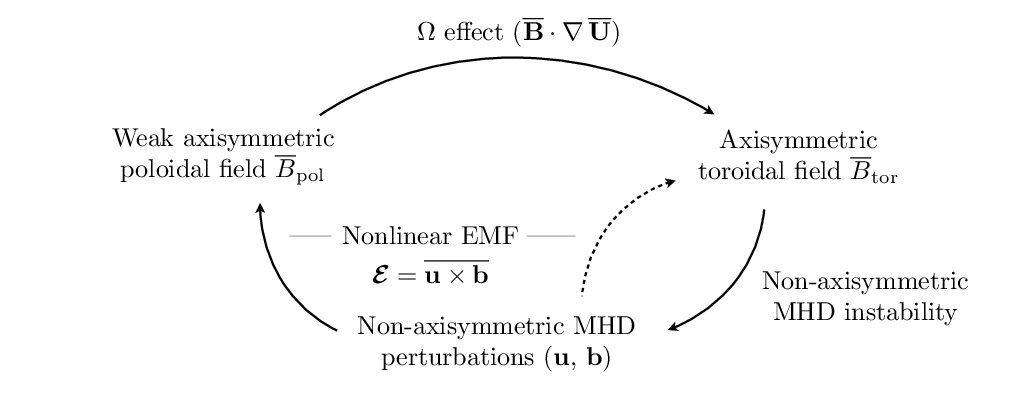}
\caption{Self-sustaining nonlinear dynamo process in
  shear flows prone to MHD instabilities. Positive nonlinear feedback
  on the axisymmetric poloidal field is essential to the whole
  mechanism, whereas nonlinear feedback on the toroidal field can be
  either positive or negative. \label{figSSP}}
\end{figure}

Trying to connect two seemingly different problems is always nice, but
analogies can also be treacherous. Is there actually a self-sustaining
nonlinear dynamo process similar to the hydrodynamic SSP ?
The answer is positive: all three dynamical mechanisms involved in the
hydrodynamic problem have a direct counterpart in the
dynamo problem \citep*{rincon07,rincon08}. To see this, consider
the equation for the axisymmetric projection of the (pristine, not
mean-field) induction equation in the Cartesian shearing sheet model,
\begin{equation}
  \label{eq:instabilitydriven}
  \dpart{\protect\meanvB}{t}=-SB_x\vec{e}_y+
\curl{\mean{\left[\left({\vU}-\vU_S\right)\times{\vB}\right]}}+\eta\Delta\meanvB~,
\end{equation}
paired with the diagram in \fig{figSSP}. 

\smallskip

\begin{enumerate}
\item The first stage of the dynamo process, analogous to the lift-up
  effect in the hydrodynamic problem, is simply the $\Omega$ effect already encountered in
\sect{largescaleOmega}. This effect is associated with the first term
on the r.h.s. of \equ{eq:instabilitydriven}. Imagine that we start with
two weak axisymmetric poloidal loops of  magnetic field (\fig{figSSPsketch},
right), just like we started with a pair of streamwise vortices in the
hydrodynamic problem. In the presence of a background shear flow, 
the $x$-component of the field is going to be slowly stretched into a
stronger azimuthal/toroidal axisymmetric field
$\meanvB_y(x,z,t)$.  Just like
the hydrodynamic lift-up, this effect is linear, and the azimuthal
component of the field only grows linearly in time in the absence
of resistive effects and any other dynamics. 
\smallskip

\item The second stage of the
dynamo process, analogous to the non-axisymmetric shear-instability of the
streaky flow in the hydrodynamic problem, is the excitation of a
linear, exponentially growing non-axisymmetric MHD instability
mediated by the slowly built-up axisymmetric azimuthal field.
In the Keplerian MRI dynamo problem, this instability
is...the MRI, which favours perturbations $\fluctvU$ and
$\fluctvB$ with a wavenumber polarisation along the magnetic field
$(\vec{k}\cdot\vB\neq 0)$. In the presence of a
predominantly azimuthal field, we see that the most unstable
perturbations are non-axisymmetric \citep{balbus92}: this is exactly
what we need here ! The situation is a bit more complex for the
(non-rotating) magnetic-buoyancy-driven dynamo, as the relevant
instability in that case is a mixture of axisymmetric magnetic
buoyancy and non-axisymmetric Kelvin-Helmoltz instability
\citep{cline03,rincon08, tobias11}. In both problems, however, the net
effect is the joint excitation of non-axisymmetric velocity field and
magnetic field fluctuations $\fluctvU(x,y,z,t)$ and $\fluctvB(x,y,z,t)$. 

\smallskip

\item Both magnetic induction and the Lorentz force play a key
  role in these two MHD instabilities, so that $\fluctvU$ and
  $\fluctvB$ are locked into a phase relationship. As they reach
  finite amplitudes, the natural cross-correlation between these
  perturbations generates a significant nonlinear axisymmetric EMF
$\meanvEMF=\mean{\fluctvU\times\fluctvB}$,
just like a Reynolds stress was
generated by streamwise-dependent perturbations in the hydrodynamic
problem. Under generic conditions, this instability-driven EMF appears
to support the poloidal axisymmetric field introduced initially
against resistive decay, providing the required closure of the dynamo
loop in \equ{eq:instabilitydriven}. This is the third stage of the
dynamo SSP. There is also a similar feedback on the toroidal
axisymmetric field, but it is not as critical to the sustainment of
the dynamo as a whole as its poloidal counterpart, since the toroidal
field can always be amplified by the $\Omega$ effect \citep[in the
MRI dynamo problem, the nonlinear feedback on the toroidal
field has in fact been found to be essentially destructive,
see][]{lesur08,riols15}.
\end{enumerate}
 
The previous description of the SSP is essentially laminar, but as we will
see later the process can also be understood in a statistical
sense at large $Re$ and $Rm$. In this respect, it is worth emphasising
that the free energy source that ultimately makes the excitation of
the two aforementioned instability-driven dynamos possible is always the
original background shear, just like in the hydrodynamic problem. In
other words, the different SSPs provide the dynamical channels by
which these diverse out-of-equilibrium shearing systems attempt to
restore equilibrium.

\subsubsection{Nonlinear dynamo cycles and subcritical bifurcations*\label{NLcycles}}
The dynamical similarities between the transition of hydrodynamic shear flows
and instability-driven dynamos translate into a very similar formal
mathematical dynamical system structure. In particular, it is now
established that transition in these two classes of problems is due
to global homoclinic and heteroclinic bifurcation mechanisms
\citep{ott02}. These bifurcations are significantly more
complex to grasp mathematically than local,
supercritical linear bifurcations such as kinematic dynamo
instabilities, and we will therefore only attempt to
describe them in relatively simple terms in what follows. For a more
in-depth discussion, see \cite{riols13} in the MRI dynamo context, 
and \cite{vanveen11,kreilos12}  in the hydrodynamic context. 

\begin{figure}
  \centering\includegraphics[width=0.9\textwidth]{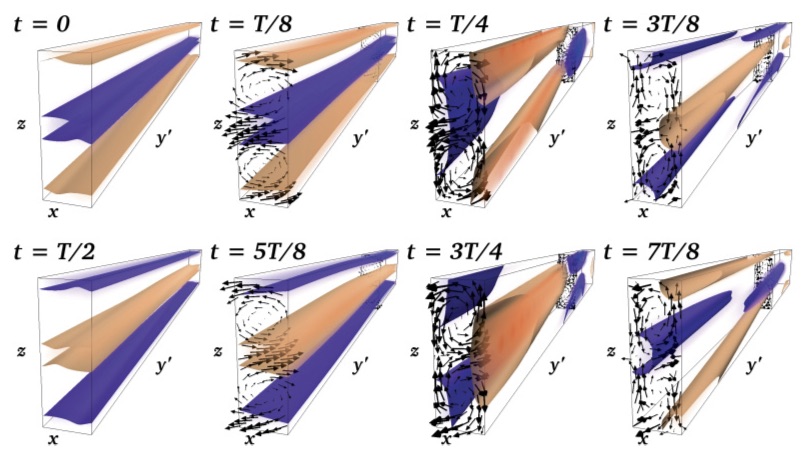}
\caption{Isosurfaces of $B$ coloured by the sign of $B_y$ at different
  stages of a simple nonlinear MRI dynamo cycle of period $T$,
  computed from the three-dimensional incompressible MHD equations in
  the Cartesian Keplerian shearing box numerical model. The
  different steps of the SSP are clearly illustrated: from $t=0$ to
  $t=T/8$, an axisymmetric azimuthal field $B_y$ with alternating
  polarities along the $z$ direction is transiently amplified via the
  $\Omega$ effect. The development of this field then supports the
  exponential growth of an MRI-unstable non-axisymmetric wave packet,
  whose nonlinear evolution results in a global field reversal
  at $T/2$ (the arrows here show non-axisymmetric MRI velocity
  perturbations in an $x-z$ meridional plane, and should not be
  confused with the
  representation of poloidal field lines in \fig{figSSPsketch}
  (right)). The whole process repeats itself in the second half of
  the cycle. As well as reversing the field, the nonlinear feedback of
  non-axisymmetric MRI perturbations taking their energy from the
  shear sustains the $(x,z)$-dependent ``large-scale'' axisymmetric
  poloidal field component against Ohmic dissipation \citep{herault11}.
\label{figherault11}}
\end{figure}

Let us focus on the MRI dynamo transition: as $Rm$ increases, the
dynamo SSP first gives rise  to self-sustained invariant nonlinear
solutions born in pairs at
saddle-node bifurcations. Depending on the details of the 
problem, such as boundary conditions, these solutions can be either
nonlinear fixed points or travelling waves \citep{rincon07}, or
nonlinear cycles/periodic orbits such as that shown in \fig{figherault11} 
\citep{herault11}. For similar
results in hydro, see e.g. \cite{nagata90,waleffe03,wedin04,viswanath07}.
More and more of these pairs of solutions come into existence 
in different parts of the phase space of the corresponding dynamical
system at larger $Rm$, each of them (labelled by $i$) with a different
saddle-node $Rm^i_{\mathrm{SN},c}$. Now, the lower branch solutions
are always linearly unstable, and their
unstable manifolds almost invariably end up colliding in phase space 
with either their own stable manifold (homoclinic case) or with the stable
manifold of a different solution (heteroclinic case) at $Rm$ usually
barely larger than the corresponding saddle-node
$Rm^i_{\mathrm{SN},c}$
\citep[][see \cite{kreilos12} for similar hydro results]{riols13}.
The outcome of such global bifurcations is an abrupt change in dynamical
complexity, including chaotic behaviour and a sensitive dependence 
on initial conditions. At transitional $Rm$ values for which the
number of pairs of invariant solutions is small, this dynamical
complexity is restricted to relatively small fractal-like bubbles of
phase space corresponding to a nonlinear ``sphere of influence'' of the
solutions, so that most initial conditions still decay to the
laminar Keplerian shear flow state. However, as $Rm$ and the number of
invariant solutions grow, an increasingly large fraction of the phase
space becomes populated by homoclinic and heteroclinic
tangles, so that any random initial condition is now much more likely 
to excite chaotic dynamics. Finally, at large $Rm$, almost any initial
condition ends up in a state of full-blown MHD turbulence.
It is this gradual overall explosion of dynamical complexity that
is captured in the intricate transition landscape of \fig{figriols13}. 

Note finally that a majority of the numerical work mentioned 
so far in this subsection, with the exception of the
magnetic-buoyancy driven dynamo of \cite{cline03}, has been conducted
in the convenient shear-periodic Cartesian shearing box framework (\sect{shearingsheet}),
which naturally begs the question of the influence of boundary
conditions on the results. It is therefore important to point out that
the subcritical MRI dynamo has also been found to be excitable in
several wall-bounded configurations such as Keplerian plane Couette
flow \citep{rincon07} and more recently quasi-Keplerian Taylor-Couette flow
\citep{guseva17}. This suggests that the subcritical SSP process is a
robust dynamical phenomenon largely independent of a particular
geometry or boundary conditions. A thorough discussion of
this aspect of the problem can be found in \cite{herault11}.

\subsubsection{$Pm$-dependence of the MRI dynamo transition*}
\begin{figure}
\centering\includegraphics[width=\textwidth]{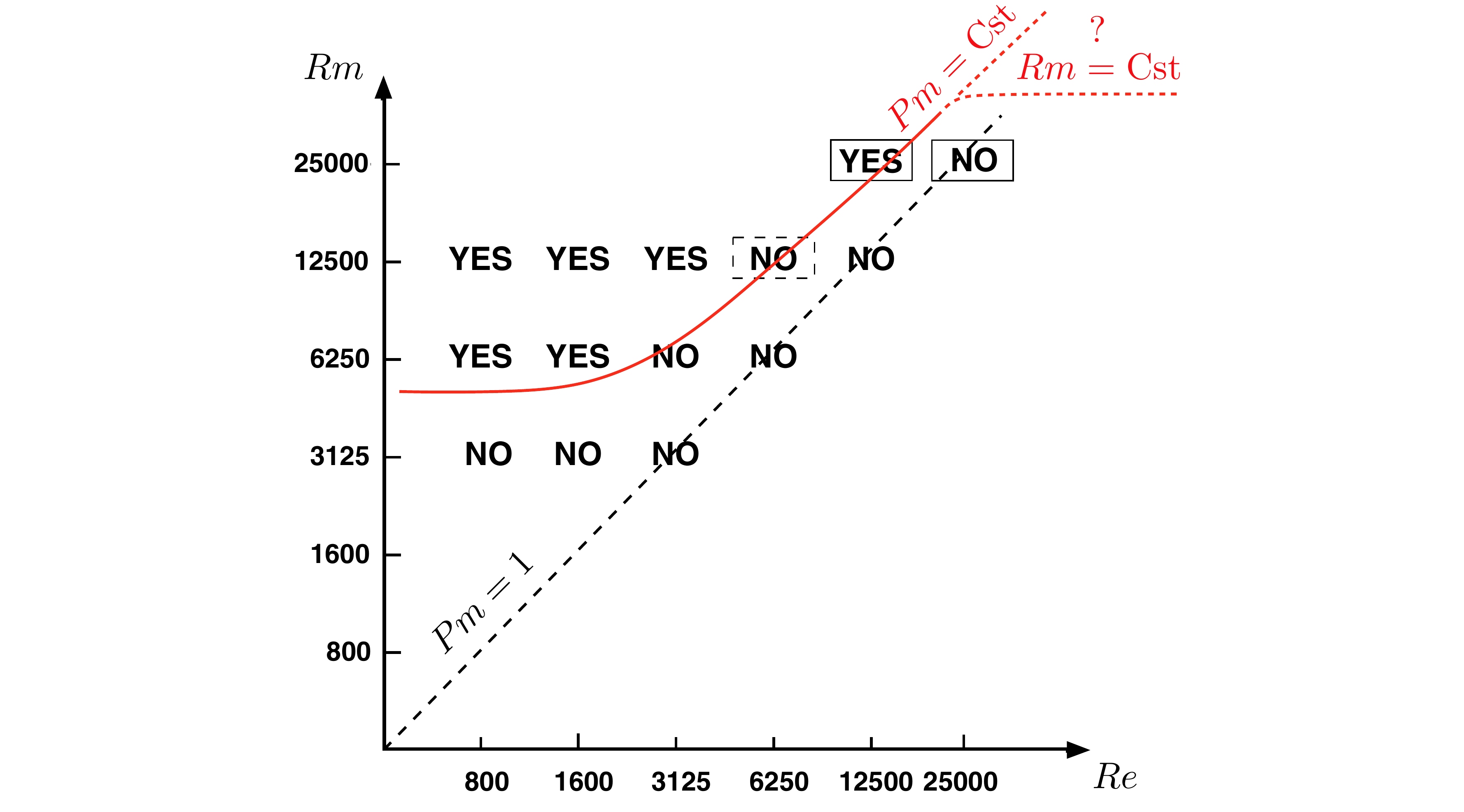}
\caption{Dependence of the MRI dynamo transition on $Re$ and $Rm$ in
  numerical simulations in the shearing box model
  \citep[adapted from][]{fromang07b}. ``Yes'' points correspond to
  simulations in which zero net-flux MRI turbulence is sustained for
  long times. The solid red line has been added to the original plot to
  outline the increase of the typical $Rm$ at which the transition
  occurs with increasing $Re$. The dashed red lines outline the
  two possible options for the actual dynamo transition diagram: a lower
  limit in $Pm$, or a lower limit in $Rm$. This question-mark red zone
  is still \textit{terra incognita}.\label{figfromangstability}}
\end{figure}

Having laid out the general principles and main features of
MHD-instability-driven dynamos, we are now in a position to
discuss several important problems at the core of current
research on the MRI dynamo and accretion-disc dynamos,
but also potentially relevant in a much broader context.
The first of these issues is a reported strong
dependence of the MRI dynamo excitation on the relative
ordering of viscous and resistive dissipative processes
\citep{fromang07b}. \Fig{figfromangstability} shows that
zero-net-flux MRI turbulence is only sustained for $Pm\gtrsim
1$ for typical Keplerian shearing box-type simulations at moderate
$Rm$ accessible to current supercomputers\footnote{$Re$ and
  $Rm$ in the MRI dynamo problem are usually constructed from the
  global shearing rate $S$ and typical vertical scale height
  $H$ of the domain, $Re=SH^2/\nu$, $Rm=SH^2/\eta$. The systematic shearing
  of non-axisymmetric fluctuations in this problem implies that dynamo
  action occurs for significantly larger values of $Rm$ than in
  other dynamo problems. This gives the misleading impression that the
  regimes investigated have asymptotically large $Rm$, while they are
  essentially transitional. Note also that the transition $Re$ and
  $Rm$ in \fig{figfromangstability} are much larger than in
  \fig{figriols13}. This difference is a consequence of the longer
  time taken to wind-up non-axisymmetric MRI-unstable fluctuations
  in the much larger azimuthal to radial/vertical aspect ratio domain
  used in the latter study \citep{herault11,riols15}.}.
This problem, of course, is reminiscent of the small-scale dynamo
transition problem (compare the general trends in
\fig{figfromangstability} and \fig{figgrowthratePm}), for which 
it is now established that the dynamo survives at low $Pm$ (\sect{lowPmsection}).
However, we have also found that the mathematical structure of these two 
dynamo transitions is quite different, so that extrapolating the low-$Pm$
small-scale dynamo existence result to the MRI dynamo requires quite a
leap of faith. As a matter of fact, even the latest brute force
spectral simulations of the MRI problem at numerical resolutions now
comparable to those at which a small-scale dynamo was found at
$Pm\simeq 0.1$ have failed to exhibit a truly low-$Pm$ MRI dynamo
\citep*{walker16}, so that it cannot currently be excluded that these
two problems have different answers.

The physical reasons underlying the $Pm$-dependence of the MRI dynamo
transition remain debated. There have for instance been suggestions
that it may already be explained at a quasilinear level through a
$Pm$-dependence of linear non-axisymmetric MRI-unstable perturbations,
or of their direct nonlinear feedback on the dynamo field
\citep{squire15MRI}. Another possible explanation put forward in the
context of a detailed analysis of nonlinear cyclic solutions 
is that the disappearance of the dynamo at increasing $Re$ and
fixed $Rm$ (and therefore decreasing $Pm$) is due to an increase of
the turbulent mixing of the magnetic field \citep{riols15}. As $Re$ is
increased, MRI-unstable non-axisymmetric velocity fluctuations excited in the
self-sustaining process are allowed to cascade down inertially into
smaller-scale fluctuations (this effect is not captured in quasilinear models).
These fluctuations then effectively act as a turbulent diffusivity on the
MRI-mediating large-scale axisymmetric field, making it harder to
sustain it. In order to recover a dynamo at larger $Re$, one therefore
also has to increase $Rm$, resulting in the seeming $Pm\simeq \mathrm{Constant}$
transition barrier of \fig{figfromangstability}. This turbulent mixing
effect has also been diagnosed in direct numerical simulations
\citep{walker17}. This second $Pm$-dependence explanation is also not
particularly surprising or original in the light of some of our
earlier discussions: for instance, the Kazantsev analysis of the
small-scale dynamo problem shows that it is the increase of the
roughness of the velocity field which makes small-scale dynamo action
harder at low $Pm$ (\sect{kazantsevregimes}). Turbulent diffusivity
and ``noisy'' velocity fields are also known to be a possible
impediment to dynamo action in low-$Pm$ liquid-metal experiments
(e.g. \cite{frick10,forest12,miralles13}, see also
\cite{ponty05,laval06,dubrulle11} for simulations and theory). There
is, however, a subtle but possibly important difference between the
MRI dynamo case and the classical scenario in
which a prescribed turbulent flow mixes the magnetic field: in the MRI
problem, the excitation of the turbulence rests on the very existence
of the dynamo field, hence turbulent diffusion/mixing is itself a
subcritical phenomenon tied to the MRI. 

If the latter explanation for the $Pm$-dependence of the MRI dynamo
transition is correct, the survival of the dynamo at low $Pm$ and
large $Rm$ should depend on whether the rate of nonlinear induction
of the large-scale component of the magnetic field mediating the MRI
exceeds the rate at which turbulent mixing erodes it. Which
mechanism is dominant asymptotically is difficult to assess:
due to the characteristics of the MRI, these two rates are \textit{a  priori}
formally both of the order $S$ (or $\Omega$) if the typical (vertical
and radial) scale of field-mixing fluctuations is comparable to that
of the dynamo field itself (in Keplerian discs,
$\Omega=v_{\mathrm{th}}/H$, where $v_{\mathrm{th}}$ is the thermal
sound speed and $H$ is the typical vertical pressure scale height, and
MRI fluctuations are typically mildly subsonic). The answer may
therefore depend on miscellaneous factors such as the problem geometry.
Several numerical studies have in fact hinted at a mild dependence 
of the apparent $Pm$ transition barrier on parameters such as
the vertical stratification or geometric aspect ratio of the system
\citep{davis10,oishi11,nauman16}, pushing it down to $Pm\simeq 0.5$. 
It is entirely possible that these parameters affect the characteristics of
velocity fluctuations and their mixing interactions with the dynamo
field, and it has in particular recently been argued that the
problem may be overcome by making the vertical dimension of the system
(and vertical scale of the dynamo field) large enough in comparison to
the other dimensions \citep{shi16,walker17}. However, definitive
numerical evidence for the MRI
dynamo at ``asymptotically'' small $Pm$ (say $\lesssim 0.1$) and large
enough $Rm$, of the kind gathered on the small-scale dynamo problem, is
still lacking for any kind of zero net-flux configuration. 
Consequently, it is not known whether the dynamo transition curve
sketched in \fig{figfromangstability} is actually 
a constant $Pm$ or constant $Rm$ line at asymptotically large $Re$ and $Rm$.
Survival of the dynamo down to $Pm=0.1$ has so far only been found in Keplerian
simulations with boundary conditions allowing for the creation of net
azimuthal magnetic flux \citep{kapyla11}. Such configurations are
however formally outside the realm of the original self-sustaining
MRI dynamo problem. We will briefly come back to this problem at the
end of the next paragraph.

\subsubsection{From the MRI dynamo to large-scale accretion-disc
  dynamos*\label{subtostat}}
In the previous paragraphs, we covered some theoretical aspects 
of instability-driven dynamo \textit{transitions} at ``mild'' $Rm$.
But what about the fully-developed, large-$Rm$ regime~? Is it possible
to formulate a physically-motivated statistical closure theory in this
regime that would describe the dynamics of the large-scale dynamo field,
such as the butterfly dynamo activity displayed in \fig{figdavis10}~? 
If so, how does it relate to the mean-field theory of large-scale
dynamos reviewed in \sect{largescalemeanfield}~?

\begin{figure}
  \centering\includegraphics[width=\textwidth]{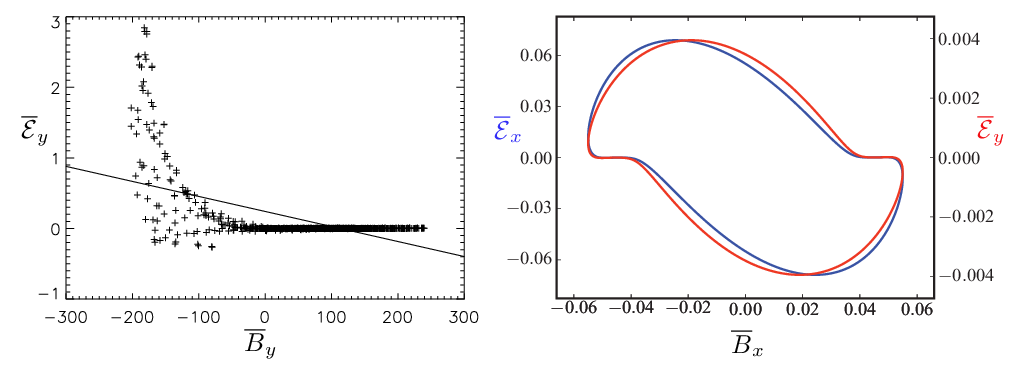}
  \caption{Phase portraits of simple transitional instability-driven
    dynamo solutions representing
    the instantaneous relationship between the azimuthally-averaged
    electromotive force and magnetic field. Left: magnetic buoyancy
    dynamo case (adapted from \cite{cline03} using the notations of this
    article). Right: nonlinear cyclic incompressible MRI dynamo
    solution in the shearing box 
    \cite[adapted from][]{herault11}.\label{figphaseportraits}}
\end{figure}

\paragraph{\textit{$\meanvEMF(\meanvB)$
  relationship in the transitional regime\label{transitional}}.}
A natural first step to approach this difficult problem is to look in
numerical simulations at the relationship
$\meanvEMF(\meanvB)$
between the large-scale axisymmetric/streamwise-independent field
$\meanvB$ and the EMF $\meanvEMF$
that feeds it, starting cautiously from the dynamically simpler
transitional regime. \Fig{figphaseportraits} shows this relationship for
relatively laminar (chaotic or just periodic) nonlinear solutions
of the magnetic-buoyancy-driven and (unstratified, incompressible) MRI
dynamo problems in this regime. Clearly, the classical linear
mean-field ansatz~(\ref{eq:meanfieldexpansion}) postulating a linear
$\meanvEMF(\meanvB)$ relationship
does not describe the dynamics of these solutions very well. Note also
that applying this ansatz here has the effect of linearizing
\equ{eq:instabilitydriven} for $\meanvB$, thereby ``degrading''
the pristine subcritical nonlinear dynamo transition into a kinematic
linear dynamo instability. The knot of the problem is that the EMF
$\meanvEMF=\mean{\fluctvU\times\fluctvB}$
sustaining instability-driven dynamos is primarily due to what we will
refer to as \textit{active} non-axisymmetric/streamwise-independent
fluctuations, i.e. fluctuations $\fluctvU$ and $\fluctvB$ that are
\textit{jointly} driven by an MHD instability and are locked into a
particular phase relationship. At large $Re$ and $Rm$,
these fluctuations cascade down into smaller-scale
slaved, \textit{passive} fluctuations that are not MRI-unstable,
however the dynamo itself does not owe its existence to this
small-scale ``turbulence'' \citep[see
e.g.][]{cline03,davies11,herault11,riols15,squire15MRI,bhat16b}. The
cross-correlation $\meanvEMF$ between the 
active fluctuation components  $\fluctvU$ and $\fluctvB$
must depend on the dynamo field $\meanvB$ that mediates the
instability somehow, but there is no obvious reason why it should be
directly proportional to $\meanvB$ in the transitional
regime. In fact, asymptotic quasi-linear calculations of the EMF in
the Keplerian MRI dynamo problem, in which $\fluctvU$ and $\fluctvB$
are simply taken to be linear MRI eigenmodes in a sinusoidal azimuthal
field $\meanB_y(z)$, already predict a complex nonlinear functional
relationship $\meanvEMF(\meanvB)$ \citep{lesur08b}.

\paragraph{\textit{Statistical regime of the unstratified MRI
    dynamo\label{statisticalunstrat}}.}
That a simple classical linear mean-field ansatz
fails to describe instability-driven dynamos at transitional $Rm$ is
not altogether particularly surprising considering that these dynamos
are neither kinematic, nor statistical/turbulent in this regime.
However, we also know that the  dynamical complexity of these dynamos
increases tremendously with $Rm$, motivating the quest for a fully
statistical description at large $Rm$. The remainder of this
discussion will focus on the Keplerian MRI
dynamo, whose statistical theory has been under most intense scrutiny
among the family of instability-driven dynamos, due to its broader
relevance to astrophysical accretion.

For the sake of clarity, we will also concentrate on the simplest
possible incompressible, unstratified Keplerian MRI dynamo problem
with boundary conditions that conserve magnetic flux, and postpone the
discussion of possible stratification effects relevant to the
accretion disc context to the next paragraph. This unstratified problem
already presents us with a significant statistical theory challenge,
in that its simulation at super-transitional $Rm$ already displays
coherent large-scale MRI dynamo activity cycles with periods much
larger than $\Omega^{-1}$, intertwined with turbulent MHD fluctuations
at all scales available down to the dissipation  scales
\citep{lesur08}. These long-lived statistical activity cycles
are much more complex than the
laminar nonlinear cyclic solutions described in \sect{NLcycles},
but their dynamics still appears to be essentially powered
by the self-sustaining nonlinear MRI dynamo process, now understood as
a large-scale MRI-driven statistical injection mechanism
\citep{riols17}. More specifically, their sustainment appears to
result from a statistical accumulation of nonlinear EMF increments
associated with an ensemble of non-axisymmetric MRI-unstable
active fluctuations growing and evolving over the shorter $\Omega^{-1}$
timescale. Importantly, smaller-scale turbulent MHD fluctuations are
slaved to this large-scale statistical dynamo SSP via a
\textit{direct} nonlinear cascade, and do not therefore appear to be 
the primordial cause of the dynamo through an inverse cascade%
\footnote{The same kind of conclusion also seems to
  apply to the dynamical origin of large-scale motions and
  regeneration cycles in the cousin problem of non-rotating
  hydrodynamic turbulent shear flows
  \citep{rawat15,farrell16}.}. In this respect, the dynamics is not
fundamentally different from that of other turbulent flows powered by
different natural instabilities, such as thermal convection.
In both cases, we find simple nonlinear solutions at relatively low
values of the control parameter (simple nonlinear cycles/saturated
convection rolls or hexagons) that become increasingly chaotic as the
latter is increased. A vigorous, well-ordered system-scale dynamics
representative of the underlying instability mechanism and its
symmetries is recovered statistically in
the fully developed turbulent regime
(MRI dynamo SSP-driven statistical cycles/buoyancy-driven
thermal winds). The main qualitative difference between the zero
net-flux MRI dynamo problem and a problem like thermal convection is
the nonlinear mathematical nature of its transition.

Having argued that the SSP should be envisioned as a statistical
process in the large-$Rm$ regime, we may ask whether some meaningful
lower-dimensional statistical theory or numerical closure
model of the large-scale dynamics and nonlinear stresses can be devised%
\footnote{Interestingly, there is a growing
  appetite for similar reduced analytical and numerical models in the
  hydrodynamic shear flow community
  \citep{barkley11,rath13,barkley15,rawat15}.}.
A two-scale statistical approach based on a separation
between non-axisymmetric MRI-unstable fluctuations evolving on the
$\Omega^{-1}$ timescale on the one hand, and the axisymmetric
dynamo MRI-mediating field evolving on a significantly larger
timescale on the other hand, makes sense at large $Rm$.
What kind of large-scale couplings should we expect in this context ?
While the $\Omega$ effect is an integral part of the MRI
dynamo SSP, different studies have shown that the off-diagonal
term that couples back the azimuthal component $\meanB_y$ of the
large-scale field  to the radial $\meanB_x$ component is not reducible
to an $\alpha$ effect in the unstratified regime, i.e.
unstratified MRI dynamo cycles are not $\alpha\Omega$ cycles
\citep{lesur08,gressel10,squire15MRI}. In fact, the net
(volume-averaged) kinetic and magnetic helicities are
numerically zero in simulations without stratification. Instead, 
mean-field projections of the dynamics of simulations using the
test-field method suggest that the mean-field version of this
dynamo can be \red{essentially described by the anisotropic diffusion
  \equs{Bxoffdiag}{Byoffdiag}.}
This is not really surprising in the sense that an off-diagonal
diffusive term $-\eta_{yx}\partial^2_z\meanB_y$  is the next-simplest
way after the $\alpha$ effect to couple back the azimuthal field to
the radial field in the mean-field framework. The interesting question
is whether such a prescription is actually functionally correct in the
large-$Rm$ regime (as shown in \fig{figphaseportraits} (right), it is
not in the transitional regime) and, if so, whether it can be derived
in a transparent, physical way.

There is still no definitive answer to these questions.
Careful numerical scrutiny of increasingly complex
dynamo cycles suggests that $\meanvEMF(\meanvB)$ is 
closer on average to a linear relationship for super-transitional
statistical cycles at large $Rm$ than for simpler transitional cycles
\citep{riols17}. The reasons underlying this behaviour are not 
understood: they may involve the particular physics of the
MRI, or may be a more universal statistical convergence effect 
connected to the improved scale-separation between the period of
statistical cycles and the typical timescale of evolution of
MRI-unstable fluctuations in the statistical regime.
As explained earlier, the statistical dynamo SSP essentially relies on an
accumulation of MRI-unstable fluctuations excited in the injection
range, not on smaller-scale inertial MHD turbulence. Therefore, a
quasilinear approach explicitly factoring in the role of linear
shearing wave packets transiently amplified by the MRI appears to provide the
best starting point for the development of self-consistent statistical
theory \citep{lesur08}. A possibly complementary way to look at this
problem in a ``mean-field'' sense is in terms of a large-scale non-helical
magnetic-shear-current-effect-driven dynamo (\sect{MHDsheardynamo})
supported by nonlinear MHD fluctuations driven by the MRI
\citep{squire15PRE,squire16JPP}. 

For completeness, let us finally mention a radically different (and
admittedly slightly exotic) potential avenue of research on this
problem. As explained in \sect{NLcycles}, we may abstractly envision the dynamics
of the MRI dynamo at large $Rm$ as an intricate game of pinball in the phase
space of the dynamical system, with bumpers consisting of many different
nonlinear cycles embedded in homoclinic and heteroclinic tangles. At
asymptotically large $Rm$, one would naturally expect some kind of
ergodic regime in which a large region of the phase space is densely
populated with such structures, motivating
an effective description of the large-scale component of the turbulent
dynamics based on the statistics of these
structures. This is the idea underlying periodic
orbit theory (\cite{cvit92}, see also \cite{cvit16} for a textbook
introduction). While this approach has not yet matured sufficiently to
become usable in practical fluid-dynamical situations, it has started
to receive significant attention in recent years in the context of the
non-rotating hydrodynamic shear flow turbulence problem 
cousin to the instability-driven dynamo problem
\citep[e.g.][]{chandler13,lucas15,budanur17}, and is therefore 
worth keeping an eye on in the future with the dynamo context in mind.

\paragraph{\textit{Large-scale dynamo processes in stratified accretion discs}.}
The MRI dynamo problem is obviously relevant to astrophysical
accretion, but accretion discs are stratified along the
$z$-direction perpendicular to their mid-plane. and likely have open
magnetic boundaries at their coronal level. What kind of impact do these
effects have on a dynamo in a Keplerian shear flow ? This problem
remains a bit of a quagmire and a detailed review of it is well beyond
the scope of this review. However, it is interesting for the sake of
completeness to discuss at a basic level some of the new physics that
appears in this context in connection with the large-scale
statistical dynamo mechanisms reviewed in \sect{largescale}. 

First, as we saw in \sect{alphastrat}, stratification facilitates the
excitation of a large-scale helical dynamo in rotating turbulence
involving vertical motions. In the context of stratified accretion discs,
nonlinear MRI dynamics becomes intertwined with magnetic-buoyancy
dynamics perpendicular to the plane of the disc, which can in turn
give rise to an $\alpha$ effect\footnote{To
  avoid any possible confusion, we emphasise that this
  mechanism is distinct from the magnetic-buoyancy-driven
  dynamo \citep{cline03} discussed earlier in \sect{instabdynamoevidence}.
  The latter does not require rotation, and it is actually unclear
  whether it can operate when $\Omega$ is of the order of $S$
  \citep[see discussion in][]{rincon08}.}, or contribute
to a $z$-migration of the activity patterns visible in \fig{figdavis10}
\citep{branden95,davis10,donnelly13}. Second, open magnetic
boundary conditions enable the build-up of a net magnetic flux
which, in a Keplerian flow, provides a means to steadily excite
turbulent MHD fluctuations through the ``classical'' linear MRI in a
mean net-flux field. This turbulence may in turn tangle this net field
and reinforce it along the lines of a classical statistical large-scale
dynamo scenario seemingly distinct from the self-sustaining MRI dynamo 
process \citep[see e.g.][]{kapyla11,gressel15}. Finally, as explained
in \sect{quenchingmodels}, such boundaries also in principle 
make it possible to circumvent the magnetic helicity conservation
constraint, with possible implications for large-scale dynamo growth.
As with the generic $\alpha$ effect problem reviewed in
\sect{largescale}, there is however still quite a lot of disagreement
and confusion surrounding the question of possible connections between
accretion-disc dynamos and magnetic helicity dynamics
\citep[e.g.][]{vishniac09,blackman12,ebrahimi14,bodo17}.

This multitude of possible dynamo-facilitating effects in discs, on
top of the unstratified MRI dynamo SSP discussed earlier, has given
rise to a variety of phenomenological explanations of the cyclic
``butterfly-type'' statistical behaviour illustrated in \fig{figdavis10}. 
The relative dynamical importance and connections between these
various effects have however been particularly difficult to assess due
to the intrinsic statistical and physical complexity of simulations,
and the jury is still out as to which one is fundamentally the driver
of cyclic dynamo activity in MHD simulations of astrophysical accretion
\citep[e.g.][]{gressel15,squire15PRE}. The personal opinion of this
author, largely informed by Occam's razor, is that the  backbone
of accretion-disc dynamos, if they exist, must be the
unstratified MRI-driven dynamo SSP. After all, the
MRI remains the only known viable, numerically demonstrable physical
mechanism by which turbulent kinetic energy can be extracted from the
shear in a Keplerian flow. This, in conjunction with the fundamentally
magnetic-field-inducing nature of this instability, strongly suggests
that it is one way or another central to the statistical growth of
magnetic energy in the accretion-disc dynamo problem.

\subsubsection{Other instability-driven and subcritical dynamos\label{otherinstab}}
\paragraph{\textit{Taylor-Spruit and magnetoshear dynamos}.}
Besides the MRI and magnetic-buoyancy-driven dynamos, there
are at least two other potential MHD-instability-driven dynamo
mechanism candidates that may be understood in terms of a subcritical
nonlinear dynamo SSP. The first one involves a non-axisymmetric
Tayler-type MHD instability \citep{tayler73,pitts85}, and might be
relevant to the magnetism of stably stratified stellar radiative zones
\citep{spruit02}. While there have now been several numerical studies
dedicated to this problem
\citep*{braithwaite06,zahn07,gellert08,jouve15}, a clear, reproducible
numerical demonstration of dynamo sustainment in the full MHD problem
is still lacking. Interestingly though, subcritical behaviour and
sensitive initial dependence on initial conditions has recently been
observed in a nonlinear mean-field model of this dynamo modelling the
nonlinear feedback of Tayler-unstable fluctuations as a
periodic $\alpha$ effect with a quadratic large-scale magnetic-field
dependence \citep{stefani19}.

The second mechanism involves non-axisymmetric 
magnetoshear instabilities in thin, stably-stratified shear layers
\citep*{cally03,miesch07}. This mechanism may in particular turn out
to be important in the solar dynamo context, as the results of a few
global numerical simulations now suggest that such instabilities
could control and perhaps even drive some aspects of the dynamo at the
tachocline \citep*{miesch07b,charbonneau14,lawson15,gilman18,plummer19}.

\paragraph{\textit{Different kinds of subcritical dynamo transitions}.}
Despite their potential astrophysical relevance, MHD-instability-driven
dynamos form an admittedly rather special class of nonlinear
subcritical dynamos. In order to provide a slightly more complete
overview of subcriticality in dynamo theory, we should therefore also
say a few words about other known instances of subcritical dynamos,
whose dynamical phenomenology appears to be quite
distinct from that of MHD instability-driven dynamos. In a vast
majority of cases, these dynamos can be understood as originating in
local subcritical pitchfork or Hopf bifurcations: in this scenario, a
three-dimensional (possibly differentially-rotating) flow excites a
kinematic dynamo above a well-defined linear $Rm_c$, but the
dynamo can survive below $Rm_c$ due to the cooperation of the
nonlinear effects of the Lorentz force. There are different ways in
which the latter can be indirectly beneficial to magnetic field
amplification, an important example of which is by counteracting the
impeding effects of the Coriolis force on thermal convection in
convection-driven dynamo problems. This kind of dynamical
phenomenology is usually associated  with bistable and
hysteretic behaviour, at it gives rise to two
subcritical weak-field and strong-field branches merging at a
saddle-node bifurcation at $Rm_{\mathrm{SN}}<Rm_c$. Subcritical dynamo
transitions like this have been discussed for many years in the
context of geodynamo and planetary dynamo problems
\citep*[e.g.][see also \cite{dormy11} for an accessible introductory
review]{roberts78,roberts88,christensen99,kuang08,goudard08,
  simitev09,morin09,sreenivasan11,dormy16,dormy18},
but have also drawn attention in the context of laboratory experiments
\citep{fuchs01,ponty07,berhanu09,monchaux09,yadav12,verma13}, and more
recently stellar magnetism \citep{browning08,morin11,gastine12,gastine13}.

\subsection{The large-$Rm$ frontier\label{largeRmfrontier}}
Let us conclude this section with a discussion of broader numerical
efforts aiming at probing and understanding the behaviour of
large-scale dynamos at very large $Rm$. As mentioned in \sect{numexplor},
the consistent generation of large-scale magnetic fields in the latest
generation of simulations of the solar dynamo, for instance, suggests that a
dynamical tipping point has been reached, or is close to being reached
in the three-dimensional numerical modelling of many large-scale dynamos.
However, the sensitivity of the results to the numerical and physical
implementation of the models also suggests that they are not yet asymptotic
by any measure and lie in a rather uncomfortable transitional parameter
range that is barely resolved. This is not entirely surprising, considering that most global
simulations to date barely exceed the small-dynamo threshold $Rm_{c,\mathrm{ssd}}\simeq 60$ at
$Pm=O(1)$ (\fig{figgrowthratePm}), and that the exploration of the $Pm<1$ region of parameter
space has only recently become possible. Our lack of theoretical
understanding of fundamental nonlinear large-scale dynamo mechanisms at $Rm\gg1$,
$St=O(1)$ is coming back to bite us here, because we cannot assess
with confidence what aspects of the dynamics are potentially missing,
or are not yet modelled accurately in the simulations.
In this respect, it is important to stress again that the solar dynamo
has merely been used in what precedes as an illustration of the general state and
limitations of current high-resolution numerical simulations of
large-scale astrophysical dynamos. Although different large-scale dynamo
mechanisms may be at work in different environments, the exact same
kinds of problems and challenges are present in the modelling of all
high $Rm$ astrophysical dynamos, including of course stellar,
galactic, and accretion-disc dynamos. Interestingly, the geodynamo likely operates
  at $Rm$ of only $O(10^3)$, a regime that simulations such as that
  shown in \fig{figgeosolardynamo} (top) are fast-approaching (albeit for
$Pm>0.05$). The problem may therefore not be that critical in this
particular context, as we may soon be in the position to adequately
resolve the full range of magnetic fluctuations excited in the Earth's
core numerically.

It remains unclear if the problem of
nonlinear, turbulent large-scale dynamo action at large $Rm$ is
amenable to a universal treatment. We have already discussed in
\sect{NLgrowth} the possibility that dynamical small-scale dynamo
fields creep up to system scales at large $Pm$ typical of the warm
interstellar medium of galaxies, and we know very little about how
saturation works at low $Pm$ typical of stellar and planetary
interiors. Besides, recall that high-resolution Cartesian simulations
of helical dynamos in periodic domains seem to be heavily dynamically
quenched at $Rm$ of the order of a few hundreds and down to $Pm=0.1$
(\sect{numobs}). A new generation of high-resolution simulations
addressing these problems is therefore strongly needed if
further progress is to be achieved. In view of all the remaining
unknowns, numerical experiments using almost any kind of geometric
and dynamical set-up all appear to be legitimate in this quest,
as long as as they can reach highly supercritical $Rm$ and produce
strong organised large-scale fields (preferably at low or large $Pm$).

\begin{figure}
 \centering\raisebox{0.0\textwidth}{\includegraphics[height=0.615\textwidth]{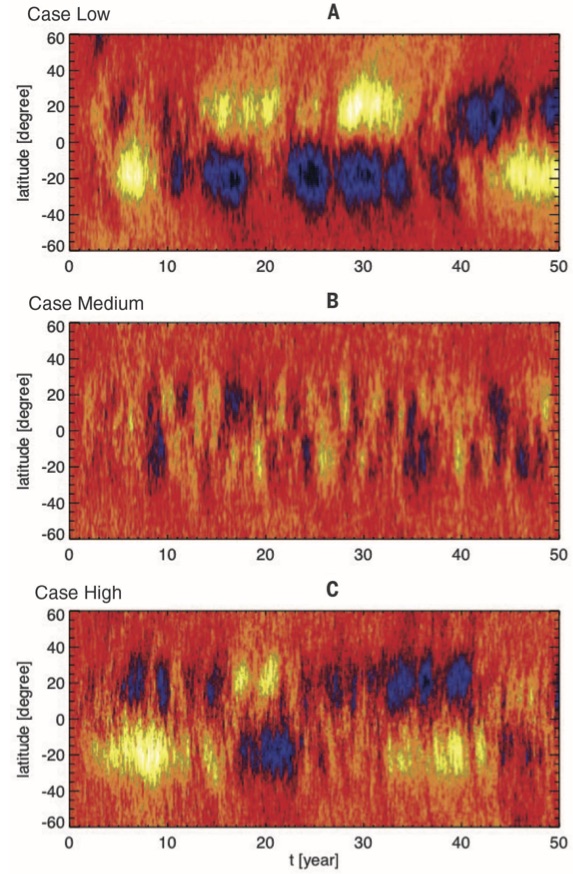}}
 \hspace{0.05\textwidth}\includegraphics[height=0.6\textwidth]{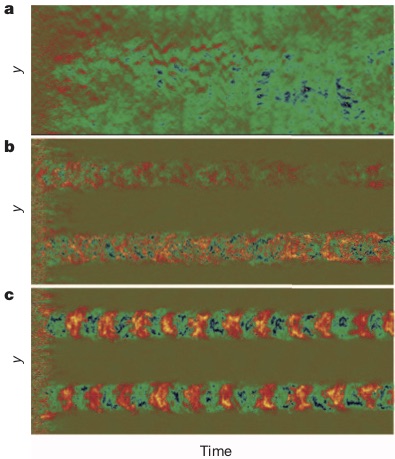}
\caption{Two seemingly very qualitatively different recoveries of
  large-scale dynamo waves in helical turbulence with shear at large
  $Rm$. Left: toroidal magnetic-field renderings in \textit{nonlinear}
  spherical-shell dynamo simulations of rotating turbulent convection
  at increasingly large $Rm$ up to 2000 \citep[adapted from][]{hotta16}. A
  wave is observed at relatively low $Rm$ (top
  panel). It is overwhelmed by small-scale fields at $Rm$ mildly
  supercritical with respect to small-scale dynamo action
  (middle panel), but re-appears at even larger $Rm$
  (lower panel). Right: magnetic-field renderings in
  2.5D \textit{kinematic} dynamo simulations of an array of GP flows
  in the presence of a large-scale sinusoidal shear $\meanU_x(y)$
  \citep[adapted from][]{tobias13} at $Rm=2500$. No clear large-scale dynamo
  activity is observed in the presence of a small-scale helical flow
  and no shear (top panel), or in the presence of a non-helical flow
  and of a strong shear (middle panel).  A large-scale dynamo wave
  only emerges at high $Rm$ if the flow is helical and the shear is
  made of the order of the turnover rate of the flow (lower
  panel).\label{hottatobias}}
\end{figure}

Several groups have started to play this game in very
different ways. For instance, a recent analysis of the brute-force
high-resolution solar dynamo simulations at $Rm=O(10^3)$,
$Pm=O(1)$ shown in \fig{figgeosolardynamo} (bottom), 
suggests that dynamically saturated small-scale fields
at $Rm>Rm_{c,\mathrm{ssd}}$ could have a beneficial effect on the development
of a large-scale dynamo field \citep[][shown in \fig{hottatobias}
left]{hotta16}. This result seems somewhat reminiscent of the
magnetic-shear-current-effect-driven dynamo discussed earlier in
\sect{MHDsheardynamo}, but \citeauthor{hotta16}
argue that this behaviour is related to the suppression or
quenching of small-scale flows by the dynamically
saturated small-scale fields. It is not entirely clear 
why and how such a suppression would facilitate the emergence
of a large-scale field but, in any case, this effect, if
 real, appears difficult to derive analytically and cannot easily be
 predicted and understood without the assistance of numerical
 simulations. It is also easy to see how different simulations 
 conducted at limited resolutions and/or based different small-scale closure
 schemes could miss small-scale dynamical effects of this
 kind, or just poorly describe them.

Another idea recently put forward on the
basis of much more idealised 2.5D Cartesian simulations at $Rm=2500$
of kinematic sheared helical dynamos driven by an array of replicated
Galloway-Proctor flow cells, is that the emergence of coherent
large-scale fields at large $Rm$ is only possible if the small-scale
dynamo (at or below the GP flow scale in this case) is somehow suppressed
\citep[][shown in \fig{hottatobias} right]{tobias13,cattaneo14}. In
these simulations, a large-scale Parker wave only emerges at high $Rm$
if there is a strong-enough large-scale shear flow that suppresses or
at least impedes the small-scale dynamo. Similar results have been
obtained at more modest
$Rm$ in 3D simulations of the same problem in the fully nonlinear regime
\citep{pongkiti16}.
For the suppression of the small-scale dynamo to be
possible, however, the shearing rate must be comparable to the
turnover rate of the eddies driving the small-scale dynamo most
efficiently. The potential problem with this ``suppression
principle'',  then, is that the turnover rate of these eddies
will always be much larger than the global rotation or shearing rate
at large $Re$ typical of astrophysical systems. It is therefore
difficult to envision how the system could avoid being infested by
small-scale fields in this regime. The proponents of the shear-suppression
principle argue that the energy of these fields should be comparable
to the relatively small kinetic energy of the small-scale eddies (in
comparison to that of the eddies at the turbulent injection scale),
but there is currently no guarantee that this is true, especially
considering the many remaining uncertainties
regarding the ultimate nonlinear fate and scale of saturation of
small-scale dynamo fields  (\sect{NLss}).

These are just a few examples of the diversity of 
activities on this problem. Another interesting ongoing thread
already discussed in \sect{quenchingmodels} is the  numerical research
on the effects of boundary conditions on helicity fluxes. And what about
the different large-scale non-helical shear dynamo mechanisms discussed in
\sect{sheardynamo} ? Could they be particularly relevant, and possibly
dominant, in large $Rm$ astrophysical conditions, if helical dynamos
are catastrophically quenched ? As frustrating as
it seems today, our earlier observation that the current generation of
simulations of large-scale MHD dynamos is probably at a ``tipping
point'' is in fact rather welcome news for numerical and theoretical
research alike, as it all but guarantees that the next ten to twenty
years of numerical research on turbulent large-scale MHD dynamos at
large $Rm$ will be very exciting.

\section{Dynamos in weakly-collisional plasmas\label{plasma}}
The previous sections have been concerned with dynamo theory 
in single-fluid MHD. However, many plasmas in the Universe
are not in this regime ! For instance, cold protoplanetary-disc
accretion involves multi-fluid effects such as ambipolar
diffusion and the Hall effect \citep{armitage10,fromang13}. Equally
importantly, there is a wide range of weakly-collisional,
high-energy astrophysical plasmas, including the ICM, protogalactic
plasmas,  hot accretion flows around black holes, and supernova or GRB
shocks, for which even a standard multi-fluid MHD description is
\textit{a priori} inappropriate. These plasmas host, and have
almost certainly dynamically amplified magnetic fields throughout
cosmic times \citep{zweibel97,kulsrud08,subramanian19}. It is
therefore essential from an astrophysical and cosmological perspective
to understand how dynamos behave in this kind of
environment. But the development of plasma dynamo theory is also
motivated by experimental prospects: several important limitations
of liquid-metal MHD experiments, not least the impossibility to reach
$Pm\geq 1$ regimes, have become painfully apparent and annoying in
recent years. Several experimental projects have recently been started
to attempt to circumvent this problem by creating new experimental
designs that use either dense (essentially fluid, but large $Pm$) laser plasmas
\citep{meinecke15,tzeferacos18}, or more dilute, variable $Pm$ plasmas
confined into metre-size spherical or cylindrical vessels \citep{forest15,plihon15}. 

The aim of this section is to provide a reasonably light introduction
to the magnetised plasma physics involved in the plasma
dynamo problem, and to discuss some possible implications of
kinetic effects on dynamos in the light of a few recently
obtained theoretical and numerical results. For the sake of briefness,
we leave out multi-fluid effects in what follows. Readers interested in the
particular topic of the Hall effect are referred to the numerical
studies of \cite{mininni05} on helical, large-scale dynamos and
\cite*{gomez10} on small-scale dynamos, and references therein.
Other multi-fluid effects, e.g. ion-neutral collisions, are known
to be at least indirectly relevant to the experimental context
and are for instance critical to the driving of dynamo flows 
in low-density plasma experiments \citep{cooper14}. We also leave 
out the important question of the generation of cosmological seed fields 
by plasma mechanisms like the Biermann battery, as these represent
a rather different and already thoroughly-reviewed area of research
\cite[e.g.][]{kulsrud08,widrow12,durrer13,subramanian19}.

\subsection{Kinetic versus fluid descriptions}
The dynamics of charged plasmas is generically described by the
Vlasov equation for ion and electron species $s=i,e$,
\begin{equation}
\label{eq:vlasov}
\dpart{f_s}{t}+\vec{v}\cdot\grad{f_s}+\left[\f{Z_se}{m_s}\left(\vE+\f{\vec{v}\times\vB}{c}\right)+\frac{\vec{F}_s}{m_s}\right]\cdot\dpart{f_s}{\vec{v}}=C_s\left[f_s\right]~,
\end{equation}
coupled to Maxwell's equations (subject to the quasi-neutrality
condition in all regimes of interest here). Here,
$f_s(\vec{r},\vec{v},t)$,  $Z_s$ and $m_s$  are respectively the
distribution function, particle charge and mass of species $s$,
$\vec{v}$ is the velocity of particles, $\vec{F}_s$ stands for an
external force acting on the plasma particles, and
$C_s\left[f_s\right]$ is a collision term. In three-dimensional
physical space, distribution functions are usually functions of three
spatial dimensions, three velocity-space dimensions, and
time. Needless to say, solving the time-dependent Vlasov-Maxwell
system in six dimensions is challenging from both mathematical
and numerical perspectives.

In the limit where collisions take place on much shorter time and
spatial scales than any other physics, the kinetic
(velocity-space) degrees of freedom are destroyed, the distribution
functions are Maxwellian, and the entire dynamics of the
plasma is described by our beloved isotropic three-dimensional fluid
MHD \equs{eq:cont}{eq:entropy} (in the single-fluid case).
In the more general magnetised, weakly-collisional or fully
collisionless case, continuity, momentum and energy equations for
each species are obtained by taking respectively the zeroth, first and
second order velocity moments of  \equ{eq:vlasov}, with the bulk
(fluid) number density, velocity, and pressure tensor of each species given by
\begin{equation}
n_s(\vec{r},t)=\int\diff^3\vec{v}f_s~,
\end{equation}
\begin{equation}
\label{eq:firstmoment}
\vU_s(\vec{r},t)=\f{1}{n_s}\int\diff^3\vec{v}\,\vec{v}f_s~,
\end{equation}
\begin{equation}
\label{eq:ptens}
\tens{P}_s(\vec{r},t)=m_s\int\diff^3\vec{v} (\vec{v}-\vU_s)(\vec{v}-\vU_s)f_s~,
\end{equation}
but the system now also has many  internal degrees of freedom
encapsulated in the non-trivial velocity-space dependences of the
distribution function solutions of \equ{eq:vlasov}. Two
types of dynamics are possible in this kind of system: fluid-like
``macroscale'' dynamics, i.e. turbulence and flow instabilities at
the outer scale of the system (convection, MRI, dynamo etc.), and kinetic
``microscale'' processes, such as collisionless damping, resonances,
velocity-space cascades, plasma oscillations, or Larmor-scale
instabilities (in magnetised plasmas). The problem in that case is
that both types of dynamics are usually strongly coupled:
if we want to understand the macroscale magnetofluid dynamics of
a collisionless plasma, we cannot simply discard the dynamics
taking place at kinetic scales. In particular, we
are not allowed anymore \textit{a priori} 
to use the standard compressible MHD equations on their own. 
At the very least, we should be supplementing them with
physically-motivated prescriptions (fluid closures) for higher-order
fluid moments (pressure tensors, heat fluxes, etc.) describing the
large-scale dynamical effects of kinetic-scale
phenomena. With no such prescription available for the problem at hand
\citep*[these closures are generically very hard to derive rigorously from kinetic
theory, see e.~g.][]{snyder97,passot07}, we have no other choice but to solve
the entire multiscale problem in an appropriate kinetic
framework. For instance, five-dimensional gyrokinetics is commonly
used to study plasma turbulence in magnetically-confined fusion
plasmas \red{\citep{krommes12,catto19}}. In the dynamo problem,
however, the spectre of anti-dynamo
theorems and the absence of a strong external guide field suggest that
the ``simplest'' available kinetic framework is the six-dimensional
Vlasov-Maxwell system (or rather 3D-3V, where 3V denotes the three
velocity-space dimensions) in its almost full glory (bar
quasi-neutrality). 

\subsection{Plasma dynamo regimes}
Before we discuss quantitative attempts to solve this seemingly
daunting problem, or rather, some small parts of it, we will try to
gain some insights into its nature and distinctive character. To keep the
complexity and algebra to a minimum, we will henceforth assume
that the plasma is composed of hot ions and electrons of equal
temperature, $T_e=T_i$. The thermal/sound speed of each
species is $\vths=\sqrt{2k_BT_s/m_s}$, their mean free path is
$\lambdamfps$, and the intra-species collision frequency is
$\nu_{ss}$. We are interested in the weakly-collisional regime $k
\lambdamfpi\ge 1$, $\omega/\nu_{ii}\ge 1$, where $k>1/L$ and $\omega$ are
the typical wavenumber and turnover rate and/or pulsation of dynamical 
fluctuations. Electrons can in principle be either fluid or
collisionless (but we will soon make them collisional to simplify
matters further).

\subsubsection{Different regimes and orderings}
There are arguably two interesting kinetic sub-regimes that we should
consider within this weakly-collisional ion ordering in the context of
the dynamo problem:
\smallskip
\begin{itemize}
\item an \textit{unmagnetised} regime in
which the magnetic field is so weak that the cyclotron frequency of
the ions $\Omega_i=Z_ieB/(m_ic)$ is \red{much smaller than any dynamical 
frequency $\omega$ in the problem ($\Omega_i,\nu_{ii}\ll\omega$), and 
their Larmor (gyration) radius $\rho_i=\vthi/\Omega_i$ is much
larger than any dynamical spatial scale $\ell$ in the problem
($\ell\ll \lambdamfpi, \rho_i$),}
\smallskip
\item a \textit{magnetised regime} in which
the ion gyration dynamics is faster than anything else
($\nu_{ii},\omega\ll\Omega_i$, $\rho_i\ll 1/k, \lambdamfpi$)~.
We will be particulary interested in the \textit{weakly-magnetised} 
case corresponding to an ion plasma beta parameter $\beta_i=4\pi n_im_i
\vthi^2/B^2$ larger than unity (understood as an order of magnitude). 
\end{itemize}
\smallskip

At first glance, dynamo action in the unmagnetised, collisionless
regime appears to be the most basic problem that we may
want to solve, and we will indeed soon find out that a dynamo in the
magnetised regime is a significantly more complex, but also
much more interesting problem. As a quick teaser, let us simply note
that  magnetised, gyrotropic plasmas (collisional
or not, as long as $\rho_s/\lambdamfps\ll 1$) have anisotropic pressure
tensors with respect to the local direction $\hatvB$ of the magnetic
field. It is the presence of these extra degrees of freedom, coupled
to the kinetic degrees of freedom in the collisionless case, that
has major implications for the overall magnetised dynamics
in this regime, including the dynamo process.

A perspicacious reader might be slightly disturbed by the very idea of
considering dynamo action in a ``magnetised'' plasma, since 
one of the goals of dynamo theory is precisely to understand how
magnetic fields grow in the first place.  It is important to
clarify at this stage that plasma magnetisation as defined here,
$\rho_i\le\lambdamfpi$, is distinct from dynamical saturation,
$B^2/(4\pi)\sim n_im_i U_i^2$, and is expected to happen well before the
latter in the process of magnetic field amplification even in
relatively subsonic flows. It is therefore perfectly legitimate  to
ask whether dynamo growth can take place in a turbulent plasma once
the latter has become magnetised in the above sense. 

Finally, because hot, low-density plasmas magnetise for tiny
sub-equipartition  fields, the magnetised regime also appears to be
the most directly relevant to the study of magnetofluid dynamics in a
variety of plasma environments ranging from low-density plasma experiments to
high-energy astrophysical plasmas such as hot accretion flows or 
the ICM. For instance, in the hot ICM, $T\sim 3\times 10^7$~K (only
$T_e$ can be directly inferred from observations, $T_i$ is usually
assumed to be of the same order), $\lambdamfpi\sim
0.1-10$~kpc,  $L\sim 100~$kpc (a few $10^{18}$~km), $\ell_0\sim
10-50$~kpc, and $\lambdamfpi\sim 0.1-10$~kpc (assuming $T_e\sim T_i$).
Protons in the ICM plasma magnetise for magnetic fields as small as
$10^{-18}$~G, while the (observed) dynamical saturation level is of
the order of a few $\mu$G (the Mach number of fluid-scale velocity
fluctuations $u_i$ inferred from observations is $M=u_i/\vthi \sim 0.3$),
corresponding to $\rho_i\sim 10^4$~km and $\beta_i\sim 10-100$
\citep{rosin11}. Hence, today's ICM is strongly magnetised in the sense
that $\rho_i/\lambdamfpi\ll 1$, but only weakly magnetised according to our
definition of weak magnetisation $\beta_i\gg 1$ ! This distinction will
turn out to be important from a kinetic stability point of view.

\subsubsection{Making compromises: the hybrid Vlasov-Maxwell model}
Collisionless dynamo problems are obviously harder to solve than their
MHD counterparts, and numerical simulations are therefore inevitably going
to play an important role in the development of their theory.
However, 3D-3V simulations of the full Vlasov-Maxwell 
problem still require a prohibitively large amount of CPU time, and
one may wonder if and what assumptions can be made that would simplify
the problem as much as possible without affecting its fundamental nature.
The most significant possible such simplification is to get rid of all the
complicated electron-scale kinetic dynamics by considering a hybrid
plasma model in which ions are collisionless, but electrons are fluid
(collisional), unmagnetised, and isothermal. In this framework, we do not need to
solve the dynamical electron Vlasov equation and instead simply adopt
a very standard Ohm's law. The Maxwell-Faraday equation then reduces to
the standard induction equation with a Hall term
\begin{equation}
  \label{eq:inducsimple}
  \dpart{\vB}{t}=\curl{\left(\vU_i\times\vB\right)}+\eta\Delta{\vB}-\f{1}{Z_in_ie}\curl{\left(\vJ\times\vB\right)}~.
\end{equation}
The presence of the magnetic diffusivity $\eta$ in this model follows
from the collisionality assumption on electrons, but could also be
postulated as a simple ``anomalous'' resistivity closure representing
unresolved dynamical phenomena taking place at electron scales
(in the remainder of this section, the term ``collisionless'' is used
in the hybrid sense, as we will always take for granted the existence of a
  small electrical resistivity associated with electron
  collisions). The isothermal electron assumption, on the other hand,
implies that an inhomogeneous baroclinic Biermann battery source term
\citep{biermann50} is absent from the r.h.s. of the induction
equation, and that the Weibel plasma instability \citep{weibel59} is
filtered out of the problem. While this hybrid model
almost certainly does not provide an entirely satistfactory
representation of any laboratory or high-energy astrophysical plasma,
and discards potentially important electromagnetic phenomena
(especially in shocked plasmas), its essential merit is in a 
drastic reduction of theoretical complexity, which leaves us with
the simplest genuine kinematic dynamo instability problem possible
in a non-fluid context. The Hall term is nonlinear and negligible if
the ratio between the magnetic and kinetic energies is small compared to
$(\ell_\eta/d_i)^2$, where $d_i$ is the ion inertial length (assuming
the magnetic energy is concentrated at $\ell_\eta$). 

\subsubsection{The 3D-3V ``small-scale'' dynamo problem}
The fact that we still have a standard induction equation
implies that the usual anti-dynamo theorems still apply, and that we
will inevitably have to deal with three spatial coordinates, as well as
three velocity-space coordinates. Even within the hybrid framework,
simulations of the problem therefore require millions of CPU hours
on latest-generation supercomputers and have only recently become
feasible. The remainder
of this section will essentially focus on the fluctuation
dynamo problem (in the MHD sense of \sect{smallscale}), which appears
to be the most directly relevant to the
generation of cosmic magnetic fields (for instance, rotation is not 
generally considered to be a very important dynamical ingredient in
galaxy clusters).

In order to keep the formulation as close as possible to the MHD case,
we will assume that a flow of collisionless plasma is driven through
the ion Vlasov \equ{eq:vlasov} by a random, isotropic, incompressible,
non-helical, $\delta$-correlated-in-time force $\vec{F}$ confined to
large fluid scales $\ell_0$. The simple question that we would like
to ask once again is whether amplification of a small (to be specified
depending on the regime considered) magnetic seed can take place in
such a system. 

\subsection{Collisionless dynamo in the unmagnetised regime}

\begin{figure}
\centering\hbox{\includegraphics[height=0.275\textheight]{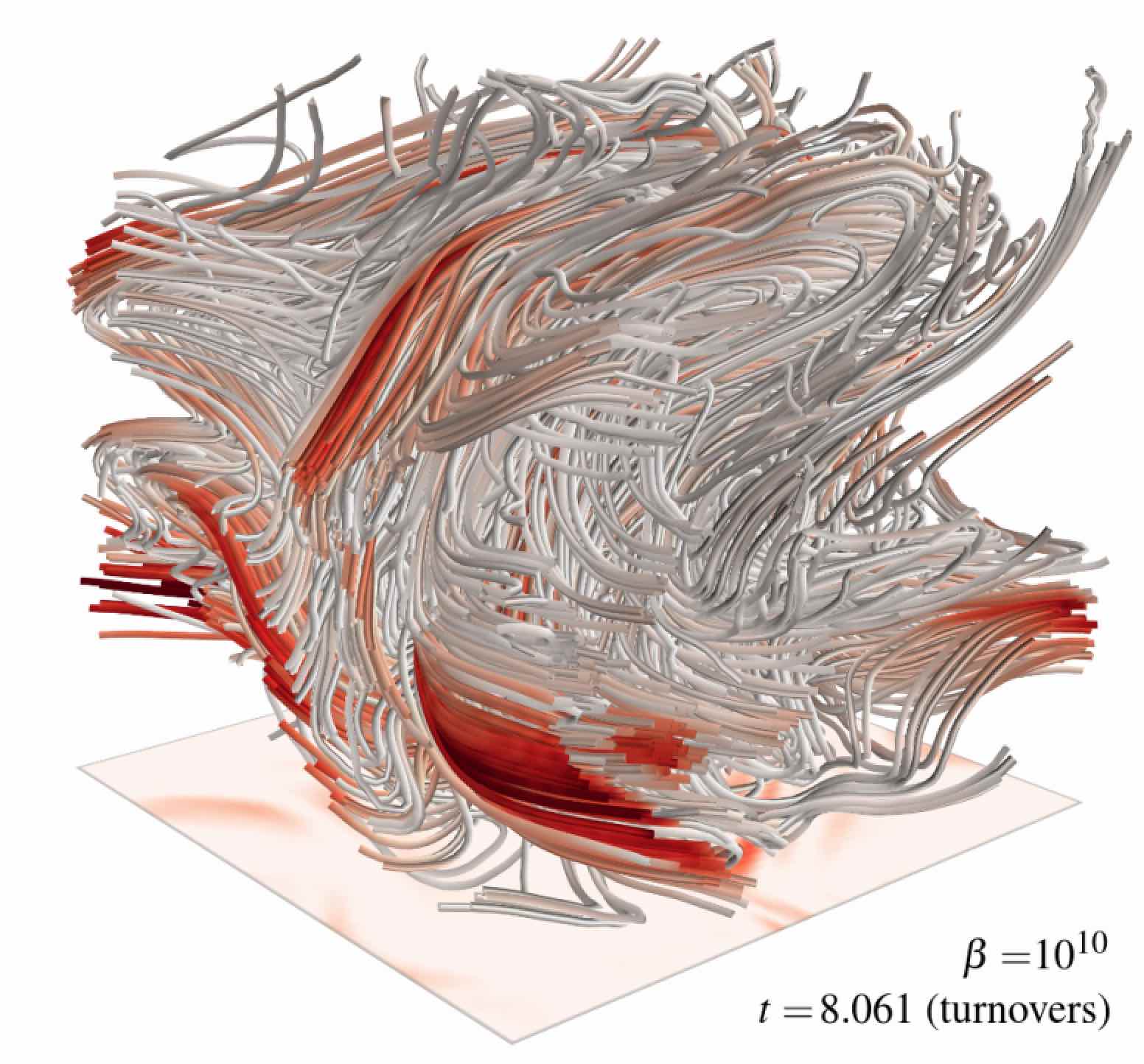}\hspace{-0.3cm}
\includegraphics[height=0.25\textheight]{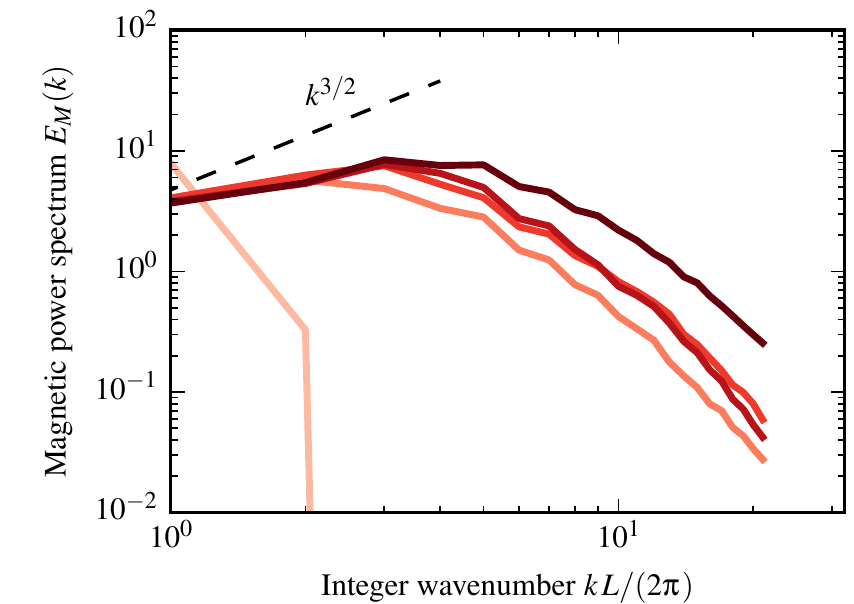}}
\caption{Left: 3D-rendering of magnetic-field lines and magnetic-field
  strength (reds represent large $B$) in a 3D-3V hybrid Vlasov-Maxwell
  simulation of the fluctuation dynamo at $Rm=1600$ in the unmagnetised
  regime with non-helical forcing at the box scale (denoted by $L$ in the
  caption). Right: time-evolution of the  magnetic power spectrum in
  the simulation (lighter colours represent earlier times). The
  Kazantsev spectrum is shown for reference  \citep[adapted
  from][]{rincon16}. \label{figVlasovBunmagnetized}}
\end{figure}

Let us first address the seemingly simpler question of fluctuation
dynamo action in the \textit{unmagnetised} regime. Imagine that we
start with a random, zero net-flux magnetic field seed (say a
primordial magnetic seed in the intergalactic plasma or the ICM)
sufficiently weak that the ion Larmor radius is larger than the 
scale $\ell_0$ at which the collisionless flow of plasma is stirred. 
In this case, no magnetised plasma dynamics can take place at
``internal'' kinetic scales, since there is no such scale in the
problem in the first place. The difference between the kinematic
dynamo problem in this regime and its MHD counterpart then boils down to
differences in the characteristics of the fluid-scale driven
velocity-field fluctuations $\vec{u}_i$ appearing in
\equ{eq:inducsimple}. In the MHD problem, the fluid
flow is a solution of the forced \textit{Navier-Stokes} \equ{eq:mom}
(without the Lorentz  force in the kinematic regime), but here
$\vec{u}_i$ must be indirectly calculated from the ion distribution
function $f_i$ solution of the forced ion \textit{Vlasov} \equ{eq:vlasov}.

The first numerical simulations of this problem \citep{rincon16}
using an Eulerian hybrid Vlasov-Maxwell code have led to the
naive re-discovery that fluid and collisionless systems generate very
different fluid-scale flows given the exact same forcing function
characteristics. This is certainly no surprise to experimental plasma
physicists, who have long been facing the issue
of forcing coherent large-scale flows of collisionless
plasma (Cary Forest, private communication). A possibly slightly
less obvious finding, though, is that this has significant
quantitative consequences for the kinematic small-scale dynamo.
In a collisionless plasma, ions can stream freely
through the system and subsonic fluid-scale ion velocity fluctuations
at wavenumber $k$ therefore phase-mix  (collisionlessly damp) on a
timescale $(k\vthi)^{-1}$ smaller by a factor Mach number than their
turnover timescale $(k u_{i,k})^{-1}$ (the Mach number of the
aforementioned simulations is $M\simeq 0.15-0.2$). In other words, the
plasma flow in the collisionless, unmagnetised regime is effectively
very viscous, decorrelates on a timescale much faster than its
turnover time, and is essentially confined to the forcing scale. This
can also be understood at the phenomenological level by noting that
the viscosity in a collisional fluid is proportional to
$\lambda_{\mathrm{mfp},i}$. If the forcing function is random, the
flow remains chaotic though, leaving open the possibility of fast
dynamo action. The situation should therefore be very similar to the
large-$Pm$ small-scale MHD dynamo regime described in \sect{zeldo}
and \sect{largePmsection} (the ``Stokes flow'' dynamo). Numerical
simulations conducted at the same $64^3$ spatial resolution 
as that of the original MHD dynamo simulations of \cite{meneguzzi81}
(multiplied by a $51^3$ resolution of the 3V-velocity space !)
have shown that an exponentially growing collisionless unmagnetised
fluctuation dynamo is indeed possible, and that the dynamo field
evolves into a folded structure with a spectrum reminiscent of the
$Pm>1$ MHD case. This is illustrated in
\fig{figVlasovBunmagnetized}, to be compared with
\fig{figMFP81small} (right) and \fig{figvizlargepm} (right) for the
MHD case\footnote{These simulations, conducted in 3D-3V over several
  fluid turnover times using 512 IBM BG/Q cores, are undoubtedly the
  \red{most-expensive-to-date} numerical simulations of a viscous flow !}.
The main difference appears to be that $Rm_c$ is close to
1500 for the unmagnetised collisionless case, compared to 60 in the
$Pm\gg 1$ MHD case. While this exact result requires independent
confirmation, the difference has been
interpreted as a consequence of the fact that the correlation time 
of the collisionless flow is significantly smaller than its turnover
time, while it is of the same order in a turbulent fluid.
Eddies  of collisionless plasma at scales $\ell_0$ (which look more like
``wobbles'' in that case) can therefore only stretch the magnetic
field in a coherent way for a very short time $\ell_0/\vthi$ before
they decorrelate. This in turn suggests that the collisionless dynamo
should be much less efficient than its MHD counterpart for a given
$Rm$, or equivalently have a substantially higher critical $Rm$ (if
$Rm$ is defined as the ratio between the typical magnetic diffusion
time and eddy turnover time). This may turn out to be significant for
dynamo experiments in weakly-collisional plasmas (larger $Rm$ are
easier to achieve in plasmas than in liquid metals though, as the
magnetic diffusivity of a plasma can be reduced substantially by
simply increasing its temperature).

\subsection{Introduction to the dynamics of magnetised plasmas}
Imagine now that an unmagnetised plasma dynamo, or a different kind of
magnetic-field generation mechanism such as a Biermann battery, or a
combination of both, has generated a field sufficiently large
that the plasma is entering the magnetised
regime $k\rho_i\sim 1$, $\omega/\Omega_i\sim 1$. As explained earlier,
the magnetic energy can still be well below equipartition with the
kinetic energy of the flow when this transition occurs, so we do not
expect the Lorentz force to become important at this stage (i.e.
we are not yet entering a dynamical regime saturated by fluid
nonlinearities). However, the new ordering of scales implies that the
internal dynamics of the plasma  (collisional or not) is going to change
drastically. Before we can address the full magnetised dynamo problem,
we therefore need to gain a basic understanding of the dynamics of
magnetised plasmas and their interaction with magnetic fields. And we
are in for a few surprises !

\subsubsection{Fluid-scale dynamics: pressure anisotropies and $\mu$-conservation\label{drift}}
Let us assume for a moment that we are well into the magnetised
regime, and that all the dynamics takes places at low fluid
wavenumbers $k_0$  at which the typical shearing or compression rate
of fluid motions (denoted by $S$ in this section) is
small in comparison to the Larmor gyration rates, $k_0 \rho_i\ll 1$,
$S/\Omega_i\ll 1$.
By taking the first velocity moment of the Vlasov \equ{eq:vlasov} for
ions in this drift-kinetics limit \citep{kulsrud83}, we can derive
an anisotropic momentum equation governing the evolution of the bulk
ion velocity,
\begin{equation}
\label{eq:anisotropicmomentum}
m_in_i\dlag{\vU_i}{t}=-\grad{\left(\pperp+\f{B^2}{8\pi}\right)}
+\div{\left[\hatvB\hatvB\left(\pperp-\ppar+\f{B^2}{4\pi}\right)\right]}~,
\end{equation}
where $\pperp=\sum_s \pperps$ and $\ppar=\sum_s\ppars$,
\begin{equation}
\label{eq:pperp}
\pperps=m_s\int\diff^3\vec{v}\f{\vperp^2}{2}f_s~,
\end{equation}
\begin{equation}
\label{eq:pparallel}
\ppars=m_s\int\diff^3\vec{v}\vpar^2f_s,
\end{equation}
$\vpar=\vec{v}\cdot\hatvB$ and
$\vec{v}_\perp=\vec{v}-\vpar\hatvB$ (in the expressions above and in
what follows, we redefine $\vec{v}$ to be the peculiar velocity of
particles in a local frame moving at the bulk species fluid velocity
$\vU_s$, and Lagrangian derivatives denote variations in that frame). 
In order to close \equ{eq:anisotropicmomentum}, we see that we 
must derive supplementary evolution equations for $\pperps$ and
$\ppars$. To do this, we take respectively the perpendicular and parallel
second-order moments of \equ{eq:vlasov} in the same drift-kinetic
limit, and combine them with the continuity equation for each species
and ideal induction \equ{eq:Bmag} for the field strength
\citep*{chew56,kulsrud83,snyder97}. The outcome of the calculation is
\begin{equation}
\label{eq:CGLtot1}
n_s B\dlag{}{t}\left(\f{\pperps}{n_s B}\right) = -\div{\left(\qperps\hatvB\right)}-\qperps\div{\hatvB}-\f{\nu_{s}}{3}(\pperps-\ppars)~,
\end{equation}
\begin{equation}
\label{eq:CGLtot2}
\f{n_s^3}{B^2}\dlag{}{t}\left(\f{\ppars
    B^2}{n_s^3}\right)=-\div{\left(\qpars\hatvB\right)}+2\qperps\div{\hatvB}-\f{2}{3}\nu_{s}(\ppars-\pperps)~,
\end{equation}
where
\begin{equation}
  \qperps=m_s\int\diff^3\vec{v}\f{\vperp^2}{2}\vpar f_s~
\end{equation}
and
\begin{equation}
\qpars=m_s\int\diff^3\vec{v}\vpar^3 f_s
\end{equation}
are the parallel fluxes of perpendicular and parallel heat
respectively. For simplicity we have used a simple Krook collision
operator, $C_s[f_s]=-\nu_{s}(f_s-f_{M,s})$, where $\nu_s$ accounts for
both intra-species and inter-species collisions
(i.e. $\nu_i=\nu_{ii}+\nu_{ie}$) and $f_{M,s}$ is a Maxwellian
with temperature $T_s=(\Tpars+2\Tperps)/3$, shifted by $\vU_s$
\citep{snyder97}. If we brutally close \equs{eq:CGLtot1}{eq:CGLtot2}
by discarding the heat-flux and collision terms, we obtain the
well-known double-adiabatic, or CGL equations \citep*[after][]{chew56},
\begin{equation}
\label{eq:CGL1}
\dlag{}{t}\left(\f{\pperps}{n_s B}\right)=0~,
\end{equation}
\begin{equation}
\label{eq:CGL2}
\dlag{}{t}\left(\f{\ppars
    B^2}{n_s^3}\right)=0~,
\end{equation}
which are arguably the simplest possible collisionless closure
for the kind of magnetised problem at hand. These equations reflect
at the macroscopic level the conservation of the magnetic moment
$\mu_s=m_s\vperp^2/2B$ and longitudinal
invariant $J=\oint\vpar(s') \diff s'$  of particles when the
(background or fluctuating) magnetic field varies on time and spatial
scales much longer than the particles Larmor gyration scales (for
$\mu_s$ conservation) and bounce timescale in magnetic mirrors (for $J$
conservation -- see \cite{boyd03}, $s'$ is a curvilinear guiding
centre coordinate of the particles).

\begin{figure}
\centering\includegraphics[width=\textwidth]{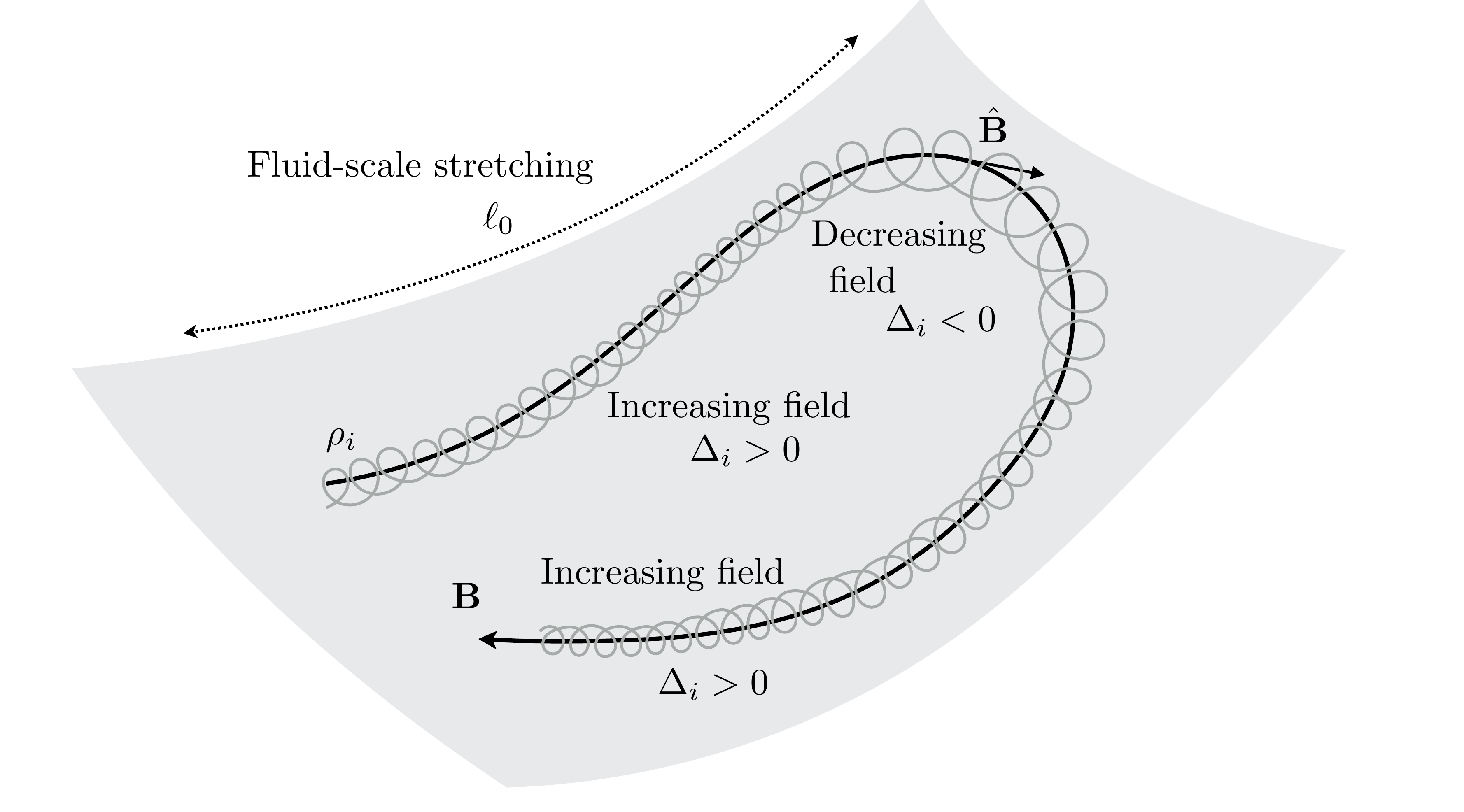}
\caption{$\mu$-conservation and pressure-anisotropy dynamics in
  magnetic folds. Note that the $\rho_i/\ell_\eta\ll 1$ configuration
  depicted here is for illustrative purposes only and is not
  guaranteed to be attainable in practice. \label{figfoldsanisotropy}}
\end{figure}

The conservation of $\mu$ and $J$ notably has important implications
for the dynamics of the pressure-anisotropy term $\pperp-\ppar$
appearing in \equ{eq:anisotropicmomentum}. Combining
\equs{eq:CGLtot1}{eq:CGLtot2} for ions in particular, we find
that
\begin{eqnarray}
\dlag{(\pperpi-\ppari)}{t} & = & (\pperpi+2\ppari)\dlag{\ln
  B}{t}-(3\ppari-\pperpi)\dlag{\ln n_i}{t} \nonumber\\ & &
-\div{\left[(\qperpi-\qpari)\hatvB\right]}-3\qperpi\div{\hatvB}
\label{eq:aniso}\\ & & 
-\nu_i(\pperpi-\ppari)~\nonumber
\end{eqnarray}
(for $T_e\sim T_i$ and strong-enough collisions, it can be shown that
electron pressure anisotropy is smaller by a factor $\sqrt{m_e/m_i}$
than ion pressure anisotropy).
In the CGL limit, this equation notably shows that any local increase
or decrease in magnetic-field strength in the process of a
large-scale magnetofluid instability such as the MRI, or a dynamo
instability, generates a corresponding local (non-dimensional)
fluid-scale pressure anisotropy
\begin{equation}
  \Delta_i\equiv\f{\pperpi-\ppari}{P_i}~,
\end{equation}
where $P_i\equiv (\ppari+2\pperpi)/3$.
\Fig{figfoldsanisotropy} illustrates how this dynamics is expected to
play out in the magnetic folds encountered in the small-scale MHD dynamo
problem in \sect{smallscale}. Positive pressure anisotropy ($\pperpi>\ppari$)
should develop in regions of strong magnetic stretching where the
magnetic field is straight and increases, and negative pressure
anisotropy ($\pperpi<\ppari$) should develop in regions of decreasing
field-strengths associated with strong magnetic curvature
\red{(energetically this happens through the so-called betatron effect
  of perpendicular acceleration or deceleration of the particles'
  gyration by the perpendicular electric field associated with a
  local magnetic field that slowly changes in time)}.

As noted by \cite{kulsrud97}, $\mu$-conservation creates an
acute problem for the growth of magnetic fields in collisionless
magnetised plasmas, as it would seem to
imply that order-of-magnitude increases in field-strength
(through the stretching of the field by fluid-scale motions) should
be accompanied by corresponding order-of-magnitude increases in the
perpendicular thermal energy of the plasma. Generating so much heating
by the mere subsonic stretching of a magnetic field in a high-$\beta$
plasma, however, looks highly implausible from an energetics
perspective. Accordingly, it has been
shown recently using variational arguments that significant
magnetic-field amplification is impossible in a strictly $\mu$-conserving
magnetised plasma \citep[\cite*{helander16}, see also numerical
results by][]{santos14}. In practice, we can see from the second term
on the r.h.s of  \equ{eq:anisotropicmomentum} that any fluctuation
$\delta \pperp$ induced by a change in magnetic-field strength $\delta B$
generates a mirror force $\mu\grad_\parallel \delta B$ acting against
the velocity field stretching the field in the first place \citep[this
mechanism plays an important part in the damping of slow
hydromagnetic waves in a collisionless magnetised
plasma,][]{barnes66}. This resistance to stretching is distinct from the fluid
Lorentz force, and is independent of the absolute magnitude of $B$. It
therefore affects the flow even when the field itself is dynamically
irrelevant. The only thing that matters here is that the
plasma is magnetised, and that $\mu$ is conserved. 

Collisions, on the other hand, break $\mu$-conservation by
kicking particles out of their Larmor orbits, and tend to relax
pressure anisotropies. The competition between the build-up of
pressure anisotropies associated with the stretching of the magnetic
field by velocity fluctuations $\fluctvU_i$, and their relaxation
by collisions, is most easily seen by taking the secondary collisional
($k\lambdamfpi\ll 1$) limit of \equ{eq:aniso} in combination with
\equ{eq:Bmag}, also discarding heat fluxes for simplicity
\citep{schekochihin10}. The result is
\begin{equation}
\label{eq:braginskiinu}
\pperpi-\ppari = 3m_in_i\nu_B\left(\hatvB\hatvB-\f{\tens{I}}{3}\right):\grad{\fluctvU_i}~,
\end{equation}
where $\nu_B=0.96 k_B T_i/(m_i\nu_{ii})$  is the so-called Braginskii
viscosity (the derivation of this exact expression requires the
introduction of a more realistic collision operator than our simple
Krook operator). This limit forms the basis for
\citeauthor{braginskii65}'s (\citeyear{braginskii65}) anisotropic
fluid MHD, and is sometimes referred to as the ``dilute'' magnetised
plasma regime in astrophysics \citep{balbus01}. 
It is easy to see that pressure-anisotropic dynamics is
important in this collisional regime too. Indeed, substituting
\equ{eq:braginskiinu} into \equ{eq:anisotropicmomentum}, we find that
the generated pressure anisotropy now acts as a (collisional) viscous
term that tends to damp (dissipate) the very motions that induce the
magnetic field. Formally though, the presence of an effective
collision term (either particular, or due to some dynamical scattering
process) relaxes the strict $\mu$-invariance relationship between the
perpendicular pressure and magnetic-field strength, and therefore
alleviates the strict impossibility of dynamo
action\footnote{Equally formally, it may be argued that collisions
  can never be entirely neglected anyway, because fluctuations of the
  distribution function are usually cascaded down to very small scales
  in velocity-space where collisions become important. This subtlety
  clearly gets in the way of our desperate quest for clarity and
  simplicity in this section, and will therefore henceforth simply be
  discarded. Readers seeking clarity on these particular matters may
  (or may not) find it in a collection of papers on
  \textit{Collisions in collisionless plasmas} recently published in
  this very journal.}.

Complementary to these results, it has also been recently realised
that the dynamical fluid-scale pressure-anisotropic response of a
magnetised plasma can interrupt the propagation of linearly-polarised,
finite-amplitude Alfv\'en waves \citep*{squire16},
and can even lead to a form of ``magneto-immutable'' Alfv\'enic MHD
turbulence characterised by minimal magnetic stretching \citep{squire19}.
At first glance, all these results are not terribly good news for dynamo
theory, for i) at this stage there is no guarantee that significant
magnetic-field growth is possible in a weakly-collisional or
collisionless turbulent plasma, ii) they all suggest that we have no
choice but to work with quite a lot of kinetic complexity if we want
to make progress on the problem. In  particular, neither  the CGL
system, nor any other closure of \equs{eq:CGLtot1}{eq:CGLtot2} without
an effective collision term, have a plasma dynamo effect in them
because they both originate in the $k\rho_i\ll 1$, $\omega/\Omega_i\ll
1$ ordering. $\mu$-conservation must be broken one way or the other,
either by finite Larmor radius (FLR) effects, collisions, or both,
for a dynamo to be possible.

\subsubsection{Kinetic-scale dynamics: pressure-anisotropy-driven instabilities}
We have just seen that $\mu$-conservation dictates the 
dynamics of pressure anisotropy in a magnetised plasma evolving on
slow, fluid scales, but that many orders of magnitude changes in the
magnetic field are impossible in a stricly $\mu$-conserving system
based on energetic considerations. Is there a way out of this
problem for dynamos in magnetised plasmas ? There is actually
an important physical consideration that we have not yet factored
in. When we try to change the magnetic-field strength
under the constraint of $\mu$-conservation, such as in the fold
depicted in \fig{figfoldsanisotropy}, we are in fact driving the
system out of equilibrium by creating velocity-space
gradients of free-energy, namely pressure anisotropy. And, just like
pushing a fluid out of equilibrium by creating large-scale
spatial gradients (of entropy, angular momentum etc.) results
in the excitation of  dynamical instabilities whose nonlinear tendency
is to drive back the system towards equilibrium, we may expect
that the generation of pressure anisotropy excites kinetic
instabilities whose effect should be to relax the former one way or
the other. Most importantly, the fluctuations associated with these
instabilities may scatter particles similarly to individual particle
collisions, breaking $\mu$-invariance and relaxing
pressure-anisotropy in the process.

Is the $\mu$-conserving, fluid-scale dynamics of a collisionless,
 magnetised plasma actually subject to pressure-anisotropy-driven
kinetic instabilities ? The answer is yes ! The most important ones
for us are  called the ion firehose, mirror and ion-cyclotron
instabilities \citep*[][see \cite{gary93}, chap. 7 for a textbook
presentation]{rosenbluth56,chandra58,parker58,
vedenov58,rudakov61,gary92,southwood93,hellinger07},
and they are now routinely detected in
\textit{in situ} measurements of different anisotropic 
heliospheric plasmas \citep[e.g.][]{hellinger06,joy06,bale09,genot09}.
To illustrate simply how such instabilities originate, and why they
are impossible to discard in the problem at hand, we do not actually need to
do any kinetic physics. Let us get back to the collisional fluid
Braginskii framework introduced above for a minute, and consider the simple
problem of incompressible Alfv\'enic oscillations of a magnetic field in the
presence of a background ion pressure anisotropy $\Delta_i$. To do
this, we perturb \equ{eq:anisotropicmomentum} and \equ{eq:induc}
linearly around such an anisotropic state with perturbations
$\delta\vU_{\perp,i}$, $\delta\vB_\perp$ with non-zero parallel wavenumber
$k_\parallel$ and $k_\perp=0$, also setting $\delta \tens{P}=0$. 
To linear order, these waves are simply magnetic-curvature perturbations 
and do not change the magnetic-field strength, so we can also set
$\delta B=0$. As can easily be guessed from the second term on the
r.h.s. of \equ{eq:anisotropicmomentum}, the dispersion relation for
this wave polarisation is just
\begin{equation}
\label{eq:firehosedisprel}
\omega^2=\left(k_\parallel\vthi\right)^2\left(\Delta_i+\f{2}{\beta_{\perp,i}}\right)~.
\end{equation}
If $\Delta_i=0$, these are just parallel shear Alfv\'en waves. The
effect of a weak background pressure anisotropy, then, is to change the
phase velocity of the waves. But if the parallel pressure is
sufficiently strong in comparison to the perpendicular one, so that
$\Delta_i<-2/\beta_{\perp,i}$, we see that the waves turn into an
instability, called the parallel firehose
instability in analogy with the
wiggly instability of a real fire/garden hose let lose at one end,
in which pressurised water is injected at the other end. Thus, in an
anisotropic plasma, a MHD process as simple as an Alfv\'en wave can
turn unstable. This is spooky ! \Equ{eq:firehosedisprel} shows that
instability is only possible if the plasma is weakly magnetised,
i.e. $\beta_{\perp,i}\ge 1$ (as $|\Delta_i|$ is typically of order
one or smaller), and therefore do not affect low-$\beta$
magnetically-confined fusion plasmas for instance. This condition
means that the local background magnetic field must be flexible
enough, or equivalently magnetic tension weak-enough, to allow the
exponential growth of magnetic-field-line wiggles.
Another significant result of this very simple analysis is that there is no
small-scale cut-off to the instability growth rate in Braginskii MHD,
i.e. the smaller the scale $1/k$ of the perturbations, the higher the
growth rate. Hence, this fluid system is catastrophically plagued by
the simple parallel firehose instability. In reality, the process is
regularised at small-scale by kinetic FLR effects that are ordered
out in the drift-kinetic \red{limit introduced in
  \sect{drift}}. For $k_\parallel \vthi\sim
\Omega_i$, or equivalently $k_\parallel\rho_i\sim 1$, the ions
streaming along field lines at their typical thermal speed start to
sample significant variations in the magnetic field before they can
complete a single Larmor orbit, i.e. the adiabatic $\mu$-invariance
is broken. This fuzziness in the field that ions see makes the
instability less effective and ultimately damps it at $\rho_i$ scales.

\begin{figure}
  \centering\includegraphics[width=\textwidth]{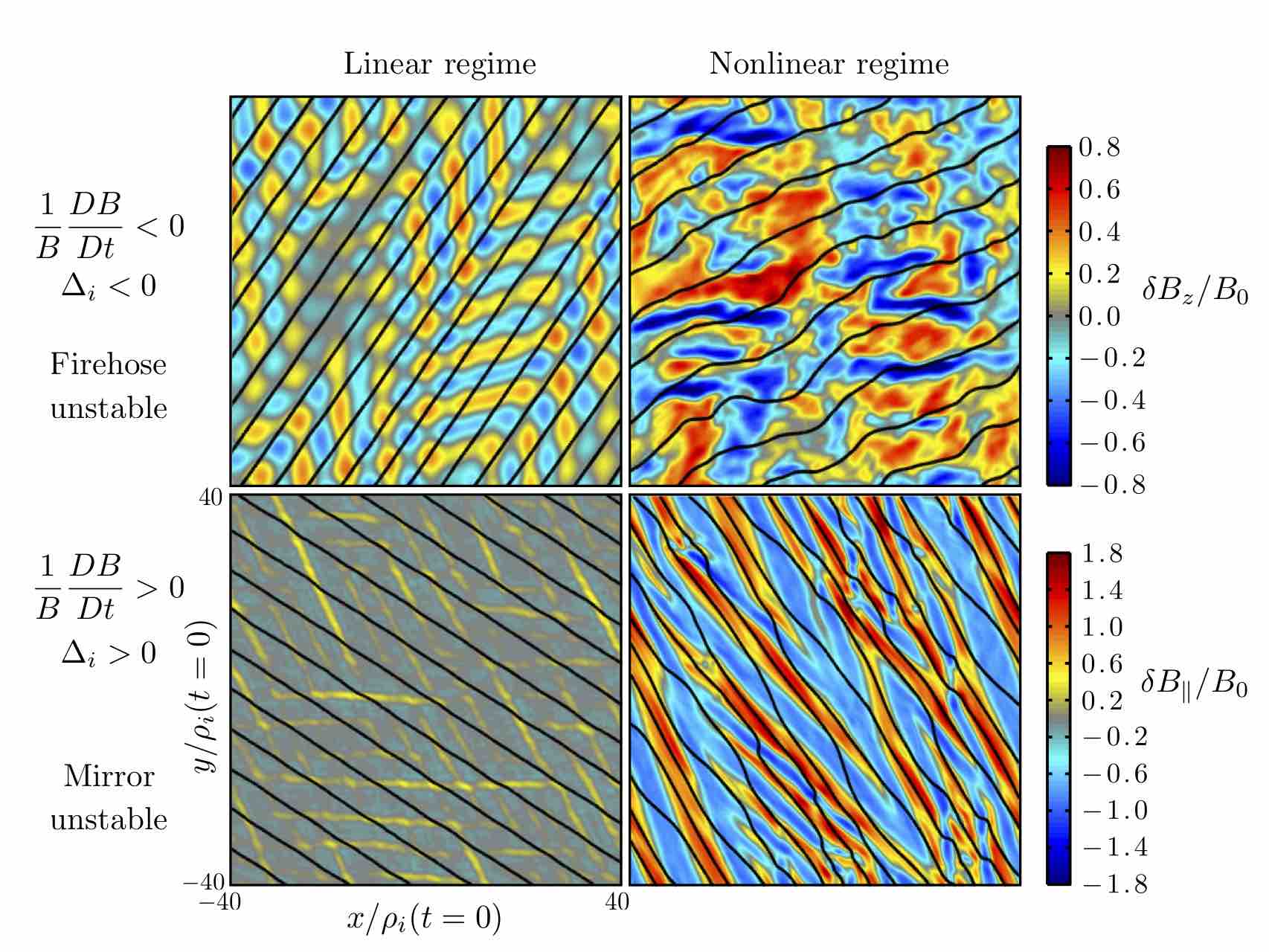}
\caption{\label{figkunzinstabilities}Snapshots of 2D hybrid-PIC simulations
  of firehose and mirror instabilities in a shearing, collisionless,
  weakly-magnetised plasma  ($\beta_i=200$). Top panels: an initially
  straight but inclined magnetic field $B_0$ in the
  $(x,y)$ plane (solid black lines) is ``unsheared'' by a large-scale
  linear shear flow $\vU_S=-Sx\,\vec{e}_y$, resulting in a decrease of 
  the magnetic-field strength and generation of negative ion pressure
  anisotropy through $\mu$-conservation. This preferentially excites
  an oblique firehose instability characterised by out-of-plane,
  perpendicular magnetic fluctuations $\delta B_z$. The left panel
  shows the linear stage of instability,
  and the right panel  the saturated stage involving
  finite-amplitude magnetic fluctuations at $\rho_i$ scales. Bottom
  panels: in a second numerical experiment, the initial inclination of
  the magnetic field in the $(x,y)$ plane is set-up in such a way
  that the shear winds up the field, resulting in an increase of
  magnetic-field strength and generation of positive ion pressure
  anisotropy. This preferentially excites an oblique mirror instability
  growing parallel magnetic-field fluctuations
  $\delta  B_\parallel$. The left panel shows the preferred
  orientation of the instability in the linear stage. The nonlinear
  stage, depicted in the right panel, consists of finite-amplitude,
  elongated mirror traps characterised by steep magnetic gradients at
  $\rho_i$ scale at their ends \citep*[adapted from][]{kunz14}.}
\end{figure}

The oblique firehose, mirror and ion-cyclotron instabilities are
physically quite different from the parallel firehose introduced
above, as they involve kinetic pressure fluctuations and in some cases
resonant particles. An in-depth analysis of these instabilities
would require us to get into the gory details of kinetic theory
and is outside the scope of these notes. Let us  therefore simply
state a few important linear results of most  immediate relevance to
the dynamo problem:
\smallskip
\begin{itemize}
\item \textit{any} pressure anisotropy, either positive or negative,
  can give rise to at least one of these instabilities. The (oblique
  or parallel) firehose instability is excited when the parallel
  pressure is larger than the perpendicular one,
  $\Delta_i\lesssim -2/\beta_i$, while the mirror
instability is excited in the opposite situation where the
perpendicular pressure exceeds the parallel one,
$\Delta_i>1/\beta_i$. Thus, the necessary instability condition
of weak magnetisation $\beta_i>1$ encountered in the parallel firehose
case extends to the mirror instability. This condition is always
satistified for sub-equipartition magnetic fields in a subsonic
flow, which is the regime we are most interested in here.
The ion-cyclotron instability has no well-defined $\beta_i$ threshold
but also lives on the $\Delta_i>0$ side of the problem parameter space. 
\smallskip
\item all these instabilities develop preferentially at scales
comparable to the ion Larmor radius, $k\rho_i=O(1)$, and
gyration frequency $|\omega|/\Omega_i=O(1)$ when $|\Delta_i|=O(1)$. 
They are therefore extremely fast in comparison to the fluid-scale
dynamics, and have the potential to break $\mu$-invariance.
\end{itemize}
\smallskip
Snapshots of magnetic perturbations in the linear regime of the
oblique firehose and mirror instabilities, developing on top of a
slowly-evolving background magnetic field driving a uniform negative
or positive ion pressure anisotropy, are shown in the left-hand panels
of \fig{figkunzinstabilities}.

\subsubsection{Saturation of kinetic instabilities in a shearing magnetised plasma*\label{satkin}}
The  theoretical analysis and numerical results presented so far show
that slowly-evolving, fluid-scale magnetic structures in a collisionless
magnetised  plasma, such as the magnetic fold in
\fig{figfoldsanisotropy}, should become mirror or ion-cyclotron
unstable in regions of
developing positive $\Delta_i$, and firehose unstable in regions 
of developing negative $\Delta_i$ \citep{schekochihin05b}. Besides,
the fact that all these instabilities grow at a very fast exponential
pace compared to the fluid timescales over which pressure
anisotropies  typically develop in a magnetised plasma implies that
large-scale plasma motions can barely start to stretch or squeeze the
magnetic field before these instabilities kick-in and saturate.
Now, are these kinetic-scale wiggles just funny, distracting, but
ultimately inoffensive, or is their back-reaction on the dynamics at
fluid scales critical to an overall plasma dynamo mechanism~? Our
earlier observation that mechanisms breaking $\mu$-invariance are
needed to break the dynamo deadlock in a collisionless magnetised
plasma suggests the latter. Besides, both theoretical analyses
\citep{schekochihin08} and the numerical results reproduced
in \fig{figkunzinstabilities} show that mirror and
firehose-unstable perturbations $\delta B$ can grow up to
strengths comparable to that of the original field on top of which 
they develop when a pressure anisotropy is dynamically generated at
fluid scales, i.e. $\delta B/B=O(1)$. This is hardly a negligible
result in the dynamo context.

\begin{figure}
\centering\includegraphics[height=0.45\textwidth]{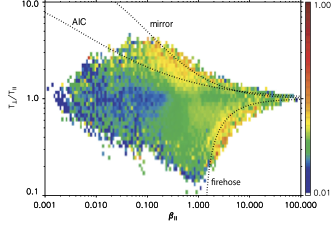}
\caption{Example of statistics of ion temperature anisotropy versus
  $\beta_i$ in the solar wind \citep{bale09}. Most recorded events at
  $\beta_i>1$ are confined into a region of parameter space seemingly
  delimited by the linear threshold of the  mirror
  ($\Delta_i=1/\beta_i$) and firehose ($\Delta_i=-2/\beta_i$)
  instabilities (dotted lines). The AIC line corresponds to a fixed
  small ion-cyclotron instability iso-growth-rate
  line.\label{SWanisotropy}}
\end{figure}

Before we can answer this key question, we therefore have no
alternative but to have a closer look at how these different
instabilities saturate in
a situation where an ion pressure anisotropy is slowly generated by 
a shearing, expansion or compression of the magnetic field at fluid
scales. Recalling the generic nonlinear tendency of instabilities to
drive back their host system towards equilibrium, and the incredibly
fast growth rate of pressure-anisotropy-driven instabilities in
comparison to fluid timescales,  it seems reasonable at the
phenomenological level to expect that their
nonlinear effect is to act as an almost instantaneous limiter on
fluid-scale pressure anisotropies, essentially pinning them
at the mirror stability threshold $\Delta_i=1/\beta_i$ in regions of
increasing magnetic-field strength, and at the firehose stability
threshold $\Delta_i\approx -2/\beta_i$ in regions of decreasing field
strength \citep[the relevance of the ion-cyclotron instability in the
problem remains an open question, but appears to depend on factors not
considered in the present discussion, such as the ratio between
electron and ion temperatures,][]{sironi15}. This phenomenological
argument is supported by statistical measurements in the solar wind
plasma, such as shown in \fig{SWanisotropy}, which suggest that
admissible plasma states are bounded by the marginal stability
threshold of the firehose and mirror instabilities \citep[note however
that pure CGL dynamics may confine the plasma to a similar region of
parameter space,][]{strumik16}. Detailed asymptotic considerations
and dedicated simulations also suggest that the distances to
instability thresholds in the saturated regime in a slowly-shearing
magnetised plasma scale as positive powers of the scale-separation
parameter $S/\Omega_i$, where $S$ is the typical magnetic-stretching
rate at fluid scales \citep{rosin11,kunz14,rincon15}. This parameter
is of course very small in all natural situations of interest
(although not necessarily in simulations).

The pressure-anisotropy delimiter argument is intuitive and appealing
in that it is somewhat physically motivated, circumvents the full
kinetic complexity of the problem, and breaks the $\mu$-invariance of
the equations. Accordingly, this approach has already been used in
several large-scale magnetofluid simulations of collisionless dynamo
and MRI as a quick and easy fluid-scale closure
\citep{sharma06,santos14}. However, it is not a foolproof alternative
 to solving the full kinetic problem either, as the latter turns out to be
 significantly more subtle in practice. A particular problem here is that
 there are actually at least two distinct dynamical ways in which the
 saturated kinetic instabilities nonlinearly back-react on the pressure
 anisotropy. The first possibility is for the unstable kinetic-scale
 mirror and firehose fluctuations to effectively screen particles from
changes in magnetic-field strength. For instance,
the early nonlinear evolution of the mirror instability is such that
particles get trapped in deepening magnetic troughs where the total
effective magnetic field, which is the sum of the original,
slowly-evolving
fluid-scale field, plus mirror-unstable magnetic perturbations,
remains constant \citep*{rincon15}. There is an analogous, albeit
simpler mechanism in the firehose case, see
\cite{schekochihin08}. Considered from a fluid-scale perspective, this
process is therefore somewhat equivalent to an effective reduction of
the stretching rate of the field by fluid eddies, and may in that
sense be roughly understood as an \textit{enhancement of the
  effective plasma viscosity} (see \cite{mogavero14} for a more
thorough discussion).  The second possibility is for finite-amplitude
magnetic mirror or firehose fluctuations at $\rho_i$ scales, such as shown in
the right panels of \fig{figkunzinstabilities}, to scatter particles.
This FLR scattering acts as to enhance the effective collisionality of the
plasma and therefore  provides a way to relax the pressure anisotropy
in which the instability
originates. Considered from a fluid-scale perspective, this process
can be seen as a \textit{reduction of the effective plasma viscosity},
as it decreases the effective mean free path of particles, to which
the viscosity is directly proportional in a regular fluid.
A complementary way to look at these seemingly very
different two possible outcomes is to revisit Braginskii's viscosity 
\equ{eq:braginskiinu} by interpreting $\nu_{ii}$ as an effective
collision frequency $\nu_{i,\mathrm{eff}}$ associated with
particle-scattering off the magnetic wiggles at $\rho_i$ scales.  The
size of the pressure anisotropy can be decreased
either by an effective reduction of the fluid-scale induction rate in the
numerator, or by an enhancement of the effective collisionality of the
plasma in the denominator. These two scenarios should have different
implications for magnetic-field growth.

There are many important subtleties and remaining uncertainties as to
how the different instabilities unfold and interact in detail, and
as to how they effectively regulate the pressure anisotropy in
turbulent magnetofluid systems. For the purpose of this
discussion, it is sufficient to mention that both types of dynamics
are observed in dedicated numerical simulations of the mirror and
firehose instabilities in a slowly-shearing, magnetised plasma
\citep*{kunz14,riquelme15}. The screening mechanism takes
precedence over the scattering mechanism in all cases in the early
stages of nonlinear evolution ($\delta B/B\ll 1$). Its main effect
is to turn the initial exponential growth of the instabilities into a
much slower secular growth taking place on a timescale comparable to the
typical fluid-scale shearing time $S^{-1}$. The transition from the secular
growth regime to the scattering  regime is not well understood and
depends on the exact instability considered, but appears to be much
faster for the firehose instability.
Note also that the nonlinear particle-scattering regime \textit{a priori}
seems to provide the most obvious way out of the difficulty to grow
magnetic fields effectively in a magnetised plasma. With
the notable exception of oblique firehose modes, the most
linearly-unstable instability modes close to threshold actually sit at
scales significantly larger than $\rho_i$ \citep{hellinger07,rosin11} 
and do not therefore break $\mu$-invariance in the linear or 
early nonlinear stages.

\subsection{Collisionless dynamo in the magnetised regime}
It should now be clear that the collisionless dynamo problem in the
magnetised regime hides much more dynamical complexity than the
already not-that-easy standard small-scale isotropic dynamo
problem. But, barring other as yet unknown unknowns, we are at least
finally all set to tackle the problem !

\subsubsection{Is dynamo possible in the magnetised regime ?}
The first important question, obviously, is whether significant
magnetic-field growth is possible at all in the magnetised regime.
Two similar sets of ``small-scale'' fluctuation dynamo simulations
have so far been performed to answer this question, a set of hybrid
Eulerian (grid) Vlasov-Maxwell simulations \citep{rincon16}
essentially probing the magnetisation transition, and a set of hybrid
PIC simulations going deeper into the magnetised regime
\citep{saintonge18}. The overall range of fluid/kinetic scale
separation covered by these simulations is approximately $0.015\leq
\rho_i/\ell_0\leq 15$. The computational costs of simulations with
significant scale separation currently prevent a thorough exploration
of the full problem parameter space. Ideally, one would also like
to perform a single simulation covering magnetic-field
growth starting from a very weak-field, unmagnetised regime, up
to equipartition. However, this is currently very difficult to achieve
due to the prohibitive scale separation between the fluid dynamo-growth
timescale and the internal kinetic timescales (the ion thermal
box-crossing time in the unmagnetised regime, and the ion-cyclotron
timescale in the magnetised regime). Consequently, the not-so-ideal
strategy adopted in all studies so far to probe
the dynamo at different levels of magnetisation and magnetic-field
strength has been to integrate for just a few turnover times several
distinct simulations initiated with seed magnetic fields of different
r.m.s. strengths corresponding to different r.m.s. $\rho_i/\ell_0$,
letting the field grow over one to three orders of magnitude at most
in each case.

\begin{figure}
\setlength{\fboxsep}{0pt}
\setlength{\fboxrule}{0.0035\textwidth}
\vspace{5pt}
\centering\raisebox{0.0035\textwidth}{\framebox{\includegraphics[height=0.393\textwidth]{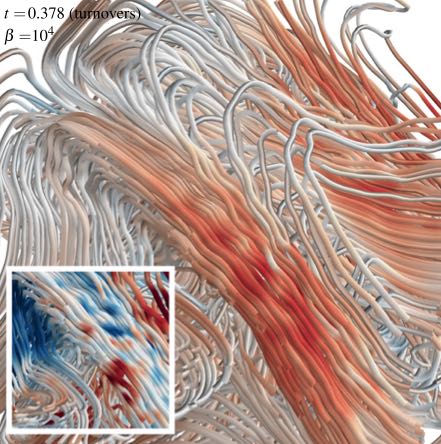}}}
\hspace{0.05\textwidth}\includegraphics[height=0.4\textwidth]{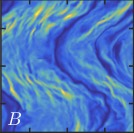}\\
\vspace{0.5cm}
\centering\includegraphics[width=0.5\textwidth]{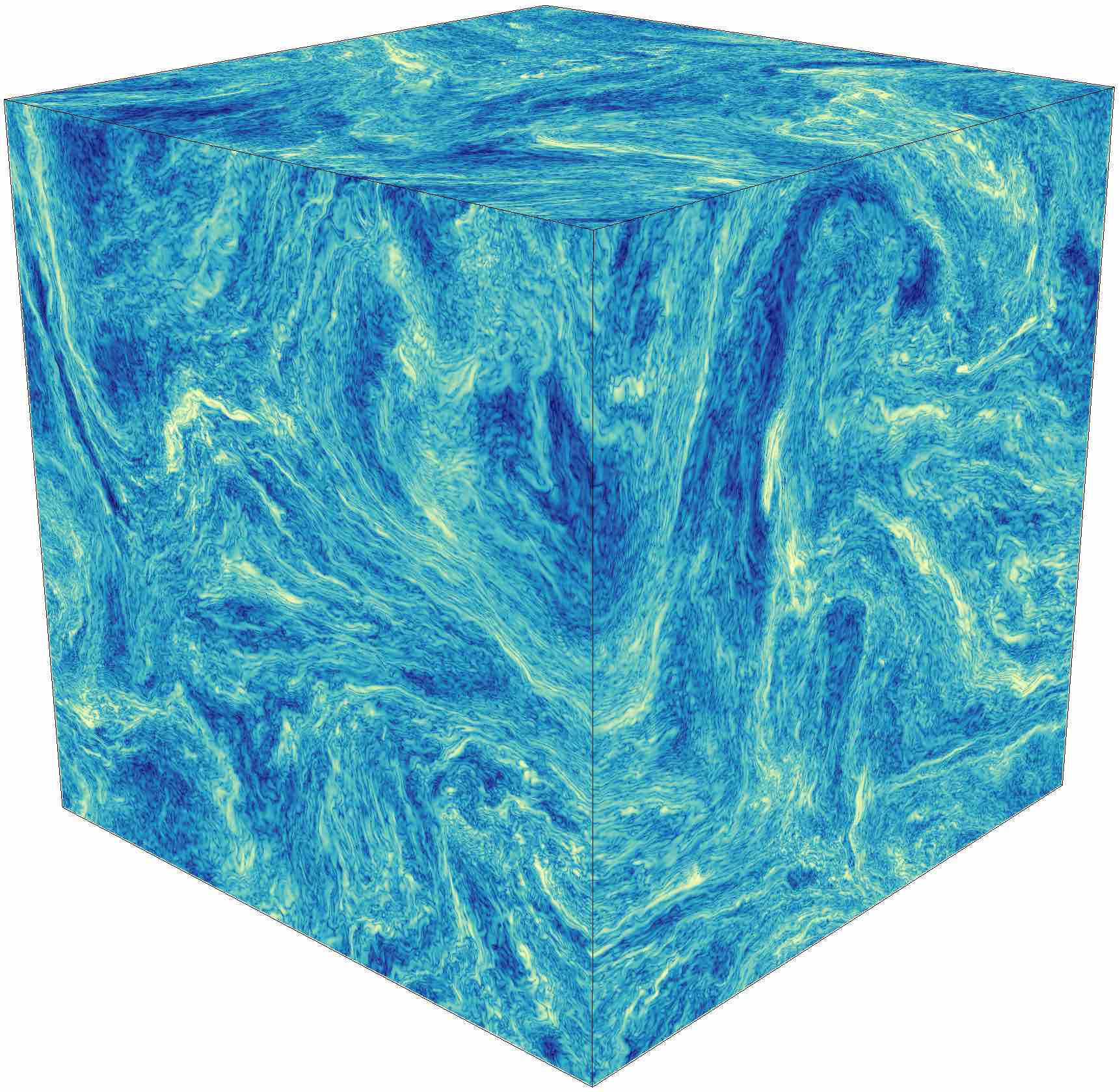}
  \caption{Top left: 3D rendering of magnetic-field lines in a 3D-3V
    hybrid Eulerian Vlasov-Maxwell simulation of small-scale dynamo
    action in the magnetised regime ($\rho_i/\ell_0=0.016$ initially). A
    collisionless plasma flow forced at the box wavenumber
    ($k_0=2\pi/\ell_0$) slowly stretches the initial magnetic-field
    seed, generating local pressure anisotropies
    (positive $\Delta_i$ in red, negative $\Delta_i$ in blue). This
    in turn excites parasitic kinetic-scale instabilities
    \citep[adapted from][]{rincon16}. The inset is a close-up showing ion
    density fluctuations in a region of $\Delta_i>0$ where the mirror
    instability is thought to be excited. Top right: 2D snapshot of
    magnetic-field strength in a
    3D-3V hybrid PIC simulation of collisionless MRI in the magnetised
    regime ($\rho_i/\ell_0=0.02$ initially). The co-development of 
    fluid-scale and kinetic-scale instabilities (mirror modes in 
    regions of increasing magnetic field) is particularly clear here: the MRI
    is responsible for the large-scale vertical sinusoidal fluctuation
    of the magnetic field, while the mirror instability generates the
    smaller-scale fluctuations in the regions where the large-scale
    magnetic field increases locally \citep[adapted from][]{kunz16}.
    Bottom: visualisation of the magnetic-field strength in a
    recent massive 3D-3V hybrid PIC simulation of collisionless plasma
    dynamo in the magnetised regime with $1008^3$ spatial resolution
    and tens of particles per cell (courtesy of St-Onge \& Kunz).}
\label{figrinconpnas}
\end{figure}

The simulations performed at small(-ish) $\rho_i/\ell_0$ exhibit an
initial build-up of positive and negative pressure anisotropies in
regions of slowly-growing and decreasing magnetic field, and the fast
excitation of fast small-scale kinetic mirror and firehose
instabilities. Strong density fluctuations associated with mirror
troughs notably develop in regions of increasing field strength, while
magnetic-field lines tend to develop a square shape in folding regions
of decreasing strength, in line with theoretical calculations of the
nonlinear development of the firehose instability (Melville \&
Schekochihin, private communication). This is illustrated
qualitatively in \fig{figrinconpnas} (top left).
More quantitative diagnostics of the excitation of kinetic-scale
instabilities in simulations can be found in the work of \cite{saintonge18}.
\fig{figrinconpnas} (top right) shows a snapshot of  hybrid-PIC
simulations by \cite*{kunz16} of the cousin three-dimensional
collisionless MRI problem, in which the local development of the
mirror instability has also been cleanly diagnosed in regions of
positive pressure anisotropy where the magnetic field is
amplified by the fluid-scale MRI (see also \cite*{schoeffler11}
for simulations of reconnection showing the firehose instability
developing in shrinking plasmoids). Finally, \fig{figrinconpnas}
(bottom) shows a vivid snapshot of the mind-blowing kinetic-fluid
multiscale dynamics at work in a very recent high-resolution hybrid-PIC
simulation by St-Onge \& Kunz of collisionless dynamo in the
magnetised regime \citep{kunz19}. These different sets of numerical
simulations therefore confirm that the general multiscale dynamical
phenomenology discussed earlier holds.

In addition to that, exponential growth of the magnetic field over
fluid timescales has been found in the Eulerian simulations of
\citeauthor{rincon16} at the magnetisation transition
($\rho_i/\ell_0>0.5$), and in the more recent PIC simulations of
\citeauthor{saintonge18} down to
$\rho_i/\ell_0 \sim 0.015$, showing that a growing magnetised-plasma
dynamo is possible. In both regimes, the field once again develops a
folded geometry, and a Kazantsev-like magnetic-energy spectrum
reminiscent of the large-$Pm$ MHD fluctuation dynamo
(\sect{largepmselection}). In the more magnetised simulations of
\citeauthor{saintonge18}, this dynamically-evolving
folded field structure is strongly entangled with a sea of
nonlinear pressure-anisotropy-driven kinetic-scale fluctuations at
$k\rho_i\sim 1$ that break $\mu$-invariance (there is no such
adiabatic invariance at the magnetisation transition where the Larmor
gyration period is of the order of the fluid turnover time). One of the
magnetised plasma dynamo simulations of \citeauthor{saintonge18} was
also integrated up to the nonlinear dynamical (equipartition)
saturation regime $M^2\beta_i=O(1)$. The statistically steady state
that they obtain also appears to be largely reminiscent of its
small-scale MHD dynamo counterpart, up to some relatively minor
differences in magnetic-field curvature stemming from the
presence of the saturated kinetic-scale instability fluctuations.

\subsubsection{How do magnetisation and kinetic effects affect dynamo growth ?*}
The global overall picture emerging so far from simulations is that a
fluctuation dynamo effect is possible in both unmagnetised and
magnetised collisionless plasmas dynamically stirred at fluid scales,
despite all the constraints described earlier. Besides, the dynamo in
both regimes appears to have a lot in common with the MHD fluctuation
dynamo at large $Pm$. If we look at the problem more
closely though, it is clear that kinetic-scale physics must play an
important role in the magnetised regime, and that these limited
preliminary numerical explorations have probably not
yet uncovered the full essence of the plasma dynamo. In any
case, it is interesting to ask both how internal plasma physics
plays out in practice in the simulations so as to make the dynamo look
MHD-like, and to what extent it may also induce new fluid-scale
dynamics compared to the MHD case.

One basic aspect of this problem, which has nothing to do with
kinetic-scale instabilities, is that the fluid-scale transport and
impedance properties of the plasma change as it becomes magnetised. More
specifically, particles are not allowed to stream freely anymore
perpendicular to magnetic-field lines once the particulate (not fluid)
Lorentz force becomes significant. As the perpendicular effective
mean free path of ions decreases, the  perpendicular viscosity of the
plasma should decrease, resulting in a cascade of kinetic energy to
smaller spatial scales with larger (perpendicular) rates of strain. An
extension of the kinetic energy spectrum towards smaller scales in the
magnetised regime is indeed observed in both the \cite{rincon16}
and \cite{saintonge18} simulation sets. The reduction in perpendicular
streaming and phase-mixing also implies that the correlation time of
the flow becomes larger in the magnetised regime, making magnetic
stretching more coherent, efficient and fluid-like. Finally, this
effect can also affect the way the plasma absorbs the mechanical
power injected at large scales. For a fixed external energy injection
rate in the system, it is possible that less of the available power is
diverted by phase-mixing into fluctuations of the distribution
function (``free-energy'' dissipated at small scales in
velocity-space) in the magnetised regime, and more into the bulk flow
of ions. A more vigorous flow could then also lead to more efficient
stretching \citep{saintonge18}. Note that none of
these effects are present in fluid MHD. The results of the
simulations of \cite{rincon16} suggest that dynamo growth becomes faster
(``self-accelerates'') as the plasma goes through the magnetisation
transition. If confirmed, this behaviour may be related to one or
several of these effects (a faster perpendicular rate of strain does
not in principle imply a faster stretching of the magnetic field
though). The more recent simulations of \cite{saintonge18}
also suggest that this may only be a transitional effect, as the
dynamo growth rate in the strongly magnetised regime seemingly
asymptotes to an MHD-like value of the order $u_{i,\mathrm{rms}}/\ell_0$.

A distinct, and perhaps also more interesting question, is that 
of the regulation of pressure-anisotropy by mirror and firehose
instabilities during the phase of magnetic-field growth in the
magnetised regime, and its consequences for the dynamo as a
whole. As this problem lends itself to a variety of pleasant
speculations, and its precise outcome may  depend on several
subtle factors, let us for a minute consider what may
be anticipated at the phenomenological level from what we know of
the saturation of these instabilities, before looking at the results of
simulations. If microscale instabilities, in particular the firehose
instability, eventually predominantly saturate by dynamically
scattering ions at an effective scattering rate $\nu_{i,\mathrm{eff}}(B)$,
we may expect that the fluid-scale viscosity $\nu$ of the plasma (not
to be confused with a scattering or collision rate) changes as
magnetisation increases, i.e. $\nu(B)\sim\vthi^2/\nu_{i,\mathrm{eff}}(B)$,
resulting in a magnetic-field-strength dependent Reynolds number
$Re_\mathrm{eff}(B)$.
In particular, if $\nu_{i,{\mathrm{eff}}}(B)$ and $Re_\mathrm{eff}(B)$
increase with increasing magnetisation, as may be the case for
instance if $\nu_{i,\mathrm{eff}}$ is somehow directly related to the ion
cyclotron frequency characteristic of magnetised-plasma
instabilities, we in principle have the conditions for a
self-accelerating dynamo loop by which magnetic-field growth
indirectly results
in the excitation of viscous-scale fluid velocity fluctuations
$u_{i,\nu}(B)\sim u_{i,0} Re^{-1/4}_\mathrm{eff}(B)$ at increasingly
smaller scales $\ell_{\nu}(B)\sim\ell_0Re^{-3/4}_\mathrm{eff}(B)$ and
with increasingly higher shearing rates
$S(B)\sim\hatvB\hatvB:\grad{\fluctvU_{i,\nu}(B)}\sim
(u_{i,0}/\ell_0)Re^{1/2}_\mathrm{eff}(B)$, which in turn should lead
to even faster magnetic growth and even smaller-scale motions etc.
(the Kolmogorov phenomenology is used for illustrative purposes).
Obviously though, how all of this plays out in practice depends
on the exact magnetic-field-strength dependence of the effective
scattering rate $\nu_{i,{\mathrm{eff}}}(B)$, fluid-scale plasma
viscosity $\nu(B)$, Reynolds numbers $Re_\mathrm{eff}(B)$ and
instantaneous stretching rate $S(B)$ , all of
which themselves depend on the details of saturation of
pressure-anisotropy-driven kinetic
instabilities. Several variants of this self-accelerating,
\textit{explosive-growth} model have been explored via zero-dimensional toy
dynamo equations using different physically motivated prescriptions for
$Re_{\mathrm{eff}}(B)$ and for the instantaneous dynamo growth rate
$\gamma=S(B)$ \citep*{schekochihin06,mogavero14,melville16}.

What do existing kinetic simulations tell us on this problem ?
\cite{saintonge18} show that the first fluid-scale effect of 
kinetic-scale instabilities in their early stage of saturation is
to effectively reduce the parallel magnetic-stretching rate
$S=\hatvB\hatvB:\grad{\fluctvU_i}$
so as to counteract the build-up of pressure anisotropy. Hence, while
the perpendicular viscosity of the plasma drops in the magnetised
regime, the parallel viscosity initially remains large, in the sense
that the effect of the early saturation of kinetic-scale instabilities
is to impede the motions that grow the field,  consistent with the
phenomenology discussed in \sect{satkin}. This feedback on the velocity
field is also a pure magnetisation effect that occurs even for
dynamically-weak magnetic fields, $M^2\beta_i\gg 1$, and has therefore
nothing to do with a dynamical feedback of the fluid-scale Lorentz
force on the flow. However, this is not the complete story either. As
they reach $\delta B/B=O(1)$, nonlinear kinetic-scale fluctuations
develop gradients at $\rho_i$ scales and start to scatter particles
effectively. This mechanism has a regulating effect
on the pressure anisotropy, but does not ``instantaneously'' pin
$|\Delta_i|$ to $O(1/\beta_i)$ marginal values in the simulations
either. The reason
is that, for very large $\beta_i$ (i.e. for dynamically-weak
fields), fluid-scale stretching can lead to significant excursions
from marginality for the mild scale separations typical of
simulations, $S/\Omega_i\sim 0.01-1$, $\rho_i/\ell_0 \sim
0.01-1$. Indeed, marginalising $\Delta_i$ requires
an effective scattering rate $\nu_{i,\mathrm{eff}}\sim S\beta_i$ (applying
the marginality conditions of pressure-anisotropy-driven instabilities
to \equ{eq:braginskiinu}), which for $\beta_i\gg 1$ and  $S/\Omega_i$
typical of the simulations is orders of magnitude larger than the
ion cyclotron frequency. Clearly, there is no physics at hand that can
make $\nu_{i,\mathrm{eff}}$ that large ! In reality, in the simulated
magnetised regime of dynamically-weak fields,  $\nu_{i,\mathrm{eff}}$
is neither proportional to $\Omega_i$, nor of
the order $\nu_{i,\mathrm{eff}}\sim S \beta_i$. Besides, the scattering
process appears to be rather non-uniform and to be regulated by firehose
fluctuations at the corners of folds. Considering that the parallel
distance between the bends of the magnetic field is $O(\ell_0)$
(\fig{figfoldsanisotropy}) and that particles essentially
free-stream along magnetic-field lines, \citeauthor{saintonge18} argue that
$\nu_{i,\mathrm{eff}}\sim\vthi/\ell_0$, resulting in the collisional
estimate $\nu_{i,\parallel}\sim\vthi\ell_0$ for the parallel viscosity,
or equivalently in an effective parallel Reynolds number of the plasma
$Re_{\mathrm{eff}\parallel}=(u_{i,\mathrm{rms}}\ell_0)/\nu_{i,\parallel}=O(M)$. Since
the velocity fluctuations forced at fluid scales in simulations are
subsonic, $Re_{\mathrm{eff},\parallel}$ is small. Altogether,
we see that the Reynolds number of the fluid-scale flow becomes strongly
anisotropic in the magnetised, kinetic-instability-prone regime, with
$Re_\perp>1\geq Re_{\mathrm{eff}\parallel}$. This result seems to
explain why the dynamo continues to behave as a $Pm\gg 1$, low-$Re$
fluctuation MHD dynamo in the magnetised regime, as only the parallel
Reynolds number matters for magnetic induction (we also recall that
$Re=O(1)$ is not a problem for a fluctuation dynamo as long as the
flow is chaotic and 3D, and $Rm$ is large-enough, see
\sect{largePmsection}). Note finally that
$Re_{\mathrm{eff}\parallel}$ in this regime of
dynamically-weak fields appears to be essentially independent of $B$,
and that only in the saturated regime of dynamically strong fields
$M^2\beta_i=O(1)$ does the effective scattering rate become of the
order $S\beta_i\sim (u_\mathrm{i,rms}/\ell_0)\beta_i$ in the simulations. This
provides a possible explanation as to why explosive dynamo growth is not
observed in the simulations.

\subsection{Uncharted plasma physics}
Obviously, none of the numerical results presented in the previous
paragraphs are yet truly asymptotic in terms of kinetic-to-fluid
scale-separations. In particular, we do not yet know what may
happen to the dynamo in the regime $S/\Omega_i= O(1/\beta_i)$,
$M^2\beta_i\gg 1$ in which a scattering rate
$\nu_{i,\mathrm{eff}}=O(\Omega_i)$ could in principle marginalise
$\Delta_i$ for dynamically-weak fields. Could the dynamo
self-accelerate ? Simulations of this regime are
unfortunately out of reach at the moment.  In order to accommodate
merely two decades of dynamically-weak field magnetised growth for
$M\sim 0.1$, we should start from  $\beta_i=O(10^4)$, corresponding
to a scale separation $S/\Omega_i=O(10^{-4})$, which is two orders of
magnitude larger than what could be painfully achieved so far.

More generally, a myriad of questions remain as to how dynamos
and magnetofluid dynamics operate in weakly-collisional magnetised
plasmas. How do pressure-anisotropy-driven instabilities saturate,
and what does their saturation imply for the dynamics of the magnetic
field in a magnetised plasma chaotically stirred at fluid scales
\citep{melville16} ? What happens when magnetic energy reaches
equipartition in a system in which the firehose instability tends to
regulate the pressure anisotropy in such a way as to cancel
magnetic tension and magnetic stretching  (see \equ{eq:firehosedisprel},
discussions in \cite{rosin11,mogavero14} and the recent work of
\cite{squire16,squire19} on the interruption of nonlinear Alfv\'en
waves and magneto-immutable Alfv\'enic turbulence in magnetised
plasmas) ? What are the properties of the dynamo in the magnetised,
collisional MHD Braginskii limit \citep{malyshkin02} ?
Are there different possible scattering and effective
transport regimes depending on whether the ion Larmor radius is
smaller or larger than the resistive scale
$\ell_\eta$ at which magnetic reversals occur \citep[see discussion
in][]{saintonge18} ? What about electron physics, reconnection and
flux-unfreezing in the
particular context of the plasma dynamo problem ? What about the
connections and transition between kinetic-scale magnetic-production
mechanisms such as the Biermann battery or the Weibel instability, and
the fluid-scale dynamo ? The full problem is phenomenal. What seems
clear at this stage is that addressing any of these questions
rigorously is going to require a lot more thinking and at least another
supercomputer architecture turnover time, as significant scale
separations (and therefore numerical resolution) between fluid and
kinetic scale are required to address them.

A final glimmer of lucidity pushes the author of these lines to concede
that pressure-anisotropy-driven kinetic instabilities in magnetised
plasmas may well turn out in the end to be no more than an exhausting
career-consuming technical distraction for a small group of theoreticians,
and that their practical effects may in the end well be reducible to
fairly simple effective parametrisations of viscosity and heat fluxes
in equally simple magnetofluid formulations. The questions of whether
this is the case and of what such closures should look like have
however not yet been
entirely settled, and should lie at the core of upcoming
investigations (see e.g. \cite*{squire17} for a very recent study of
this kind of question for the cousin collisionless MRI problem).

\section{A subjective outlook for the future\label{conclusion}}
What is the future up to ? Since Larmor planted the first fluid dynamo
seed a century ago, dynamo theory has grown solid roots and many
branches with colourful foliage, so as to become a venerable tree in
the vast forest of physics. But is it ever going to reach the canopy ? 

\subsection{Mathematical theory}
As many of the discussions in the previous sections have shown, most
dynamo theories, and their applications, still suffer from
major limitations. At the most fundamental level, the MHD dynamo
problem at large $Rm$, just like turbulence theory or QCD, is
generically non-perturbative, whereas the available mathematical
techniques encountered in these notes are essentially perturbative. 
In other words, we are faced with a hard theoretical physics problem,
whose solution in regimes relevant to most situations in nature
requires new insights into a much broader class of mathematics
and physics problems. 

The upside of this for current and future generations of
students and researchers is that the dynamo problem, inasmuch as we
want to consider it as a single big physics problem, is far from
solved. The downside is that further progress at the fundamental
physical and mathematical level is hard
to envision at the moment: field-theoretic approaches to turbulence
problems have had their day in the past, but have so far not delivered 
big game-changing results in non-perturbative regimes. While some
numerical techniques now appear to be moderately successful at
describing the effective large-scale dynamics of large-scale dynamos,
no existing closure procedure can be justified entirely from first
principles in the physical regimes of most astrophysical, geophysical, or
experimental interest. Theoretical developments rooted in the idea
that nonlinear coherent structures underlie the statistical dynamics,
such as instanton calculations or periodic orbit theory, are certainly
worth pursuing but have not yet made a significant difference either. 
The theoretical state of affairs is not entirely hopeless though, as
the current deadlock may be rapidly broken should significant new
developments on any classical turbulence problem, or even simpler
non-perturbative problems, appear. \red{As will be shortly discussed, 
future numerical explorations may also provide us with important new
insights into how to further develop the analytical theory.}

\subsection{Experiments and observations}
Can twenty-first century experiments and astronomical observations
come to the rescue~? The prospects for the next decade or so are quite
interesting. New experimental designs using dense laser plasmas
\citep{meinecke15} or more dilute plasmas confined into metre-size
vessels \citep{forest15} hold great promises for instance, one of the
most important being the exploration of the dissipative parameter
space of dynamos in a way that is impossible with liquid-metal
experiments. Some of these experiments are also less constrained
geometrically than earlier ones and are therefore likely to provide 
a propitious environment to significantly advance the study of dynamos
in homogeneous, isotropic fluid or plasma turbulence. There is also
some room for excitement on the astronomical side. For instance, a new
high-resolution window into cosmic magnetism, from the intergalactic
medium to galaxies, is about to be opened thanks to the construction 
of the gigantic SKA
radio-observatory\footnote{\url{https://www.skatelescope.org}}.
New observational programmes of visible-light and infrared
spectro-polarimetric measurements of magnetic fields in rotating
stars, proto-stars and planet-forming systems are also underway, which
can provide some significant phenomenological trends and constraints
on stellar dynamos accross the Hertzsprung-Russell progression, and
magnetic dynamics in young accreting systems%
\footnote{\url{http://www.cfht.hawaii.edu/en/projets/SPIRou/science.php}}.

All of this being acknowledged, it is however also clear that forthcoming
observational programmes and experiments have their own strong
limitations, both in terms of spatial, temporal and
magnetic-field strength resolution, and
accessible parameter regimes. The intrinsic complexity of experimental
and astrophysical systems also implies that we only have limited control
over what we can measure. In particular, dynamos in such systems are
unlikely to be devoid of interference by parasitic physical and
dynamical processes, which may limit their potential to provide strong
constraints on any particular fundamental dynamo mechanism. In fact, it
is extremely challenging to envision and conceive a set of
experimental or astronomical observational procedures that could probe
the depths of any particular large $Rm$ turbulent dynamo mechanism, or
cleanly discriminate between different theoretical models. Note that
the fundamental theoretical concern here is not to explain or
predict any particular astronomical observation or experimental
measurement with a sufficiently sophisticated effective statistical or
dynamical model, but to obtain better real-life constraints on the
underlying first-principle physical mechanisms, so as to lift as many
theoretical degeneracies and ambiguities as possible. This question is
fundamentally distinct from that of applied data fitting and
phenomenological modelling.

\subsection{The privileged position of numerics}
Just like theoretical physicists usually operate within the
limits of available mathematics, it would be unfair to blame
astronomers and experimentalists for the intrinsic complexity and 
uncontrollable circumstances of their
favourite object of study. Similar limitations impede the
study of virtually any physical process in the Universe, not just
dynamos. What the lack of strong experimental and observational
constraints on theory seemingly implies, however, is that
theoretical research in the field is bound to remain relatively
speculative for many years.

As heretical as it may seem to colleagues whose physics compass is
first and foremost directed by measurements performed over actual
physical systems, it is the opinion of this author that numerical
simulations in their various available forms likely provide the most
powerful tool to mitigate this
problem both in the present and future. Brute force high-resolution
simulations such as those shown in \fig{figgeosolardynamo}
are definitely needed in our arsenal to make further progress on
planetary and astrophysical dynamos, if only because they can provide
a more convenient and cleaner virtual alternative to laboratory experiments
(if carefully conducted). Some important fundamental theoretical questions
underlying dynamo theory, such as the precise role of fast reconnection on
MHD dynamos and turbulence, or the multiscale dynamics of
weakly-collisional plasma dynamos, definitely require simulations
with tens of billions of grid points or degrees of freedom to be
addressed properly, as vividly illustrated by \fig{plasmoids}
and \fig{figrinconpnas} (bottom). We are not yet routinely performing
simulations in this territory, but should get
there in the forthcoming years. On the other hand, it is important to
stress that advanced numerical machineries and huge
resolutions are almost certainly not required to get to the heart of
the matter regarding some fundamental dynamo questions, such as
how basic large-scale dynamo mechanisms operate physically in flows
with finite correlation times,
how they interact with small-scale dynamos, or how they saturate. 
As illustrated by the many numerical examples provided in these notes,
carefully controlled numerical experiments at moderate resolutions
have already illuminated many different dynamo problems, and they
undoubtedly still have a lot of untapped potential to inform our study
of many of the open questions raised in this text. One may even argue
in the light of the results presented in this review that
the main numerical challenge that we have to face is not so much that
we will never be able to simulate the Sun, the Hydra A cluster, or the
Milky Way in their full glory because of a lack of resolution, but
that it is surprisingly difficult to get computers to answer many
basic questions that could fundamentally enrich our understanding of
dynamos beyond our limited existing theoretical paradigms.

\section*{Acknowledgements}

This work has immensely benefited from many conversations with Alex
Schekochihin, Steve Cowley, Gordon Ogilvie, Michael Proctor, Carlo
Cossu, Geoffroy Lesur, Pierre-Yves Longaretti, Francesco Califano,
Antoine Riols, Tarek Yousef, and Matt Kunz.
I am also very grateful to all the mathematicians, experimental and/or
theoretical physicists, geophysicists, astrophysicists and plasma
physicists with whom I have had the opportunity to learn, discuss, and
also sometimes argue about dynamos over the years. Any attempt at an
exhaustive recollection of the people involved would most probably
fail, but let me single out Michel Rieutord, Katia Ferri\`ere,
Jean-Paul Zahn, Nigel Weiss, Juri Toomre, Steve Tobias, Axel
Brandenburg and Jean-Fran\c{c}ois Pinton for triggering
my interest in dynamos back in my graduate and postdoctoral years.
Many thanks too to Frank Stefani, Jono Squire and Nathana\"el
Schaeffer for providing some very useful feedback on the first version of
the complete draft. I would finally like to thank the LOC and SOC of
the 2017 Les Houches plasma physics school \textit{``From laboratories to
  astrophysics: the expanding universe of plasma physics''} for giving
me the opportunity to present a dynamo tutorial, and the local staff
of the Ecole de Physique for their hospitality.

\appendix
\section{Some good reads\label{biblio}}
As a supplement to these notes, here is a selection of books and
reviews available to readers interested in dipping their toes deeper
into MHD and plasma dynamo flows. 

\subsection{MHD, astrophysical fluid dynamics and plasma physics textbooks}
There is a wide diversity of MHD, AFD and plasma physics texts
available. The first chapters of \citeauthor{roberts67}'s
(\citeyear{roberts67}) \textit{\usebibentry{roberts67}{title}}
are one of the most tidy expositions of the fundamentals of MHD
available. \citeauthor{ogilvie16}'s (\citeyear{ogilvie16}) recently published
lecture notes provides a concise, clean and clear introduction to ideal
MHD and \textit{\usebibentry{ogilvie16}{title}} (strongly recommended for
students and researchers alike). \citeauthor{kulsrud05}'s
(\citeyear{kulsrud05}) \textit{\usebibentry{kulsrud05}{title}}
textbook provides an introduction to the range of mathematical models
describing the dynamics of fluid and collisionless plasmas. The reader
is also referred to the Varenna school notes on
\textit{\usebibentry{kulsrud83}{title}} by the
same author for a particularly clear derivation of drift-kinetics,
kinetic MHD, and a discussion of the CGL approximation \citep{kulsrud83}. 
Another solid and accessible reference on fundamental plasma physics is
\cite{boyd03}.

\subsection{Dynamo theory books and reviews}
A practical reference providing quick introductions and pointers to
research on many of the topics discussed in this review, as well as
interesting accounts of the early development of dynamo research, is
\citeauthor{gubbins07}'s (\citeyear{gubbins07})
\textit{\usebibentry{gubbins07}{title}} (see
also \cite{molokov07} for a broader historical perspective on the
development of MHD). The essence of dynamo theory is distilled in
\citeauthor{vainshtein72}'s (\citeyear{vainshtein72}) historical
\textit{\usebibentry{vainshtein72}{title}}. The first
sections of this review notably introduce the small-scale
versus large-scale dynamo problems in enlighting terms. 
\citeauthor{moffatt77}'s (\citeyear{moffatt77})
\textit{\usebibentry{moffatt77}{title}} offer an interesting and
concise introduction to his classic \citeyear{moffatt78}
\textit{\usebibentry{moffatt78}{title}} book on the theory of
mean-field electrodynamics. The other classic textbook on mean-field
theory is \citeauthor{krause80}'s (\citeyear{krause80})
\textit{\usebibentry{krause80}{title}}.
An entire ``yellow'' book, \textit{\usebibentry{childress95}{title}}
by \cite{childress95} is also dedicated to the mathematical theory of
fast dynamos. \citeauthor{kulsrud05}'s (\citeyear{kulsrud05})
plasma physics textbook mentioned above is one of the very few
non-specialised monographs that discusses turbulent small-scale
dynamos. \red{A new textbook on \textit{Self-Exciting fluid
dynamos} by \cite{moffatt19} has been published as 
this manuscript was finalized.}

As far as general contemporary reviews of dynamo theory research
are concerned, \citeauthor{branden05}'s (\citeyear{branden05})
\textit{\usebibentry{branden05}{title}} remains
the most exhaustive and detailed resource available,
although it is probably not the easiest entry point in
the field. \citeauthor{dormy07}'s (\citeyear{dormy07})
\textit{\usebibentry{dormy07}{title}} and the collection of
Les Houches summer school lecture notes on dynamos edited by
\cite{cardin08} provide quality introductions to
many aspects of the problem, \red{and so do the Peyresq lecture notes
  of \cite{fauve03}.}

On more specialised topics, \citeauthor{schekochihin07b}'s \citeyear{schekochihin07b}
review on \textit{\usebibentry{schekochihin07b}{title}} includes a
very clear introduction to the linear and nonlinear phenomenologies of
small-scale dynamos, and also the first accessible (albeit
rather speculative) discussion of the weakly-collisional dynamo
problem. Other interesting essays on small-scale dynamo
theory include the work of \cite{vincenzi02} on the Kazantsev model, a
chapter by \cite*{tobias11b} on MHD dynamos in the collective book
\textit{\usebibentry{tobias11b}{booktitle}}, and a few 
sections of \citeauthor{davidson13}'s (\citeyear{davidson13}) book
reviewing the current state of research on
\textit{\usebibentry{davidson13}{title}}.
Accessible reviews of large-scale dynamo theory, including diverse
perspectives on nonlinear effects and interactions between large- and
small-scale dynamos, can be found in
\cite{hughes10b,branden18,hughes18}. The seemingly inextricable
problem of nonlinear saturation of large-scale dynamos
is specifically discussed in two reviews by \cite{proctor03} and
\cite*{diamond05}. A relatively recent perspective on the development
and purpose of test-field modelling of large-scale dynamos has been
written by \cite{branden09b}. Finally, an early discussion of the
phenomenology of instability-driven dynamos and their connection to
hydrodynamic self-sutaining processes in shear flows can be found in
\cite{rincon08}.

\subsection{Astrophysical and planetary dynamo reviews}
Readers more specifically interested in relatively recent developments
on astrophysical and planetary dynamo problems may notably want to
consult the reviews by \cite{christensen10,jones11,roberts13} on geo-
and planetary dynamos,
\cite{charbonneau10,miesch12,charbonneau14,brun17,branden18}
on solar and stellar dynamos, \cite{shukurov07,branden15} on galactic
dynamos, \cite{kulsrud08,widrow12,durrer13,subramanian19} on cosmic
and primordial magnetic fields, and \cite{federrath16} on dynamos in
highly compressible astrophysical flows.

\bibliographystyle{jpp}

\bibliography{dynamo}

\end{document}